\documentclass{lmcs}
\renewcommand{\paragraph}[1]{\textit{#1}.\quad}


\usepackage{subfiles}
\usepackage{xr}
\usepackage{currfile}

\makeatletter
\def\endthebibliography{%
  \def\@noitemerr{\@latex@warning{Empty `thebibliography' environment}}%
  \endlist
}
\makeatother

\ifdefined\docroot\else
 \let\docroot\currfilepath
\fi

\newcommand\IfDocRootTF[2]{%
 \ifcurrfilepath{\docroot}{#1}{#2}%
}
\newcommand\IfDocRootT[1]{\IfDocRootTF{#1}{}}
\newcommand\IfDocRootF[1]{\IfDocRootTF{}{#1}}

\IfFileExists{../preamble.tex}{\usepackage{amssymb}
\usepackage{amsmath}
\usepackage{amsthm}
\usepackage[utf8]{inputenc}
\usepackage{stmaryrd}
\renewcommand{\sfdefault}{cmss}
\usepackage[T1]{fontenc}
\usepackage{xcolor}
\IfFileExists{../proof-dashed.sty}{\usepackage{../proof-dashed}}{\usepackage{proof-dashed}}
\usepackage{graphicx}
\usepackage{enumitem}
\usepackage{listings}
\usepackage{hyperref}
\usepackage{aliascnt}
\usepackage{mathtools}
\usepackage{microtype}
\usepackage{cleveref} 

\usepackage{marginnote}
\usepackage[textsize=footnotesize,linecolor=teal,bordercolor=teal,%
backgroundcolor=yellow!10]{todonotes} 

}{}
\IfFileExists{../macros.tex}{
\newtheorem{notation}{Notation}

\newcommand{\then}{\Rightarrow}
\newcommand{\prid}[1]{(\mathit{#1})}
\newcommand{\did}[2]{\raisebox{1.5pt}{%
$\scriptstyle\left\lfloor\!^\text{\raisebox{-1pt}{$\mathsf{#1}$}}|_%
\textsc{$\mathsf{#2}$}\!\right\rceil$}}
\newcommand{\chanto}[3]{#1[#2 \rangle #3]}
\newcommand{\pid}[1]{\mathsf{#1}}
\newcommand{\prc}[2]{\pid{#1}[\role{#2}]}
\newcommand{\pids}[1]{\til{\pid{#1}}}
\newcommand{\env}{D}
\newcommand{\aenv}{\mathbb{D}}
\newcommand{\renvsymb}{\blacktriangleright}
\newcommand{\renv}[3]{#1,#2 \renvsymb #3}

\newcommand{\NI}{\noindent}
\newcommand{\co}[1]{\overline{#1}}
\newcommand{\m}[1]{\text{\normalfont \texttt{#1}}}
\newcommand{\lfn}[1]{\texttt{#1}}
\newcommand{\auxfn}[1]{\mathbf{#1}}
\newcommand{\ctx}[1]{\mathcal #1}

\newcommand{\init}{\auxfn{init}}
\newcommand{\locs}{\auxfn{locs}}
\newcommand{\evalfn}{\auxfn{eval}}
\newcommand{\pco}{\auxfn{pco}}
\newcommand{\cofn}{\auxfn{co}}
\newcommand{\closed}{\auxfn{closed}}
\newcommand{\supfn}{\auxfn{sup}}
\newcommand{\tmerge}{\underset{\leftarrow}{\sqcup}}
\newcommand{\troot}[1]{#1}
\newcommand{\emptyseq}{\varepsilon}
\newcommand{\emptyfunc}{\emptyset}

\makeatletter
\def\oversortoftilde#1{\mathop{\vbox{\m@th\ialign{##\crcr\noalign{\kern0\p@}%
      \sortoftildefill\crcr\noalign{\kern1\p@\nointerlineskip}%
      $\hfil\displaystyle{#1}\hfil$\crcr}}}\limits}

\def\sortoftildefill{$\m@th \setbox\z@\hbox{$\braceld$}%
  \scalebox{.7}[.4]{$\braceld$}%
  \leaders\vrule\hfill%
  \scalebox{.7}[.4]{\hbox{$\braceru$}}$}

\makeatother
\newcommand{\wtil}[1]{\oversortoftilde{#1}}
\newcommand{\til}[1]{\tilde{#1}}
\newcommand{\prf}[1]{\mathcal{#1}}
\newcommand{\prfcase}{\textbf{Case}}
\newcommand{\DD}{\prf D}
\newcommand{\EE}{\prf E}
\newcommand{\FF}{\prf F}
\newcommand{\GG}{\prf G}
\newcommand{\HH}{\prf H}

\newcommand{\ASET}[1]{\{#1\}}

\newcommand{\gram}{::=}
\newcommand{\Div}{|}

\newcounter{linenos}
\newcommand{\LN}{\stepcounter{linenos}\arabic{linenos}.}
\newcommand{\ResetLN}{\setcounter{linenos}{0}}

\newcommand{\smallpar}[1]{\smallskip \noindent \emph{#1}}
\newcommand{\boldpar}[1]{\smallskip \noindent \textbf{#1}}

\newcommand{\proc}[3]{#1.\pid{#2}[\role{#3}]}

\newcommand{\role}[1]{\mathtt{#1}}
\newcommand{\roles}[1]{\til{\role{#1}}}
\newcommand{\send}[6]{\overline{#1}^{#2 #3}(#4);(#5\pp #6)}
\newcommand{\sendp}[4]{\overline{#1}[#2]!#3(#4)}
\newcommand{\recv}[4]{{#1}^{#2 #3}(#4)}
\newcommand{\recDef}[3]{\m{def}\; #1 = #2 \;\m{in}\; #3}
\newcommand{\recCall}[1]{#1}
\newcommand{\inact}{\mathbf{0}}
\newcommand{\wait}[1]{\m{wait}\, #1}
\newcommand{\close}[1]{\m{close}\, #1}
\newcommand{\Case}[4]{{#1}^{#2}\mathsf{case}(#3,#4)}
\newcommand{\inl}[2]{{#1}^{#2}.\m{inl}}
\newcommand{\inr}[2]{{#1}^{#2}.\m{inr}}
\newcommand{\res}[1]{(\boldsymbol\nu #1)\,}
\newcommand{\fwd}[2]{#1\leftrightarrow #2}
\newcommand{\pp}{\ \boldsymbol{|}\ }
\newcommand{\ppr}{\boldsymbol{|}\ \;}
\newcommand{\ppx}[1]{\ \;\boldsymbol{|}_{#1}\ \;}

\newcommand{\seq}{\vdash}
\newcommand{\hlseq}[1]{\seq {#1}}
\newcommand{\hlseqmin}[1]{\seq_{\m{min}} \hl{#1}}
\newcommand{\gseq}{\vDash}
\newcommand{\pseq}{::\ \seq}
\newcommand{\pgseq}{::\ \gseq}
\newcommand{\pair}[2]{{#1\!:}\ #2}
\newcommand{\dual}[1]{#1^{\bot}}
\newcommand{\epp}[2]{\proj{#1}{#2}}
\newcommand{\gepp}[2]{\epp{#1}{}^{#2}}

\newcommand{\tensor}{\otimes}
\newcommand{\tensort}[1]{\otimes^{\hl{#1}}}
\newcommand{\parr}{\bindnasrepma}
\newcommand{\parrt}[1]{\parr^{\hl{#1}}}
\newcommand{\with}{\binampersand}
\newcommand{\witht}[1]{{\binampersand}^{\hl{#1}}}
\newcommand{\one}{1}
\newcommand{\mix}[1]{\{#1\}}
\newcommand{\gentensor}[3]{#1\tensort{#2} #3}
\newcommand{\genparr}[3]{#1\parrt{#2} #3}
\newcommand{\genoplus}[3]{#1\oplust{#2} #3}
\newcommand{\genwith}[3]{#1\witht{#2} #3}
\newcommand{\swapG}{\simeq_{\m G}}
\newcommand{\equivG}{\equiv_{\m G}}
\newcommand{\swapT}{\simeq_{\m T}}
\newcommand{\oplust}[1]{\oplus^{\hl{#1}}}
\newcommand{\gnext}{\auxfn{next}}
\newcommand{\grecs}{\auxfn{recs}}

\newcommand{\stype}[5]{#1:#2 \langle #3, #4, #5 \rangle }

\newcommand{\tstring}{\mathbf{str}}
\newcommand{\tint}{\mathbf{int}}
\newcommand{\tunit}{\mathbf{unit}}
\newcommand{\tbool}{\mathbf{bool}}
\newcommand{\taddr}{\mathbf{addr}}
\newcommand{\tdate}{\mathbf{date}}
\newcommand{\tbyte}{\mathbf{byte}}
\newcommand{\tbytes}{\mathbf{bytes}}

\newcommand{\Locations}{\ctx{L}}
\newcommand{\Sessions}{\ctx{K}}
\newcommand{\SessionQueues}{\ctx{Q}}
\newcommand{\Roles}{\ctx{A}}
\newcommand{\Val}{\mathit{Val}}
\newcommand{\Var}{\mathit{Var}}
\newcommand{\Pids}{\ctx{P}}
\newcommand{\Trees}{\ctx{T}}
\newcommand{\Queues}{\ctx{M}}
\newcommand{\Operations}{\ctx{O}}
\newcommand{\Paths}{\m{Paths}}
\newcommand{\Labels}{\textit{Lab}}

\newcommand{\dom}{\auxfn{dom}}
\newcommand{\eval}[1]{[\![#1]\!]}
\newcommand{\jpath}[1]{\underline{#1}}
\newcommand{\carr}[1]{\langle #1 \rangle}
\newcommand{\equeues}{\auxfn{qs}}
\newcommand{\estate}{\auxfn{st}}
\newcommand{\eifs}{\m{ifc}}
\newcommand{\comto}{\raisebox{-.1em}{\resizebox{1.5em}{!}{\(\rightarrowtriangle\)}}}
\newcommand{\startwith}{\raisebox{-.1em}{\resizebox{1.5em}{!}{\(\leftrightarrowtriangle\)}}}
\newcommand{\com}[3]{{#1\!:\!}\ #2 \; \comto \; #3}
\newcommand{\cond}[3]{\m{if} \; {#1} \; \{#2\} \; \m{else} \;
\{#3\}}
\newcommand{\gencond}{\cond{\pid p.e}{C_1}{C_2}}
\newcommand{\upd}{\neq}
\newcommand{\seff}{\blacktriangleleft}
\newcommand{\enq}{\m{enq}}
\newcommand{\deq}{\m{deq}}
\newcommand{\swapC}{\simeq_{\m C}}
\newcommand{\equivC}{\equiv_{\m C}}

\newcommand{\gcom}[3]{#1 \; \comto \; #2.#3}
\newcommand{\grecv}[3]{#1 \rangle #2.#3}
\newcommand{\psv}[1]{ \raisebox{.3ex}{${}_{#1}$} }
\newcommand{\gchoice}[3]{\oplus #1\psv{#2}.#3}
\newcommand{\gbranch}[3]{\& \psv{#1}#2.#3}
\newcommand{\gapply}[3]{\prescript{#1}{#2}{\downarrow}#3}
\newcommand{\gapplygen}[1]{%
  \prescript{\role A \rangle \role B}{o}{\downarrow}#1}
\newcommand{\gend}{\m{end}}
\newcommand{\gengrecDef}{\m{rec}\ \mathbf{t}.G}
\newcommand{\gengrecCall}{\mathbf{t}}
\newcommand{\lto}[1]{\mathrel{\stackrel{{\;\;#1\;\;}}{\mbox{\rightarrowfill}}}}
\newcommand{\longto}{\quad \to \quad}
\newcommand{\toglobal}{\rightsquigarrow}
\newcommand{\gtypes}{\m{gt}}
\newcommand{\mayto}{\to^{?}}
\newcommand{\manyto}{\Rightarrow}
\newcommand{\lgbox}[1]{\fcolorbox{lightgray}{white}{#1}}

\newcommand{\initg}[4]{#1\langle #2 \Div #3 \Div #4 \rangle}
\newcommand{\serviceTyping}[4]{\initg{#1}{#2}{#3}{#4}}
\newcommand{\geninitg}{\til l:\initg{G}{\role A}{\til{\role B}}{\til{\role C}}}

\newcommand{\gengbranch}{\gcom{\role A}{\role B}{
  \{ \op{o}_i(U_i) ; G_i \}_{i} }}
\newcommand{\gengbranchI}{\gcom{\role A}{\role B}{
  \{ \op{o}_i(U_i) ; G_i \}_{i\in I} }}

\newcommand{\gengcom}{\gcom{\role A}{\role B}{\op{o}(U); G}}

\newcommand{\lsend}[2]{\oplus\hspace{0.3mm}\role{#1}.#2}
\newcommand{\genlsend}{\lsend{A}{\{\op{o}_i(U_i) ; T_i\}_{i\in I}}}
\newcommand{\genlsendB}{\lsend{B}{\{\op{o}_i(U_i) ; T_i\}_{i\in I}}}
\newcommand{\lrecv}[2]{\&\role{#1}.#2}
\newcommand{\genlrecv}{\lrecv{A}{\{\op{o}_i(U_i);T_i\}_{i \in I}}}
\newcommand{\genlrecvPrime}{\lrecv{A}{\{\op{o}_i(U_i);T'_i\}_{i \in I}}}
\newcommand{\genlrecDef}{\m{rec}\ \mathbf{t}.T}
\newcommand{\genlrecCall}{\mathbf{t}}

\newcommand{\btype}{b}
\newcommand{\btypeop}{\m{bte}}
\newcommand{\genbtypeop}{\btypeop(\ \role A, \til m\ )}

\newcommand{\proj}[2]{\left\llbracket #1 \right\rrbracket_{#2}}
\newcommand{\genproj}[1]{\proj{#1}{\pid r}}
\newcommand{\group}[2]{\left\lfloor#1\right\rfloor_{#2}}
\newcommand{\CC}{\mathcal{C}}

\newcommand{\e}{,}
\newcommand{\cut}{\bullet}
\definecolor{light-gray}{gray}{0.90}
\setlength{\fboxsep}{0.3mm}

\newcommand{\hl}[1]{{\colorbox{light-gray}{$#1$}}}

\definecolor{light-gray1}{gray}{0.90}
\definecolor{light-gray2}{gray}{0.80}
\definecolor{light-gray3}{gray}{0.85}
\newcommand{\hlone}[1]{\colorbox{light-gray1}{$#1$}}
\newcommand{\hltwo}[1]{\colorbox{light-gray2}{$#1$}}
\newcommand{\hlthree}[1]{\colorbox{light-gray3}{$#1$}}
\newcommand{\fn}{\auxfn{fn}}
\newcommand{\fp}{\auxfn{fp}}
\newcommand{\pn}{\auxfn{pn}}
\newcommand{\auxRoles}{\auxfn{roles}}
\newcommand{\DEFEQ}{\stackrel{\mbox{\scriptsize def}}{=}}  
\newcommand{\ComCom}{(\rightarrow\rightarrow)}  
\newcommand{\ComChoice}{(\rightarrow\!\oplus)}  
\newcommand{\ChoiceChoice}{(\oplus\oplus)}  
\newcommand{\ChoiceCom}{(\oplus\rightarrow)}
\newcommand{\tcsymb}{\triangleleft}
\newcommand{\tcopy}[2]{\tcsymb(\, #1,#2\, )}
\newcommand{\trcsymb}{\Yleft}
\newcommand{\op}[1]{#1}
\newcommand{\trootcopy}[2]{\trcsymb(\, #1,#2\, )}
\newcommand{\gencom}{\com{k}{\pid p[\role A].e}{\pid q[\role B].\op{o}(x)}}
\newcommand{\gencomext}{\com{k}{\pid p[\role A].e}{\pid q[\role B].
                                            \{\op{o}(x);\inact\}}}
\newcommand{\gencomj}{\com{k}{\pid p[\role A].e}{\pid q[\role B].\op{o}_j}}
\newcommand{\gensend}{\com{k}{\pid p[\role A].e}{\role B.\op{o}}}
\newcommand{\gensendj}{\com{k}{\pid p[\role A].e}{\role B.\op{o}_j}}
\newcommand{\genrecv}{\com{k}{\role A}{\pid q[\role B].\op{o}(x)}}
\newcommand{\genbranch}{\com{k}{\role A}{\pid q[\role B].\{\op{o}_i(x_i);C_i\}_i}} 
\newcommand{\genbranchI}{\com{k}{\role A}{\pid q[\role B].\{\op{o}_i(x_i);C_i\}_{i\in I} }}
\newcommand{\genbranchIJ}{\com{k}{\role A}{\pid q[\role B].
                                    \{\op{o}_j(x_j);C_j\}_{j\in I\cup J}}}
\newcommand{\metastart}[3]{{#1:}\ #2 \; \startwith \; #3}
\newcommand{\deltastart}[2]{\m{start}\ #1:\ #2}
\newcommand{\start}[3]{\m{start}\ \metastart{#1}{#2}{#3}}
\newcommand{\req}[3]{\m{req}\ \metastart{#1}{#2}{#3}}
\newcommand{\acc}[2]{\m{acc}\ {#1:}\ #2}
\newcommand{\genstart}{\start{k}{\pid p[\role A]}{\wtil{l.\pid q[\role B]}}}
\newcommand{\gendeltastart}{\deltastart{k}{\wtil{l.\prc pA}}}
\newcommand{\fresh}[1]{\#{#1}}
\newcommand{\genacc}{\acc{k}{\wtil{l.\pid q[\role B]}}}
\newcommand{\genaccC}{\acc{k}{\wtil{l.\pid q[\role C]}}}
\newcommand{\genacci}{\acc{k}{\wtil{l_i.\pid q_i[\role B_i]}}}
\newcommand{\genreq}{\req{k}{\pid p[\role A]}{\wtil{l.\role B}}}
\newcommand{\gendel}{\com{k}{\pid p[\role A].k'[\role C]}{\pid q[\role B].\op{o}(k'[\role C])}}
\newcommand{\gendelS}{\com{k}{\pid p[\role A]}{\pid q[\role B].\op{o}\langle k'[\role C] \rangle}}

\newcommand{\prcdot}{\,\raisebox{.2em}{{\fontsize{14pt}{0em}\selectfont.}}\,}
\newcommand{\qfrom}[1]{\;\m{from}\;#1}
\newcommand{\qto}[1]{\;\m{to}\;#1}
\newcommand{\jsrv}[2]{\left\langle #1 \right\rangle_{#2}}
\newcommand{\jprc}[3]{#1 \prcdot #2 \prcdot #3}
\newcommand{\jpr}[2]{#1 \prcdot #2}
\newcommand{\oneway}[2]{#1(#2)}
\newcommand{\rr}[4]{#1(#2)(#3)\{#4\}}
\newcommand{\rrfrom}[5]{#1(#2)(#3)\qfrom{#4}\;\{#5\}}
\newcommand{\notify}[3]{#1@#2(#3)}
\newcommand{\solicit}[4]{#1@#2(#3)(#4)}
\newcommand{\cq}[1]{\nu\rangle #1}
\newcommand{\choice}[2]{\left[#1\right]\left\{#2\right\}}
\newcommand{\new}{\m{new}}
\newcommand{\equivD}{\equiv_{\m D}}
\newcommand{\strBhv}{\mathfrak{B}}

\newcommand{\espc}{\hspace{-3.4pt}}
\newcommand{\enc}[1]{
  \setlength\fboxsep{1pt}
	\setlength{\fboxrule}{.5pt}
	\fbox{$#1$}
}
\newcommand{\genenc}[1]{\enc{#1}^{\hspace{1pt}\raisebox{-3pt}
{\scriptsize$\Gamma$}\!\!}}

\newcommand{\geneenc}[1]{\eenc{#1}{\!\Gamma}}
\newcommand{\eenc}[2]{\langle\hspace{-.3em}\langle#1\rangle
\hspace{-.3em}\rangle^{#2}}

\newcommand{\filter}[2]{\left.#1\right|_{#2}}

\newcommand{\jbisim}{\approx}
\newcommand{\Seq}{\mathop{\vphantom{\sum}\mathchoice
  {\vcenter{\hbox{\huge $\odot$}}} {\vcenter{\hbox{\Large
  $\odot$}}}{\mathrm{\odot}}{\mathrm{\odot}}}\displaylimits}
  
\newcommand{\bigtcopy}{\mathop{\vphantom{\sum}\mathchoice
  {\vcenter{\hbox{\huge $\triangleleft$}}} {\vcenter{\hbox{\Large $\trianglele
  ft$}}}{\mathrm{\triangleleft}}{\mathrm{\triangleleft}}}\displaylimits}

\newcommand{\addCode}[1]{{\color{blue!99}#1}}

\newcommand{\fulltextwidth}{\rule{\linewidth}{0pt}\\}
\newcommand{\fulltextwidthRule}{\rule{.88\linewidth}{0pt}\\}

\usepackage{marginnote}
\newcommand{\fabrizio}[1]{\todo{{\bf F:} {#1}}}
\usepackage{mdframed} 
\newcommand{\note}[1]{\noindent\begin{mdframed}
[hidealllines=true,backgroundcolor=lime]#1\end{mdframed}}
\newcommand{\remove}[1]{{\color{red!70!black}#1}}
\newcommand{\newstuff}[1]{{\color{blue!90!black}#1}}

\newcommand{\citeneeded}{\alert{[CIT.?]}}

\newcommand{\lub}{\triangledown}
\newcommand{\genlsendBmin}{\lsend{B}{\{o(U) ; T\}}}
\newcommand{\genlrecvPrimemin}{\lrecv{A}{\{o(U);T'\}}}
\newcommand{\genencp}[1]{\enc{#1}^{\hspace{1pt}\raisebox{-3pt}
{\scriptsize$\Gamma'$}\!\!}}

\newcommand{\pluseq}{\;\text{+=}\;}
\newcommand{\asgn}{\;\text{:=}\;}
\newcommand{\aupd}{\asgn\ \aenv}
\let\emptyset\varnothing

\makeatletter
\renewcommand{\@todonotes@drawMarginNoteWithLine}{%
\begin{tikzpicture}[remember picture, overlay, baseline=-0.75ex]%
    \node [coordinate] (inText) {};%
\end{tikzpicture}%
\marginnote[{
    \@todonotes@drawMarginNote%
    \@todonotes@drawLineToLeftMargin%
}]{
    \@todonotes@drawMarginNote%
    \@todonotes@drawLineToRightMargin%
}%
}
\makeatother

\newcommand{\hll}[1]{\color{magenta}#1}
\crefname{figure}{Figure}{Figures}
\crefname{table}{Table}{Tables}
\crefname{section}{\S}{\S}
\crefname{line}{Line}{Lines}
\crefname{theorem}{Theorem}{Theorems}
\theoremstyle{definition}
\newtheorem{example}{Example}
\crefname{example}{Example}{Examples}
\theoremstyle{definition}
\newtheorem{definition}{Definition}
\crefname{definition}{Definition}{Definitions}
\newtheorem{theorem}{Theorem}
\newtheorem{corollary}{Corollary}
\newtheorem{lemma}{Lemma}
\crefname{lemma}{Lemma}{Lemmas}
\let\proof\relax
\NewDocumentEnvironment{proof}{O{}}{\paragraph{Proof #1}}{\hfill$\square$}
\crefname{proof}{Proof}{Proofs}
\newtheorem{proposition}{Proposition}
\newtheorem{remark}{Remark}

\newcounter{mycounter}
\newcommand{\showcounter}[1]{\mbox{\textcircled{$\scriptstyle#1$}}}
\newcommand{\thiscounter}{\;\stepcounter{mycounter}\showcounter{\themycounter}\;}

\newcommand{\case}[1]{\item[]\textbf{Case} \hl{\mbox{#1}} \newline }
\newcommand{\step}[1]{\item[]\ilab{#1} }
\newcommand{\ecase}[1]{\item[]\textbf{Case} \hl{\mbox{#1}} \newline}
\newcommand{\Def}{D}
\newcommand{\Thm}{T}
\newcommand{\ilab}[1]{(\text{{#1}})}
 
\newcommand{\pnum}[1]{\raisebox{-1pt}{\large \textcircled{\raisebox{1.1pt} {\scriptsize #1}}}}

\newcommand{\smallpnum}[1]{\raisebox{-2pt}{\Large \textcircled{\raisebox
{1pt} {\hspace{-.5pt}\scriptsize #1}}}}

\newcommand{\startlab}[3]{\tau }
\newcommand{\pstartlab}[3]{\tau }

\newcommand{\dep}{\env}

\newsavebox{\mybox}
\newcommand{\tmpScaleVar}{}
\newenvironment{scaledListing}[1]
  {\renewcommand{\tmpScaleVar}{#1}\begin{lrbox}{\mybox}}
  {\end{lrbox}\scalebox{\tmpScaleVar}{\usebox{\mybox}}}}{}

\newcommand{\IFSubFileBiblio}{
  \IfDocRootT{\bibliographystyle{alpha}}
  \IfDocRootT{\bibliography{biblio}}
}

\begin{document}

\title{Applied Choreographies}
\titlecomment{This work was partially supported by \ldots}

\author[S. Giallorenzo]{Saverio Giallorenzo}
\address{Universit\`a di Bologna, Italy and INRIA, France}
\email{saverio.giallorenzo@gmail.com}
\author[F. Montesi]{Fabrizio Montesi}
\address{University of Southern Denmark, Denmark}
\email{fmontesi@imada.sdu.dk}
\author[M. Gabbrielli]{Maurizio Gabbrielli}
\address{Universit\`a di Bologna, Italy and INRIA, France}
\email{maurizio.gabbrielli@unibo.it}

\begin{abstract}
Choreographic Programming is a correct-by-construction paradigm where a
compilation procedure synthesises deadlock-free, concurrent, and distributed
communicating processes from global, declarative descriptions of communications,
called choreographies.
Previous work used choreographies for the 
synthesis of programs. Alas, there is no formalisation that provides a chain of
correctness from choreographies to their implementations. This problem
originates from the gap between existing theoretical models, which abstract
communications using channel names (à la CCS/\(\pi\)-calculus), and their
implementations, which use low-level mechanisms for message routing.
As a solution, we propose the theoretical framework of Applied Choreographies.
In the framework, developers write choreographies in a language that follows the
standard syntax and name-based communication semantics of previous works. Then,
they use a compilation procedure to transform a choreography into a low-level,
implementation-adherent calculus of Service-Oriented Computing (SOC). To manage
the complexity of the compilation, we divide its formalisation and proof in
three stages, respectively dealing with:
\emph{a}) the translation of name-based communications into their SOC
equivalents (namely, using correlation mechanisms based on message data);
\emph{b}) the projection of a choreography into a composition of partial,
single-participant choreographies (towards their translation into SOC
processes);
\emph{c}) the translation of partial choreographies and the distribution of
choreography-level state into SOC processes.
We provide results of behavioural correspondence for each stage. Thus, given a
choreography specification, we guarantee to synthesise its faithful and
deadlock-free service-oriented implementation.

\end{abstract}

\maketitle


\newcommand{\mainDir}{.}

\section{Introduction}
\label{sec:introduction}

\paragraph{Background}
Concurrent, distributed software applications have become a crucial asset of
our society: messaging, governance, healthcare, and transportation are just
some of the contexts recently revolutionised by distributed applications. A
hallmark of those applications is that their global
behaviour, usually referred to as \emph{protocol}, emerges from the
interaction of programs, also called \emph{endpoints}, that run in
parallel and rely on message passing to communicate and coordinate their
actions~\cite{DistributedSystems}.
Developers strive to correctly implement separate endpoints that, when put
together, will enact the expected protocols. If endpoints fail to follow their
protocols, the distributed system can block or misbehave --- e.g., due to
deadlocks~\cite{deadlocks} or race conditions~\cite{raceConditions}. Ensuring
that all endpoints play their respective parts correctly---i.e., that they
follow their intended protocols---is difficult due to the inherent
non-determinism of distributed programs running in parallel~\cite{o18}.

Since the early days of distributed computing, designers and developers
introduced and used tools to describe the order of interactions among the
endpoints of a system, like security protocol notation~\cite{needham1978using},
Message Sequence Charts~\cite{MSC} and UML Sequence Diagrams~\cite{UML}. The
common denominator of these tools is to present a global description of the
sequence of messages in the system, an information difficult to infer (due to
the complexity of interleaved communications) from the specified behaviours of
the endpoints. Even for simple systems with a fixed number of participants,
algorithms for extracting this information have exponential
complexity~\cite{CLM17}.

Recognising the usefulness of these approaches, in the early 2000s the W3C
assembled a Working Group tasked with the definition of a standard for
describing interactions among Web Services. This resulted in the Web Services
Choreography Description Language (WS-CDL)~\cite{wscdl}.
%
A WS-CDL artefact is a ``choreography'', which specifies the observable
behaviours of all the endpoints involved in the system of interest, formalising
from a global viewpoint the ordering conditions and constraints that regulate
the exchange of messages.

\begin{example}\label{example:intro}
We illustrate the choreographic approach with a representative example. We use
the example to also introduce the syntax of choreographies used in the
remainder of the paper. The example describes a simple business scenario among
a client process $\pid c$, a seller service located at $l_{\role S}$ and a
bank service located at $l_{\role B}$. Locations ($l$) are abstractions of
network addresses, or URIs, which identify where services can be contacted to
interact with them.
%

\vspace{.5em}
\begin{minipage}{.47\textwidth}
\begin{scaledListing}{.9}
\begin{lstlisting}[lineskip={.5em},
numbers=left,mathescape,backgroundcolor=\color{white}]
$\label{cd:start}\start{k}{\pid c[\role C]}{l_{\role S}.\pid s[\role S],
                            l_{\role B}.\pid b[\role B]};$
$\label{cd:interaction}\com{k}{\pid c[\role C].product}
                              {\pid s[\role S].\op{buy}(\ x\ )};$
$\label{cd:3}\com{k}{\pid s[\role S].\lfn{mk\_order}(\ x\ )}{
    \pid b[\role B].\op{reqPay}(\ order\ )};$
$\label{cd:4}\com{k}{\pid c[\role C].cc}{\pid b[\role B].\op{sendCC}(\ cc\ )};$
\end{lstlisting}
\end{scaledListing}
\vspace{1.5em}
\end{minipage}
\hspace{.05\textwidth}
\begin{minipage}{.47\textwidth}
\begin{scaledListing}{.9}
\begin{lstlisting}[firstnumber=5,lineskip={.5em},
numbers=left,mathescape,backgroundcolor=\color{white}]
$\label{cd:5}\m{if}\ \pid b.\lfn{confirm\_pay}(\ cc,\ order\ ) \{$ 
 $\com{k}{\pid b[\role B]}{\pid c[\role C].\op{ok}()};\
 \com{k}{\pid b[\role B]}{\pid s[\role S].\op{ok}()}$
$\}\ \m{else}\ \{$
 $\com{k}{\pid b[\role B]}{\pid c[\role C].\op{ko}()};\
 \com{k}{\pid b[\role B]}{\pid s[\role S].\op{ko}()}$
$\}$
\end{lstlisting}
\end{scaledListing}
\end{minipage}
\vspace{.5em}
%

%
At \cref{cd:start}, the client $\pid c$ asks the seller and the bank services to
create two new processes, respectively $\pid s$ and $\pid b$. The three
processes $\pid c$, $\pid s$, and $\pid b$ can communicate over a private multiparty
session $k$, intended as in Multiparty Session Types~\cite{CDYP2015}: 
each process owns a statically-defined
\emph{role} in the session, which identifies a message queue that the process uses to receive
messages asynchronously. For simplicity, at \cref{cd:start}, we assign role
$\role C$ to process $\pid c$, $\role S$ to $\pid s$, and $\role B$ to $\pid b$.
As usual, processes have local states and run concurrently.
All communications in the rest of the choreography now take place over session $k$, as indicated by the 
prefix ``$k:$'' in the other lines.
At \cref{cd:interaction}, the client $\pid c$ invokes
operation $\op{buy}$ of the seller $\pid s$ with the name of a $product$ it
wishes to buy, which the seller stores in its local variable $x$. At 
\cref{cd:3}, the seller uses
its internal function $\lfn{mk\_order}$ to prepare an order
(e.g., compute the price of the product) and sends it to the bank on operation
$\op{openTx}$, for opening a payment transaction. At \cref{cd:4}, the client
sends its credit card information $cc$ to the bank on operation $\op{pay}$.
Then, at \cref{cd:5}, the bank makes an internal choice on whether the payment
can be performed (with the internal function $\lfn{close\_tx}$, which takes the
local variables $cc$ and $order$ as parameters). The bank then notifies the
client and the seller of the final outcome, by invoking them both either on
operation $\op{ok}$ or $\op{ko}$.
\end{example}

The advantage of choreographies is their clarity: they specify the intended
global behaviour of a communicating system unambiguously. For this reason, since
the inception of WS-CDL, choreographies have been adopted also in other
practical applications, like the Business Process Model and Notation by the
Object Management Group~\cite{BPMN} and Testable
Architecture~\cite{savara:website}. In general, choreographies come with the
promise of enhancing the correctness of systems, since they equip programmers
with precise specifications of the communications that a system should enact.
This promise motivated a fruitful line of research in the areas of process
calculi and programming languages, which rotates around the question: 
``\emph{Can we use choreographies to prove that a concurrent program will
execute the right communications?}''


Inspired by this question, two development methodologies have emerged based on
choreographies. In the first, called Choreographic Programming~\cite{M13:phd},
programs are choreographies as that in our Example~\ref{example:intro}. The
idea is that the choreography defines both the internal computation performed
by processes and the communications among them. Then, a
correct-by-construction implementation (typically given in terms of a process
calculus) can be automatically synthesised~\cite{CHY12,CM13}. In the second
methodology, choreographies are used to describe protocols, which abstract
away from internal computation. The aim of this second methodology is to
verify that each process, written manually (in
contrast to being automatically synthesised, as in choreographic programming),
implements correctly its role in the protocols that it participates in.
Multiparty Session Types~\cite{HYC16} is representative of this methodology.

Both methodologies are based on the same general idea: for each endpoint
described in a choreography, we can \emph{project} a definition of its local
behaviour using a procedure known as EndPoint Projection (EPP). In
choreographic programming, this yields the local implementation of each
endpoint. For multiparty session types, this yields a type, e.g., used to check
that a process implements its role in a protocol correctly. The key technical
result that one needs to prove then is that the projection yields a set of
endpoint terms (programs or types) that, when executed in parallel, implement
exactly the communications described in the original choreography. This is
typically called the EndPoint Projection Theorem (or EPP Theorem, for short).

The model of Compositional Choreographies~\cite{MY13} unifies the two
methodologies, with the aim of combining their advantages. In that model,
programmers can describe parts of a system as in choreographic programming and
other parts as independent process terms. The model uses multiparty session
types to check that the composition of the independent process terms with the
projections of choreographic programs will behave correctly. What made
the unification of the two approaches possible is the strong
operational correspondence guaranteed by the EPP.

\paragraph{Motivation}
The main application area for choreographies so far is that of
Service-Oriented Computing (SOC), as in web services~\cite{wscdl} or
microservices~\cite{DGLMMMS16,n15}. Implementing communications in this setting is
non-trivial, since services must be loosely coupled and thus we cannot assume
the presence of any particular common middleware. However, in all previous
definitions of EPP, both the choreography language and the target language
abstract from how real-world frameworks support
communications~\cite{QZCY07,LGMZ08,CHY12,CM13,CMS17}, by modelling message
exchange through synchronisations on \emph{names} (as in CCS and the
$\pi$-calculus~\cite{M80,MPW92}).
As a consequence, the implementations of choreographic
frameworks~\cite{chor:website,aiocj:website,NY14} significantly depart from
their respective formalisations~\cite{CM13,DGGLM15,HYC16} (a common aspect of
implementing process calculi, cf.~\cite{CLM05,HYH08}). In particular,
implementations realise the creation of new channels and message routing with
additional data structures and message exchanges~\cite{M13:phd,DGLMG14} that
are absent in their formalisations. The specific communication mechanism used
in these implementations is message correlation; correlation is the reference
communication support in SOC, and is supported by mainstream technologies
(e.g., WS-BPEL~\cite{bpel}, Java/JMS, C\#/.NET).
The gap between formalisations and implementations can compromise the
correctness guarantee of choreographies.
Thus we ask: 

\noindent
\begin{center}
\emph{How can we formalise the implementation of communications in
choreographies?}
\end{center}
\noindent

A satisfactory answer should preserve the correctness guarantees down to the
level of how communications are concretely implemented.
Defining such a model is challenging: we wish to retain the typical clarity of
choreography languages, yet we need enough details to (formally) reason on how
communications are realised at the lower level. Ideally, the complexity of
implementing communications should not leak into the choreographic programming
model exposed to programmers, and should just be a ``detail'' that we can forget
about with confidence. Building this confidence is the main aim of this article.

\paragraph{Contributions and Outline}
We tackle our question by developing the framework of Applied Choreographies.
Our framework consists of three calculi, which enjoy a tight series of
correspondences.

The first calculus, called Frontend Choreographies (FC), is meant to be the
programming model exposed to programmers and is presented in
\cref{sec:choreography_language}. FC is a straightforward reformulation of the
standard calculus of Compositional Choreographies~\cite{MY13}, which we adopt
to show that our approach applies to both the methodology of choreographic
programming and that of multiparty session types. In particular,
communications are based on name synchronisation, as in standard process
calculi.

The second calculus, called Backend Choreographies (BC), has the same syntax of
FC but a different semantics: instead of using abstract name synchronisation, BC
models and keeps track of the data structures needed to implement concrete
correlation-based communications (\cref{sec_aEC}). While more involved than FC,
BC is agnostic wrt the specific technology used for correlation.


The third is a process calculus of distributed executable code, based on a
standard formal model for Service-Oriented Computing~\cite{MC11}, called Dynamic
Correlation Calculus (DCC) (\cref{sec:implementation_model}). DCC models both
data distribution and how concrete communications are implemented. Given its
low-level scope, DCC does not capture all the abstraction of choreographies.

Our main contribution is the definition of a behaviour-preserving compiler from
Frontend Choreographies to DCC distributed services. The compiler uses Backend
Choreographies as intermediate representation. This is the first correctness
result of an end-to-end translation from standard choreographies to programs
based on a real-world communication mechanism.

More concretely, our compiler includes two transitional stages (FC-to-BC and EPP)
toward the Compilation:
\begin{description}

  \item[FC-to-BC] generates the data structures needed to support the execution
    of a source FC program using message correlation (\cref{sub:encodig}).
    Essentially, we obtain a Backend choreography which is operationally
    correspondent to its source Frontend;
  
  \item[EPP] transforms a choreography that describes the behaviours of many
  participants into a set of modules, called \emph{endpoint choreographies},
  each describing the behaviour of a single participant. More precisely, the
  procedure is an endomorphism (\cref{sub:endpoint_projection}) that transforms
  a source choreography---whether FC or BC does not matter, since they share the
  same syntax---into a set of (endpoint) choreographies whose syntax is
  restricted to only partial actions (i.e., belonging to one of the two ends of
  a communication);

  \item[Compilation] takes in the BC data structures obtained from the FC-to-BC
  conversion and the endpoint choreographies obtained from the EPP
  transformation and synthesises a correct implementation of the
  source FC program as a distributed system of DCC services.

\end{description}

Starting from FC proves that our development is adequate. Programmers can use
high-level programming primitives and semantics as found in previous works on
choreographies---with state-of-the-art features like
asynchronous communications~\cite{CM13} and modular
development~\cite{MY13}---while our compilation procedure tackles the
heavy-lifting of producing correct service-oriented implementations.

We conclude our proposal discussing related and future work in
\cref{sec:related} and report in the Appendix auxiliary technical material and
the proofs of our results.

This paper integrates and extends material from~\cite{GMG18}, which presented
the main ideas behind the Applied Choreographies framework. The extensions in
this work include:
\emph{a}) full formal definitions (syntax and semantics of all three calculi);
\emph{b}) detailed examples for each main component of the work (the three
calculi, the typing system, the three stages of compilation) to illustrate their
relevant characteristics and features;
\emph{c}) full proofs of the formal properties guaranteed by the framework (in
\cref{sec:appendix}, to avoid breaking the flow of the reader with details of
the technical development). Besides the previous points, this version contains
an extended, revised, and refined presentation of all the contents presented
in~\cite{GMG18}

\IFSubFileBiblio

\section{Frontend Choreographies}
\label{sec:choreography_language}

We present Frontend Choreographies (FC), the language model intended for
programmers.

%

Before giving the formal syntax of FC, we first describe the intuition behind
its key components. The following table displays the symbols that we are going
to use, along with their names and domains.

\[\begin{array}{l@{\hspace{2em}}c@{\hspace{2em}}c}
  \mathbf{Name} & \mathbf{Symbols} & \mathbf{Domain}
  \\
  \hline
  \\
  \mathit{Choreographies} & C_1, C_2 & -
  \\
  \mathit{Processes} & \pid{p},\pid{q} & \Pids
  \\
  \mathit{Operations} & \op{o_1}, \op{o_2} & \Operations
  \\
  \mathit{Variables} & x, y & \Var
  \\
  \mathit{Sessions} & k_1, k_2 & \Sessions
  \\
  \mathit{Roles} & \role A, \role B & \Roles
  \\
  \mathit{Locations} & l_1, l_2 & \Locations
\end{array}\]\vspace{1em}

FC programs are choreographies, as in \cref{example:intro}, denoted by $C$.
A choreography describes the behaviour of some 
processes. Processes, denoted $\pid p,\pid q \in \Pids$, are intended as usual: they are independent 
execution units running concurrently and equipped with local variables, denoted $x \in \Var$.

Processes communicate by exchanging messages. A message consists of
two elements: \emph{i}) a payload, representing the data exchanged between two
processes; and \emph{ii}) an operation, which is a label used by the receiver to determine what 
it should do with the message---in object-oriented programming, these labels are 
called method names~\cite{PIERCE}; in service-oriented computing, labels are typically called 
operations as in here. Operations are denoted $\op{o} \in \Operations$.

Message exchanges happen through a session, denoted $k \in \Sessions$, which acts as a 
communication channel. Sessions in FC are behaviourally typed \cite{HLVCCDMPRT16}. Intuitively, a 
session is an instantiation of a protocol, where each process is responsible for implementing the 
actions of a role defined in the protocol. We denote roles with
$\role A, \role B \in \Roles$.

A process can create new processes and sessions at runtime by invoking service processes (services 
for short).
Services are always available at fixed locations, denoted $l \in \Locations$, meaning that they can 
be used multiple times (in process calculus terms, they act as replicated processes \cite{SW01}).

%
%

FC supports modular development by allowing choreographies, say $C$ and $C'$, to be composed 
in parallel, written $C \pp C'$. A parallel composition of choreographies is also a 
choreography, which can thus be used in further parallel compositions. Composing two choreographies 
in parallel allows the processes in the two choreographies to interact over shared location and 
session names.

We distinguish between two kinds of statements inside of a choreography: complete and partial 
actions. A complete action is internal to the system defined by the choreography, and thus does not 
have any external dependency. By contrast, a partial action defines the behaviour of some processes 
that need to interact with another choreography in order to be executed. Therefore, a choreography 
containing partial actions needs to be composed with other choreographies that provide compatible
partial actions.

To exemplify the distinction between complete and partial actions, we consider
the case of a single communication between two processes.

\[
\begin{array}{@{\hspace{2em}}c@{\hspace{2em}}|@{\hspace{2em}}c@{\hspace{2em}}}
\emph{Complete interaction} & \emph{Composed partial actions}
\\[.1em]\hline\\
\com{k}{\pid c[\role C].product}{\pid s[\role S].\op{buy}(\ x\ )}
&
  \begin{array}l
  \com{k}{\pid c[\role C].product}{\role S.\op{buy}}
  \\|\\
  \com{k}{\role C}{\pid s[\role S].\op{buy}(\ x\ )}  
  \end{array}
\end{array}
\]\vspace{1em}

Above, on the left we have the communication statement as seen at
\cref{cd:interaction} of \cref{example:intro}. This is a complete action: it
defines exactly all the processes that should interact ($\pid c$ and $\pid
s$). On the right, we implement the same action as the parallel composition of
two choreographies with partial actions: a send action by process $\pid c$ to
role $\role S$ over session $k$ (left of the parallel) and a reception by
process $\pid s$ from a role $\role C$ (right of the parallel) over the same
session $k$. More specifically, we read the send action (top of the parallel)
as ``process $\pid c$ sends a message as role $\role C$ with payload $product$
for operation $\op{buy}$ to the process playing role $\role S$ on session
$k$''. Dually, we read the receive action (bottom of the parallel) as
``process $\pid s$ receives a message for role $\role S$ and operation
$\op{buy}$ over session $k$ and stores the payload in variable $x$''. The
compatible roles, session, and operation used in the two partial actions make
them compliant. Thus, the choreography on the left is operationally equivalent
to the one on the right. Observe that partial actions do not mention the name
of the process on the other end---for example, the send action by process
$\pid c$ does not specify that it wishes to communicate with process $\pid s$
precisely. This mechanism supports some information hiding: a partial action
in a choreography can interact with partial actions in other choreographies
independently of the process names used in the latter. Expressions and
variables used by senders and receivers are also kept local to statements that
define local actions.

\subsection{Syntax of Frontend Choreographies}\label{sub:fc_syntax}

We present the formal syntax of FC, displayed in
\cref{fig:cc_syntax}.
%
In the remainder, we use the symbol $\sim$ over an
element to indicate an ordered set of elements of its kind, e.g., $\pids p$
indicates an ordered set of processes $\pid p_1, \ldots, \pid p_n$.

\begin{figure}
{\arraycolsep=1pt\def\arraystretch{1.3}
$$
\begin{array}{lr@{\quad}l@{\qquad}l@{\qquad}c@{\qquad}l@{\qquad}l}
C & \gram & \eta ; C 										& \prid{seq}
	& \Div &
		\cond{\pid p.e}{C_1}{C_2} 						& \prid{cond}
	\\
	& \Div & C_1 \pp C_2 									& \prid{par}
	& \Div &
	 	\genbranchI														& \prid{recv}
	\\
	& \Div & \inact												& \prid{inact}
	& \Div &
	 	\recDef{X}{C^\prime}{C}								& \prid{rec}
	\\
	& \Div & \recCall{X}									& \prid{call}
	& \Div &
		\genacc ; C														& \prid{accept}
	\\[1em]
\eta & \gram & \gensend									& \prid{send}
	& \Div &
		\genstart 														& \prid{start}
	\\
	& \Div & \gencom											&  \prid{com}
	& \Div &
  \genreq																	& \prid{req}
\end{array}
$$
}
\caption{Frontend Choreographies --- syntax.}
\label{fig:cc_syntax}
\end{figure}

\paragraph{Complete Actions} In term $\prid{start}$, process $\pid p$ creates a new session $k$ together with processes $\pids q$ ($\pids q$ is assumed
non-empty). Process $\pid p$, called \emph{active process}, is already running,
whereas each process $\pid q$ in $\wtil{l.\pid q}$, called \emph{service
process}, is dynamically created at the respective service location $l$. Each process is
annotated with the role it plays in the new session $k$.
Term $\prid{com}$ reads: on session $k$, process $\pid p$ sends to process $\pid
q$ a message for its operation $\op{o}$; the message carries the evaluation of
expression $e$ on the local state of $\pid p$, whilst $x$ is the variable where $\pid q$ will store the content of the message. We leave the guest language for writing local expressions ($e$) unspecified, and assume that it consists of terms for accessing local variables ($x$) and implementing standard computations based on those (e.g., arithmetics).

\paragraph{Partial Actions}
A choreography can use partial actions to interact with other choreographies
composed in parallel. Therefore, partial actions describe the behaviour of
processes that wish to synchronise with ``external'' participants. Concretely,
these external participants will be processes and/or services whose behaviours
are defined in other choreographies composed in parallel. In term
$\prid{req}$, process $\pid p$ requests some external services, respectively
located at $\til l$, to create a new session $k$ and some new external
processes. Role annotations follow the same intuition as in term
$\prid{start}$: in the new session $k$, $\pid p$ will play $\role A$ and each
new external process $\pid q_i$ will play the respective role $\role B_i$.

Term $\prid{acc}$ is the dual of $\prid{req}$ and defines a choreography
module that provides the implementation of some service processes.
We assume that $\prid{acc}$ terms are always at the top level, to capture that
choreography modules are always available. By top level, we mean that the term
is not preceded by another term in a sequential composition $\prid{seq}$.

In term $\prid{send}$, process $\pid p$ sends a message to an external process
that plays $\role B$ in session $k$. Dually, in term $\prid{recv}$, process
$\pid q$ receives a message for one of the operations $o_i$ from an external
process playing role $\role A$ in session $k$, and then proceeds with the
corresponding continuation. In the remainder, we omit curly brackets in
$\prid{recv}$ terms when they have only one operation, i.e., $\com{k}{\role
A}{\pid q[\role B].\op{o}(x);C }$ is an abbreviation of $\com{k}{\role A}{\pid
q[\role B].\{\op{o}(x);C\} }$.

\paragraph{Other Terms} Term $\prid{seq}$ is sequential composition. In a
conditional $\prid{cond}$, process $\pid p$ evaluates a condition $e$ in its
local state to choose between the continuations $C_1$ and $C_2$. Term
$\prid{par}$ is standard parallel composition, which allows partial actions in
two choreographies $C_1$ and $C_2$ to interact. Respectively, terms
$\prid{def}$, $\prid{call}$, and $\prid {inact}$ model the definition of
recursive procedures, procedure calls, and inaction.

Some terms bind identifiers in continuations---the choreography that follows
them in a sequential composition. In terms $\prid{start}$ and $\prid{acc}$,
the session identifier $k$ and the process identifiers $\pids q$ are bound (as
they are freshly created). In terms $\prid{com}$ and $\prid{recv}$, the
variables used by the receiver to store the message are bound ($x$ and all the
$x_i$, respectively). In term $\prid{req}$, the session identifier $k$ is
bound. Finally, in term $\prid{def}$, the procedure identifier $X$ is bound.
In the remainder, we omit $\inact$ or irrelevant variables (e.g., in
communications with empty messages).
Terms $\prid{com}$, $\prid{send}$, and $\prid{recv}$ include role annotations
only for clarity reasons; roles in such terms can be inferred, as shown
in~\cite{M13:phd}.

\begin{example}
\label{example:super} 
\begin{figure}
\begin{minipage}{.59\textwidth}
\begin{scaledListing}{.9}
\begin{lstlisting}[numbers=left,mathescape,lineskip={.5em},
      backgroundcolor=\color{white}]
$\label{lst:startk}\start{k}{\pid c[\role C]}{l_{\role S}.\pid s[\role S],
                            l_{\role B}.\pid b[\role B]};$
$\com{k}{\pid c[\role C].buyReq}
                                {\pid s[\role S].\op{buy}(\ x\ )};$
$\label{lst:startkp}\addCode{\req{k'}{\pid s[\role S]}{l_{\role D}.\role D}};$
$\addCode{\com{k'}{\pid s[\role S].\lfn{mk\_shipping}(\ x\ )}{\role D.\op{quoteShipping}}};$
$\addCode{\com{k'}{\role D}{\pid s[\role S]}{.\op{shippingCosts}(\ y\ )};}$
$\com{k}{\pid s[\role S].\addCode{\lfn{mk\_order}(\ x,\ y\ )}}
  {\pid b[\role B].\op{reqPay}(\ order\ )};$
$\com{k}{\pid c[\role C].cc}{\pid b[\role B].\op{sendPay}(\ cc\ )};$
\end{lstlisting}
\end{scaledListing}
\end{minipage}
\begin{minipage}{.40\textwidth}
\begin{scaledListing}{.9}
\begin{lstlisting}[firstnumber=8,numbers=left,mathescape,lineskip={.5em},
backgroundcolor=\color{white}]
$\m{if}\ \pid b.\lfn{confirm\_pay}(\ cc,\ order\ ) \{$
 $\com{k}{\pid b[\role B]}{\pid c[\role C].\op{ok}()};\
 \com{k}{\pid b[\role B]}{\pid q[\role S].\op{ok}()};$
 $\addCode{\com{k'}{\pid s[\role S]}{\role D.\op{sendShipping}}}$
$\}\ \m{else}\ \{$
 $\com{k}{\pid b[\role B]}{\pid c[\role C].\op{ko}()};\
 \com{k}{\pid b[\role B]}{\pid q[\role S].\op{ko}()};$
 $\addCode{\com{k'}{\pid s[\role S]}{\role D.\op{abortShipping}}}$
$\}$
\end{lstlisting}
\end{scaledListing}
\end{minipage}
\caption{Choreography $C_1$, extension of \cref{example:intro}.} 
\label{fig:FC_example}
\begin{minipage}{.59\textwidth}
\begin{scaledListing}{.9}
\begin{lstlisting}[numbers=left,mathescape,lineskip={.5em},
backgroundcolor=\color{white}]
$\label{lst:startkpc}\addCode{\acc{k'}{l_{\role D}.\pid d[\role D]};}$
$\addCode{\com{k'}{\role S}{\pid d[\role D].\op{quoteShipping}(\ pkg\ )};}$
$\addCode{\com{k'}{\pid d[\role D].\lfn{quote}(\ pkg\ )}{\role S.\op{shippingCosts}};}$
\end{lstlisting}
\end{scaledListing}
\end{minipage}\hspace{5px}
\begin{minipage}{.39\textwidth}
\begin{scaledListing}{.9}
\begin{lstlisting}[firstnumber=4,numbers=left,mathescape,lineskip={.5em},
backgroundcolor=\color{white}]
$\addCode{\com{k'}{\role S}{\pid d[\role D]}.\{}$
 $\addCode{\op{sendShipping}(),}$
 $\addCode{\op{abortShipping}() \hspace{1em} \}}$
\end{lstlisting}
\end{scaledListing}
\end{minipage}
  \caption{Choreography $C_2$, compliant choreography to \cref{fig:FC_example}.}
  \label{fig:FC_example_complementary}
\end{figure}
In \cref{fig:FC_example}, we extend (in \addCode{blue}) the behaviour of the
seller of \cref{example:intro} to use an external module. In the updated code,
the seller contacts an external service for the delivery of the product: the
seller receives a request $buyReq$ from the buyer. The request contains the
wanted product and the delivery address (Line 2). Next, the seller creates a
new session $k'$ with an external delivery process (Line 3) and sends to the
latter the shipping information of the product, e.g., the origin and
destination addresses (Line 4). At Line 5, the seller receives the shipping
costs, which it adds to the costs of the order at the bank (Line 6). At Lines
11 and 14, the seller notifies the delivery process if it shall ship the
product or not.
Let us call $C_1$ the code above. We report in
\cref{fig:FC_example_complementary} the module $C_2$ of a compliant delivery
service for $C_1$.
%
%
%
We obtain a working system by composing the two choreographies in parallel: $C_1 \pp C_2$.

\end{example}

\subsection{Semantics of Frontend Choreographies}
\label{sec:fc_semantics}
We give an operational semantics for FC in terms of reductions of the form
$\env,C \to \env',C'$, where $\env$ is a deployment. Deployments keep track
of: the local states of processes (the values of their local variables); and
the messages in transit in sessions, which we use to model asynchronous
communications.
%
%
In the following, we formalise our notion of deployment and we present
our reduction semantics.

\subsubsection{Frontend Deployments}
%
In the remainder, we adopt as a convention, when indicating a Frontend
Choreographies program, its shortened form ``Frontend choreography'' (lowercase
c) or simply ``choreography'' when the context clearly associates it to Frontend
Choreographies. We also use the shortened form ``Frontend deployment'' to
indicate a Frontend Choreographies deployment.

Each pair of roles in a session has two dedicated asynchronous message queues
that they can use to exchange messages, one for each direction. Formally, let
$\SessionQueues = \Sessions \times \Roles \times \Roles$ be the set of all
\emph{queue identifiers}; we write $k[\role A\rangle\role B] \in
\SessionQueues$ to identify the queue from role $\role A$ to role $\role B$ in
session $k$.

A \emph{deployment} $\env$ is an
overloaded partial function defined by cases as the sum of two partial
functions, $f_s: \Pids
\rightharpoonup \Var
\rightharpoonup \Val $ and $f_q: \SessionQueues \rightharpoonup Seq(\Operations \times
\Val )$ (their domains and co-domains are disjoint):

\[\env( z ) = \begin{cases}
  f_s(z) & \mbox{ if } z \in \Pids \\ 
  f_q(z) & \mbox{ if } z \in \SessionQueues
\end{cases}
\]\vspace{1em}

Function $f_s$ maps a process $\pid p$ to its state. A state is a partial
function from variables $x,y \in \Var$ to values $v \in \Val$.
Function $f_q$ stores the queues used in sessions. Each queue is a sequence of
messages $\til m = m_1 :: \ldots :: m_n \ | \ \emptyseq$ ($\emptyseq$ is the
empty queue), where each message $m = (\op{o}, v) \in \Operations \times \Val$
contains the operation $\op{o}$ for which the message is intended and the
payload $v$.

Deployments are a runtime concept: programmers do not need to define them, just
as they normally do not explicitly give an initial state for their programs in
other language models. Formally, we assume that choreographies without free
session names start execution with a \emph{default deployment} that contains
empty process states. Let
$\fp(C)$ return the set of free process names in $C$. Then, we formally define a
default deployment as follows.

\begin{definition}[Default Deployment\label{def:default_deployment}] Let $C$ be
a choreography without free session names. Then, the default deployment $\env$
for $C$ is defined as the function that maps all free process names in $C$ to
empty states (we write $\emptyfunc$ for the empty partial function from $\Var$
to
$\Val$):

\[
\env = \big[ \pid p \mapsto \emptyfunc \mid \pid p \in \fp(C) \big]
\]

\end{definition}

Intuitively, $\env$ is a default deployment for a choreography without free
session names $C$ if \emph{i}) $\env$ is defined for all and only the
processes that appear free in $C$ and \emph{ii}) the state of these processes is
empty.

\subsubsection{Frontend Deployment Transitions}
In our semantics, choreographic actions have effects on the state of a
system --- deployments change during execution. At the same time, a deployment
also determines which choreographic actions can be performed. For example, a
communication from role $\role A$ to role $\role B$ over session $k$ requires
a queue $k[\role A\rangle\role B]$ to exist in the deployment of the system.

We formalise the notion of which choreographic actions are allowed by a
deployment and their effects using transitions of the form
$\renv{\env}{\delta}{\env'}$, read ``the deployment $\env$ allows for the
execution of $\delta$ and becomes $\env'$ as result''. The following grammar 
defines $\delta$ actions.

\[\begin{array}{lrl@{\qquad}l}
  \delta & \gram  & \genstart & \textit{(session start)}     \\[2pt]
         & \Div   & \gensend  & \textit{(send in session)}   \\[2pt]
         & \Div   & \com{k}{\role A} {\pid q[\role B].\op{o}(x)} 
                          & \textit{(receive in session)}     
\end{array}\]\vspace{1em}

\begin{figure}
\input{figures/sos_chor_seff}
\end{figure}

The rules defining $\renv{\env}{\delta}{\env'}$ are given in \cref{fig:seff_semantics}.

\boldpar{Rule $\protect\did{\env}{Start}$}
states that the creation of a new session $k$ between an existing process
$\pid p$ and new processes $\pids q$ results in updating the deployment with:
a new (empty) state for each of the new processes $\pid q$ in $\pids q$
($\big[\pid q \mapsto \emptyset \mid \pid q \in \pids q \big]$); and a new
(empty) queue between each pair of distinct roles in the session
($\big[\chanto{k}{\role C}{\role E} \mapsto \emptyseq
        \mid \{ \role C, \role E \} \subseteq \{ \role A, \roles B \} \big]$).
				
\boldpar{Rule $\protect\did{\env}{Send}$}
models the effect of a send action.  In the first premise, we use the
auxiliary function $\evalfn$ to evaluate the local expression $e$ in the state
of process $\pid p$, obtaining the value $v$ to use as message payload. Then,
in the conclusion, we add a message $(\op{o},v)$ --- where $\op o$ is the
operation used to label the message --- to the tail of the queue
$\chanto{k}{\role A}{\role B}$, i.e., the queue expected to contain messages
sent by $\role A$ to $\role B$ in session $k$. We assume that function
$\evalfn$ always terminates --- in practice, this can be obtained by using
timeouts.

\boldpar{Rule $\protect\did{\env}{Recv}$}
models the effect of a reception. First, in the premise, we look up the head
of the message queue between sender and receiver, i.e., $(\op{o},v)$. Then, in
the conclusion, we remove the message from the queue ($\big[ \chanto{k}{\role
A}{\role B} \mapsto \til m \big]$) and update the state of the receiver at the
variable used to store the message ($\big[ \pid q \mapsto \env( \pid q )[ x
\mapsto v ] \big]$).

\subsubsection{Reductions}
Using deployment transitions, we can now define the rules for reductions
$\env,C \to \env',C'$. We call a configuration $\env,C$ a \emph{running
choreography}. The reduction relation $\to$ for FC is the smallest relation
closed under the rules given in \cref{fig:cc_semantics_sos_full}.

\begin{figure}
\input{figures/chor_semantics_sos_full}
\end{figure}

\boldpar{Rule $\protect\did{C}{Start}$}
creates a new session, by ensuring that both the new session name $k'$ and new
processes $\pids r$ are fresh wrt $\env$ ($\env\#k',\pids r$). We use the
fresh names in the continuation $C$, by using a standard substitution
$C[k'/k][\pids r/\pids q]$.

\boldpar{Rule $\protect\did{C}{Send}$} reduces a send action, if the
deployment permits it: $\renv{\env}{\gensend}{\env'}$.

\boldpar{Rule $\protect\did{C}{Recv}$} reduces a message reception, if the
deployment permits the reception of a message on one of the branches in the
receive term ($j \in I$). Recalling the corresponding rule $\did{\env}{Recv}$,
this can happen only if the deployment $\env$ has a message for operation
$o_j$ in the queue $\chanto{k}{\role A}{\role B}$.

\begin{figure}
$$
	\begin{array}{c}
		\recDef{X}{C^\prime}{\inact} \equivC \inact
		\qquad \qquad
		C \pp C^\prime \equivC C^\prime \pp C
		\qquad \qquad
		(C_1 \pp C_2) \pp C_3 \equivC C_1 \pp (C_2 \pp C_3)
		\\[.7em]
		\recDef{X}{C^\prime}{C[X]}
	 \equivC
		\recDef{X}{C^\prime}{C[C^\prime]}
		\\[.5em]
		\gencom;C \equivC \gensend;
			\com{k}{\role A}{\pid q[\role B].\{\op{o}(x);C\}}
	\end{array}
$$
	\caption{Frontend Choreographies --- structural congruence $\equivC$}
	\label{fig:chor_struct_eq}
\end{figure}
\begin{figure}
$$
	\begin{array}{c}
		\infer[\did{CS}{EtaEta}]
		{\eta;\eta^\prime \quad \swapC \quad \eta^\prime;\eta}
		{\pn(\eta) \cap \pn(\eta^\prime) = \emptyset}
		\qquad
		\infer[\did{CS}{EtaCnd}]
		{
			\begin{array}{l}
			\cond{\pid p.e}{\eta;C_1}{\eta;C_2} 
			\\ \hspace{2em} \swapC \quad \eta; \cond{\pid p.e}{C_1}{C_2}	
			\end{array}
		}
		{ \pid p \not \in \pn(\eta) }
		\\[15pt]
		\infer[\did{CS}{EtaRcv}]
		{
			\begin{array}{l}
			\com{k}{\role A}{\pid q[\role B].\{\op{o}_i(x_i); \eta; C_i\}_{i\in I}}
			\quad \swapC \quad
			\eta ; \com{k}{\role A}{\pid q[\role B].\{\op{o}_i(x_i); C_i\}_{i\in I}}
			\end{array}
		}
		{\pid q \not \in \pn(\eta)}
		\\[10pt]
		\quad
				\infer[\did{CS}{RcvRcv}]
		{
			\begin{array}{l}
				\com{k}{\role A}{\pid p[\role B].\{\op{o}_i(x_i); 
				\com{k^\prime}{\role C}{\pid q[\role D].\{\op{o}^\prime_{ij}(x'_{ij});
					C_{ij}\}_{j\in J}}\}_{i\in I}}
			\\ \hspace{2em} \swapC \quad
			\com{k^\prime}{\role C}{\pid q[\role D].\{\op{o}^\prime_j(x'_j); 
				\com{k}{\role A}{\pid p[\role B].\{\op{o}_{ij}(x_{ij});
					C_{ij}\}_{i\in I}}\}_{j\in J}}
			\end{array}
		}
		{\pid p \neq \pid q}
		\\[15pt]
		\infer[\did{CS}{CndCnd}]
		{
			\begin{array}{l}
				\cond{\pid p.e}
				{\cond{\pid q.e^\prime}{C_1}{C_2}}
				{\cond{\pid q.e^\prime}{C^\prime_1}{C^\prime_2}}
			\\ \hspace{2em} \swapC \quad \cond{\pid q.e^\prime}
				{\cond{\pid p.e}{C_1}{C^\prime_1}}
				{\cond{\pid p.e}{C_2}{C\prime_2}}
			\end{array}
		}
		{
			\pid p \neq \pid q
		}
		\\[15pt]
		\infer[\did{CS}{RcvCnd}]
		{
			\begin{array}{l}
				\com{k}{\role A}{\pid p[\role B].\{\op{o}_i(x_i); 
				\cond{\pid q.e}{C_{i1}}{C_{i2}}\}_{i\in I}}
			\\ \hspace{1em} \swapC \quad
			\cond{\pid q.e}{
				\com{k}{\role A}{\pid p[\role B].\{\op{o}_i(x_i); C_{i1}\}_{i\in I}}
			}{
				\com{k}{\role A}{\pid p[\role B].\{\op{o}_i(x_i); C_{i2}\}_{i \in I}}
			}
			\end{array}
		}
		{\pid p \neq \pid q}
	\end{array}
$$
\caption{Frontend Choreographies --- swap relation $\swapC$.}
\label{fig:chor_swap}
\end{figure}

\boldpar{Rule $\protect{\did{C}{Eq}}$} closes $\to$ under the congruences
$\equivC$ and $\swapC$. Structural congruence $\equivC$, reported in
\cref{fig:chor_struct_eq}, is the smallest congruence supporting
$\alpha$-conversion, recursion unfolding, and commutativity and associativity
of parallel composition.
The swap relation $\swapC$, reported in \cref{fig:chor_swap}, is the smallest
congruence able to exchange the order of non-interfering concurrent actions.
For example, provided $\pn$ returns the set of process names, Rule
$\did{CS}{EtaEta}$ swaps two communications respectively enacted by completely
disjoint processes.
  
Rule $\did{C}{Eq}$ also enables the reduction of complete communications on
$\prid{com}$ terms---see the last equivalence in \cref{fig:chor_struct_eq},
which unfolds a complete communication term into the two corresponding send
and receive terms.

\boldpar{Rule $\protect{\did{C}{PStart}}$} starts a new session by
synchronising a partial choreography that requests to start a session with
other choreographies that can accept the request. The premise of the rule $\{
\wtil{l.\role B} \} =
\biguplus_i \{\wtil{l_i.\role B_i}\}_i$, where $\biguplus$ indicates the
disjoint union of the list of located roles, requires that in the accepting
choreographies the list of locations and their supported roles match the
corresponding list of the request. The rest of the rule is similar to
$\did{C}{Start}$. Here it is convenient that deployment transitions are
specified by a separate set of rules, since the effect of starting a session
using partial actions is equivalent to that of using a complete start term.
The choreographies accepting the request remain available for subsequent
reuses.

Finally, rules $\did{C}{Cond}$,$\did{C}{Ctx}$, and $\did{C}{Par}$ are standard
and respectively model guarded conditionals, recursion, and parallel
composition.

\begin{example}
\label{ex:async} The interplay between $\swapC$ and rule $\did{C}{Send}$
yields an elegant formalisation of asynchronous behaviour for choreographies
that, differently from previous work~\cite{CM13}, does not require a labelled
transition system and ad-hoc reduction rules. Consider Line 10 in
\cref{example:super}, reported below.
$$
C\ \DEFEQ \ {\color{red}\com{k}{\pid b[\role B]}{\pid c[\role C].\op{ok}()}};\
 {\color{blue}\com{k}{\pid b[\role B]}{\pid q[\role S].\op{ok}()}}
$$

We can reduce $C$ as follows (for brevity, we omit deployments):
$$
\begin{array}{rl@{\qquad}l} 
C \to & 
  {\color{red}\com{k}{{\role B}}{\pid c[\role C].\op{ok}()}};\
    {\color{blue}\com{k}{\pid b[\role B]}{\pid s[\role S].\op{ok}()}} 
      & \text{by } \did{C}{Eq} \text{ with } \mathcal{R}\ =\ \equivC 
      \text{ and } \did{C}{Send}
  \\[5pt]
  \to & 
  {\color{blue}\com{k}{\role B}{\pid s[\role S].\op{ok}()}};\
     {\color{red}\com{k}{\role B}{\pid c[\role C].\op{ok}()}} 
       & \text{by } \did{C}{Eq} \text{ with } \mathcal{R}\ =\ \swapC \text{ and }\did{C}{Send}
\end{array}
$$

In this case, process $\pid s$ may receive its message before process $\pid
c$, due to asynchronous message passing (the sending actions for process $\pid
b$ are non-blocking).
\end{example}

\IFSubFileBiblio

\section{Typing}
\label{sec:typing}

In this section, we define our typing discipline for the Frontend Choreographies. Our
typing checks the behaviour of sessions against protocols, given as Multiparty
Session Types~\cite{HYC08,CDYP2015}.
Interestingly, we retain the same syntax of traditional Multiparty Session
Types yet we ensure that correct initial deployments do not corrupt at runtime
due to inconsistencies on states and message queues.
%
%

In \cref{sec:types_and_type_projection} we present the types that abstract
choreographies, called global types. We define the syntax of global types and
we introduce local types. The latter are abstract descriptions of the behaviour
of single processes, used for type checking. We also formalise how, from a
global type, we obtain a set of related local types by means of a projection
procedure. In \cref{sec:type_checking} we formalise the environment and the
rules of our type discipline. In \cref{sec:runtime_typing}, we consider the
typing of running choreographies. We illustrate why and how a choreography and
its companion deployment can become inconsistent and we present a runtime
typing extension to avoid inconsistencies. Finally, in
\cref{ssub:runtime_examples}, we present two comprehensive examples to clarify
the relationship between types and running choreographies, and in
\cref{sec:types_properites} we formalise the properties guaranteed by our
typing system. 

\subsection{Types and Type Projection}
\label{sec:types_and_type_projection}
\paragraph{Global and Local types}
As in standard Multiparty Session Types, we use \emph{global types} to
represent protocols from a global viewpoint and \emph{local types} to describe
the behaviour of each participant.
Our type system checks that a set of local types, each abstracting the
behaviour of a process in a choreography, coherently follows a global
type.
We report in \cref{fig:chor_typing_syntax} the syntax of global types $G$
and local types $T$.

\begin{figure}
$$
	\begin{array}{l@{\qquad}l}
		\textup{\textit{Global Types}} &
		\begin{array}{lrl@{\hspace{20pt}}l}	
			G & \gram & \gengbranch & \prid{communication}
			\\[3pt]
				&	\Div 	& \gengrecDef \quad \Div \quad \genlrecCall & \prid{recursion}
			\\[3pt] 
				& \Div 	& \gend & \prid{end}
		\end{array}
		\\[35pt]
		\textup{\textit{Local Types}} &
		\begin{array}{lrl@{\hspace{36pt}}l} 
			T & \gram & \lsend{\role A}{\{\op{o}_i(U_i) ; T_i\}_{i}} & \prid{send}
				\\[3pt]
				& \Div & \lrecv{\role A}{\{\op{o}_i(U_i) ; T_i\}_{i}}	& \prid{receive}
				\\[3pt]
				& \Div & \genlrecDef \quad \Div \quad \genlrecCall & \prid{recursion}
				\\[3pt]
				& \Div & \gend & \prid{end}
		\end{array}
		\\[50pt]
		\textup{\textit{Sort Types}} &
		\begin{array}{lrl@{\qquad}l}
			U & \gram & \tunit \quad \Div \quad \tint \quad \Div \quad  \tbool \quad \Div \quad \tstring \quad \Div \quad \ldots
		\end{array}
	\end{array}
$$
\caption{Frontend Choreographies --- Syntax for Global and Local Types.}
\label{fig:chor_typing_syntax}
\end{figure}

A global type $\gengbranch$ abstracts a communication, where $\role A$ can send
to $\role B$ a message on any of the operations $\op{o}_i$ and continue with the
respective continuation $G_i$.
%
%
A carried type $U$ types the value exchanged in the message.
In local types, $!\role A.\{\op{o}_i(U_i) ; T_i\}_{i}$ abstracts the sending of
a message of type $U_i$ to role $\role A$ on one of the operations $\op{o}_i$,
with continuation $T_i$. Dually, $?\role A.\{\op{o}_i(U_i) ; T_i\}_{i}$
abstracts the offering of an input choice among the operations $\op{o}_i$, with
continuation $T_i$. The other terms for recursion and end of types are
standard. As done for FC, also in types we omit curly brackets when outputs and
inputs comprise only one operation.

As an example, we report in \cref{fig:FC_typing_example} two global types, $G_1$ and $G_2$, that abstract
the choreographies presented
in~\cref{fig:FC_example,fig:FC_example_complementary}. In particular, $G_1$
types session $k$, created at locations $(l_\role S, l_\role B)$ ---
\cref{lst:startk} of \cref{fig:FC_example} --- and $G_2$ types session $k'$,
created at location $(l_\role D)$ --- request at \cref{lst:startkp} of
\cref{fig:FC_example}, accept at \cref{lst:startkpc} of
\cref{fig:FC_example_complementary}.
We also write operations followed by empty
parentheses when the type of their message $U$ is $\tunit$.

\begin{figure}
\begin{minipage}{.49\textwidth}
	\(\def\arraystretch{1.2}
	\begin{array}{ll}
	G_1 = & \gcom{\role C}{\role S}{\op{buy}(\tstring)}; 			\\
			& \gcom{\role S}{\role C}{\op{reqPay}(\tint)}; 			\\
			& \gcom{\role C}{\role B}{\op{sendPay}(\tstring)};	\\
			& \gcom{\role B}{\role C}{ \{ 											\\
			& \quad\begin{array}{l}
					\op{ok}(); \gcom{\role B}{\role S}{\op{ok}()},
					\\[{1pt}]
					\op{ko}(); \gcom{\role B}{\role S}{\op{ko}()}
				\end{array} 																			\\
			& \}}; \gend
	\end{array}
	\)
\end{minipage}
\begin{minipage}{.49\textwidth}
	\(\def\arraystretch{1.45}
	\begin{array}{ll}
	G_2 = & \gcom{\role S}{\role D}{\op{quoteShipping}(\tstring)}; \\
			& \gcom{\role D}{\role S}{\op{shippingCost}(\tint)}; 			\\
			& \gcom{\role S}{\role D}{ \{ 														\\
			& \quad\begin{array}{l}
					\op{sendShipping}(),
					\\[{1pt}]
					\op{abortShipping}();
				\end{array} 																						\\
			& \}}; \gend
	\end{array}
	\)
\end{minipage}
\caption{\label{fig:FC_typing_example}
Global types $G_1$ (left) and $G_2$ (right) abstract
the respective choreographies presented
in~\cref{fig:FC_example,fig:FC_example_complementary}.}
\end{figure}

\smallpar{Type Projection.}
To relate global types to the behaviour of processes in choreographies, we
project a global type $G$ onto a set of local types, each corresponding to the
behaviour of a single role. We report in \cref{fig:chor_typing_epp} the
projection of global types, defined following~\cite {MY13}. $\epp{G}{\role A}$
denotes the projection of $G$ onto the role $\role A$.
Intuitively, $\epp{G}{\role A}$ gives an encoding of the local actions expected
by role $\role A$ in the global type $G$.
When projecting a communication, we require the local behaviour of all roles
not involved in it to be merged with the merging operator $\sqcup$. Like
in~\cite{MY13}, $T \sqcup T'$ is isomorphic to $T$ and $T'$ up to branching,
where all branches of $T$ or $T'$ with distinct operations are also
included, formally

$$
T \ \sqcup \ T' =
\begin{cases}
T  & \mbox{if } T = T'
\\
\lrecv{\role A}{}\left\{
\begin{array}{l}
	\{\ \op{o}_h(U_h) ; T_h\ \}_{h \in I \setminus J} \quad \cup \\
	\{\ \op{o}_h(U_h) ; T'_h\ \}_{h \in J \setminus I} \quad \cup \\
	\{\ \op{o}_h(U_h) ; T_h \sqcup T'_h\ \}_{h \in J \cap I}
\end{array}
\right\} &
\begin{array}{l}
\mbox{if } T = \lrecv{\role A}{\{\op{o}_i(U_i) ; T_i\}_{i \in I}} \\ 
\mbox{and } T' = \lrecv{\role A}{\{\op{o}_j(U_j) ; T'_j\}_{j \in J}}	
\end{array}
\end{cases}
$$

\begin{figure}
{$$
\begin{array}{l@{\quad}c@{\quad}l}
	\epp{\gcom{\role B}{\role C}{\{ \op{o}_i(U_i) ; G_i \}_{i} }}{\role A} & = &
	\begin{cases}
		\lsend{\role C}{\{\op{o}_i(U_i) ; \epp{G_i}{\role C}\}_{i}}
			& \mbox{if } \role A = \role B
		\\
		\lrecv{\role B}{\{\op{o}_i(U_i) ; \epp{G_i}{\role C}\}_{i}} 
			& \mbox{if } \role A = \role C
		\\
		\bigsqcup_i \epp{G_i}{\role A} & \mbox{otherwise}
	\end{cases}
	\\[20pt]
	\epp{\gengrecDef}{\role A} & = &
	\begin{cases}
		\m{rec}\;\mathbf{t}. \epp{G}{\role A} & \mbox{if } \role A \in G
		\\
		\gend & \mbox{otherwise}
	\end{cases}
	\\[15pt]
	\epp{\mathbf{t}}{\role A} & = & \mathbf{t}
	\\[15pt]
	\epp{\gend}{\role A} & = & \gend
\end{array}
$$}
\caption{Frontend Choreographies --- Global Type Projection.}
\label{fig:chor_typing_epp}
\end{figure}

\subsection{Type checking}
\label{sec:type_checking}

Now that we defined the relation between global and local types, we can
proceed to present our system that guarantees that sessions in choreographies
follow their types.

\subsubsection{Environments}\label{sub:environments} We define our typing environments $\Gamma,
\Gamma', \ldots$ as reported in \cref{fig:cc_typing_environments}.

\begin{figure}
$$
\begin{array}{c}
	\begin{array}{rl@{\qquad}l}
		\Gamma \gram 
				& \emptyset 														& \prid{empty\ environment} 
				\\[3pt]
	\Div 	& \Gamma, \pair{\pid p.x}{U}						& \prid{variable}
				\\[3pt]
	\Div 	& \Gamma, \pair{X}{\Gamma} 							& \prid{definition}
				\\[3pt]
	\Div 	& \Gamma, \pair{k[\role A]}{T} 					& \prid{local\ session}
				\\[3pt]
	\Div 	& \Gamma, \pair{\pid p}{k[\role A]} 		& \prid{ownership}
				\\[3pt]
	\Div 	& \Gamma, \geninitg											&	\prid{service}	
	\end{array}
\end{array}
$$
\caption{Frontend Choreographies --- Typing Environments.}
\label{fig:cc_typing_environments}
\end{figure}

%

The typing of \emph{variables} denote that a process $\pid p$ has in its
state a variable $x$ of type $U$. We assume that we can write $\Gamma,
\pair{\pid p.x}{U}$ only if either $x$ has not been typed yet in $\Gamma$ or
it is already associated with the same type \(U\) (formally let \(\{u,u'\} \in U\), 
if $u = u'$ then $\Gamma, \pair{\pid p.x}{u}, \pair{\pid p.x}{u'} = \Gamma,
\pair{\pid p.x}{u}$).
We assume a similar convention for all the identifiers in $\Gamma$ except for
\emph{service} typings, whose rule for set inclusion is detailed at the end
of this section.
The typing of \emph{definition} of recursive procedures associates a
procedure identifier $X$ to a typing environment $\Gamma$.
A \emph{local session} typing $\pair{k[\role A]}{T}$ states that role $\role
A$ in session $k$ follows the local type $T$. 
%
An \emph{ownership} typing $\pair{\pid p}{k[\role A]}$ states that process
$\pid p$ owns the role $\role A$ in session $k$. Hence, each process can
participate in multiple sessions, but can play only one role in each session.
A service typing $\geninitg$ types with a global type $G$ all sessions created
by contacting the services at the locations $\til l$. In the typing,
\begin{itemize}
	\item $\role A$ is the role that the active process (the starter) should play;
	\item $\roles B$ are the roles respectively played by each service process at $l$ in
		$\til l$ --- we assume that each $l$ plays a unique role, so the lengths of 
		$\roles B$ and $\til l$ are the same;
	\item $\roles C$ are the roles implemented by the choreography that we are
	typing --- we assume $\roles C \subseteq \roles B$, i.e., that $\roles C$
	contain a subset of the roles in $\roles B$, ordered following the order in
	$\roles B$ (as of \cref{par:list_subset_predicate}).
\end{itemize}

Regarding set inclusion of service typings, when we write $\Gamma = \Gamma',
\pair{\til l}{\serviceTyping{G}{\role A}{\roles B}{\roles C}}$ we assume that:

\begin{itemize}
	\item $\{\role A, \roles B\} = \auxRoles(G)$, where function $\auxRoles$
		returns the set of roles in $G$;

	\item the locations $\til l$ are ordered lexicographically;

	\item the locations in $\til l$ do not appear in any other service
		typing in $\Gamma$;

	\item that either:
		\begin{itemize}
			
			\item $\til l$ does not appear in $\Gamma'$ and the resulting $\Gamma$
			includes it, formally $\til l \not \in \dom(\Gamma')$ and $\Gamma = 
			\Gamma', \ \pair{\til l}{\serviceTyping{G}{\role A}{\roles B}{\roles C}}$;

      \item $\til l$ appears in $\Gamma'$, such that $\Gamma' =
			\Gamma'',\pair{\til l}{\serviceTyping{G}{\role A}{\roles B}{\roles D}}$, and
			$\{\roles C\} \cap \{\roles D\} = \emptyset$, i.e., the roles in $\roles
			C$ do not appear in $\roles D$. The resulting $\Gamma$ includes in the
			service typing  of $\til l$ the merged list of roles in $\roles C$ and
			$\roles D$, following the lexicographic order in $\roles B$. We write the
			merge as $\roles D \bowtie_{\roles B} \roles C$ (see
			\cref{par:ordered_join_operator}) and $\Gamma = \Gamma'', \pair
			{\til
			l}{\serviceTyping{G}{\role A}{\roles B}{\roles D \bowtie_{\roles B}
			\roles C}}$.

		\end{itemize}

\end{itemize}

We underline that the annotation $\roles C$ in service typings plays two
important parts: it enables the composition of choreographies and it ensures that
only one choreography implements a specific role. This is mirrored in the
composition $\Gamma = \Gamma', \pair{\til l}{\serviceTyping{G}{\role A}
{\roles B}{\roles C}}$ where, if $\Gamma'$ already contains the typing for
some roles $\roles D$ in $\til l$, $\Gamma$ will contain the additional roles
defined in $\roles C$ (provided $\roles D$ and $\roles C$ contain distinct roles).


\subsubsection{Typing Judgements and Rules}
\begin{figure}
$$
\begin{array}{c}
\infer[\did{T}{Start}]
{
	\Gamma, \til l:\initg{G}{\role A}{\til{\role B}}{\til{\role B}}
	\hlseq{\genstart ; C}
}
{
	\Gamma,\til l:\initg{G}{\role A}{\til{\role B}}{\til{\role
	B}},
	\init(\ \wtil{\prc rC},k,G\ )
	\hlseq {C}
	&
	\wtil{\prc rC} = \prc pA, \wtil{\prc qB}
	&
	\til{\pid q} \not \in \Gamma
}
\\[12pt]
\infer[\did{T}{Req}]
{
	\Gamma \hlseq{\genreq ; C}
}
{
	\Gamma, \pid p:k[\role A], k[\role A]:\epp{G}{\role A} \hlseq{C}
	&
	\til l:\initg{G}{\role A}{\til{\role B}}{\emptyset} \in \Gamma
}
\\[10pt]
\infer[\did{T}{Acc}]
{
	\Gamma, \til l':\initg{G}{\role A}{\til{\role B}}{\til{\role C}} \hlseq{
		\genaccC ; C	}
}
{
	\til l \subseteq \til l'
	&
	\Gamma, \til l':\initg{G}{\role A}{\til{\role B}}{\emptyset},
	\init(\ \wtil{\pid q[\role C]},k,G\ ) \hlseq{C}
	&
	\pids q \not \in \Gamma
}
\\[12pt]
\infer[\did{T}{Com}]
{
	\Gamma, 
	\pair{k[\role A]}{\genlsendB},
	\pair{k[\role B]}{\genlrecvPrime}
	\hlseq{
		\com{k}{\pid p[\role A].e}{\pid q[\role B].\op{o}_j(x)};C}
}
{
	\Gamma \seq \pair{\pid p}{k[\role A]},\pair{\pid q}{k[\role B]}
	&
	j \in I
	&
	\Gamma \seq \pair{\pid p.e}{U_j}
	&
	\Gamma, \pair{\pid q.x}{U_j}, 
		\pair{k[\role A]}{T_j}, 
		\pair{k[\role B]}{T'_j}
		\hlseq{C}
}
\\[12pt]
\infer[\did{T}{Send}]
{
	\Gamma, 
	\pair{k[\role A]}{\genlsendB}
	\hlseq{\gensendj; C}
}
{
	j \in I
	&
	\Gamma \seq \pair{\pid p}{k[\role A]}
	&
	\Gamma \seq \pair{\pid p.e}{U_j}
	&
	\Gamma, \pair{k[\role A]}{T_j} \hlseq{C }
}
\\[12pt]
\infer[\did{T}{Recv}]
{
	\Gamma, 
		\pair{k[\role B]}{\lrecv{A}{\{\op{o}_i(U_i);T_i\}_{i \in I}}}
	\hlseq{\genbranchIJ}
}
{
	\Gamma \seq \pair{\pid q}{k[\role B]}
	&
	\forall j\in I. \ \Gamma, \pair{\pid q.x_j}{U_j}, 
	\pair{k[\role B]}{T_j}
	\hlseq{C_j}
}
\\[12pt]
\infer[\did{T}{Cond}]
{\Gamma \hlseq{\gencond}}
{
	\Gamma \seq \pair{\pid p.e}{\tbool}
	&
	\Gamma \hlseq{C_1}
	&
	\Gamma \hlseq{C_2}
}
\\[12pt]
\infer[\did{T}{Def}]
{
	\Gamma \hlseq{\recDef{X}{C'}{C}}
}
{
	\Gamma,\pair{X}{\Gamma'} \hlseq{C}
	&
	\Gamma', \pair{X}{\Gamma'} \hlseq C'
	&
	\Gamma' |_{\locs} \subseteq \Gamma
}
\\[10pt]
\infer[\did{T}{Par}]
{\Gamma_1, \Gamma_2 \hlseq {C_1 \pp C_2}}
{
	\Gamma_1 \hlseq {C_1}
	&
	\Gamma_2 \hlseq {C_2}
}
\hfill
\infer[\did{T}{End}]
{\Gamma \hlseq{\inact}}
{
\m{end}(\Gamma)
}
\hfill
\infer[\did{T}{Call}]
{
	\Gamma, \Gamma', \pair{X}{\Gamma''} \hlseq{X}
}
{
	\Gamma'' \subseteq \Gamma'
	&
	\m{end}(\Gamma)
}
\hfill
\end{array}
$$
\caption{Frontend Choreographies --- Typing Rules.}
\label{fig:typing_rules_full}
\end{figure}
%
A judgement $\Gamma \hlseq{C}$ states that the choreography $C$ follows the
specifications given in $\Gamma$.
We comment the typing rules reported in \cref{fig:typing_rules_full}.
%

Rule $\did{T}{Start}$ types a session start. In the first premise, the service
typing $\til l:\initg{G}{\role A} {\til {\role B}}{\til{\role B}}$ checks that
the continuation implements all the roles in protocol $G$. The function $\init$
assembles the typing environment that correctly types --- with the appropriate
ownerships and local typings --- the freshly-started session $k$, given the
global type $G$ and the processes in $\pids p$, each playing its corresponding role
in $\roles B$. Formally,
$$
\init(\ \wtil{\pid p[\role A]},k,G\ ) =
\left\{\ \pid q: k[\role B],\ k[\role B]: \epp{G}{\role B} 
	\;|\; \pid q[\role B] \in \left\{\wtil{\pid p[\role A]}\right\}\ \right\}.
$$
where the type of each process $\pid p \in \pids p$ playing role $\role B \in
\roles B$ is the local type projection $\epp{G}{\role B}$ of the global type
$G$. In $\did{T}{Start}$, we abuse the notation $\pids q \not \in \Gamma$ to
check that all freshly created processes in $\pids q$ do not appear in $\Gamma$
(i.e., there is no variable or ownership typings in $\Gamma$ associated with any
process in $\pids q$).

Rule $\did{T}{Req}$ types $\prid{req}$ terms and is similar to $\did{T}{Start}$,
although it only performs the checks for the process $\pid p$, playing role
$\role A$, that requests the creation of the new session $k$. Dually,
$\did{T}{Acc}$ mirrors Rule $\did{T}{Start}$ and $\til{l} \subseteq \til{l'}$
checks that (following \cref{par:list_subset_predicate}) the list of
locations of the service typing in $\Gamma$ includes the locations in the
$\prid{acc}$ term. In the premise, we type the continuation \(C\) with \(\til
l':\initg{G}{\role A}{\til{\role B}}{\emptyset}\) since \(\prid{acc}\)
terms can only appear at top level in choreographies.

Rule $\did{T}{Com}$ types a complete communication. From left to right the
premises check that:
\begin{enumerate}

	\item the sender $\pid p$ and the receiver $\pid q$ own their respective
	roles in the session;
	\item since $j \in I$:
	\begin{itemize}
		\item operation $\op{o}_j$ can be effectively selected by the sender,
		according to its local type;
	
		\item similarly, $\op{o}_j$ is among the operations offered by the
		receiver, according to its local type;
	
	\end{itemize}

	\item the expression of the sender ($e$) has the type\footnote{The
		judgement $\seq v: U$ reads as ``value $v$ has type $U$''.} $U_j$,
		expected by the protocol;

	\item the resulting environment $\Gamma, \pair{\pid q.x}{U_j}, 
	\pair{k[\role A]}{T_j}, \pair{k[\role B]}{T'_j}$ correctly types the
	continuation $C$, in particular that:

	\begin{itemize}

	\item the receiver $\pid q$ correctly uses the reception variable $x$
	in $C$;

	\item processes $\pid p$ and $\pid q$ proceed according to their local types,
	respectively $T_j$ and $T'_j$.

	\end{itemize}

\end{enumerate}

Rules $\did{T}{Send}$ and $\did{T}{Recv}$ share part of the checks commented
for $\did{T}{Com}$ and judge the respective partial terms $\prid {send}$ and
$\prid{recv}$. Note that, as in standard multiparty session types, the local
typing of the branching process $\pid q$ is contravariant wrt the branches in
the choreography, i.e., the rule $\did{T}{Recv}$ checks that the operations
supported by the typing $o_i, i \in I$ are at least a subset of the actual
operations $o_j, j \in I \cup J$ provided in the $\prid{recv}$ term.
%

Rule $\did{T}{Cond}$ checks that the expression of a conditional has a
compatible type ($\tbool$) and that both branches $C_1$ and $C_2$ are
correctly typed by $\Gamma$.

Rule $\did{T}{Def}$ checks procedure definitions. Here, function $|_{\locs}$
applied to an environment $\Gamma$ returns all service typings in it. In the
rule we write $\Gamma' |_{\locs} \subseteq \Gamma$ to check that the body of
the recursive procedure does not introduce unexpected services, i.e., services
that are not present at top level.

In rule $\did{T}{Par}$ we extend the set inclusion for $\Gamma_1,\Gamma_2$
point-wise to the identifiers in $\Gamma$ to merge typings and to check that
choreographies executing in parallel do not implement overlapping roles at
locations.

In rule $\did{T}{End}$, the predicate $\gend(\Gamma)$ holds if the protocols
for all sessions in $\Gamma$ have terminated (i.e., all local typings have
type $\gend$).

$\did{T}{Call}$ checks a procedure call. The premise $\Gamma'' \subseteq
\Gamma$ checks that procedure $X$ does not introduce unexpected typings (and,
by extension, behaviours) wrt the active sessions contained in $\Gamma'$. The
premise $\gend(\Gamma)$ makes sure that the remaining sessions in the typing
environment have all terminated.

\subsection{Runtime Typing}
\label{sec:runtime_typing}
To prove that well-typed FC programs never go wrong, we need to pay attention
to how their deployments evolve at runtime. For example, in rule
$\did{C}{Send}$, the deployment $\env$ must contain the proper queue where the
sender can deliver its message: a remarkable difference wrt previous works on
choreographies, where such conditions do not exist and choreographies can
always continue execution (see,
e.g.,~\cite{QZCY07,choOrchCorrespondence,CHY12,CM13}). 

To guarantee that well-typed FC programs never go wrong, we must guarantee that
their companion deployments evolve in a consistent way. We address this issue by
extending our typing discipline to check runtime states.

\smallpar{Wrong Deployments.}
We want to rule out ``wrong'' deployments. 
Intuitively, we say that a deployment is wrong wrt a choreography if e.g.,
%
	 processes have undefined variables that are used in the choreography or
	 a message queue does not contain messages as expected by the protocol of the
session in which it is used.
%

Wrong deployments may cause unpredictable executions or faulty behaviours,
e.g., deadlocks. We illustrate the consequences of having wrong deployments
with this simple running choreography:

\[
 \env,\com{k}{\pid p[\role A].y}{\pid q[\role B].\op{o}(x)};\inact
\]

\begin{itemize}[topsep=5pt,itemsep=-2pt,partopsep=1ex,parsep=1ex]
\item \emph{(uninitialised variables)} assume that $\env$ is such that
the state of process $\pid p$ in $\env$, $\env(\pid p)$, does not contain a
value for variable $y$; then the condition $\evalfn(\ y, \env(\pid p)\ )$
given in rule $\did{\env}{Send}$ is undefined and rule $\did{C}{Com}$ cannot
be applied, causing the choreography to get stuck.



\item \emph{(protocol violations)} assume that $\env(\ \chanto{k}{\role
A}{\role B}\ ) = (\op{o}', v)$ where $\op{o} \neq \op{o}'$. Namely, that
\emph{i}) in session $k$ process $\pid q$ (playing role $\role B$) has a
message in its receiving queue from process $\pid p$ (playing role $\role A$)
and \emph{ii}) the operation of the message is $\op{o}'$, different from operation
$\op{o}$ expected in the choreography. If we let the choreography reduce
following the previous point, it ends up deadlocked. After the reduction, the
queue used by $\pid p$ contains in its head the message $(\op{o}', v)$ and
we cannot apply rule $\did{C}{Recv}$, as it expects to find a message for $\op{o}$
at that position.
\end{itemize}

To avoid these outcomes, we extend our type system to prove that, given a
well-typed choreography and a non-wrong companion deployment, our semantics
never produces wrong deployments. Note that this development is transparent to
programmers, since default deployments are trivially never wrong.

\paragraph{Runtime Global Types}
To capture asynchrony and partial runtime
states, we extend the syntax of global types with:

$$
\begin{array}{rl@{\qquad}l}
	G \gram & \ldots
	\\[2pt]
	\Div & \gchoice{\role A}{\role B}{\{\op{o}_i(U_i)\}};G 
		& \prid{global\ choice} 
	\\[2pt]
	\Div & \gbranch{\role A}{\role B}{\{\op{o}_i(U_i);G_i\}_{i \in I}} 
		& \prid{global\ branch}
	\\[2pt]
	\Div & \grecv{\role A}{\role B}{\op{o}(U)};G 
		& \prid{global\ buffer}
\end{array}
$$

Global choice and branch are the equivalent of a complete communication
$\gengcom$ where: $\gchoice{\role A}{\role B}{\{\op{o}_i(U_i)\}};G$ means that
role $\role A$ can choose to send a message to role $\role B$ on operation $
\op{o}_i$ with type $U_i$, proceeding with continuation $G$; while $
\gbranch{\role A}{\role B}{\{\op{o}_i(U_i);G_i\}_{i \in I}}$ means that $\role B$
can receive a message from $\role A$ on any operation $\op{o}_i, i \in I$,
proceeding with the related continuation $G_i$. 

When the choice performed by $\role A$ is applied to the branch controlled by
$\role B$, we obtain term $\grecv{\role A}{\role B}{\op{o}(U)}$, which marks
that $\role A$ has sent the message but $\role B$ still has to consume it.

\paragraph{Semantics of Global Types}
To express the (abstract) execution of protocols, we give a semantics for
global types. Formally, $G \to G'$ is the smallest relation on the
recursion-unfolding of global types satisfying the rules in
\cref{fig:cc_global_types_semantics}.

\begin{figure}
$$
\begin{array}{c}
  \infer[\did{G}{Send}]
  {
    \gchoice{\role A}{\role B}{\{\op{o}_i(U_i)\}; G} 
    \quad \to \quad
    G'
  }
  {
    o \in \bigcup_i\{o_i\}
    &
    G' = \gapply{
      \role A \rangle \role B
    }{
      o
    }{
      G
    }
  }
  \qquad
  \infer[\did{G}{Recv}]
  {
    \grecv{\role A}{\role B}{\op{o}(U)}; G 
    \quad \to \quad
    G
  }
  {\ }
  \\[10pt]
  \infer[\did{G}{Eq}]
  {
    G \longto G'
  }
  {
    \mathcal{R} \in \{\equivG, \swapG\}
    &
    G \ \mathcal{R}\ G_1
    &
    G_1 \to G_1'
    &
    G_1' \ \mathcal{R} \ G'
  }
\end{array}
$$
{\color{gray} \hrule}
\begin{center}
\footnotesize Reduction Rules.
\end{center}
\vspace{.5em}
  $$
  \begin{array}{rcl}
    \gcom{\role A}{\role B}{\{o_i(U_i);G_i\}} & \equivG &
    \gchoice{\role A}{\role B}{\{o_i(U_i)\}};
      \gbranch{\role A}{\role B}{\{o_i(U_i);G_i\}}
    \\[10pt]
    G[\gengrecDef'] & \equivG & 
    G\left[\ G'[\gengrecDef'/\gengrecCall]\ \right]
  \end{array}
  $$
{\color{gray} \hrule}
\begin{center}
\mbox{\footnotesize Structural Congruence.}
\end{center}
\vspace{.5em}
  $$
  \begin{array}c
    \infer[\did{GS}{ChoCho}]
    {
      \gchoice{\role A}{\role B}{\{o_i(U_i)\}};
      \gchoice{\role C}{\role D}{\{o_j(U_j)\}} 
      \swapG
      \gchoice{\role C}{\role D}{\{o_j(U_j)\}};
      \gchoice{\role A}{\role B}{\{o_i(U_i)\}}
    }{
      \role A \neq \role C \vee \role B \neq \role D
    }
    \\[5pt]
    \infer[\did{GS}{BrcBrc}]{
      \gbranch{\role A}{\role B}{\{o_i(U_i);
        \gbranch{\role C}{\role D}{\{o_j(U_j);G_{ij}\}}\}}
      \swapG
      \gbranch{\role C}{\role D}{\{o_j(U_j);
        \gbranch{\role A}{\role B}{\{o_i(U_i);G_{ij}\}}\}}
    }{
      \role A \neq \role C \vee \role B \neq \role D
    }
    \\[5pt]
    \infer[\did{GS}{ChoBrc}]{
      \gchoice{\role A}{\role B}{\{o_i(U_i)\}};
      \gbranch{\role C}{\role D}{\{o_j(U_j);G_j\}}
      \swapG
      \gbranch{\role C}{\role D}{\{o_j(U_j);
        \gchoice{\role A}{\role B}{\{o_i(U_i)\}};G_j\}}
    }{
      \role A \neq \role D
    }
    \\[5pt]
    \infer[\did{GS}{BufBrc}]{
      \grecv{\role A}{\role B}{o(U)}; 
      \gbranch{\role C}{\role D}{\{o_j(U_j);G_j\}}
      \swapG
      \gbranch{\role C}{\role D}{\{o_j(U_j);
        \grecv{\role A}{\role B}{o(U)};G_j\}}
    }{
      \role A \neq \role C \vee \role B \neq \role D
    }
    \\[5pt]
    \infer[\did{GS}{BufBuf}]{
      \grecv{\role A}{\role B}{o(U)};
      \grecv{\role C}{\role D}{o'(U')}
      \swapG
      \grecv{\role C}{\role D}{o'(U')};
      \grecv{\role A}{\role B}{o(U)}
    }{
      \role A \neq \role C \vee \role B \neq \role D
    }
    \\[5pt]
    \infer[\did{GS}{ChoBuf}]{
      \gchoice{\role A}{\role B}{\{o_i(U_i)\}};
      \grecv{\role C}{\role D}{o(U)}
      \swapG
      \grecv{\role C}{\role D}{o(U)};
      \gchoice{\role A}{\role B}{\{o_i(U_i)\}}
    }{
      \role A \neq \role D
    }
  \end{array}
$$
{\color{gray} \hrule}
\noindent\begin{center}
{\footnotesize Swap Relation.}
\end{center}
\vspace{.5em}
  $$
  \begin{array}c
  \begin{array}{ccll}
    \gapply{%
      \role A\rangle\role B}
      {o_j}
      {\gbranch{\role A}{\role B}{\{o_i(U_i);G_i\}_{i\in I}}} & = &
        \grecv{\role A}{\role B}{o_j(U_j)};G_j & \mbox{ if } j \in I
  \\[5pt]
  \gapply{%
    \role A\rangle\role B}
    {o_j}
    {\gbranch{\role C}{\role D}{\{o_i(U_i);G_i\}_{i\in I}}} & = &
     \gbranch{\role C}{\role D}{\{o_i(U_i);
      \gapply{\role A\rangle \role B}{o_j}{G_i}\}}
      & \mbox{ if } \role A \neq \role C \vee \role B \neq \role D
  \end{array}
  \\[25pt]
  \gapplygen{ \grecv{\role C}{\role D}{o(U)};G } = 
    \grecv{\role C}{\role D}{o(U)};\gapplygen G
  \qquad
  \gapplygen{ \gchoice{\role C}{\role D}{\{o_i(U_i)\}};G } = 
    \gchoice{\role C}{\role D}{\{o_i(U_i)\}};\gapplygen G  
  \\[5pt]
  \gapplygen{\gengrecDef} = \gengrecDef 
  \hfill
  \gapplygen{\gengrecCall} = \gengrecCall
  \hfill
  \gapplygen{\gend} = \gend
  \end{array}
  $$
{\color{gray} \hrule}
\noindent\begin{center}
{\footnotesize Application Function.}
\end{center}
\caption{Global types --- Semantics.}
\label{fig:cc_global_types_semantics}
\end{figure}

Rule $\did{G}{Send}$ allows the sending of a message from a $
\prid{global\ choice}$. The continuation $G'$ is obtained from the application of the sending
to the corresponding $\prid{global\ branch}$, with function $\gapply{\role A
\rangle \role B}{o_i}{G}$ that transforms the related branch in $G$ into a 
$\prid{global\ buffer}$ on the selected operation $\op{o_i}$, followed by the
respective continuation $G_i$.

The actual reception of the message is executed in rule $\did{G}{Recv}$. In
$\did{G}{Eq}$ we model the splitting of complete communications and recursion
unfolding with the structural equivalence $\equivG$, the smallest congruence
defined by the rules in \cref{fig:cc_global_types_semantics}. To capture the
semantics of asynchronous message delivery, we define the swap relation
$\swapG$ as the smallest congruence defined by the rules in
\cref{fig:cc_global_types_semantics}. Both congruences are similar to what
presented for choreographies in \cref{sec:fc_semantics}. Note that
rules $\did{GS}{ChoBrc}$ and $\did{GS}{ChoBuf}$ enable the swapping of
choice terms with receptions, as long as the swap preserves the causal
consistency between operations (i.e., we do not swap a sending that is causally
dependent from a reception on the same role).

\paragraph{Runtime Type checking and Typing Rules}
We extend the typing rules given in the previous section to check runtime
terms. The extension consists in
\emph{i}) new terms for $\Gamma$, 
and \emph{ii}) the introduction of rule $\did{T}{\env C}$ to type runtime
choreographies.
We extend the grammar of typing environments with
$$
\begin{array}{rl@{\qquad}l}
\Gamma \gram & \ldots 														 \\[2pt]
\Div & \Gamma, \pid p@l & \prid{location}					 \\[2pt]
\Div & \Gamma,\chanto{k}{\role A}{\role B}: T 
												& \prid{buffer}	
\end{array}
$$
where $\Gamma,\pid p@l$ states that process $\pid p$ runs at location $l$ and a
\emph{buffer} typing $\chanto{k}{\role A}{\role B}: T$ types the messages in
the queue where the process implementing role $\role B$ in session $k$
receives messages from role $\role A$. We extend to buffer typings the
assumption for set inclusion stated for standard elements in $\Gamma$. For
location typings, we assume that we can write $\Gamma,\pid p@l$ only if $\pid
p@l \not \in \Gamma$. This formalises the requirement that a process can
appear only in one choreography (e.g., given the choreography $C = C_1 \pp
C_2$ process $\pid p \in \pn(C)$ appears either in $C_1$ or in $C_2$) and
that it is associated only to one location.


\begin{figure}
$$
\begin{array}c
	\begin{array}{lcl}
	\epp{\gcom{\role C}{\role D}{\{\op{o}_i(U_i);G_i\}}}{\role B}^\role A & = &
	\begin {cases}
		\gend & \mbox{if } \role C = \role A\ \wedge\ \role D = \role B
		\\
		\bigsqcup_i\epp{G_i}{\role B}^\role A & \mbox{otherwise}
	\end{cases}
		\\[15pt]
	\epp{\grecv{\role C}{\role D}{\op{o}(U);G}}{\role B}^\role A & = &
	\begin{cases}
		\lrecv{\role A}{\op{o}(U)};\epp{G}{\role B}^\role A 
			& \mbox{if } \role C = \role A \wedge\ \role D = \role B
		\\
		\epp{G}{\role B}^\role A & \mbox{otherwise}
	\end{cases}
	\\[15pt]
	\epp{\mathbf{t}}{\role B}^{\role A} =
	\epp{\gend}{\role B}^\role A =
	\epp{\gengrecDef}{\role B}^\role A & = & \gend
	\end{array}
\end{array}
$$
\caption{Frontend Choreographies --- Buffer Type Projection.}
\label{fig:chor_typing_epp_buffer}
\end{figure}

To relate the typings of queues to the buffer types expected by the protocol
of sessions, we define the \emph{buffer type projection} $\epp {G} {\role
B}^\role A$, which follows the rules in \cref{fig:chor_typing_epp_buffer} and
returns the expected buffer type of role $\role B$ from $\role A$ in $G$.
$\epp{G}{\role B}^\role A$ extracts from $G$ the partial receptions of the
form $\grecv{\role A} {\role B}{\op{o}(U)}$, translating them to local types
of the form $\lrecv{\role A}{\op{o}(U)}$.
Below, we report the rule that extends global type projection for global
buffers.
$$
\epp{\grecv{\role A}{\role B}{o(U)};G}{\role C} = 
\begin{cases}
	\lrecv{\role A}{o(U) ; \epp{G}{\role C}} & \mbox{if } \role C = \role B
	\\
	\epp{G}{\role C} & \mbox{otherwise}
\end{cases}
$$
Note that we do not need to extend the projection to $\prid{global\ choice}$
and $\prid{global\ branch}$. Indeed, in our setting we consider only running
global types that are evolution of a global type, hence global choices and
branches are always balanced. Given a running global type $G$, we can always
obtain an equivalent ($\swapG$, $\equivG$)  global type $G'$ which is absent
from $\prid{global\ choice}$ and $\prid{global\ branch}$ terms. We call a
running global type \emph{canonic} if it contains no $\prid{global\ choice}$
and $\prid{global\ branch}$ terms. When writing projections of global types we
assume $G$ to be in canonic form.

Finally, we extend our typing discipline with a new rule $\did{T}{\env C}$
that checks for coherence among types, choreographies, and deployments. To
define $\did{T}{\env C}$, we formalise a predicate, called
\emph{partial coherence}\footnote{Partial because it accounts for missing
typings of roles implemented by external partial choreographies.} and denoted
$\pco (\Gamma)$, that holds if and only if, for all sessions $k$, the local
and buffer typings of $k$ follow (are projections of) the same global type $G$.

%

\begin{definition}[Partial Coherence]\label{def:partial_coherence} We write
$\pco(\Gamma)$ when, for all sessions $k$ in $\Gamma$, there exists a global
type $G$ such that
$$
\forall\; \pair{k[\role B]}{T} \in \Gamma,\ T = \epp{G}{\role B}\
\quad \wedge \quad \forall\; \role A \in \auxRoles(G) \setminus \{\role B\},\
	\Gamma \seq \pair{\chanto{k}{\role A}{\role B}}{\epp{G}{\role B}^\role A}
$$
\end{definition} 

Rule $\did{T}{\env C}$ is defined as:

$$
\infer[\did{T}{\env C}]
{
	\Gamma \hlseq{\env,C}
}
{
	\pco(\Gamma) 
		&
	\Gamma \hlseq \env
		&
	\Gamma \hlseq C
}
$$

where a judgement $\Gamma \hlseq{\env,C}$ states that $C$ and $\env$ are
coherent according to $\Gamma$ and all sessions in $\Gamma$ are coherent.
$\Gamma$ is an abstraction between $\env$ and $C$ and guarantees that $\env$ cannot go wrong. Formally

\begin{definition}[Deployment Judgements]\label{def:deployment_judgements}
$$ \Gamma \hlseq{\env} \iff \begin{cases}
			(1) \ \forall\ \pair{\pid p.x}{U} \in \Gamma,\; \env( \pid p ).x : U
			\\
			(2) \ \forall\ \chanto{k}{\role A}{\role B}: T \in \Gamma \wedge
			\env(\chanto{k}{\role A}{\role B}) = \til m, \ \btypeop(\role A, \til m
			) = T
	\end{cases}
$$
\end{definition}
We comment the checks performed by $\Gamma \hlseq \env$:
%
%
$(1)$ checks that, for each typing $\pair{\pid p.x}{U}$ in $\Gamma$, $\env$
associates $x$, in the state of process $\pid p$, to a value of type
$U$;
%
$(2)$ uses buffer types to check that the typing of a message queue in $\Gamma$
is correct wrt to the actual sequence of messages stored by that queue in
$\env$. We extract the type of a queue $\til m$, i.e., the sequence of message
receptions from a role $\role A$, with function $\btypeop(\role A,
\til m)$. Formally,

\begin{definition}[Buffer Type Extraction]\label{def:bte}
Let $\seq \pair{v_i}{U_i}$, $i \in [1,n]$ and $\til m =
(\op{o}_1,v_1)::\cdots::(\op{o}_n,v_n)$ then $\btypeop( \ \role A, \til m \ ) \ = \
\lrecv{\role A}{\op{o}_1( U_1 )}; \ \cdots; \ \lrecv{\role A}{\op{o}_n( U_n )}$.
\end{definition}

\subsection{Runtime Examples, Typing and Reductions} 
\label{ssub:runtime_examples}

In this section, we present two running examples that illustrate the
relationship between global types and choreographies. First we report
a basic case where a session starts and two processes exchange a message. Then
we consider a started session and comment the asynchronous delivery of messages.

\begin{table}
\resizebox{\columnwidth}{!}{%
\(%
\begin{array}{c|c|c|c}
& \mbox{Typing Environment} & \mbox{Choreography} & \mbox{Deployment}
\\[.5em] \hline \hline \rule{0pt}{5.5em}
\thiscounter
  &
\begin{array}{lll}
  G \hspace{.8em} & = &
  \gcom{\role A}{\role B}{\op{pass}( \tstring )};
  \\ &&
  \gcom{\role B}{\role C}{\op{fwd}( \tstring )};
  \\ &&
  \gend
\\[.5em]
k[\role A] & = & \lsend{\role B}{\op{pass}(\tstring)};\gend
\\
k[\role B] & = & \lrecv{\role A}{\op{pass}(\tstring)};\\&&
                 \lsend{\role C}{\op{fwd}(\tstring)};\gend
\\
k[\role C] & = & \lrecv{\role B}{\op{fwd}(\tstring)};\gend
\end{array}
&
\begin{array}{ll}
  C' = &
  \com{k}{\pid a[\role A].\texttt{"ok"}}{\pid b[\role B].\op{first}(\ x\ )};
  \\ &
  \com{k}{\pid b[\role B].x}{\pid c[\role C].\op{second}(\ x\ )}
\end{array}
&
\env'
\\[4em] \hline
\rule{0pt}{1.5em}
&
G \to G' \mbox{ by } \did{G}{Eq}, \did{G}{Send} & 
C' \to C'' \mbox{ by } \did{C}{Eq}, \did{C}{Send} & 
\hspace{-.5em}\begin{array}c
\delta = \com{k}{\pid a[\role A].\texttt{"ok"}}{\role B.\op{pass}}
\\
\renv{\env'}{\delta}{\env''} \mbox{ by } \did{\env}{Send}
\end{array}
\\[.5em] \hline \rule{0pt}{5.5em}
  \thiscounter
&
\begin{array}{lll}
  G' \hspace{.8em} & = &
  \hl{\grecv{\role A}{\role B}{\op{pass}(\tstring)}};
  \\ &&
  \gcom{\role B}{\role C}{\op{fwd}( \tstring )};
  \\ &&
  \gend
\\[.5em]
\hl{k[\role A]} & = & \hl{\gend}
\\
k[\role B] & = & \lrecv{\role A}{\op{pass}(\tstring)};\\&&
                 \lsend{\role C}{\op{fwd}(\tstring)};\gend
\\
k[\role C] & = & \lrecv{\role B}{\op{fwd}(\tstring)};\gend
\\
\hl{\chanto{k}{\role A}{\role B}} & = & \hl{\lrecv{\role A}{\op{pass}
(\tstring)}}
\end{array}
&
\begin{array}{ll}
  C'' = &
  \hl{\com{k}{\role A}{\pid b[\role B].\op{pass}(\ x\ )}};
  \\ &
  \com{k}{\pid b[\role B].x}{\pid c[\role C].\op{fwd}(\ x\ )}
\end{array}
&
\hl{\env''(\chanto{k}{\role A}{\role B})} = \hl{(\op{pass}, \texttt{"ok"})}
\\[5em] \hline \rule{0pt}{1.5em}
&
G' \to G'' \mbox{ by } \did{G}{Recv} & 
C'' \to C''' \mbox{ by } \did{C}{Recv} & 
\hspace{-.5em}\begin{array}c
\delta' = \com{k}{\role A}{\pid{b}[\role B].\op{pass}(x)}
\\
\renv{\env''}{\delta'}{\env'''} \mbox{ by } \did{\env}{Recv} 
\end{array}
\\[.5em] \hline \rule{0pt}{4em}
  \thiscounter
&
\begin{array}{lll}
  G'' \hspace{.8em} & = &
  \hl{\gcom{\role B}{\role C}{\op{fwd}( \tstring )}};
  \\ &&
  \gend
\\[.5em]
k[\role A] & = & \gend
\\
\hl{k[\role B]} & = & \hl{\lsend{\role C}{\op{fwd}(\tstring)};\gend}
\\
k[\role C] & = & \lrecv{\role B}{\op{fwd}(\tstring)};\gend
\end{array}
&
\begin{array}{ll}
  C''' = &
  \hl{\com{k}{\pid b[\role B].x}{\pid c[\role C].\op{fwd}(\ x\ )}}
\end{array}
&
\hl{\env'''(\pid b).x} = \hl{\texttt{"ok"}}
\end{array}%
\)%
}
\caption{
\label{tab:example_reductions}
Example of message delivery on elements of interest of
choreography $C'$ (second column), its companion deployment $\env'$ (third
column), and their typing environment (first column). 
}
\end{table}
\begin{example}[Start and Message Delivery]\label{es:start_msg_delivery}
We consider a running choreography $C,\env$ and a global type $G$ such
that $\env$ is a default deployment (cf. \cref{def:default_deployment}) and
$$
\begin{array}{c@{\qquad}c}
	\begin{array}{ll}
	C = &
	\start{k}{\pid a[\role A]}{
		l_{\role B}.\pid b[\role B],l_{\role C}.\pid c[\role C]};
	\\ &
	\com{k}{\pid a[\role A].\texttt{"ok"}}{\pid b[\role B].\op{pass}(\ x\ )};
	\\ &
	\com{k}{\pid b[\role B].x}{\pid c[\role C].\op{fwd}(\ x\ )}
	\end{array}
	&
	\begin{array}{ll}
	G = &
	\gcom{\role A}{\role B}{\op{pass}( \tstring )};
	\\ &
	\gcom{\role B}{\role C}{\op{fwd}( \tstring )};
	\\ &
	\gend
	\end{array}
\end{array}
$$
The global type $G$ is used in the typing environment $\Gamma$ to check
$C,\env$, formally the service typing 
$
	\pair{l_{\role B},l_{\role C}}{
		\serviceTyping{G}{\ \role A\ }{\ \role B,\role C\ }{\ \role B,\role C\ }
	}
$ belongs to $\Gamma$ and $\Gamma \hlseq{C,\env}$.

Now, we let $\env,C$ reduce to $\env',C'$ following rules $\did{C}{Start}$ and
$\did{\env}{Start}$ so that $\env$ contains the data and queues needed to
support interactions on session $k$. Finally, we report in~\cref{tab:example_reductions}:

\begin{itemize}
	\item left column, the main elements in the typing environment
	$\Gamma$, i.e., the evolution of the type $G$. To show how partial coherence
	(\cref{def:partial_coherence}) holds, we report also the local and buffer
	types of $\role A$, $\role B$, and $\role C$ projected from $G$ following
	global type projection $\epp{G}{\role A}$ for local types (see
	\cref{fig:chor_typing_epp}) and buffer type projection $\epp{G}{\role
	A}^{\role B}$ (see \cref{fig:chor_typing_epp_buffer}) for buffer types. For
	brevity, we omit to report empty buffer types such as $\chanto{k}{\role
	A}{\role B} =	\gend$; 
	\item middle column, the reduction of choreography $C$;
	\item right column, the main changes in $\env$.
\end{itemize}
To ease the reading of the example, we highlight in $\hl{\
\mbox{\rule{0pt}{.7em}grey}\ }$ the elements that have been changed by the
reduction. To keep our example brief, we only report the reduction (sending
and reception) of the first interaction in $C$, namely $\com{k}{\pid a[\role
A].\texttt{"ok"}} {\pid b[\role B].\op{pass}(\ x\ )}$.

In~\cref{tab:example_reductions}, row \showcounter{1} shows on the left column
the original type $G$ and the global type projection onto the local types of
roles $\role A$, $\role B$, and $\role C$; in the next two columns we reported
for completeness the reductions $C'$ and $\env'$.
Next, we let the running choreography reduce, applying rules $\did{C}{Eq}$, $
\did{C}{Send}$, and $\did{\env}{Send}$ to let process $\pid a$ deliver its
message in the queue $\chanto{k}{\role A}{\role B}$ of process $\pid b$. We also
let $G$ reduce to $G'$ with rule $\did{G}{Send}$.
In row \showcounter{2} we report the result of the reductions. In the left
column, $G'$ indicates that role $\role A$ has sent a message to $\role B$,
which should consume it in the next step. This is also mirrored by the buffer
projection, where the buffer typing $\chanto{k}{\role A}{\role B}$ is
$\lrecv{\role A}{\op{pass}}$. The deployment $\env''$ contains the actual
message sent by $\pid a$ in the queue owned by $\pid b$. The reduced
choreography is still well-typed as, applying function $\btypeop( \ \role A,
\env''( \chanto{k}{\role A}{\role B}) \ )$ on the interested queue, we obtain
the same local type of the buffer typing $\chanto{k}{\role A}{\role B}$.
Finally, we let the running choreography and the global type reduce again,
allowing process $\pid b$ to consume the message. We show the result of the
reductions in row \showcounter{3}, where in deployment $\env'''$ we can find
that the value of the message has been assigned to the receiving variable $x$
of $\pid b$.
\end{example}

\setcounter{mycounter}{0}
\begin{table}
\resizebox{\columnwidth}{!}{%
\(%
\begin{array}{c|c|c|c}
& \mbox{Typing Environment} & \mbox{Choreography} & \mbox{Deployment}
  \\[.5em] \hline\hline \rule{0pt}{5em}
\thiscounter
  &
\begin{array}{lll}
  G \hspace{.8em} & = &
  \gcom{\role A}{\role B}{\op{first}};
  \\ &&
  \gcom{\role A}{\role B}{\op{second}};
  \\ &&
  \gend
\\[.5em]
k[\role A] & = & \lsend{\role B}{\op{first}};\\&&
                 \lsend{\role B}{\op{second}};
                 \gend
\\
k[\role B] & = & \lrecv{\role A}{\op{first}};\\&&
                 \lrecv{\role A}{\op{second}};\gend
\end{array}
&
\begin{array}{ll}
  C = &
  \com{k}{\pid a[\role A]}{\pid b[\role B].\op{first}};
  \\ &
  \com{k}{\pid a[\role A]}{\pid b[\role B].\op{second}}
\end{array}
&
\env
\\[4em] \hline
\rule{0pt}{4em}
\showcounter{A}
&
\!\begin{array}l
G \to G' \mbox{ by } \did{G}{Eq}, \did{G}{Send} \mbox{i.e.,}
\\
\!\begin{array}{lcl}
G \equivG G_1 & = & \hl{\gchoice{\role A}{\role B}{\op{first}}};\\&&
          \hl{\gbranch{\role A}{\role B}{\op{first}}};\\&&
          \gcom{\role A}{\role B}{\op{second}};\\&& 
          \gend
\end{array}
\\
G_1 \to G'_1 \mbox{ and } G'_1 \equivG G'
\end{array}
& 
C \to C' \mbox{ by } \did{C}{Eq}, \did{C}{Send} & 
\hspace{-.5em}\begin{array}c
\delta = \com{k}{\pid a[\role A]}{\role B.\op{first}}
\\
\renv{\env}{\delta}{\env'} \mbox{ by } \did{\env}{Send}
\end{array}
\\[.5em] \hline \rule{0pt}{5.5em}
  \thiscounter
&
\begin{array}{lll}
  G' \hspace{.8em} & = &
  \hl{\grecv{\role A}{\role B}{\op{first}}};
  \\ &&
  \hl{\gchoice{\role A}{\role B}{\op{second}}}; \\ &&
  \hl{\gbranch{\role A}{\role B}{\op{second}}};
  \\ &&
  \gend
\\[.5em]
\hl{k[\role A]} & = & \hl{\lsend{\role B}{\op{second}};\gend}
\\
k[\role B] & = & \lrecv{\role A}{\op{first}};\\&&
                 \lrecv{\role A}{\op{second}};\gend
\\
\hl{\chanto{k}{\role A}{\role B}} & = & 
  \hl{\lrecv{\role A}{\op{first}()}};\gend
\end{array}
&
\begin{array}{ll}
  C' = &
  \hl{\com{k}{\role A}{\pid b[\role B].\op{first}}};
  \\ &
  \hl{\com{k}{\pid a[\role A]}{\role B.\op{second}}}; \\ &
  \hl{\com{k}{\role A}{\pid b[\role B].\op{second}}}
\end{array}
&
\hl{\env'(\chanto{k}{\role A}{\role B})} = \hl{(\op{first}, \_)}
\\[5em] \hline \rule{0pt}{4em}
\showcounter{B}
&
\!\begin{array}l
G' \to G'' \mbox{ by } \did{G}{Eq}, \did{G}{Send} \mbox{i.e.,}
\\
\!\begin{array}{lcl}
G' \swapG G_2 & = & 
          \hl{\gchoice{\role A}{\role B}{\op{second}}};\\&&
          \hl{\grecv{\role A}{\role B}{\op{first}}};\\&&
          \gbranch{\role A}{\role B}{\op{second}};\\&&
          \gend
\end{array}
\\
G_2 \to G'_2 \mbox{ and } G'_2 \swapG G''  
\end{array}
& 
C' \to C'' \mbox{ by } \did{C}{Eq}, \did{C}{Send} & 
\hspace{-.5em}\begin{array}c
\delta' = \com{k}{\role A}{\pid{b}[\role B].\op{second}}
\\
\renv{\env'}{\delta'}{\env''} \mbox{ by } \did{\env}{Send} 
\end{array}
\\[.5em] \hline \rule{0pt}{6.5em}
  \thiscounter
&
\begin{array}{lll}
  G'' \hspace{.8em} & = &
  \grecv{\role A}{\role B}{\op{first}}; \\ &&
  \hl{\grecv{\role A}{\role B}{\op{second}}}; \\ &&
  \gend
\\[.5em]
k[\role A] & = & \gend
\\
\hl{k[\role B]} & = & 
  \lrecv{\role A}{\op{first}};\\&&
  \hl{\lrecv{\role A}{\op{second}}};\\&&
  \gend
\\
\hl{\chanto{k}{\role A}{\role B}} & = & 
  \lrecv{\role A}{\op{first}}\\&&
  \hl{\lrecv{\role A}{\op{second}}}\\&&
  \gend
\end{array}
&
\begin{array}{ll}
  C'' = &
  \com{k}{\role A}{\pid b[\role B].\op{first}};
  \\ &
  \hl{\com{k}{\role A}{\pid b[\role B].\op{second}}}
\end{array}
&
\!\begin{array}{lll}
\hl{\env''(\chanto{k}{\role A}{\role B})} & = &
  (\op{first}, \_):: \\&& 
  \hl{(\op{second}, \_)}  
\end{array}
\end{array}%
\)%
}
\caption{
\label{tab:example_reductions_2}
Example of asynchrony and effects on elements of interest of choreography $C$
(second column), its companion deployment $\env$ (third column), and their
typing environment (first column). }
\end{table}
\begin{example}[Asynchronous Message Delivery]\label{eg:async_msg_delivery}
In this example, we consider a well-typed running choreography $\Gamma
\seq{\env,C}$ where $C$ and its correspondent reduced global type
$G$ are:
$$
\begin{array}{c@{\qquad}c}
	\begin{array}{ll}
	C = &
	\com{k}{\pid a[\role A]}{\pid b[\role B].\op{first}()};
	\\ &
	\com{k}{\pid a[\role A]}{\pid b[\role B].\op{second}()}
	\end{array}
	&
	\begin{array}{ll}
	G = &
	\gcom{\role A}{\role B}{\op{first}( \tunit )};
	\\ &
	\gcom{\role A}{\role B}{\op{second}( \tunit )};
	\\ &
	\gend
	\end{array}
\end{array}
$$
We keep the same conventions on notation defined in the previous example with
the addition of omitting round parenthesis for void values. We report in
\cref{tab:example_reductions_2} a possible sequence of reduction. Following the
previous example, we use row \showcounter{1} to summarise the status of (from
left to right) the typing environment $\Gamma$, the choreography $C$, and its
companion deployment $\env$. 

In row \showcounter{A} we report the main elements involved in the reduction.
In the left-most cell of the raw, the global type $G_1$ is structurally
equivalent ($\equivG$) to $G$ and that appears in rule $\did{G}{Eq}$ to split
the complete communication $\gcom{\role A}{\role B}{\op{first}()}$ into its
equivalent $\gchoice{\role A}{\role B}{\op{first}()};\gbranch{\role A}{\role
B}{\op{first}()}$. Then $G_1$ reduces to $G_1'$ with rule $\did{G}{Send}$
and, as of rule $\did{G}{Eq}$, we take $G'$ as structurally equivalent to
$G_1'$, as shown in row \showcounter{2}, $G'$ splits the complete
communication $\gcom{\role A}{\role B}{\op{second}()}$ into its equivalent
$\gchoice{\role A}{\role B}{\op{second}()};\gbranch{\role A}{\role
B}{\op{second}()}$. The reduction of $C$ mirrors that of $G$: it splits the
complete communication on operation $\op{first}$, consumes the sending, and
finally splits the other complete communication on operation $\op{second}$,
resulting in $C'$ (row \showcounter{2}). The sending is applied on $\env$
which contains the related message in queue $\chanto{k}{\role A}{\role B}$ in
its reductum $\env'$.

Then, in row \showcounter{B} we allow the delivery of operation $\op{second}$.
This illustrates how asynchrony works at both levels of global types and
choreographies. As before, we start from the left-most cell in the row. First we
consider $G_2$, which is swap-equivalent to $G'$, after applying to it rule
$\did{GS}{ChoBuf}$. This brings on top the $\prid{global\ choice}$ on operation
$\op{second}$. Then $G_2$ reduces to $G_2'$ with rule $\did{G}{Send}$ and, as of
rule $\did{G}{Eq}$, we take $G'' = G_1'$. The reduction on $C',D'$ is similar to
that of $G'$.
\end{example}

\subsection{Properties}
\label{sec:types_properites}

We close this section with the main guarantees of our type system.

First, our semantics preserves well-typedness:
\begin{theorem}[Subject Reduction\label{thm:paper_subj_red}]
$\Gamma \hlseq{\env,C}$ and 
$\env,C \to \env',C'$ imply
$\Gamma' \hlseq{\env',C'}$ for some $\Gamma'$.
\end{theorem}

We report in \cref{sec:proof_of_typing} the proof of
\cref{thm:paper_subj_red}.

We now relate $\Gamma$ and $\Gamma'$ to prove that the behaviours of sessions in
a well-typed choreography follow their respective types. We denote $\epp{G}{k}$
the projection of a global type $G$ for a session $k$ and let $\epp{G}{k}$ be
the set of local and buffer typings as obtained by the projection of $G$ on each
of its roles:

\begin{definition}[Global Type Projection]
\vspace{-.5em}
$$
	\epp{G}{k} = \{\
		\pair{k[\role A]}{\epp{G}{\role A}} \ | \ \role A \in \auxRoles(G)
	\ \},
	\{\
		\pair{\chanto{k}{\role A}{\role B}}{\epp{G}{\role B}^{\role A}}
		\ | \ \role A \in \auxRoles(G), \role B \in \auxRoles(G) \setminus \{\role A\}
	\ \}
$$
\end{definition}

We say that a reduction is ``at session $k$'' if it is obtained by consuming a
communication term for session $k$ (as in~\cite{HYC08}), and we write
$k\not\in \Gamma$ when $k$ does not appear in any local typing in $\Gamma$. Then we
have:
\begin{theorem}[Session Fidelity\label{thm:session_fidelity}]
Let $\Gamma,\Gamma_k \hlseq{\env,C}$, $k \not \in \Gamma$. Then, $\env,C \to
\env',C'$ with a redex at session $k$ implies that, for some $G$ and $\Gamma'$,
$k \not \in \Gamma'$, (i) $\Gamma_k \subseteq \epp{G}{k}$, (ii) $G \to G'$,
(iii) $\Gamma_k' \subseteq \epp{G'}{k}$, and (iv) $\Gamma',\Gamma_k' \hlseq{\env',C'}$.
\end{theorem}
\cref{thm:session_fidelity} states that all communications on sessions
follow the expected protocols ($\Gamma'$ may differ from $\Gamma$ for the
instantiation of a new variable). The proof of \cref{thm:session_fidelity} is
reported in \cref{sec:proof_of_typing}.

Finally, we present the definition of the coherence predicate $\cofn$:

\begin{definition}[Coherence\label{def:coherence}]
$\cofn(\Gamma)$ holds iff $\forall\ k \in \Gamma\ ,\ \exists\ G$ s.t. 
\begin{itemize}
	\item $\til l:\initg{G}{\role A}{\til {\role B}}{\til{\role C}} \in \Gamma \
	 	 \wedge\ \roles C = \roles B$ and
 	 \item $\forall\ \role A \in \auxRoles(G)$,
	 	 	$\pair{k[\role A]}{T} \in \Gamma 
	 	 	\quad \wedge \quad 
	 	 	T = \epp{G}{\role A} 
	 	 	\quad \wedge \quad
	 	 	\forall\ \role B \in \auxRoles(G)\setminus\{\role A\},\ 
	 	 		\Gamma \seq \chanto{k}{\role B}{\role A} : \epp{G}{\role A}^{\role
	 	 		B}$
\end{itemize}
\end{definition}
Coherence extends partial coherence to check that
\emph{i}) all needed services to start new sessions are present and
\emph{ii}) all the roles in every open session are correctly implemented by
some processes. 

Coherent and well-typed systems are deadlock-free, as stated by
\cref{thm:deadlock-freedom}.
\begin{theorem}[Deadlock-freedom\label{thm:deadlock-freedom}]
$\Gamma \hlseq{\env,C}$ and $\cofn(\Gamma)$ imply that either (i) $C \equivC
\inact$ or (ii) there exist $\env'$ and 
$C'$ such that $\env,C \to \env',C'$.
\end{theorem}

We report the proof of \cref{thm:deadlock-freedom} in
\cref{sec:proof_of_deadlock_freedom}.

\IFSubFileBiblio

\section{Backend Choreographies}
\label{sec:implementation_model}

We now present \emph{Backend Choreographies} (BC).
The syntax of programs in BC is the same as that of FC\@. Also the two
semantics are close, except that FC models communication over named
channels while BC formalises message exchange based on message correlation, as
found in Service-Oriented Computing (SOC)~\cite{bpel}.
Formally, thanks to the separation between choreographic programs and
deployments presented in FC, we can let FC and BC share a large fragment of
semantics rules, while the significant differences between the two
semantics of message exchange---name-based for FC, correlation-based for
BC---are isolated within their specific deployments and deployment
transitions.

The structure and semantics of the Backend deployments $\aenv$ is one of
our major contributions: it formalises, at the level of choreographies, how to
implement sessions using the communication mechanism of message correlation
typical of SOC systems.

In the following, we first informally introduce correlation-based message
exchange, then we formalise data and queues in (the deployment of the)
Backend Choreographies, and finally we formalise correlation-based message exchange
in the semantics of deployment transitions in BC.

%

\begin{figure}
	\includegraphics[width=\textwidth]{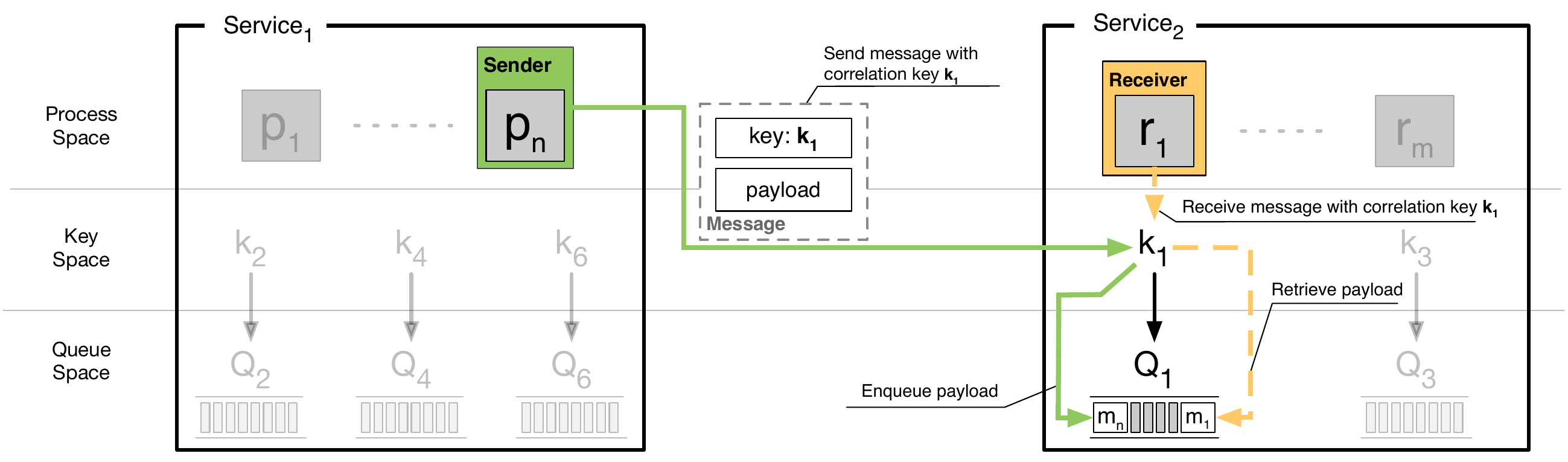}
	\caption{Depiction of correlation-based message exchange in SOC.}
	\label{fig:correlation}
\end{figure}

\subsection{Correlation-based Communication} 
\label{sec_aEC}

Processes in SOC run within services and communicate asynchronously. To realise
asynchronous communication, services provide an unbound number of
first-in-first-out message queues that processes interact with. The interaction
happens from processes that associate a message insertion/retrieval action with a
\emph{correlation key}, which uniquely identifies the queue subject of the action.
Concretely, a correlation key corresponds to a set of data that the service
associates to a specific queue.

Processes retrieve messages from the queues of their enclosing service. This is
represented in (the right side of) \cref{fig:correlation} by process $\pid r_1$,
which wants to consume a message received on queue $\mathsf{Q}_1$, associated to
the correlation key $\mathsf{k}_1$. The request is satisfied by the service,
which delivers message $\mathsf{m}_1$ to $\pid \mathsf{r}_1$, also removing the
interested message from the head of queue $\mathsf{Q}_1$. 
The complement of the action above is message insertion. Any process (within the
queue-enclosing service and remote) can insert data into a queue by sending a
message to the service owning the queue. That message must associate the payload
to be inserted with the correlation key that identifies the queue within the
service.
Concretely, when a service receives a message from the network, it inspects its
content, looking for a valid correlation key, i.e., one that points to any of
its queues. If a queue can be found, the message is enqueued in its tail. In
\cref{fig:correlation}, this is represented by data $\mathsf{k}_1$ marked by the
attribute \textsf{key} in the message sent by process $\pid p_\mathsf{n}$ (of
$\pid{Service}_1$) to $\pid{Service}_2$. At reception, $\pid{Service}_2$:
\begin{enumerate}
	\item checks for the presence of the attribute \textsf{key};
	\item extracts the corresponding key $\mathsf{k}_1$;
	\item finds the queue $\mathsf{Q}_1$, pointed by $\mathsf{k}_1$;
	\item enqueues the received payload in $\mathsf{Q}_1$ as message
	$\mathsf{m}_\mathsf{n}$.
\end{enumerate}
As depicted in \cref{fig:correlation}, messages in SOC contain correlation keys
as either part their payload or in some separate header. As in~\cite{MC11}, also
here we abstract from such details.
To summarise, two processes can communicate over correlation-based messaging if:
\emph{i}) the sender knows the (location of the) service where the addressee is
running and \emph{ii}) the sender and the addressee know the key corresponding
to a queue in the addressee service.
After having presented the mechanism of correlation for message exchange, we can
proceed to explain how we model SOC systems in BC.

\smallpar{Data and Process state.}
%
Data in SOC is structured following a tree-like format, e.g.,
XML~\cite{XML} or JSON~\cite{JSON}.
In BC, we use trees to represent both the payload of messages and the state
of running processes (as in, e.g., BPEL~\cite{bpel} and Jolie~\cite{MGZ14}).

Formally, we consider rooted trees $t \in \Trees$, where $\Trees = \Val \cup
\Locations \cup Set(\Labels \times \Trees)$ and
\[ t \gram \quad v \quad \Div \quad l \quad \Div \quad \{\
\pair{\jpath{x_1}}{t_1}, \dots, \pair{\jpath{x_n}}{t_n} \ \} \]
i.e., a tree (node) is either a value $v$, a location $l$, or a set of ordered
pairs of edge labels $\jpath{x},\jpath{y} \in \mathit{\Labels}$ and tree nodes.
We assume tree nodes to be values or locations only in leaves.
Now we can define BC variables as paths on trees (the latter, we remind,
represents state of processes) as sequences of labels $x,y \in Seq(\Labels)$
such that $x \gram \jpath{x}.x \ \Div \ \emptyseq$, where $\emptyseq$ is the
empty sequence, which we often omit for brevity. When writing paths in their
extended form, e.g., $\jpath{x}.\jpath{y}.\jpath{z}.\emptyseq$, we often use the
abbreviation $\jpath{x.y.z}$.

In addition, we define two operators to handle trees: path application
and deep copy.
The path-application operator $x(t)$ is used to access the sub-nodes pointed
by path $x$ in tree $t$. Intuitively, $x(t)$ returns either the value, the
location or the sub-tree pointed by path $x$ in $t$. If $x$ is not present in
$t$, \(x(t)\) returns an empty set of ordered pairs label-tree. Formally,
\[
\jpath{x}.x( t ) = \begin{cases}
x(\ \jpath{x}.\emptyseq(t)\ ) & \mbox{if } x \neq \emptyseq
\\
t' & \mbox{if } 
	x = \emptyseq
	\mbox { and } 
	t = \{\ \pair{\jpath{x}}{t'}, \dots, \pair{\jpath{x_n}}{t_n} \ \} 
\\
\emptyset & \mbox{otherwise}
\end{cases}
\]
The deep-copy operator $t \tcopy{x}{t'}$ is a (total) replacement operator
that returns the tree obtained by replacing in $t$ the sub-tree rooted in
$x(t)$ with $t'$. If $x$ is not present in $t$, $t \tcopy{x}{t'}$ adds the
smallest chain of empty nodes to $t$ such that it stores $t'$ under path $x$.
Formally,
\[
t \tcopy{\jpath{x}.x}{t'} = \begin{cases} 
\emptyset \tcopy{\jpath{x}.x}{t'} & \mbox{if } t \in \Val \cup \Locations
\\
(\ t \ \setminus \ \{\ \pair{\jpath{x}}{\jpath{x}(t)}\ \}\ ) \cup 
\{\ \pair{\jpath{x}}{t'}\ \} &
\mbox{if } t \not \in \Val \cup \Locations \mbox{ and } x = \emptyseq
\\
(\ t \ \setminus \ \{\ \pair{\jpath{x}}{\jpath{x}(t)}\ \}\ ) \cup 
\{\ \pair{\jpath{x}}{\jpath{x}(t)\tcopy{x}{t'}}\ \} &
\mbox{otherwise}
\end{cases}
\]


\subsection{Backend Deployments, Transition Rules, and BC Semantics}
In addition to the convention of using the terms ``Frontend
choreography/deployment'' to indicate a Frontend Choreographies
program/deployment, in the reminder we adopt the same convention for ``Backend
choreographies'' and ``Backend deployments''. We use the term ``choreography''
alone, when the context makes it clear if we refer to Backend or Frontend ones.

We can now define the notion of deployment for BC, denoted $\aenv$, which
includes:
\begin{itemize}
	\item the locality of processes;
	\item queues, pointed by a combination of a location and a correlation key;
	\item the state of processes.
\end{itemize}

Formally, $\aenv$ is an overloaded partial function defined by cases as the sum
of three partial functions
$g_l: \Locations \rightharpoonup Set( \Pids )$, $g_m: (\Locations \times \Trees)
\rightharpoonup Seq( \Operations \times \Trees )$, and $g_s: \Pids
\rightharpoonup \Trees$. The domains and co-domains of the functions are
disjoint, hence:
\[
\aenv( z ) = \begin{cases}
	g_l( z ) & \mbox{ if } z \in \Locations, \\
	g_m( z ) & \mbox{ if } z \in (\Locations \times \Trees), \\
	g_s( z ) & \mbox{ otherwise}
\end{cases}
\]
Function $g_l$ maps a location to the set of processes running in the service at
that location. Given a location $l$, we read $\aenv(l) = \{\pid p_1, \ldots,
\pid p_n\}$ as ``the processes $\pid p_1, \ldots, \pid p_n$ are running at the
location $l$'' (we assume each process $\pid p$ to run at most at one location).
Function $g_m$ maps a couple location-tree to a message queue. This reflects
message correlation as informally described above, where a queue resides in a
service, i.e., at its location, and is pointed by a correlation key. Given a
couple $l:t$, we read $\aenv(l:t) = \til m$ as ``the queue $\til m$ resides in a
service at location $l$ and is pointed by correlation key $t$''. The queue $\til
m$ is a sequence of messages
$\til m ::= m_1::\cdots::m_n \ | \ \emptyseq$
and a message of the queue is $m ::= (\op{o},t)$, where $t$ is the payload of
the message and $\op{o}$ is the operation on which the message was received.
Pairing operation labels with message payloads is typical of SOC
implementations in general. Indeed, while not essential for the correct delivery
of messages, operation labels are used by processes to program external choices
(for instance, a process expecting to receive a message on either of two
mutually-exclusive operations, e.g., to continue or exit a loop). The case
applies also to BC, where we preserve the association between payload and
operations \((o,t)\)---similarly to FC \((o,v)\).
Function $g_s$ maps a process to its local state. Given a
process $\pid p$, the notation $\aenv(\pid p) = t$ means that $\pid p$
has local state $t$.

%
%

\paragraph{Backend Deployment Transitions}
In BC, we replace the deployment transitions of FC with the rules defining
$\renv{\aenv}{\delta}{\aenv'}$, reported in \cref{fig:seff_semantics_AC}. We
comment them in the following.

\begin{figure}
\small
	{
	$$
	\hspace{-10pt}\begin{array}{c}
		\infer[\did{\aenv}{Start}]
	{
		\renv{\aenv}{\genstart}{
		%
		\aenv'
		} 
	}
	{
		\pid p \in \aenv(l)
		&
		\renv{\aenv}{\supfn(\ \{\ l.\pid p[\role A],\wtil{l.\pid q.[\role B]}\ \}\ )}{\aenv'}
	}
	\\[15pt]
	\infer[\did{\aenv}{Sup}]{
		\renv{\aenv}{\supfn(\ \{\ l_i.\pid q_i[\role B_i]\ \}_{i \in I}\ )}{
		\aenv'''\big[\
			\pid q_h \mapsto \{ \jpath{k} : t\}
		\ \big]_{h \in \{2,\ldots, n\}} \showcounter{8}
		}
	}{
	\begin{array}c
		\pid q_1 \in \aenv \ \showcounter{1}
		\hspace{2em}
			j \in I \setminus \{i\}
		\hspace{2em}
			\jpath{B_i.l}( t ) = l_i \showcounter{2}
		\hspace{2em}
			\jpath{B_i.B_j}(t) = t_{ij} \showcounter{3}
		\hspace{2em}
		l_j:t_{ij} \not \in \aenv \showcounter{4}
		\\[.5em]
			\aenv' = \aenv\big[\ l_i \mapsto \aenv(l_i) \cup \{\pid q_i\} \ \big] 
			\showcounter{5}
		\hspace{1em}
			\aenv'' = \aenv'\big[\ l_i:t_{ij} \mapsto \emptyseq\ \big] \showcounter{6}
		\hspace{1em}
			\aenv''' = \aenv''\big[\ \pid q_1 \mapsto \aenv''(\pid q_1)
			\tcopy{\jpath{k}}{t}\ \big] \showcounter{7}
	\end{array}
	}
	\\[15pt]
	\infer[\did{\aenv}{Send}]
		{
			\renv{\aenv}{\gensend}{
				\aenv\left[\
					l:t_c \mapsto
							\aenv(l:t_c) :: (\op{o}, t_m)
						\ \right]
			}
		}
		{
			l = \jpath{k.B.l}(\ \aenv(\pid p) \ )
			\qquad
			t_c = \jpath{k.A.B}(\ \aenv(\pid p) \ )	
			\quad
			t_m = \evalfn(e, \aenv(\pid p))
		}
	\\[15pt]
	\infer[\did{\aenv}{Recv}]
	{
		\renv{\aenv}{\genrecv}{
		\aenv'\left[\
			\pid q \mapsto
					\aenv'(\pid q) \tcopy{\jpath{x}}{t_m}
		\ \right]
		}
	}
	{
		t_c = \jpath{k.A.B}(\ \aenv(\pid q) \ )
		&
		\pid q \in \aenv(l)
		&
		\aenv( l:t_c ) = (\op{o},t_m) :: \til m
		&
		\aenv' = \aenv[\ l:t_c \mapsto \til m \ ]
	}
	\end{array}
	$$}
 	\caption{Backend Choreographies --- Deployment transitions.}
	\label{fig:seff_semantics_AC}
\end{figure}

Rule $\did{\aenv}{Start}$ simply retrieves the location of process $\pid p$
(the one that requested the creation of session $k$) and uses rule
\did{\aenv}{Sup} to obtain the new deployment $\aenv'$ that supports
interactions over session $k$. Namely, $\aenv'$ is an updated version of $\aenv$
with: \emph{i}) the newly created processes for session $k$ and \emph{ii}) the
queues used by the new processes and $\pid p$ to communicate over session $k$.
In addition, in $\aenv'$, \emph{iii}) the new processes and $\pid p$ contain in
their states a structure, rooted in $\jpath{k}$ and called \emph{session
descriptor}, that includes all the information (correlation keys and the
locations of all involved processes) to support correlation-based communication
in session $k$.
Formally, this is done by rule \did{\aenv}{Sup} where we \showcounter{1}
retrieve the starter process, here called $\pid q_1$, which is the only process
already present in $\aenv$. Then, given a tree $t$, we ensure it is a proper
session descriptor for session $k$, i.e., that:
\begin{itemize}
	
	\item[\showcounter{2}] $t$ contains the location $l_i$ of each process,
	represented by its role in the session $\role B_i$, under path
	$\jpath{B_i.l}$;
	\item[\showcounter{3}] $t$ contains a correlation key $t_{ij}$ for each
	ordered couple of roles $\role B_i$, $\role B_j$ under path $\jpath{B_i.B_j}$,
	such that \showcounter{4} there is no queue in $\aenv$ at location $l_j$
	pointed by correlation key $t_{ij}$;
\end{itemize}
Finally, we assemble the update of $\aenv$ in four steps:
\begin{itemize}
	\item[\showcounter{5}] first, we obtain $\aenv'$ by adding in $\aenv$ the
	processes $\pid q_2, \ldots, \pid q_n$ at their respective locations;
	\item[\showcounter{6}] second, we obtain $\aenv''$ by adding to $\aenv'$ an
	empty queue $\emptyseq$ for each couple $l_j:t_{ij}$;
	\item[\showcounter{7}] third, we obtain $\aenv'''$ from $\aenv''$ by
	storing in the state of (the starter) process $\pid q_1$ the session
	descriptor $t$ under path $\jpath{k}$;
	\item[\showcounter{8}] finally, we update $\aenv'''$ such that each new
	created process ($\pid q_2, \ldots, \pid q_n$) has in its state just the
	session descriptor $t$ rooted under path $\jpath{k}$.
\end{itemize}

We deliberately define in \did{\aenv}{Sup} the session descriptor $t$ with a
set of constrains on data, rather than with a procedure to obtain the data
for correlation. In this way, our model is general enough to capture
different methodologies for creating correlation keys (e.g., UUIDs or API
keys).
%

Rule $\protect\did{\aenv}{Send}$
models the sending of a message. We comment the premises. From left to right,
the first gets the location $l$ of the receiver $\role B$ from the state of
the sender $\pid p$; the second retrieves the correlation key in the state of
$\pid p$ (playing role $\role A$) to send messages to role $\role B$; the
third evaluates the expression $e$ of the sender $\pid p$ using its local
state to get a value $t_m$. Function $\evalfn$ evaluates expressions in a
process state, traversing its paths and performing local computation. We
highlight that, since in BC we preserve the syntax of Fronted Choreographies, we
make two assumptions: that expressions (e.g., $e$ in $\did{\aenv}{Send}$) are
defined on $\Var$iables and that $\evalfn$ in BC automatically maps variables
$x$, $y$, $z$ into the respective paths $\jpath{x}.\emptyseq$,
$\jpath{y}.\emptyseq$, and $\jpath{z}.\emptyseq$, used to access process
states in $\aenv$.
Finally, in the conclusion of the rule, we add the message $(\op{o},t_m)$ in
the queue pointed by $l:t_c$ that we found via correlation.

Rule $\protect\did{\aenv}{Recv}$
models a reception. From left to right, the first premise finds the
correlation key $t_c$ for the queue that $\pid q$ (playing role $\role B$)
should use to receive from $\role A$ in session $k$. The second premise
retrieves the location $l$ of $\pid q$. The third accesses the queue pointed
by $l:t_c$ and retrieves message $(\op{o},t_{m})$. The last premise updates
$\aenv$ to $\aenv'$ removing $(\op{o},t_{m})$ from the interested queue.
Dually to rule $\did{\aenv}{Send}$, where $\evalfn$ maps variables into paths,
in the conclusion of rule $\did{\aenv}{Recv}$ we map $x$, i.e., the intended
variable that should store the payload $t_m$ in the state of $\pid q$, into
path $\jpath{x}.\emptyseq$.

\subsection{Encoding Frontend Choreographies to Backend Choreographies and Properties} 
\label{sub:encodig}

Now that we presented Backend Choreographies, we can proceed with our main
intent of defining a compilation procedure from high-level FC programs to
low-level services. 
Here, we tackle the transition from FC programs to their intermediate
representation toward SOC systems as Backend Choreographies. 
Specifically, we translate FC programs that use the abstract mechanism of
communication over names, into BC programs that use the concrete mechanism of
correlation-based communication. We prove our translation correct, i.e., that
our encoding guarantees an operational correspondence between the semantics of a
Frontend choreography and its Backend encoding.

Formally, since choreographies in BC have the same syntax of FC ones, we can
translate FC runtime terms $\env, C$ to BC runtime terms by encoding the FC
deployment $\env$ to an appropriate Backend deployment. Notably, BC deployments
contain more information wrt FC deployments. We extract these data
from $\Gamma$, the typing environment of $\env,C$. 


\begin{definition}[Encoding FC in BC\label{def:bc_encoding}]
Let $\Gamma \hlseq{\env,C}$ and $\geneenc{\env}$ be defined by the algorithm in
\cref{fig:algo_encoding_deployment}. Then, the Backend encoding of $\env,C$ is
defined as $\geneenc{\env},C$.
\end{definition}

\begin{figure}

\begin{lstlisting}[xleftmargin=.3\textwidth,mathescape=true,numbers=left,numberblanklines=false] 
$\geneenc{\env}$ = $\hspace{.3em} \aenv \asgn \emptyfunc$
       
       foreach $\pid p@l$ in $\Gamma$
          $\aenv \aupd \left[\ l \mapsto \aenv(l) \cup \{\pid p\} \ \right]$
          $\aenv \aupd \left[\ \pid p \mapsto \emptyset \ \right]$

       foreach $\pair{\pid p.x}{U}$ in $\Gamma$
          $\aenv \aupd [\; \pid p \mapsto \aenv(\pid p) \tcopy{\jpath{x}}{\env( \pid p )( x )} \;]$

       foreach $\{\  \pair{\pid p}{k[\role A]}\, \ \pair{\pid q}{k[\role B]},\
       \pid q@l \}$ in $\Gamma$
          $t \asgn \auxfn{fresh}(\aenv,l)$
          $\aenv \aupd [\ \pair{l}{t} \mapsto \env(\chanto{k}{\role A}{\role B}) \ ]$
          $\aenv \aupd [\ \pid p \mapsto \aenv(\pid p) \tcopy{\jpath{k.A.B}}{t} \ ]$
          $\aenv \aupd [\ \pid q \mapsto \aenv(\pid q) \tcopy{\jpath{k.A.B}}{t} \ ]$
          $\aenv \aupd [\ \pid p \mapsto \aenv(\pid p) \tcopy{\jpath{k.B.l}}{l} \ ]$
          $\aenv \aupd [\ \pid q \mapsto \aenv(\pid q) \tcopy{\jpath{k.B.l}}{l} \ ]$

       return $\aenv$
\end{lstlisting}
  \caption{Encoding Algorithm from Frontend to Backend Deployments.}
  \label{fig:algo_encoding_deployment}
\end{figure}

What the algorithm $\geneenc{\env}$ does is:
\begin{enumerate}
  \item include in $\aenv$ all (located) processes present in $\env$ (and typed
  in $\Gamma$);
  \item translate the state (i.e., the association \(\Var\)iable-\(\Val\)ue) of each
  process in $\env$ to its correspondent tree-shaped state in $\aenv$;
  \item for each ongoing session in $\env$, set the proper correlation keys and
  queues in $\aenv$ and, for each queue, import and translate its related
  messages.
\end{enumerate}

More precisely, in the algorithm defined in \cref{fig:algo_encoding_deployment}
at Line 1, we create a new Backend deployment $\aenv$ and assign to it the
totally undefined function ($\emptyfunc$); $\aenv$ is an empty Backend deployment
that will be later refined via the updates on $\aenv$ at Lines 3--16. Then,

\begin{itemize}

  \item \emph{Lines 3--5}, for each located process $\pid p@l$ in $\Gamma$,
  we update the locations of $\aenv$ to contain $\pid p$ at location $l$ (Line
  4) and we include process $\pid p$ in $\aenv$, associating to it an empty
     state, i.e., the empty tree $\emptyset$ (Line 5);

  \item \emph{Lines 7--8}, for each variable $x$ (typed in $\Gamma$) of a
  process $\pid p$, we update the state of process $\pid p$ in $\aenv$ to
  include the association of $x$ to its value in the state $\env(\pid p)$. As
  done in rules $\did{\aenv}{Send}$ and $\did{\aenv}{Recv}$, we map FC
  variables $x \in \mathit{Var}$ into BC paths $\jpath{x} \in
  Seq(\mathit{Lab})$;

  \item \emph{Lines 10--16}, follow the same principles to support
  correlation-based exchanges as formalised in rule $\did{\aenv}{Sup}$; for
  each couple of processes $\pid p, \pid q$, respectively playing distinct
  roles $\role A$ and $\role B$ in a session $k$, with $\pid q$ located at $l$:

  \begin{itemize}

    \item \emph{Line 11}, we obtain a fresh correlation key $t$ with
    auxiliary function $\auxfn{fresh}$. The latter takes deployment $\aenv$ and
    location $l$ as input and returns a correlation key which is fresh among
    the keys associated to location $l$ in $\aenv$. Formally $t$ is such that
    $\pair{l}{t} \not \in \dom(\aenv)$;

    \item \emph{Line 12}, we associate correlation key $t$ with location $l$
    in $\aenv$ and make it point the corresponding queue of messages from
    role $\role A$ to role $\role B$ in $\env$ (accessed with triple
    $\chanto{k} {\role A}{\role B}$). Note that we can directly copy message
    queues from $\env$ into $\aenv$. Indeed, while message queues in $\env$
    and $\aenv$ are respectively of type $Seq(\Operations \times
    \Val)$ and $Seq(\Operations \times \Trees)$, by definition
    $\Trees$ subsumes $\Val$;

    \item \emph{Line 13--14}, we include in the state of processes $\pid p$
    (Line 13) and $\pid q$ (Line 14) correlation key $t$, storing it under
    path $\jpath{k.A.B}$;

    \item \emph{Line 15--16}, we include in the state of processes $\pid p$
    (Line 15) and $\pid q$ (Line 16) the location of role $\role B$ under
    path $\jpath{k.B.l}$.

  \end{itemize}

\end{itemize}

The encoding from FC to BC guarantees a strong operational correspondence.

\begin{theorem}[Operational Correspondence (FC $\leftrightarrow$ BC)]
\label{thm:encoding_operational_correspondence}
Let $\Gamma \seq \env,C$. Then:
  \begin{enumerate}
    \item (Completeness) $\env,C \to \env',C'$ implies $\geneenc{\env},C \to 
    \eenc{\env'}{\!\Gamma'},C'$ for some $\Gamma'$ s.t. $\Gamma' \hlseq{\env',C'}$;
    \item (Soundness) $\geneenc{\env},C \to \aenv,C'$ implies $\env,C \to
    \env',C'$ and $\aenv = \eenc{\env'}{\Gamma'}$ for some $\Gamma'$ s.t.
    $\Gamma' \hlseq {\env',C'}$.
  \end{enumerate}
\end{theorem}

\begin{proof}[Sketch of \cref{thm:encoding_operational_correspondence}]
We sketch the proof of Theorem~\ref{thm:encoding_operational_correspondence},
analysing its two parts: \emph{(Completeness)} and \emph{(Soundness)}.
The proof of Completeness is by induction on the derivation of $\env,C$. The
main observation is that the encoded system $\geneenc{\env},C$ mimics $\env,C$
by applying the same semantic rules on $C$ and the corresponding deployment
transitions (e.g., respectively defied by rules $\did{\env}{Send}$ and
$\did{\aenv}{Send}$).
Let $\aenv'$ be the Backend environment obtained from the reduction $
\geneenc{\env},C \to \aenv,C'$ on rule $\did{C}{Start}$. Since
\cref{fig:algo_encoding_deployment} and rule $\did{\aenv}{Sup}$ (on which rule
$\did{\aenv}{Start}$ relies) implement the same principles of
\(\geneenc{\env},\) we know that $\jpath{k.A.B}(\aenv)$ and $\jpath{k.A.B}(\
\eenc{\env'}{\Gamma'}\ )$ will be the same, except possibly for \emph{i}) the
location of processes and \emph{ii}) trees of correlation keys corresponding to
the same paths. Concretely, item \emph{i}) derives from the fact that $\Gamma$
and $\Gamma'$ can disagree on the location of the same process $\pid p$, and
item \emph{ii}) is caused by the random generation of correlation keys, for
which, considering a correlation key rooted in $\jpath{k.A.B}$ of a process
$\pid p$, the trees obtained from $\jpath{k.A.B}(\ \aenv(\pid p)\ )$ and $
\jpath{k.A.B}(\ \eenc{\env'}{\Gamma'}(\pid p)\ )$ may differ. However, these
discrepancies do not constitute a problem, since both locations and correlation
keys are used consistently in their respective deployments, which are thus
interchangeable.

We can extend the same observation also for Soundness, which is proved by
induction on the derivation of $\geneenc{\env},C$.
\end{proof}


\IFSubFileBiblio

\section{Dynamic Correlation Calculus}
\label{sec:implementation_model}

In this section, we introduce the Dynamic Correlation Calculus (DCC), the target
language of our compilation. 

DCC is a straightforward extension of a previous proposal called Correlation
Calculus~\cite{MC11}, which is a process calculus that formalises
service-oriented, correlation-based communications. Indeed, while we
started this work considering CC as the target language of our compilation, we
found it limited for our purposes: in CC each process receives from only one
message queue, while we need processes to be able to select receptions from
multiple queues (as in our Backend deployments). Hence, we defined DCC as an
extension of CC with the support for the dynamic creation and selection of
queues in processes.

We deem DCC a choice that fits the practical motivations of this work thanks to
its closeness to the implementation languages/frameworks listed below, which
casts a good outlook on the affordability of future implementations of our
theoretical results.
First, CC formalises the semantics of message exchange of Jolie, a
service-oriented programming language~\cite{MGZ14}. Thus CC specifications are
directly translatable into Jolie executable programs. This is not the case for
our DCC code, as Jolie lacks the primitives to let processes create and select
queues. Fortunately, the distance between CC and DCC is close enough so that
supporting the extended features in Jolie would entail minimal change, i.e., the
inclusion of the syntactic primitives for queue creation and
selection\footnote{The \(\prid{newque}\) and \(\mathtt{from}\) and
\(\mathtt{to}\) particles in \cref{fig:jc_syntax}.} and the implementation of
the associated semantics---a direct extension of the one-process-one-queue
semantics of the current implementation.
Second, the service-oriented language BPEL~\cite{bpel} lets processes
create and receive from multiple queues, making DCC a useful reference for
BPEL-based implementations.
Third, besides service-oriented languages, DCC abstracts real-world
message-exchange models where processes can interact with multiple message
queues. This is the case, e.g., for some versions of the actor
model~\cite{agha85} where one actor can be associated with many
queues/mailboxes~\cite{haller07} and in some popular message-exchange
middlewares~\cite{vinoski06,videla12}, which are suitable alternatives to the
implementation targets above.

\begin{figure}
$$
\begin{array}c
	\begin{array}{l@{\qquad}l}
		\textit{Services} & 
		\begin{array}{r@{\qquad}l@{\qquad}l}
		S\ \gram & \jsrv{\strBhv, P, M}{l} 	& \prid{srv} 
		\\
			\Div 	& S \pp S' 					& \prid{net}
		\end{array}
		\\[15pt]
		\textit{Start Behaviour} & 
		\begin{array}{r@{\qquad}l@{\qquad}l}
		\strBhv \gram & !(x);B 					& \prid{acpt} 
		\\
				 \Div & \inact 					& \prid{inact}
		\end{array}
		\\[15pt]
		\textit{Processes} &
		\begin{array}{r@{\qquad}l@{\qquad}l}
		P\ \gram & \jpr{B}{t}  		& \prid{prcs} 
		\\
				\Div & P \pp P' 				& \prid{par}
		\end{array}
	\end{array}
	\\[45pt]
	\textit{Behaviours}\hfill\\[1pt]
	\begin{array}{rl@{\qquad}l}
		B \gram &
		?@e_1( e_2 );B  & \prid{reqst}
			\\[1pt]
		\Div & \op{o}(x) \qfrom e; B & \prid{input} 
		\\[1pt]
		\Div & \recDef{X}{B'}{B} & \prid{def}
		\\[1pt]
		\Div & \cq{x};B & \prid{newque}
		\\[1pt]
		\Div & x = e; B & \prid{assign}
	\end{array}
	\quad
	\begin{array}{rl@{\qquad}l}
		\Div & \sum_{i}\choice{o_i(x_i) \qfrom e}{B_i} & \prid {choice}
		\\[1pt]
		\Div & \op{o}@e_1( e_2 ) \qto	e_3; B & \prid{output}
		\\[1pt]
		\Div & \cond{e}{B_1}{B_2} & \prid{cond}
		\\[1pt]
		\Div & \inact & \prid{inact}
		\\[1pt]
		\Div & \mbox{X} & \prid{call}
	\end{array}
\end{array}
$$
	\caption{Dynamic Correlation Calculus --- Syntax.}
	\label{fig:jc_syntax}
\end{figure} 

\paragraph{Syntax}
We now introduce the syntax of DCC, which we report in \cref{fig:jc_syntax} and
which comprises two layers: \emph{Services}, ranged over by $S$, and
\emph{Processes}, ranged over by $P$.

In the syntax of services, term $\prid{srv}$ is a service, located at $l$,
with a \emph{Start Behaviour} $\strBhv$ and running processes $P$ (both described
later on) and a queue map $M$. The queue map is a partial function $M: \Trees
\rightharpoonup \mathit{Seq}(\Operations \times \Trees)$ that, similarly to
function $g_m$ in Backend deployments, associates a correlation key $t$ to a
message queue. We model messages like in BC where a message is a couple
$(o,t)$, $o$ being the operation on which the message has been received, and
$t$ the payload of the message. Services are composed in parallel in term
$\prid{net}$.

Concerning behaviours, in DCC we distinguish between start behaviours and
process behaviours. Process behaviours define the general behaviour of
processes in DCC, as described later on. Start behaviours use term $!(x)$ to
indicate the availability of a service to generate new local processes on
request. At runtime, the start behaviour $\strBhv$ of a service is activated by the
reception of a dedicated message that triggers the creation of a new process.
The new process has (process) behaviour $B$, which is defined in $\strBhv$ after
the $!(x)$ term, and an empty state. The content of the request message is
stored in the state of the newly created process, under the bound path $x$. As
in Backend Choreographies, also in DCC paths are used to access process states.

Finally, processes $\prid{prc}$ in DCC consists of a behaviour $B$ and a
state $t$ and can be composed in parallel $\prid{par}$. Process states $t$
are trees and, in \emph{Behaviours}, operations ($\op{o}$), procedures ($X$),
paths ($x$), and expressions ($e$, evaluated at runtime on the state of the
enclosing process) are all the same as defined for Backend Choreographies
(\cref{sec_aEC}).
Terms $\prid{input}$ and $\prid{output}$ model communications. In
$\prid{input}$, the process stores under $x$ a message $\mathtt{from}$ the
head of the queue correlating with $e$ and received on operation $\op{o}$.
Dually, term $\prid{output}$ sends a message on operation $\op{o}$. The three
expressions in the term define: $e_1$, the location of the service where the
addressee is running; $e_2$, the content of the message; $e_3$, the
key that correlates with the receiving queue of the addressee.
Term $\prid{choice}$ is an $\prid{input}$-choice: when one of the inputs can
receive a message from the queue correlating with $e$ on operation
$\op{o}_i$, it discards all other inputs and executes the continuation $B_i$.
Term $\prid{reqst}$ is the dual of $\prid{acpt}$ and asks the service located
at $e_1$ to spawn a new process, passing to it the message in $e_2$.
Term $\prid{newque}$ models the creation of a new queue that correlates with
a unique correlation key (in the service hosting the running process). The
correlation key is stored under path $x$ in the state of the process, for
later access. The remaining terms are standard.


\begin{figure}
\scalebox{.89}[.89]{%
\(
\hspace{-15pt}\begin{array}{c}
	\infer[\did{DCC}{Assign}]
	{ \jpr{x = e\ ;B}{t} \quad \to \quad \jpr{B}{t\tcopy{x}{t'}}}
	{ t' = \evalfn(x,t) }
	\qquad
	\infer[\did{DCC}{Ctx}]
	{
		\jpr{\recDef{X}{B_1}{B}}{t}
		\quad \to \quad
		\jpr{\recDef{X}{B_1}{B'}}{t'}
	}
	{
		\jpr{B}{t} \to \jpr{B'}{t'}
	}
	\\[15pt]
	\infer[\did{DCC}{Cond}]
	{
		\jpr{\cond{e}{B_1}{B_2}}{t}
			\quad \to \quad
		\jpr{B_i}{t}
	}
	{
		i = 1 \mbox{ if } \evalfn( e, t ) = \mbox{true}, i = 2 \mbox{ otherwise}
	}
	\qquad
	\infer[\did{DCC}{PEq}]
	{ \jsrv{\strBhv,\ P,\ M}l \quad \to \quad \jsrv{\strBhv,\ P',\ M}l }
	{ P \equivD P_1 \pp P_2 & P_1 \to P_1' & P_1' \pp P_2 \equivD P'}
	\\[15pt]
	\infer[\did{DCC}{Newque}]
	{
		\jsrv{\strBhv,\ \jpr{B}{t} \pp P,\ M}l
			\quad \to \quad 
		\jsrv{\strBhv,\ \jpr{B}{t \tcopy{x}{t_c}} \pp P,\ M'}l
	}
	{
		B = \cq{x};B
		&
		t_c \not \in \dom(M)
		&
		M' = M[t_c \mapsto \emptyseq]
	}
	\\[15pt]
	\infer[\did{DCC}{Recv}]
	{
		\jsrv{\strBhv,\ \jpr{B}{t} \pp P,\ M}l 
		\quad \to \quad 
		\jsrv{\strBhv,\ \jpr{B_j}{t\tcopy{x_j}{t_m}} \pp P ,\ M[t_c \mapsto \til m]}l
	}
	{
		\begin{array}c
		B \in \{\; o_j(x_j) \qfrom e;B_j 
		\; , \; 
						  \sum_{i \in I}\choice{o_i(x_i) \qfrom e}{B_i}  
	  \; \}
	  \\[2pt]
	  j \in I
	  \qquad
	  t_c = \evalfn(e,t)
	  \qquad
	  M(t_c) = (o_j,t_m)::\til m	
		\end{array}
	}
	\\[15pt]
	\infer[\did{DCC}{InSend}]
	{
		\jsrv{\strBhv,\ \jpr{B}{t} \pp P, M}{l}
		\quad \to \quad 
		\jsrv{\strBhv,\ \jpr{B'}{t} \pp P, M[t_c \mapsto M(t_c)::(o,t_m)]}{l}
	}
	{
	\begin{array}c
		B = \notify{\op{o}}{e_1}{e_2} \qto e_3;B'
		\qquad
		\evalfn( e_1, t ) = l
		\\[2pt]
		\evalfn( e_3, t ) = t_c
		\qquad
		\evalfn( e_2, t ) = t_m
		\qquad
		t_c \in \dom(M)
	\end{array}
	}
	\\[15pt]
	\infer[\did{DCC}{InStart}]
	{
		\jsrv{!(x);B',\ \jpr{B}{t} \pp P,\ M}{l}
		\quad \to \quad
		\jsrv{!(x);B',\ Q \pp \jpr{B''}{t} \pp P,\ M}{l}
	}
	{
		B = \notify{?}{e_1}{e_2};B''
		&
		\evalfn( e_1, t ) = l
		&
		Q = \jpr{B'}{\emptyset\tcopy{x}{\evalfn( e_2, t )} }
	}
	\\[15pt]
	\infer[\did{DCC}{Send}]
	{
		\jsrv{\strBhv, \jpr{B}{t} \pp P,\ M}{l} 
		\pp 
		\jsrv{\strBhv', P',\ M'}{l'}
		\quad \to \quad
		 \jsrv{\strBhv, \jpr{B''}{t} \pp P,\ M}{l} \pp 
		 \jsrv{\strBhv', P', M''}{l'}
	}
	{
		\begin{array}c
		B = \notify{\op{o}}{e_1}{e_2} \qto e_3; B''
		\qquad
		\evalfn( e_1, t ) = l'
		\qquad
		\evalfn( e_3, t ) = t_c
		\\[2pt]
		\evalfn(	e_2, t ) = t_m
		\qquad
		t_c \in \dom(M')
		\qquad
		M'' = M'[t_c \mapsto M'(t_c)::(o,t_m)]	
		\end{array}
	}
	\\[15pt]
	\infer[\did{DCC}{Start}]
	{
		\begin{array}l
		\jsrv{\strBhv,\ \jpr{B}{t} \pp P, M}{l}
		\pp 
		\jsrv{\strBhv',\ P',\ M'}{l'} 
		 \quad \to \quad 
		\jsrv{\strBhv,\ \jpr{B''}{t} \pp P,\ M}{l}	
		\pp 
		\jsrv{\strBhv', Q \pp P',\ M'}{l'} 
		\end{array}
	}
	{
		B = \notify{?}{e_1}{e_2};B''
		&
		\strBhv' = !(x);B'
		&
		\evalfn( e_1, t ) = l'
		&
		Q = \jpr{B'}{\emptyset\tcopy{x}{\evalfn( e_2, t )}}
	}
	\\[15pt]
	\infer[\did{DCC}{SPar}]
	{
		S \pp S_1 \quad \to \quad S' \pp S_1
	}
	{
		S  \to  S'
	}
	\qquad
	\infer[\did{DCC}{SEq}]
	{
		S \quad \to \quad S'
	}
	{
		S \equivD S_1
		&
		S_1  \to  S_1'
		&
		S_1' \equivD S'
	}
\end{array}
\)
}
\vspace{2em}
{\color{gray}\hrule}
\vspace{2em}
\[
	\begin{array}{c}
	\jpr{\recDef{X}{B}{\inact}}{t} \; \equivD \; \jpr{\inact}{t}
	\qquad
	P \pp P' \; \equivD \; P' \pp P
	\qquad
	(P_1 \pp P_2) \pp P_3 \; \equivD \; P_1 \pp (P_2 \pp P_3)
	\\[5pt]
	P \quad \equivD \; P \pp \jpr{\inact}{t}
	\qquad
	\jpr{\recDef{X}{B}{X}}{t} \; \equivD \; \jpr{\recDef{X}{B}{B/X}}{t}
	\qquad
	S \pp S' \; \equivD \; S' \pp S
	\\[5pt]
	(S_1 \pp S_2) \pp S_3 \; \equivD \; S_1 \pp (S_2 \pp S_3)
	\end{array}
\]%
\caption{Dynamic Correlation Calculus --- Semantics.}
\label{fig:js_semantics_full}
\end{figure}

\paragraph{Semantics}
In \cref{fig:js_semantics_full}, we report the rules defining the semantics
of DCC, a relation $\to$ closed under a (standard) structural congruence
$\equivD$ that supports commutativity and associativity of parallel
composition. We comment the rules.

Rules $\did{DCC}{Assign}$, $\did{DCC}{Ctx}$, and $\did{DCC}{Cond}$ are
standard for, respectively, assignments, procedure definition, and condition
evaluation. Rule $\did{DCC}{PEq}$ uses equivalence $\equivD$ on DCC processes
to describe parallel execution and recursion. The rules of $\equivD$ are
reported in the lower part of \cref{fig:js_semantics_full}.

Rule $\did{DCC}{Newque}$ adds to $M$ an empty queue ($\emptyseq$) correlating
with a randomly generated key $t_c$. The key is stored under path $x$ of the
process that requested the creation of the queue. As in rule
\did{\aenv}{Sup} of Backend Choreographies (see \cref{sec_aEC}), we do not
impose a structure for correlation keys, yet we require that they are
distinct within their service.

Rule $\did{DCC}{Recv}$ models message reception. Since both $\prid{input}$
and $\prid{choice}$ define receptions of messages, we consider both cases in
the rule. Indeed, the first premise of the rule captures the presence of either
an $\prid{input}$---with shape $o_j(x) \qfrom e$---or a
$\prid{choice}$---with shape $\sum_{i \in I}\choice{o_i(x_i) \qfrom e}{B_i}$. In both
cases, we obtain the correlation key of the receiving queue from the
evaluation of expression $e$ against the state of the receiving process
($t$). Then, we inspect queue map $M$ and check if it has a message in its
head received on operation $o_j$. If this holds, the rule removes the message
from the queue and stores the payload ($t_m$) under path $x_j$ in the state
of the process.

Regarding message delivery, in DCC, there are two output actions: \emph{i})
$\prid{output}$ used by a process to communicate with another one and \emph{ii})
$\prid{reqst}$ used by a process to require the creation of a new process in a
service. Since in DCC communications can happen within the same service or
between two services, we describe two sets of rules, either for internal and
inter-service message delivery.

We start from the easier case of internal delivery, defined by rules
$\did{DCC}{InSend}$ and $\did{DCC}{InStart}$. In rule $\did{DCC}{InSend}$ a
process $\jpr{B}{t}$ sends a message into a queue of its hosting service. This
is illustrated by the second premise of the rule where the location $l$,
corresponding to the evaluation of expression $e_1$ against the state of the
sender process, is the same of its hosting service. As expected, correlation key
$t_c$ must point an actual queue of the service. This is checked by the last
premise, which requires $t_c$ to be in the domain of queue map $M$. In the
conclusion of the rule, we update the content of the queue pointed by $t_c$
including message $(o,t_m)$ in its tail. In rule $\did{DCC}{InStart}$ a service
accepts the request to create a new process from one of its local processes. In
the conclusion of the rule, we find the newly created process $Q$. The behaviour
of the new process corresponds to the one associated with the $\prid{acpt}$ term
of the service ($B'$). The state of the new process is empty ($\emptyset$)
except for the inclusion of the payload of the request, stored under path $x$
and obtained from the evaluation of $e_2$ against $t$.

Message delivery between two services is defined by rules $\did{DCC}{Send}$
and $\did{DCC}{Start}$. The two rules are similar to their respective
internal cases, except for requiring the location defined by the sender
(i.e., the one obtained from the evaluation of expression $e_1$ against the
state $t$ of the sender process) to match that of the receiving service.

The last two rules in \cref{fig:js_semantics_full} are $\did{DCC}{SPar}$ and
$\did{DCC}{SEq}$ and define the (parallel) execution of networks of services.

\IFSubFileBiblio

\section{Compiling Frontend Choreographies into DCC processes}
\label{sec:compiler}

We now present our main result: the correct compilation of high-level Frontend
Choreographies into low-level DCC networks of services (and processes).
\begin{figure}
	\includegraphics[width=\textwidth]{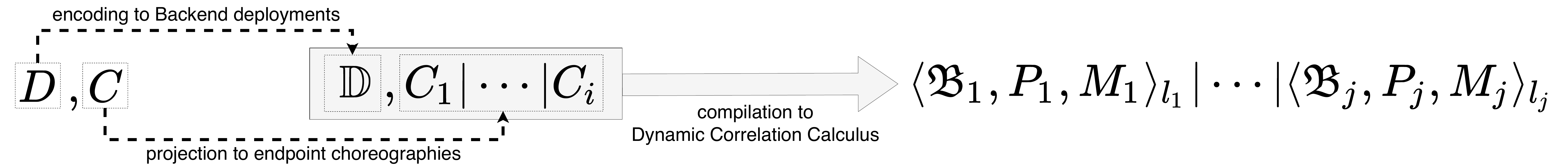}
	\caption{Scheme of compilation from Frontend Choreographies to Dynamic Correlation Calculus.}
	\label{fig:formal_scheme}
\end{figure}
We depict in \cref{fig:formal_scheme} a schematic representation of the stages
involved in the compilation from FC to DCC programs. Concretely, given an FC
program $\env, C$ and its typing environment $\Gamma$, our compilation procedure
consists of three stages:

\begin{description}

	\item[FC-to-BC] the encoding, defined in \cref{sub:encodig}, of the
	Frontend deployment $\env$ to a correspondent Backend deployment $\aenv =
	\geneenc{\env}$. This stage is depicted in \cref{fig:formal_scheme} with the
	relation \(D \dashrightarrow \mathbb{D}\);

	\item[EPP] the projection of the choreography $C$ into a
	parallel composition of partial choreographies (i.e., containing only actions
	concerning one participant), each defining the behaviour of a single active or
	service process in $C$. This stage is called \emph{Endpoint
	Projection}, it is presented in \cref{sub:endpoint_projection}, and it
	is depicted in \cref{fig:formal_scheme} with the relation \(C \dashrightarrow
	C_1|\cdots|C_i\);

	\item[Compilation] the compilation of the composition of the results of the
	previous stages---essentially, an endpoint Backend choreography---into a
	network of corresponding DCC services and their located processes. We present
	the compilation in \cref{sub:BC_to_DCC}. This stage is depicted in
	\cref{fig:formal_scheme} with the relation \(\mathbb{D},C_1|\cdots|C_i
	\Rightarrow\langle\strBhv_1,P_1,M_1\rangle_{l_1}|\cdots|\langle\strBhv_j,P_j,M_j\rangle_{l_j}\)
	(the left part highlighted in grey to help readability);

\end{description}

The division in three stages makes the definition of the compilation process,
and its related checks for correctness, simpler. In particular, they ease the
extraction of the behaviour of a single process (\textbf{EPP})
from the source Frontend choreography and of its state
(\textbf{FC-to-BC}) from the source Frontend deployment.
In the remainder of this section, we detail the projection stage
(\textbf{EPP}), we define how we pair the outputs of
\textbf{FC-to-BC} and \textbf{EPP} and the properties of that
pairing, and we present the \textbf{Compilation} stage and the
properties of our main contribution.


\subsection{Endpoint Projection (EPP)} 
\label{sub:endpoint_projection}
Given a choreography\footnote{Since the EPP acts on the syntax and FC and BC
share the same syntax, distinguishing between them here is irrelevant.} $C$, its
Endpoint Projection (EPP), denoted $\epp{C}{}$, returns an
operationally-equivalent composition of \emph{Endpoint choreographies}.
Intuitively, an Endpoint choreography is a choreography that does not contain
complete actions---i.e., terms $\prid{start}$ and $\prid{com}$---and that
describes the behaviour of a single process. We remind that a choreography can
contain two kinds of processes: \emph{active processes} which are already
running, and \emph{service processes} which accept requests to create new active
processes at their respective associated location $l$. As detailed later on, our
EPP procedure projects Endpoint choreographies on all processes, both active and
service ones.

Our definition of EPP is an adaptation of that presented in~\cite{MY13} and
it is divided into two components:

\begin{itemize}

	\item a \emph{process projection} that derives the Endpoint choreography of
	a single process $\pid p$ from a given choreography $C$, written
	$\epp{C}{\pid p}$;

	\item the actual EPP of a given choreography $C$, which results into the
	parallel composition of:

	\begin{itemize}
		\item the process projections of all active processes in $C$;

		\item the process projections of all service processes in $C$, with the
		exception that we merge into the same Endpoint choreography all process
		projections of service processes that accept requests at the same
		location.

	\end{itemize}
	
\end{itemize}

In the next paragraphs, first we present the process projection and next the actual Endpoint Projection.

\paragraph{Process Projection}
Let us start the definition of process projection by formalising Endpoint choreographies.

\begin{definition}[Endpoint Choreographies]
Given a (Frontend/Backend) choreography $C$. If either:
\begin{itemize}
	\item $C = \acc{k}{l.\pid q[\role B]};C'$, and $\pid q$ is the only free
	process name in $C'$;
	\item $C$ has only one free process name.
\end{itemize}
then $C$ is an Endpoint choreography.
\end{definition}

The process projection of a subject process $\pid p$ in a choreography $C$,
written $\proj{C}{\pid p}$, returns the Endpoint choreography obtained
following the rules defined in \cref{fig:cc_projection}.

\begin{figure}
	$$
	\begin{array}{lcl}
		\genproj{\genstart;C}  & = &
			\begin{cases}
				\genreq ; \genproj{C} & 
					\mbox{if } \pid r = \pid p
				\\
				\acc{k}{l.\pid r[\role C]} ; \genproj{C} &
					\mbox{if } l.\pid r[\role C] \in \{\wtil{l.\pid q[\role B]}\}
				\\
				\genproj{C} & \mbox{otherwise}
			\end{cases}
			\\[20pt]
		\genproj{\gencom ; C} & = &
			\begin{cases}
				\gensend ; \genproj{C} & 
					\mbox{if } \pid r = \pid p
				\\
				\genrecv ; \genproj{C} &
					\mbox{if } \pid r = \pid q
				\\
				\genproj{C} & \mbox{otherwise}
			\end{cases}
	\\[20pt]
	\genproj{\genacc;C} & = &
 		\begin{cases}
			\acc{k}{l.\pid r[\role C]; \genproj{C}} &
				\mbox{if } l.\pid r[\role C] \in \{\wtil{l.\pid q[\role B]}\}
			\\
			\genproj{C} & \mbox{otherwise}
		\end{cases}
	\\[15pt]
	\genproj{\genreq ; C} & = &
			\begin{cases}
				\genreq ; \genproj{C} & \mbox{if } \pid r = \pid p
				\\
				\genproj{C} & \mbox{otherwise}
			\end{cases}
		\\[15pt]
		\genproj{\gensend ; C} & = &
			\begin{cases}
				\gensend ; \genproj{C} & \mbox{if } \pid r = \pid p
				\\
				\genproj{C} & \mbox{otherwise}
			\end{cases}
		\\[15pt]
				\genproj{\recDef{X}{C^\prime}{C}} & = &
				\recDef{X_{\pid r}}{
						\proj{\ C'[X_{\pid r}/X]\ }{\pid r}
					}{
						\proj{\ C[X_{\pid r}/X]\ }{\pid r}
					}
			\\[10pt]
			\genproj{X} & = & \recCall{X}
			\\[5pt]
			\genproj{\gencond} & = &
				\begin{cases}
					\cond{\pid p.e}{\epp{C_1}{\pid r}}{\epp{C_2}{\pid r}} 
					& \mbox{if } \pid r = \pid p
					\\
					\genproj{C_1} \sqcup \genproj{C_2} & \mbox{otherwise}
				\end{cases}
			\\[15pt]
			\genproj{\genbranchI} & = &
				\begin{cases}
					\com{k}{\role A}{\pid q[\role B].
						\{\op{o}_i(x_i);\epp{C_i}{\pid r}\}_{i\in I}} 	
							& \mbox{if } \pid r = \pid q
					\\
					\bigsqcup_{i\in I} \genproj{C_i} & \mbox{otherwise}
				\end{cases}
			\\[15pt]
			\genproj{C_1 \pp C_2} & = & \genproj{C_1} \pp \genproj{C_2}
			\\[15pt]
			\genproj{\inact} & = & \inact
		\end{array}
$$
	\caption{Frontend Choreographies --- process projection.}
	\label{fig:cc_projection}
\end{figure}

Process projection follows the structure of the source choreography. We
briefly comment the rules in \cref{fig:cc_projection}, from top to bottom.

We start with the complete actions $\prid{start}$ and $\prid{com}$ which, if
participated by the subject process, are projected into proper partial terms.
When projecting a $\prid{start}$ action, if the subject process is the active
process $\pid p$, we project a $\prid{req}$. If otherwise the subject process
is one of the service processes in $\pids q$, we project an (always-available)
$\prid{accept}$. Similarly, when projecting a $\prid{com}$ action, if the
subject process is the sender or the receiver in the interaction, we
respectively project a $\prid{send}$ or a $\prid{recv}$. Partial actions
$\prid{accept}$, $\prid{req}$, and $\prid{send}$ are projected verbatim, except
for $\prid{accept}$ terms, which define the availability of only the subject
process.

When projecting a $\prid{rec}$ term, we project both the body of the
procedure ($C'$) and the choreography $C$. This is safe even if $\pid r$ does
not take part into the body of $X$, indeed, in that case, the projection of
$C'$ is just an $\prid{inact}$ term. As a consequence, we can safely project
$\prid{call}$ terms verbatim.

The projections of conditionals and receptions are peculiar. Indeed, we
project a conditional verbatim if the subject process evaluates the
condition; for all other processes, we merge their behaviours with the
merging (partial commutative) operator $\sqcup$, defined by the rules
reported in Appendix (\cref{fig:merging_function}). $C \sqcup C'$ is defined
only for Endpoint choreographies and returns a choreography isomorphic to $C$
and $C'$ up to receptions, where all receptions with distinct operations are
also included. We use $\sqcup$ also in the projection of $\prid{recv}$ terms,
where we require the behaviour of all processes not receiving the message to
be merged.

When projecting two choreographies in parallel we return the parallel
composition of their respective projections, while $\prid{inact}$ is projected
verbatim.

Finally, we draw attention on the definition of the rule of process projection
for $\prid{rec}$ terms. Indeed, applying a na\"ive rule like
\[
	\genproj{\recDef{X}{C^\prime}{C}} \quad = \quad
	\recDef{X}{
			\proj{C'}{\pid r}
		}{
			\proj{C}{\pid r}
		}
\]
in the EPP would yeld more than one procedure with the same identifier, which
could prevent the obtained projection from being typable as, according to the
typing rules defined in \cref{sec:typing}, we cannot have in $\Gamma$ two
definition typings on the same identifier. To tackle the issue, the rule for
$\prid{rec}$ terms in \cref{fig:cc_projection} guarantees the coherent
definition and usage of process-unique identifiers through renaming. The
renaming is safe as, by assumption, we consider well-sorted choreographies where
definitions always precede recursive calls.

We conclude the paragraph with the formal definition of process projection.

\begin{definition}[Process Projection]\label{def:process_projection}
	$\proj{C}{\pid r}$ is a partial homomorphism from (Frontend/Backend)
	choreographies to Endpoint Choreographies, inductively defined by the rules in
	\cref{fig:cc_projection}.
\end{definition}

\paragraph{Endpoint Projection}
We can now proceed to define our Endpoint Projection.

In the definition below, we use the grouping operator $\group{C}{l}$, which
returns the set of all service processes accepting requests at location $l$.
We report in Appendix (\cref{fig:grouping}) the rules that inductively
define $\group{C}{l}$.

\begin{definition}[Endpoint Projection\label{def:epp}] Let $C$ be a
(Frontend/Backend) choreography. The endpoint projection of $C$, denoted by
$\proj{C}{}$, is defined as:
$$
\proj{C}{} =
	\underbrace{
		\raisebox{-2.5em}{}
		\prod\limits_{\pid p \ \in \ \fp(C)} \proj{C}{\pid p}
		}_{(i)} 
	\pp
	\underbrace{
	\raisebox{-2.5em}{}
	\prod_{l} \left( 
		\bigsqcup_{ \pid p \ \in \ \group{C}{l}} \proj{C}{\pid p} 
	\right)}_{(ii)}
$$
\end{definition}

Commenting \cref{def:epp}, the EPP of a choreography is the parallel composition
of two kinds of Endpoint choreographies: (\emph{i}) Endpoint choreographies that
are the process projection of active processes $\pid p \in \fp(C)$ and
(\emph{ii}) Endpoint choreographies that are the merge ($\sqcup$) of the process
projections of all service processes available at the same location $l$, i.e.,
$\pid p \in \group{C}{l}$.

\begin{example}
As an example of Endpoint Projection, let $C$ be the choreography at Lines
5--9 of \cref{example:intro} (for convenience, we report the mentioned
snippet of code grayed-out in the lower part of \cref{fig:example_epp}). The EPP of $C$,
$\epp{C}{}$, is the parallel composition of the process projections of
processes $\pid c$, $\pid s$, and $\pid b$, i.e., respectively $\epp{C}{\pid
c}$, $\epp {C} {\pid s}$, and $\epp {C}{\pid b}$. As per \cref{def:epp},
$\epp{C}{} = \epp{C}{\pid c}
\pp
\epp{C}{\pid s} \pp \epp{C}{\pid b}$.

\begin{figure}
	\centering
\[
	\hspace{-5pt}\begin{array}{l@{\hspace{5em}}l}
		\epp{C}{\pid b} =
		\begin{array}{l}
				\cond{\pid b.\lfn{confirm\_pay}(cc,\ order)}{
					\\[2pt] 
						\quad \com{k}{\prc bB}{\role C.\op{ok}};\;
						\com{k}{\prc bB}{\role S.\op{ok}} 
					\\[2pt] 
				}{
					\\[2pt] \quad \com{k}{\prc bB}{\role C.\op{ko}};\;
					\com{k}{\prc bB}{\role S.\op{ko}}
					\\[2pt]
				} 
		\end{array}
		&
		\begin{array}l
			\\
			\epp{C}{\pid c} = \com{k}{\role B}{\prc cC.\{ \; \op{ok}(),\; \op{ko}() \; \}}
			\\[20pt]
			\epp{C}{\pid s} = \com{k}{\role B}{\prc sS.\{ \; \op{ok}(),\; \op{ko}() \; \}}
		\end{array}
	\end{array}
\]
{\color{gray} \hrule}
{\color{black!66}
\[
C = \begin{array}l
\m{if}\ \pid b.\lfn{confirm\_pay}(\ cc,\ order\ ) \{ \\
 \quad \com{k}{\pid b[\role B]}{\pid c[\role C].\op{ok}()};\
 \com{k}{\pid b[\role B]}{\pid s[\role S].\op{ok}()}
\\ \}\ \m{else}\ \{\\\
 \quad \com{k}{\pid b[\role B]}{\pid c[\role C].\op{ko}()};\
 \com{k}{\pid b[\role B]}{\pid s[\role S].\op{ko}()}
\\ \}
\end{array}
\]
}

\caption{Example of Endpoint Projection (top-half) of Lines 5--9 of \cref{example:intro}
(lower-half).}
\label{fig:example_epp}
\end{figure}

We report in the top half of \cref{fig:example_epp} the projections
$\epp{C}{\pid c}$, $\epp {C} {\pid s}$, and $\epp {C}{\pid b}$. The example is
useful to illustrate that the projection of the conditional is homomorphic on
the process ($\pid b$) that evaluates it. The projection of a $\prid {com}$
term results into a partial $\prid{send}$ for the sender---as in the two
branches of the conditional in $\epp{C}{\pid b}$---and a partial $\prid
{recv}$ for the receiver---as in $\epp{C}{\pid c}$ and $\epp{C}{\pid s}$.
Note that the EPP merges branching behaviours: in $\epp{C}{\pid c}$ and
$\epp{C}{\pid s}$ the two complete communications are merged into a partial
reception on either operation $\op{ok}$ or $\op{ko}$.
\end{example}

\subsection{Properties}
\label{sec:epp_properties}

We conclude this section presenting the guarantees provided by the Endpoint
Projection wrt to the source Frontend choreography, as formalised in
\cref{thm:epp}. Before presenting \cref{thm:epp}, introduce the notion of
\emph{pruning} (as defined in \cite{CHY12}), where $\prec$ specifies an
asymmetric relation between two choreographies $C$ and $C'$, written $C \prec
C'$, in which $C$ prunes some unused accepts and receptions of $C'$. To give a
formal definition to our pruning relation, we present the two concepts of
subtyping of typing environments and minimal typing system. Below we just give
the intuition on both concepts, which are formalised in the Appendix:

\begin{itemize}

	\item given two typing environments $\Gamma$ and $\Gamma'$,
	$\Gamma$ is a subtype of $\Gamma'$, written $\Gamma \prec \Gamma'$, if
	$\Gamma$ is identical to $\Gamma'$ up to \emph{i}) some local and global
	types that are more constrained in $\Gamma$ than in $\Gamma'$ and \emph{ii})
	some service typings present in $\Gamma'$ and not present in $\Gamma$. We
	report the formal definition of $\Gamma \prec \Gamma'$ in
	\cref{def:environment_subtyping},

	\item the minimal typing system $\Gamma \hlseqmin{C}$ uses the
	minimal global and local types to type sessions and services in $C$. We
	report in \cref{sub:minimal_typing} the formal definition of minimal typing.

\end{itemize}

We can finally formalise the pruning relation.

\begin{definition}[Pruning\label{def:pruning}]
 Let $\Gamma\hlseqmin{C}$ and $\Gamma'\hlseqmin{C'}$, if $\Gamma \prec
 \Gamma'$ then $C$ prunes $C'$ under $\Gamma$, written $\Gamma \hlseqmin{C} \prec C'$, or $C \prec C'$ for short.
\end{definition}

The shortened form $C \prec C'$ is similar to \cite{CHY12}, where, as here, it
does not lose precision since it is always possible to reconstruct appropriate
typings. The pruning of $C'$ by $C$ means that $C$ omits unused inputs and
service processes present in $C'$. The $\prec$ relation is thus a strong
bisimulation since $C \prec C'$ means that the two choreographies have precisely
the same observable behaviours, except for the receive actions at pruned
receptions and unused available service processes.

We can now write the statement of our EPP Theorem.

\begin{theorem}[EPP Theorem\label{thm:epp}]
Let $\env,C$ be a well-typed Frontend choreography. Then,
\begin{enumerate}
\item (Well-typedness) $\env, \epp{C}{}$ is well-typed.
\item (Completeness) $\env,C \to \env',C'$ implies $\env,\epp{C}{} \to
	\env',C''$ and $\epp{C'}{} \prec C''$.
\item (Soundness) $\env,\epp{C}{} \to \env',C''$ implies $\env,C \to
\env',C'$ and $\epp{C'}{} \prec C''$.
\end{enumerate}
\end{theorem}

We report in \cref{sec:proof_of_epp} the proof of \cref{thm:epp}.






\subsection{From Backend Endpoint Choreographies to DCC (Compilation)}\label{sub:BC_to_DCC}

This is the last stage of our compilation process, where, given a parallel
composition of Backend Endpoint choreographies, we obtain a network of DCC
services that faithfully follow the semantics of the source choreography.

Given a Backend deployment  $\aenv$, a parallel composition of endpoint choreographies $C$, and a typing environment $\Gamma$, we write $\genenc{\aenv,C}$ to indicate the compilation of $\aenv,C$ under $\Gamma$ into DCC.
 
To formally define $\genenc{\aenv,C}$, we use some auxiliary functions:
\begin{itemize}
	\item $\filter{C}{l}$ returns the endpoint choreography in $C$
	correspondent to the service process accepting requests at location $l$
	(e.g., $\filter{C}{l} = \acc {k}{l.\prc pA};C''$);
	\item $\filter{C}{\pid p}$ returns the endpoint choreography in $C$
		correspondent to process $\pid p$;
	\item $\genenc{C}$, given a single endpoint choreography $C$ and a typing environment $\Gamma$, compiles $C$ to DCC, using the
	rules in \cref{fig:encoding};
	\item $l \in \Gamma$, a predicate satisfied if, according to $\Gamma$,
	location $l$ contains or can spawn processes;
	\item $\filter{\aenv}{l}$ returns the partial function of type $\Trees
	\rightharpoonup \mathit{Seq}(\Operations\times\Trees)$ that corresponds to the
	projection of function $g_m$ in $\aenv$ with location $l$ fixed. Formally, for
	each $t$ such that $\aenv(l:t) = \til m$, $\filter{\aenv}{l}(t) = \til m$.
\end{itemize} 
%
%
%
%

%
\begin{definition}[Compilation\label{def:compilation}]
Let $\aenv$ be a Backend deployment, $C$ a parallel composition of endpoint
choreographies, and given the typing environment $\Gamma$
\[
\genenc{\aenv,C} \; =
\hspace{-5pt}\begin{array}l
	\;\prod\limits_{
		\begin{array}c
			l \in \Gamma
		\end{array}}
	\hspace{-5pt}
	\jsrv{
		\genenc{\filter{C}{l}},
		\prod\limits_{\pid p \ \in \ \aenv(l)}
			\jpr{\genenc{\filter{C}{\pid p}}}
				{\aenv(\pid p)}
		\ ,\
		{\filter{\aenv}{l}}
	}{\!\!l}
\end{array}
\]
\end{definition}
%
Intuitively, for each service $\jsrv{\strBhv,P,M}l$ in the compiled network:
\emph{i}) the start behaviour $\strBhv$ is the compilation of the endpoint
choreography in $C$ accepting the creation of processes at location $l$;
\emph{ii}) $P$ is the parallel composition of the compilation of all active
processes located at $l$, equipped with their respective states according to
$\aenv$; \emph{iii}) $M$ is the set of queues in $\aenv$ corresponding to
location $l$.

\begin{figure}
$$
\hspace{-5pt}
\begin{array}{l}
	\text{Let } \pid p@l' \in \Gamma, \
	
	\genenc{
		\req{k}{\pid p[\role A]}{\wtil{l.\role B}};C} \quad = \quad
			\auxfn{start}(k, l'.\role A, \wtil{l.\role B}); \genenc{C} 
	\\[5pt]

	\mbox{
	\small$
	\auxfn{start}(k, l_\role A.\role A, \wtil{l_\role B.\role B}) =
	\underbrace{
		\Seq\limits_{
			\vphantom{\raisebox{-5pt}{C}}\substack{ \role I \in \{\role A, 
			\til{\role B}\}}
		}	\jpath{k.I.l} = l_{\role I}\ ;
	}_{s_1}
	\underbrace{
		\Seq\limits_{\vphantom{\raisebox{-5pt}{C}}\role I \in \{\til{\role B}\}}
			\left(	\begin{array}l
				\cq{\jpath{k.I.A}} \ ; \\ 
				\notify{?}{\jpath{k.I.l}}{\jpath{k}} \ ; \\ 
				\oneway{\op{sync}}{\jpath{k}} \qfrom \jpath{k.I.A}	
			\end{array}
			\right);
	}_{s_2}
	\underbrace{
		\Seq\limits_{\vphantom{\raisebox{-5pt}{C}}\role{I} \in \{\til{\role B}\}}
			\notify{\op{start}}{\jpath{k.I.l}}{\jpath{k}} \qto \jpath{k.A.I}
	}_{s_3}
	$}

	\\[35pt]

	\text{Let } l \in \til l, \til l \in \Gamma, \

		\genenc{\acc{k}{l.\pid q[\role B]}; C} \hspace{3em} = \quad
		\auxfn{accept}(\ k,\ \role B, \Gamma(\til l) \ );
		\genenc{C}\ , 

	\\[5pt]
		\mbox{\small$
		\auxfn{accept}(k, \role B,\initg{G}{\role A}{\til{\role C}}{\roles D}) = \
			\underbrace{\vphantom{\raisebox{-15pt}{C}}
				\oneway{!}{\jpath{k}} \ ;
			}_{a_1}
			\underbrace{\vphantom{\raisebox{-15pt}{C}}
				\Seq\limits_{	\role I \in \{\role A,\til{\role C}\}\setminus
			\{\role B\}}
					\Big( \cq{\jpath{k.I.B}} \Big) ;
			}_{a_2}
			\underbrace{\vphantom{\raisebox{-15pt}{C}}
				\notify{\op{sync}}{\jpath{k.A.l}}{\jpath{k}} \qto{\jpath{k.B.A}};
			}_{a_3}
			\underbrace{\vphantom{\raisebox{-15pt}{C}}
				\oneway{\op{start}}{\jpath{k}} \qfrom{\jpath{k.A.B}}
			}_{a_4}
		$}
	\\[35pt]
		
		\begin{array}{l@{\hspace{58pt}}ll}
				
			\genenc{\gensend;C} & = &
		 		\notify{\op{o}}{\jpath{k.B.l}}{e} \qto{\jpath{k.A.B}}; \genenc{C}
		
		\\[15pt]

			\genenc{\genbranchI} & = & \sum\limits_{i \in I}\choice{\oneway{\op{o}_i}{x_i} 
						\qfrom{\jpath{k.A.B}}}{\genenc{C_i}\ }

		\\[15pt]

			\genenc{\cond{\pid p.e}{C_1}{C_2}} & = & \cond{e}{\genenc{C_1}\ }{\genenc{C_2}\ }

		\\[8pt]
		
		 	\genenc{\recDef{X}{C'}{C}} & = & \recDef{X}{\genenc{C'}}{\genenc{C}}

		\\[5pt]
			
			\genenc{X} & = & X

	 	\\[5pt]

			\genenc{\inact} & = & \inact
		\end{array}
		
	\end{array}
	$$
\caption{Compiler from Endpoint Choreographies to DCC.}
\label{fig:encoding}
\end{figure}
We comment the rules in \cref{fig:encoding}, where the notation $\Seq$ 
is the sequence of behaviours $\Seq_ {i \in [1,n]}(B_i) = B_1 ; \ldots
; B_n$.

\paragraph{Requests}
Function $\m{start}$ defines the compilation of $\prid{req}$ terms. Function
$\m{start}$ compiles $\prid{req}$ terms to create the queues and a part of the
session descriptor for the starter (this is similar to what rule
\did{\aenv}{Sup} does in Backend deployment transitions, \cref{sec_aEC}).
%
%
Given a session identifier $k$, the located role of the starter ($l_\role
A.\role A$), and the other located roles in the session ($\wtil {l_\role
B.\role B}$), $\m{start}$ returns the DCC code that: 
\begin{itemize}
	\item[$s_1$] includes in the session descriptor all the locations of the
processes involved in the session;
	\item[$s_2$] for each role, except for the starter,
		\begin{itemize}
			\item creates the key and the correlated queue that the current role will use in the session to communicate with the starter;
			\item requests the creation of the service process that will play the current role in the session;
			\item waits on the reserved operation $\mathit{sync}$ to receive the correlation data for the session defined by the newly created process.
		\end{itemize}
	\item[$s_3$] sends to the newly created processes the complete session
	descriptor obtained after the reception (in the $\mathit{sync}$ step) of all
	correlation keys.
\end{itemize}

%

\paragraph{Accepts}
$\prid {acc}$ terms define the start behaviour of a spawned process at a
location.
Given a session identifier $k$, the role $\role B$ of the service process, and
the service typing $\initg{G}{\role A}{\til{\role C}}{\roles D}$ of the
location, function $\m{accept}$
%
compiles the code that: ($a_1$) accepts the request to spawn a process,
($a_2$) creates its queues and keys, updates the session descriptor
received from the starter, and sends it back to the latter ($a_3$). Finally with ($a_4$) the new process waits to start the session.


\paragraph{Other terms}
A $\prid{send}$ term compiles to a DCC $\prid{output}$ term. Notably,
the compiled code contains the same elements used by the semantics of BC to
implement correlation, i.e., the location of the receiver ($\jpath{k.B.l}$) and
the key that correlates with its queue ($\jpath{k.A.B}$).
Similarly, $\prid{recv}$ compiles to $\prid{choice}$, which defines the path
($\jpath{k.A.B}$) of the key correlating with the receiving queue.
%



\begin{example}
As an example of compilation, we compile the first two lines of the choreography
$C$ in \cref{example:intro}, considering a deployment $\aenv$ and a typing
environment $\Gamma$ such that \(\Gamma\hlseq{\aenv,C}\).
$$
	\genenc{\aenv,\epp{C}{}} = 
		\jsrv{\inact,P_{\pid c}}{l_{\role C}} 	\pp
		\jsrv{\strBhv_{\role S},\inact}{l_{\role S}} \pp
		\jsrv{\strBhv_{\role B},\inact}{l_{\role B}}
$$
where
$$P_{\pid c} = \left\{
\begin{array}{l}
\jpath{k.S.l} = l_{\role S}; 												\quad \jpath{k.B.l}=l_{\role B};
\quad \cq{\jpath{k.S.C}};																	\quad?@\jpath{k.S.l}( \jpath{k} );
\quad \op{sync}(\jpath{k})\qfrom{\jpath{k.S.C}};								\\[2pt]
\cq{\jpath{k.B.C}};																	\quad?@\jpath{k.B.l}( \jpath{k} );
\quad \op{sync}(\jpath{k})\qfrom{\jpath{k.B.C}};								\quad
\op{start}@\jpath{k.S.l}(\jpath{k}) \qto{\jpath{k.C.S}};	\\[2pt]
\op{start}@\jpath{k.B.l}(\jpath{k}) \qto{\jpath{k.C.B}};	\quad
{\color{gray}\texttt{ /* end of start-request */ }}	\\[2pt]
\op{buy}@\jpath{k.S.l}(product) \qto{\jpath{k.C.S}};			\quad
\ldots
\end{array}\right.$$

and

\[\strBhv_{\role S} = \left\{
\begin{array}{l}
!(\jpath{k}); 																			\quad
\cq{\jpath{k.C.S}};																	\quad
\cq{\jpath{k.B.S}};																	\quad
\op{sync}@\jpath{k.C.l}(\jpath{k}) \qto{\jpath{k.S.C}}; 	\\[2pt]
\op{start}(\jpath{k}) \qfrom{\jpath{k.C.S}}; 						\quad
{\color{gray}\texttt{ /* end of accept */ }}				\quad
\op{buy}(x) \qfrom{\jpath{k.C.S}}; 											\quad
\ldots
\end{array}\right.\]

We omit to report $\strBhv_{\role B}$, which is similar to $\strBhv_{\role S}$.
\end{example}

\subsection{Properties of Applied Choreographies} 
We conclude this section by presenting our main result, i.e., a compiler from
Frontend choreographies to DCC networks and its properties.

In our definition, we use the term \emph{projectable} to indicate that, given a choreography $C$, we can obtain its projection $\epp{C}{}$. Formally

\begin{definition}[Projectable Choreography]
	Let $C$ be a choreography, we call $C$ \emph{projectable} if there is a choreography $C$ such that $C' = \epp{C}{}$.
\end{definition}

\cref{thm:applied_choreographies} defines our result, for which, given a
well-typed, projectable Frontend choreography, we can obtain its correct
implementation as a DCC network. Such result is obtained by merging the
properties of the stages \textbf{FC-to-BC}(\cref{sub:encodig}),
\textbf{EPP}(\cref{sub:endpoint_projection}), and
\textbf{Compilation}(\cref{sub:BC_to_DCC}).

\begin{theorem}[Applied Choreographies\label{thm:applied_choreographies}] Let
$\env,C$ be a Frontend choreography where $C$ is projectable and $\Gamma
\hlseq{\env, C}$ for some $\Gamma$. Then:
\begin{enumerate}[topsep=-10pt,itemsep=-5pt,partopsep=1ex,parsep=1ex]
 \item (Completeness)
 $\env, C \to \env',C'$ implies \\
	\[\begin{array}{ccccc}
			\genenc{\geneenc{\env},\epp{C}{}} \to^+ \enc{\eenc{\env'}{\Gamma'},C''}^{\Gamma'}
			& \mbox{ and } &
			\epp{C'}{} \prec C''
			& \mbox{ and } &
			\mbox{for some } \Gamma',\ \Gamma' \hlseq{\env',C'}	
	\end{array}\]
 \item (Soundness)
 $\genenc{\geneenc{\env},\epp{C}{}} \to^* S$ implies 
	\[\begin{array}{ccccccc}
 			\env,C \to^* \env', C'
			& \mbox{ and } &
			S \to^* \enc{\eenc{\env'}{\Gamma'},C''}^{\Gamma'}
			& \mbox{ and } &
			\epp{C'}{} \prec C''
			& \mbox{ and } &
			\mbox{for some } \Gamma',\ \Gamma' \hlseq{\env',C'}	
	\end{array}\]
\end{enumerate}
\end{theorem}

We report in \cref{sub:proof_of_fc_to_dcc_compilation} the proof of
\cref{thm:applied_choreographies}. 

By \cref{thm:deadlock-freedom} and \cref{thm:applied_choreographies},
deadlock-freedom is preserved from well-typed choreographies to their final
translation in DCC\@. We say that a network $S$ in DCC is deadlock-free if it
is either a composition of services with terminated running processes or it can
reduce.

\begin{corollary} $\Gamma \hlseq{\env,C}$ and $\cofn(\Gamma)$ imply that
$\enc{\env,\epp{C}{}}^{\Gamma}$ is deadlock-free.
\end{corollary}

\IFSubFileBiblio

\section{Related Work and Discussion}
\label{sec:related}

This is the first work that formalises how we can use choreographies in the
setting of a practical communication mechanism used in Service-Oriented
Computing (SOC), i.e., message correlation. Previous formal choreography
languages specify only an EPP procedure towards a calculus based on name
synchronisation, leaving the design of its concrete support to implementors.
Chor~\cite{chor:website} and AIOCJ~\cite{aiocj:website} are the respective
implementations of the models found in~\cite{CM13} and~\cite{DGGLM15}.
However, the implementations of their EPP depart significantly from their
respective formalisations, since they are based on message correlation instead
of name synchronisation.
This means that there is no proof that the implementation strategies followed in
these languages correctly supports synchronisation on names.
%
%
Implementations of other frameworks based on sessions share similar
issues~\cite{HYH08,HNYDH13,NY14}. Our work gives the first correctness result
for the compilation of choreographies to a language close to real-world
implementations. More in general, our results are a useful reference to
formalise the implementation of session-based languages. In the
future, this line of work may pave the way to establishing certified
choreography compilation.
We believe that our approach can be easily applied to many models that use choreographies and 
sessions (or channel-based communications), including those designed around 
(variants of) the $\pi$-calculus~\cite{CHY12,CM13,MY13,HYC16} and those based on linear 
logic~\cite{CMS17,CMSY17}.

Our development shows that it is possible to keep a simple language model as
frontend, allowing developers to abstract from how sessions are concretely
implemented. Nevertheless, our Frontend Choreographies are expressive, as
illustrated by our examples, and recent studies have shown that choreography
languages such as this are Turing complete~\cite{CM16}. There are many works
that investigate how to introduce different features to choreographies, which
we have not studied here and leave to future work. Examples include nested
protocols~\cite{DH12}, asynchronous two-way exchanges~\cite{CMS17}, and
general recursion~\cite{CM17}. These features are orthogonal to our
development, so their inclusion should be straightforward. A more interesting
feature to add may be session delegation for choreographies~\cite{CM13,HYC16}.
Delegation allows to transfer the responsibility to continue a session from a
process to another. Introducing delegation in FC is straightforward, since we
can just import the development from~\cite{CM13,MY13}. Implementing it in BC
and DCC would be more involved, but not difficult: delegating a role in a
session translates to moving the content of a queue from a process to another,
and ensuring that future messages reach the new process. The mechanisms to
achieve the latter part have been investigated in~\cite{HYH08}, which use
retransmission protocols. Formalising these ``middleware'' protocols and
proving that they preserve the intended semantics of FC could be an
interesting future work.

%

%

In the semantics of BC, we abstract from how correlation keys are generated.
With this loose definition we capture several implementations, provided they
satisfy the requirement of uniqueness of keys (wrt to locations). As future
work, we plan to implement a language, based on our framework, able to support
custom procedures for the generation of correlation keys (e.g., from database
queries, cookies, etc.).

\IFSubFileBiblio

\section{Conclusion}
\label{sec:conclusion}

In this paper, we presented our framework of Applied Choreographies, which
includes three calculi: a high-level choreographic language intended for
developers, an intermediate-representation choreographic language, and a
low-level, close-to-implementation distributed calculus. We equip our framework
with a tight series of behavioural correspondences, so that we guarantee that
low-level distributed programs compiled from high-level sources faithfully
follow their source specifications. By pairing our compilation with a type
system and static checks that guarantee the absence of deadlocks in high-level
choreographies, we obtain that the compiled distributed systems are
deadlock-free. Specifically, we target Service-Oriented distributed systems that
communicate over correlation mechanisms.

Besides the contribution above, Applied Choreographies introduce a novel
semantics for choreographies that abstracts the features of choreographies
(message passing, creation of new sessions and processes) from their
implementation (and the related complexity).
To this end we \emph{i}) equip choreographies with a global deployment and
\emph{ii}) define a separate semantics of effects on deployments.
This separation allows us to compose our semantics of choreographies with other
definitions of deployment and effects so that we have a straightforward way to
capture different communication semantics (e.g., synchronous, asynchronous with
buffers) and implementations (e.g., distributed objects~\cite{ChinC91}). 
The notion of deployments let us formalise how choreographies can go wrong (see
\cref{sec:runtime_typing}) and show that the theory of session types is useful
not only to type communications on choreographies (\cite{CM13,MY13}) but also to
check the correctness of deployments.
%
It is worth noting that, except for the declaration of locations, Applied
Choreographies has the same types and syntax from previous
works~\cite{CM13,MY13}, hence developers have only to specify protocols and
choreographies and do not need to deal with deployment information or
correlation data.

We have already mentioned some short-term future work in the previous section.
More long term projects include the investigation of compilation to other target
languages/communication mechanisms besides correlation-based ones-orientation,
for instance those found in Erlang and Scala+Akka. Clearly this would be a major
development, since the actor-based concurrency and message passing of these
languages are substantially different from that based on correlation, considered
in this paper.
%
%
Another ambitious goal is the application of our research to the Internet of
Things (IoT) setting. IoT promotes the communication among heterogeneous
entities---which use a wide range of communication media and data
protocols--whose integration result in a cumbersome low level programming
activity. Indeed, to achieve a higher degree of
interoperability~\cite{HICSS18,GGLZ19} propose the use of high-level,
service-oriented languages for communication technology integration in IoT
systems. In particular, an extension of Jolie is introduced~\cite{HICSS18,GGLZ19}
which natively integrates the two most adopted protocols for IoT communication
(CoAP and MQTT). We plan to take this approach further by developing a suitable
version of Applied Choreographies, specifically designed for IoT applications,
which can then be compiled to the Jolie extension mentioned above. This would
allow one to import in the IoT field the correct-by-construction approach
through the formal correctness of compilation that we have developed in this
paper.

\IFSubFileBiblio

\bibliographystyle{alpha}
\bibliography{biblio}

\newcommand{\etalchar}[1]{$^{#1}$}
\begin{thebibliography}{{{AIO}}16}

\bibitem[Agh85]{agha85}
Gul~A Agha.
\newblock Actors: A model of concurrent computation in distributed systems.
\newblock Technical report, MASSACHUSETTS INST OF TECH CAMBRIDGE ARTIFICIAL
  INTELLIGENCE LAB, 1985.

\bibitem[{{AIO}}16]{aiocj:website}
{{AIOCJ} Team}.
\newblock {AIOCJ framework}, 2016.
\newblock \url{http://www.cs.unibo.it/projects/jolie/aiocj.html}.

\bibitem[B{\etalchar{+}}14]{JSON}
Tim Bray et~al.
\newblock The javascript object notation (json) data interchange format, 2014.

\bibitem[BGG{\etalchar{+}}06]{choOrchCorrespondence}
Nadia Busi, Roberto Gorrieri, Claudio Guidi, Roberto Lucchi, and Gianluigi
  Zavattaro.
\newblock Choreography and orchestration conformance for system design.
\newblock In {\em {COORDINATION}}, pages 63--81. Springer, 2006.

\bibitem[BPSM{\etalchar{+}}98]{XML}
Tim Bray, Jean Paoli, C~Michael Sperberg-McQueen, Eve Maler, and Fran{\c{c}}ois
  Yergeau.
\newblock Extensible markup language (xml).
\newblock {\em W3C Recommendation REC-xml-19980210}, 16, 1998.

\bibitem[CC91]{ChinC91}
Roger~S. Chin and Samuel~T. Chanson.
\newblock Distributed object-based programming systems.
\newblock {\em {ACM} Computing Surveys}, 23(1):91--124, 1991.

\bibitem[CD88]{DistributedSystems}
George~F. Coulouris and Jean Dollimore.
\newblock {\em Distributed Systems: Concepts and Design}.
\newblock Addison-Wesley Longman Publishing Co., Inc., Boston, MA, USA, 1988.

\bibitem[CDCYP15]{CDYP2015}
Mario Coppo, Mariangiola Dezani-Ciancaglini, Nobuko Yoshida, and Luca Padovani.
\newblock Global progress for dynamically interleaved multiparty sessions.
\newblock {\em MSCS}, 760:1--65, 2015.

\bibitem[CES71]{deadlocks}
Edward~G Coffman, Melanie Elphick, and Arie Shoshani.
\newblock System deadlocks.
\newblock {\em ACM Computing Surveys (CSUR)}, 3(2):67--78, 1971.

\bibitem[{Cho}16]{chor:website}
{Chor Team}.
\newblock {Chor Programming Language}, 2016.
\newblock \url{http://www.chor-lang.org/}.

\bibitem[CHY12]{CHY12}
Marco Carbone, Kohei Honda, and Nobuko Yoshida.
\newblock Structured communication-centered programming for web services.
\newblock {\em ACM Transactions on Programming Languages and Systems (TOPLAS)},
  34(2):1--78, 2012.

\bibitem[CLM05]{CLM05}
Samuele Carpineti, Cosimo Laneve, and Paolo Milazzo.
\newblock Bopi - {A} distributed machine for experimenting web services
  technologies.
\newblock In {\em {ACSD}}, pages 202--211. IEEE, 2005.

\bibitem[CLM17]{CLM17}
Lu{\'{\i}}s Cruz{-}Filipe, Kim~S. Larsen, and Fabrizio Montesi.
\newblock The paths to choreography extraction.
\newblock In {\em FoSSaCS}, volume 10203 of {\em Lecture Notes in Computer
  Science}, pages 424--440, 2017.

\bibitem[CM13]{CM13}
Marco Carbone and Fabrizio Montesi.
\newblock Deadlock-freedom-by-design: multiparty asynchronous global
  programming.
\newblock In {\em POPL}, pages 263--274, 2013.

\bibitem[CM16]{CM16}
Lu{\'{\i}}s Cruz{-}Filipe and Fabrizio Montesi.
\newblock A core model for choreographic programming.
\newblock In {\em {FACS}}, volume 10231 of {\em Lecture Notes in Computer
  Science}, pages 17--35, 2016.

\bibitem[CM17]{CM17}
Lu{\'{\i}}s Cruz{-}Filipe and Fabrizio Montesi.
\newblock Procedural choreographic programming.
\newblock In {\em {FORTE}}, volume 10321 of {\em Lecture Notes in Computer
  Science}, pages 92--107. Springer, 2017.

\bibitem[CMS17]{CMS17}
Marco Carbone, Fabrizio Montesi, and Carsten Sch{\"u}rmann.
\newblock Choreographies, logically.
\newblock {\em Distributed Computing}, pages 1--17, 2017.
\newblock Also: CONCUR, pages 47--62, 2014.

\bibitem[CMSY17]{CMSY17}
Marco Carbone, Fabrizio Montesi, Carsten Sch{\"{u}}rmann, and Nobuko Yoshida.
\newblock Multiparty session types as coherence proofs.
\newblock {\em Acta Informatica}, 54(3):243--269, 2017.

\bibitem[DGG{\etalchar{+}}15]{DGGLM15}
Mila {Dalla Preda}, Maurizio Gabbrielli, Saverio Giallorenzo, Ivan Lanese, and
  Jacopo Mauro.
\newblock Dynamic choreographies.
\newblock In {\em {COORDINATION}}, pages 67--82. Springer, 2015.

\bibitem[DGL{\etalchar{+}}14]{DGLMG14}
Mila {Dalla Preda}, Saverio Giallorenzo, Ivan Lanese, Jacopo Mauro, and
  Maurizio Gabbrielli.
\newblock {AIOCJ:} {A} choreographic framework for safe adaptive distributed
  applications.
\newblock In {\em {SLE}}, pages 161--170, 2014.

\bibitem[DGL{\etalchar{+}}17]{DGLMMMS16}
Nicola Dragoni, Saverio Giallorenzo, Alberto~Lluch Lafuente, Manuel Mazzara,
  Fabrizio Montesi, Ruslan Mustafin, and Larisa Safina.
\newblock Microservices: yesterday, today, and tomorrow.
\newblock In {\em Present and ulterior software engineering}, pages 195--216.
  Springer, 2017.

\bibitem[DH12]{DH12}
Romain Demangeon and Kohei Honda.
\newblock Nested protocols in session types.
\newblock In {\em CONCUR}, pages 272--286, 2012.

\bibitem[GGLZ18]{HICSS18}
Maurizio Gabbrielli, Saverio Giallorenzo, Ivan Lanese, and Stefano~Pio Zingaro.
\newblock A language-based approach for interoperability of iot platforms.
\newblock In {\em 51st Hawaii International Conference on System Sciences,
  {HICSS} 2018, Hilton Waikoloa Village, Hawaii, USA, January 4-7, 2018}, 2018.
\newblock {T}o appear.

\bibitem[GGLZ19]{GGLZ19}
Maurizio Gabbrielli, Saverio Giallorenzo, Ivan Lanese, and Stefano~Pio Zingaro.
\newblock Linguistic abstractions for interoperability of iot platforms.
\newblock In {\em Towards Integrated Web, Mobile, and IoT Technology}, pages
  83--114. Springer, 2019.

\bibitem[GH05]{GH05}
Simon Gay and Malcolm Hole.
\newblock Subtyping for session types in the pi calculus.
\newblock {\em Acta Informatica}, 42(2-3):191--225, November 2005.

\bibitem[GMG18]{GMG18}
Saverio Giallorenzo, Fabrizio Montesi, and Maurizio Gabbrielli.
\newblock Applied choreographies.
\newblock In {\em Formal Techniques for Distributed Objects, Components, and
  Systems - 38th {IFIP} {WG} 6.1 International Conference, {FORTE} 2018, Held
  as Part of the 13th International Federated Conference on Distributed
  Computing Techniques, DisCoTec 2018, Madrid, Spain, June 18-21, 2018,
  Proceedings}, pages 21--40. Springer, 2018.

\bibitem[HLV{\etalchar{+}}16]{HLVCCDMPRT16}
Hans H{\"u}ttel, Ivan Lanese, Vasco~T Vasconcelos, Lu{\'\i}s Caires, Marco
  Carbone, Pierre-Malo Deni{\'e}lou, Dimitris Mostrous, Luca Padovani,
  Ant{\'o}nio Ravara, Emilio Tuosto, et~al.
\newblock Foundations of session types and behavioural contracts.
\newblock {\em ACM Computing Surveys (CSUR)}, 49(1):1--36, 2016.

\bibitem[HNY{\etalchar{+}}13]{HNYDH13}
Raymond Hu, Rumyana Neykova, Nobuko Yoshida, Romain Demangeon, and Kohei Honda.
\newblock Practical interruptible conversations.
\newblock In {\em {RV}}, pages 130--148, 2013.

\bibitem[HO07]{haller07}
Philipp Haller and Martin Odersky.
\newblock Actors that unify threads and events.
\newblock In {\em Coordination Models and Languages}, pages 171--190. Springer,
  2007.

\bibitem[HYC08]{HYC08}
K.~Honda, N.~Yoshida, and M.~Carbone.
\newblock Multiparty asynchronous session types.
\newblock In {\em Proc. of POPL}, volume 43(1), pages 273--284. ACM, 2008.

\bibitem[HYC16]{HYC16}
Kohei Honda, Nobuko Yoshida, and Marco Carbone.
\newblock Multiparty asynchronous session types.
\newblock {\em Journal of the ACM (JACM)}, 63(1):1--67, 2016.

\bibitem[HYH08]{HYH08}
Raymond Hu, Nobuko Yoshida, and Kohei Honda.
\newblock Session-based distributed programming in java.
\newblock In {\em ECOOP}, pages 516--541, 2008.

\bibitem[{Int}96]{MSC}
{International Telecommunication Union}.
\newblock Recommendation \mbox{Z.120}: Message sequence chart, 1996.

\bibitem[{JBo}13]{savara:website}
{JBoss Community}.
\newblock Savara, 2013.
\newblock \url{http://www.jboss.org/savara/}.

\bibitem[LGMZ08]{LGMZ08}
I.~Lanese, C.~Guidi, F.~Montesi, and G.~Zavattaro.
\newblock Bridging the gap between interaction- and process-oriented
  choreographies.
\newblock In {\em {SEFM}}, pages 323--332. {IEEE}, 2008.

\bibitem[MC11]{MC11}
Fabrizio Montesi and Marco Carbone.
\newblock Programming services with correlation sets.
\newblock In {\em {ICSOC}}, pages 125--141, 2011.

\bibitem[MGZ14]{MGZ14}
Fabrizio Montesi, Claudio Guidi, and Gianluigi Zavattaro.
\newblock Service-oriented programming with {J}olie.
\newblock In {\em Web Services Foundations}, pages 81--107. 2014.

\bibitem[Mil80]{M80}
Robin Milner.
\newblock {\em A Calculus of Communicating Systems}, volume~92 of {\em {LNCS}}.
\newblock Springer, 1980.

\bibitem[Mon13]{M13:phd}
Fabrizio Montesi.
\newblock {\em Choreographic Programming}.
\newblock Ph.{D}. thesis, IT University of Copenhagen, 2013.
\newblock
  \url{http://www.fabriziomontesi.com/files/choreographic_programming.pdf}.

\bibitem[MPW92]{MPW92}
Robin Milner, Joachim Parrow, and David Walker.
\newblock A calculus of mobile processes, {I and II}.
\newblock {\em Information and Computation}, 100(1):1--40,41--77, September
  1992.

\bibitem[MY13]{MY13}
Fabrizio Montesi and Nobuko Yoshida.
\newblock Compositional choreographies.
\newblock In {\em CONCUR}, pages 425--439, 2013.

\bibitem[New15]{n15}
Sam Newman.
\newblock {\em Building microservices: designing fine-grained systems},
  chapter~4.
\newblock O'Reilly Media, Inc., 2015.

\bibitem[NM92]{raceConditions}
Robert~HB Netzer and Barton~P Miller.
\newblock What are race conditions? some issues and formalizations.
\newblock {\em ACM Letters on Programming Languages and Systems (LOPLAS)},
  1(1):74--88, 1992.

\bibitem[NS78]{needham1978using}
Roger~M Needham and Michael~D Schroeder.
\newblock Using encryption for authentication in large networks of computers.
\newblock {\em Communications of the ACM}, 21(12):993--999, 1978.

\bibitem[NY14]{NY14}
Rumyana Neykova and Nobuko Yoshida.
\newblock Multiparty session actors.
\newblock In {\em {COORDINATION}}, pages 131--146, 2014.

\bibitem[OAS07]{bpel}
OASIS.
\newblock {WS-BPEL}.
\newblock \url{http://docs.oasis-open.org/wsbpel/2.0/wsbpel-v2.0.html}, 2007.

\bibitem[O'H18]{o18}
Peter O'Hearn.
\newblock Experience developing and deploying concurrency analysis at facebook.
\newblock In {\em International Static Analysis Symposium}, pages 56--70.
  Springer, 2018.

\bibitem[OMG04]{UML}
OMG.
\newblock Unified modelling language, version 2.0, 2004.

\bibitem[{OMG}11]{BPMN}
{OMG}.
\newblock {B}usiness {P}rocess {M}odel and {N}otation.
\newblock \url{http://www.omg.org/spec/BPMN/2.0/}, 2011.

\bibitem[Pie02]{PIERCE}
Benjamin~C. Pierce.
\newblock {\em Types and Programming Languages}.
\newblock MIT Press, MA, USA, 2002.

\bibitem[QZCY07]{QZCY07}
Zongyan Qiu, Xiangpeng Zhao, Chao Cai, and Hongli Yang.
\newblock Towards the theoretical foundation of choreography.
\newblock In {\em WWW}, pages 973--982. IEEE Computer Society Press, 2007.

\bibitem[SW01]{SW01}
D.~Sangiorgi and D.~Walker.
\newblock {\em The \(\pi\)-calculus: a Theory of Mobile Processes}.
\newblock Cambridge University Press, 2001.

\bibitem[Vin06]{vinoski06}
Steve Vinoski.
\newblock Advanced message queuing protocol.
\newblock {\em IEEE Internet Computing}, 10(6), 2006.

\bibitem[VW12]{videla12}
Alvaro Videla and Jason~JW Williams.
\newblock {\em RabbitMQ in action: distributed messaging for everyone}.
\newblock Manning, 2012.

\bibitem[{W3C}04]{wscdl}
{W3C WS-CDL Working Group}.
\newblock {WS-CDL} version 1.0, 2004.
\newblock \url{http://www.w3.org/TR/2004/WD-ws-cdl-10-20040427/}.

\end{thebibliography}

\newpage

\appendix
 
\section{Additional Material} 

\subsection{Typing}

\begin{definition}[List Subset\label{par:list_subset_predicate}]
Let $\emptyseq$ be the empty list and $\til N$, $\til M$ be two lists of
elements $n$ of the kind $\til N \gram \emptyseq \ | \ n,\til N'$, the predicate
$\til N \subseteq \til M$ holds if $\til N = \til M = \emptyseq$ or, assuming
$\til N = n, \til N'$ and $\til M = m,\til M'$ either $n = m$ and $\til N'
\subseteq \til M'$ or $\til N \subseteq
\til M'$.
\end{definition}

\begin{definition}[Ordered Join Operator\label{par:ordered_join_operator}]
Let $\til N$, $\til L$, and $\til M$ be three lists of elements as defined
in \cref{par:list_subset_predicate}, the ordered-join operator $\til N
\bowtie_{\til L} \til N$ is defined as
$$
\begin{array}{lll} 
  \til N \bowtie_{\emptyseq} \til M & = & \emptyseq
  \\
  \til N \bowtie_{l, \til L} \til M & = & 
  \begin{cases}
    \til N \bowtie_{\til L} \til M &
    \mbox{if } l \not \in {\til N} \cup {\til M}
    \\
    l, \til {N'} \bowtie_{\til L} \til M &
    \mbox{if } \til N = l, \til {N'}
    \\
    l, \til N \bowtie_{\til L} \til {M'} &
    \mbox{if } \til M = l, \til {M'}
  \end{cases}
\end{array}
$$
\end{definition}

\subsection{Compiling Frontend Choreographies into DCC Processes} 

\newpage

\begin{figure}[!ht]
	\[
	\begin{array}{rcl}
		\begin{array}l
		\acc{k}{l.\pid p[\role A]}; C_1
		\; \sqcup \\
		\acc{k}{l.\pid q[\role A]};C_2 
		\end{array}
		& = & \acc{k}{l.\pid p[\role A]} ; (C_1 \sqcup C_2)
		\\[3em]
		\!\begin{array}l
			\genreq; C_1
			\; \sqcup \\[2pt]
			\req{k}{\prc{q}{A}}{\wtil{l.\role B}}; C_2
		\end{array}
		& = & \genreq;(C_1 \sqcup C_2)
		\\[3em]
		\begin{array}l
			\gensend; C_1
			\; \sqcup \\
			\com{k}{\pid q[\role A].e}{\role B.o}; C_2
		\end{array}
		& =	& \gensend; (C_1 \sqcup C_2)
		\\[3em]
		\begin{array}l
		\com{k}{\role A}{\prc{p}{B}.\{\ o_i(x_i); C_i \ \}_{i \in I} }
		\; \sqcup \\[2pt]
		\com{k}{\role A}{\prc{q}{B}.\{\ o_j(x_j); C'_j \ \}_{j \in J} }
		\end{array}
		& = & 
		\com{k}{\role A}{\prc{p}{B}.\left\{
		\begin{array}{ll}
		& \{\ o_i(x_i) ; C_i\ \}_{i \in I \setminus J}\\[2pt]
		\cup & \{\ o_i(x_i) ; C'_i\ \}_{i \in J \setminus I}\\[2pt]
		\cup & \{\ o_i(x_i) ; C_i \sqcup C'_i\ \}_{i \in I\cap J}
		\end{array}
		\right\}}
		\\[3em]
		\!\begin{array}l
			\cond{\pid p.e}{C_1}{C_1'} \; \sqcup \\[2pt]
			\cond{\pid q.e}{C_2}{C_2'}
		\end{array}
		& = &
		\cond{\pid p.e}{C_1 \sqcup C_2}{C_1' \sqcup C_2'}
		\\[3em]
		\begin{array}l
		\recDef{X}{C_1'}{C_1} \; \sqcup \\
		\recDef{Y}{C_2'}{C_2}	
		\end{array}
		& = &\recDef{X}{C_1'\sqcup C'_2}{C_1 \sqcup C_2}
		\\[3em]
		X \; \sqcup \; Y & = & X
		\\[2em]
		\inact \; \sqcup \; \inact & = & \inact
	\end{array}
	\]
	\caption{Merging Function}
	\label{fig:merging_function}
\end{figure}

\newpage 

\begin{figure}[!ht]
\[
\begin{array}{lcl}
	\group{
	\start{k}{\pid p[\role D]}
		{\wtil{l.\pid q[\role B]}};C}{l} \quad & = & 
			\group{\genacc;C}{l}
\\[1em]
	\group{\genacc;C}{l} & = &
	\begin{cases}
		\{\pid r\} \cup \group{C}{l} 
			& \mbox{if } l.\pid r[\role A] \in \{\ \wtil{l.\pid q[\role B]}\ \}
			\\
			\group{C}{l} & \mbox{otherwise}
	\end{cases}
	\\[2em]
		\group{\eta;C}{l} & = & \group{C}{l} \quad \mbox{if } \eta \neq \prid{start}
	\\[1em]
		\group{\cond{\pid p.e}{C_1}{C_2}}{l} & = &
		\group{C_1}{l} \cup \group{C_2}{l}
	\\[1em]
		\group{\recDef{X}{C'}{C}}{l} & = &
		\group{C'}{l} \cup \group{C}{l}
	\\[1em]
	\group{X}{l}  & = & \emptyset 
	\\[1em]
	\group{\inact}{l} & = &	\emptyset
	\\[1em]
	\group{C_1 \pp C_2}{l} & = & \group{C_1}{l} \cup \group{C_2}{l}
\end{array}
\]
\caption{Service Grouping}
\label{fig:grouping}
\end{figure}

\newpage

\section{Proofs}
\label{sec:appendix}
\newcommand{\proofsDir}{proofs}

\subsection{Proofs of Subject Reduction and Session Fidelity}
\label{sec:proof_of_typing}

In order to prove Subject Reduction (\cref{thm:paper_subj_red}), we prove the
stronger result of Typing Soundness, defined in \cref{thm:sub_red}. We use
\cref{thm:sub_red} to also prove Session Fidelity (\cref{thm:session_fidelity}).

In order to define and prove \cref{thm:sub_red}, we provide additional
definitions and lemmas, in particular:

\begin{itemize}
	
	\item we define an annotated semantics for FC (\cref
	{sub:ac_annotated_semantics}) to track reductions on sessions;
	
	\item we define subtyping (\cref{sub:local_subtyping}) for local types and for
	typing environments. On these definitions we prove lemmas used to relate
	evolutions of the typing environment wrt reductions in choreographies;

	\item we define an annotated semantics for global types
	(\cref{sub:reductions_for_global_types}) and prove
	\cref{lemma:session_fidelity}, guaranteeing that global types and local types
	in the typing environment evolve accordingly.

\end{itemize}

Finally, we proceed to prove Typing Soundness
(\cref{ssub:proof_of_typing_soundness}) and consequently Subject Reduction and
Session Fidelity.

\subsubsection{FC Annotated Semantics}
\label{sub:ac_annotated_semantics}

We define the semantics of annotated FCs by marking transitions with the name
of the session whose term has reduced. We annotate other reductions as $\tau$.
We range over annotated labels with 
$$\beta \quad \gram \quad \com{k}{\role A}{\role B.o} \quad \Div \quad k:\grecv
{\role A} {\role B} {o(x)} \quad \Div \quad \tau$$

We report the annotated semantics of FC in \cref
{fig:cc_annotated_semantics}. Intuitively, we mark reductions over a
session $k$ with $\com{k}{\role A}{\role B.o}$ for message sends ($\did{C}
{Send}$ and $\did{C}{Com}$) and $k:\grecv
{\role A} {\role B} {o(x)}$ for receptions ($\did{C}{Recv}$).

\begin{figure}
\begin{displaymath}
\begin{array}{c}
\infer[\did{C}{Start}]
{
	\dep,\ \genstart; C
	\quad \lto {\startlab{k'}{\pid p}{\pids r}} \quad
	\dep',\
	C[k'/k][\pids r/\pids q] 
}
{
	\env\fresh{k',\pids r}
	&
	\delta = \start{k'}{\pid p[\role A]}{\wtil{l.\pid q[\role B]}}	
	&
	\renv{\dep}{\delta}{\dep'}
}
\\[10pt]
\infer[\did{C}{Send}]
{
	\dep,\ \eta; C
	\quad \lto {\com{k}{\role A}{\role B.o}} \quad
	\dep',\ C
}{
	\eta = \gensend
	&
	\renv{\dep}{\eta}{\dep'}
}
\\[10pt]
\infer[\did{C}{Recv}]
{
	\dep,\ \genbranchI
	\quad \lto {k:\grecv{\role A} {\role B}{o_j(x_j)} } \quad
	\dep',\ C_j
}{
	j \in I
	&
	\renv{\dep}{\com{k}{\role A}{\pid q[\role B].o_j(x_j)}}{\dep'}
}
\\[10pt]
\infer[\did{C}{Cond}]
{
	\dep,\ \cond{\pid p.e}{C_1}{C_2}
	\quad\lto{\tau}\quad
	\dep,\ C_i
}
{
	i = 1 \text{ if } \evalfn(\ e, \env(\pid p)\ ) = \mathtt{ true },
	i = 2 \text{ otherwise}
}
\\[10pt]
\infer[\did{C}{Ctx}]
{
	\dep, \recDef{X}{C_2}{C_1} \quad \lto \beta \quad
	\dep^\prime, \recDef{X}{C_2}{C^\prime_1}
}
{
	\dep, C_1 \lto \beta \dep^\prime, C^\prime_1
}
\\[10pt]
\infer[\did{C}{Eq}]
{
	\dep, C \quad \lto \beta \quad \dep^\prime, C'
}
{
	\ctx{R} \in \{\,\equiv\,,\,\swapC\,\}
	&
	C \,\ctx{R}\, C_1
	&
	\dep, C_1 \lto \beta \dep^\prime, C_1'
	&
	C_1 \,\ctx{R}\, C'
}
\\[5pt]
\infer[\did{C}{Par}]
{
	\dep, C_1\pp C_2
	\quad \lto \beta \quad
	\dep', C'_1 \pp C_2
}
{
	\dep, C_1
	\quad \lto \beta \quad
	\dep', C'_1
}
\\[10pt]
\infer[\did{C}{PStart}]
{
\begin{array}l
	\dep,
	\req{k}{\pid p[\role A]}
	{\wtil{l.\role B}}; C
	\pp
	\prod_i
	\big(
	\genacci; C_i
	\big)
	\lto{\pstartlab{k'}{\pid p}{\pids r_1,\cdots,\pids r_n}}
	\\ \hspace{1em}
	\dep',\ 
	C[k'/k] \pp \prod_i\big(\ C_i[k'/k][\pids r_i /\pids q_i] \ \big)
	\pp
	\prod_i
	\big(
	\genacci; C_i
	\big)
\end{array}
}
{
	i \in \{1,\dots,n\}
	\hfill
	\env\fresh{k',\pids r}
	\hfill
	\{ \wtil{l.\role B} \} = \biguplus_i \{\wtil{l_i.\role B_i}\}
	\hfill
	\{\pids r\} = \bigcup_i \{ \pids r_i \}
	\\[2pt]
	\delta = \start{k'}{\pid p[\role A]}{\wtil{l_1.\pid r_1[\role
	B_1]},\dots,\wtil{l_n.\pid r_n[\role B_n]}}
	\qquad
	\renv{\dep}{\delta}{\dep'}
}
\end{array}
\end{displaymath}
\caption{Fronted Choreographies --- annotated semantics.}
\label{fig:cc_annotated_semantics}
\end{figure}




\subsubsection{Local Types and Typing Environment Subtyping} 
\label{sub:local_subtyping}

We define a subtyping relation on local types
following~\cite{GH05,CHY12,MY13}. We write the subtyping relation as $T'
\prec T$, which intuitively indicates that $T'$ is more constrained than $T$
in its behaviour. Note that, like in~\cite{CHY12,MY13}, the input type is
covariant and the output type is contravariant for this relation.

\begin{definition}[Local Subtyping\label{def:local_subtyping}] We define the
subtyping relation between local types as $T' \prec T$, which is the smallest
relation over closed local types, satisfying the rules
$$
\begin{array}c
\fulltextwidth
\infer[\did{SubT}{Eq}]
	{T \prec T'}
	{T'' \prec T' & T \approx T''} 
\qquad
\infer[\did{SubT}{Send}]
	{ !\role A.\{o_i(U_i) ; T_i\}_{i \in I} \ \prec \ !\role A.\{o_i(U'_i) ;
		T'_i\}_ {i \in J} } {J \subseteq I & \forall\ i \in J\ |\ T_i \prec T'_i \
		\wedge \ U_i \prec U'_i}
\\[5pt]
\infer[\did{SubT}{Recv}]
	{
		?\role A.\{o_i(U_i) ; T_i\}_{i \in I} \ \prec \ 
		?\role A.\{o_i(U'_i) ;	T'_i\}_ {i \in J}}
	{I \subseteq J & \forall\ i \in I\ |\ T_i \prec T'_i \ \wedge \ U_i \prec
	U'_i}
\qquad
\infer[\did{SubT}{Val}]
{
	U \prec U
}
{}
\qquad
\infer[\did{SubT}{End}]
	{ \gend \prec T }
	{ \gend \approx T }
\end{array}
$$	
\end{definition}

In rule $\did{SubT}{Eq}$, $T \prec T'$ if there exists a local type $T''$,
subtype of $T'$, such that $T \approx T''$, i.e., $T''$ approximates $T$,
$\approx$ being the standard tree isomorphism on recursive types.

Although not directly relevant in the current proof, we also define the
subtyping for global types $G \prec G'$, which intuitively follows that of
local ones. Subtyping for global types is used in the definition of
Environment subtyping. The relation between subtyping of Environments and of
global types (in service typings) will become relevant when proving
properties of our Endpoint Projection (see \cref{sec:proof_of_epp}). Our
definition of subtyping for global types follows~\cite{MY13}.

\begin{definition}[Global Subtyping\label{def:global_subtyping}]
  $G \prec G'$ is the smallest relation over closed global types
  satisfying the rules below
  $$
    \begin{array}c
      
    \infer[\did{SubG}{Com}]
    {
      \gcom{\role A}{\role B}{\{o_i(U_i);G_i\}_{i \in I}} \prec
      \gcom{\role A}{\role B}{\{o_j(U'_j);G'_j\}_{j \in J}}
    }
    {
      I \subseteq J
      &
      \forall\ i \in I,\ G_i \prec G'_i \ \wedge U_i \prec U_i'
    }
    \\[10pt]
    \infer[\did{SubG}{Recv}]
    {
      \grecv{\role A}{\role B}{o(U);G} \prec
      \grecv{\role A}{\role B}{o(U');G'}
    }
    {
      U \prec U'
      &
      G \prec G'
    }
    \\[10pt]
    \infer[\did{SubG}{Eq}]
    {
      G \prec G'
    }
    {
      G'' \prec G'
      &
      ( G'' \approx G \ \vee \ G'' \swapG G )
    }
    \qquad
    \infer[\did{SubG}{End}]
    { \gend \prec G }
    { \gend \approx G }
    \end{array}
  $$

\end{definition}

Finally, we define a subtyping relation between Typing Environments.
Intuitively $\Gamma \prec \Gamma'$ means that $\Gamma'$ and $\Gamma$ are
identical Typing Environments up to \emph{a}) some local and global types
that are more constrained in $\Gamma$ --- i.e., subtypes of a correspondent
global/local type --- than in $\Gamma'$ and \emph{b}) some service typings
not present in $\Gamma$.

\begin{definition}[Typing Environment Subtyping\label
{def:environment_subtyping}] Let $\Gamma$ and $\Gamma'$ be two typing
  environments, where $\Gamma' = \Gamma'',\Gamma_{l}\ $, for which $\dom(\Gamma) = \dom(\Gamma'')$ and $\Gamma_l$ contains only service typings. Then, $\Gamma \prec \Gamma'$ if and only if
  \[
		\begin{array}{lll}
      (\mathit{i}) & \forall\ \pair{\pid p.x}{U} \in \Gamma, 
        & \Gamma' \hlseq{ \pair{\pid p.x}{U} }
      \\
      (\mathit{ii}) & \forall\ \pair{X}{\Gamma_x} \in \Gamma,
        & \Gamma' \hlseq{ \pair{X}{\Gamma_x} }
      \\
      (\mathit{iii}) & \forall\ \pair{\pid p}{k[\role A]} \in \Gamma, 
        & \Gamma' \hlseq{ \pair{\pid p}{k[\role A]} }
      \\
      (\mathit{iv}) & \forall\ \pid p@l \in \Gamma,
        & \Gamma' \hlseq{ \pid p@l }
      \\
      (\mathit{v}) & \forall\ \pair{\chanto{k}{\role A}{\role B}}{T} \in
        \Gamma, & \Gamma' \hlseq{ \pair{\chanto{k}{\role A}{\role B}}{T} }
      \\
      (\mathit{vi}) & \forall\ \pair{k[\role A]}{T} \in \Gamma, 
        & \Gamma' \hlseq{ \pair{k[\role A]}{T'} } \mbox{ and } T \prec T'
      \\
      (\mathit{vii}) & \forall\ \pair{\til l}{\serviceTyping{G}{\role
      A}{\roles B}{\roles C}} \in \Gamma, & \Gamma' \hlseq{\pair{\til
      l}{\serviceTyping{G'}{\role A}{\roles B}{\roles C}}} \mbox{ and } G
      \prec G'
    \end{array}
  \]
\end{definition}

Commenting the definition, the subtyping relation for typing environments
states that an environment $\Gamma$ is a subtype of an environment $\Gamma'$
if

\begin{itemize}
  
  \item they type the same variables (\emph{i}), procedure definitions (\emph{ii}), role ownerships (\emph{iii}), process locations (\emph{iv}), and buffers (\emph{v}) and they agree on their judgements;

  \item they type the same local sessions (\emph{vi}) and the local type in $\Gamma$ is a subtype of the local type in $\Gamma'$;

  \item if they type the same service (\emph{vii}) (note that $\Gamma'$ is allowed to have additional service typings wrt $\Gamma$) and the global type in $\Gamma$ is a subtype of the global type in $\Gamma'$.

\end{itemize}

In \cref{lem:subsumption} we prove that if $\Gamma \prec \Gamma'$ and
$\Gamma$ types a running choreography $\dep,C$ also $\Gamma'$ types that
choreography.

\begin{lemma}[Subsumption\label{lem:subsumption}]
Let $\Gamma \prec \Gamma'$ and $\Gamma \hlseq {\dep,C}$ for some $\dep, C$
then $\Gamma' \hlseq{\dep,C}$.
\end{lemma}

\begin{proof}
	The proof is immediate by \cref{def:local_subtyping} and rules
	$\did{T}{Recv}$, $\did {T}{Send}$, and $\did {T}{Com}$. Intuitively, the
	lemma holds since the local typings in $\Gamma'$ allow for additional,
	unused actions in $\dep,C$.
\end{proof}

We also prove \cref{lem:up_to_buffer} which guarantees that the typing of
choreographies ($C$) is invariant wrt buffer types.

\begin{lemma}[Buffer types invariance\label{lem:up_to_buffer}] Let $\Gamma =
\Gamma',\Gamma_b$ where $\Gamma_b$ contains only buffer typings. If $\Gamma'
\hlseq C$ then $\Gamma \hlseq C$.
\end{lemma}

\begin{proof} 
	Trivial from the definition of rule $\did{T}{\dep C}$ and $\Gamma \hlseq{C}$
	for which buffer typings affect only predicate $\pco$ and the typing of
	deployments.
\end{proof}

\subsubsection{Reductions for Global Types}
\label{sub:reductions_for_global_types}

We annotate the reductions of global types with labels $$\gamma \quad \gram
\quad \gcom{\role A}{\role B}{o} \qquad \Div \qquad \grecv{\role A}{\role
B}{o}$$ and report below the correspondent annotated semantics.

$$
\hspace{-5pt}\begin{array}c
  \infer[\did{G}{Send}]
  {
    \gchoice{\role A}{\role B}{\{\op{o}_i(U_i)\}; G} 
    \quad
		\lto{\gcom{\role A}{\role B}{o}}
    \quad
    G'
  }
  {
    o \in \bigcup_i\{o_i\}
    &
    G' = \gapply{
      \role A \rangle \role B
    }{
      o
    }{
      G
    }
  }
  \qquad
  \infer[\did{G}{Recv}]
  {
    \grecv{\role A}{\role B}{\op{o}(U)}; G 
    \quad
		\lto{ \grecv{\role A}{\role B}{o} }
    \quad
    G
  }
  {\ }
  \\[10pt]
  \infer[\did{G}{Eq}]
  {
    G \lto{\gamma} G'
  }
  {
    \mathcal{R} \in \{\equivG, \swapG\}
    &
    G \ \mathcal{R}\ G_1
    &
    G_1 \lto{\gamma} G_1'
    &
    G_1' \ \mathcal{R} \ G'
  }
\end{array}
$$

In \cref{lem:proj_sub} we account for the fact that any output reduction at the
level of global types can constrain the projected local types of the roles not
involved in the reduction. Indeed, referring to rule $\did{G}{Send}$, the output
operation chooses one of the available continuations $G'$ and discards all the
others. Therefore the local types of the other roles not involved in the
reduction can be constrained by the removal of the unused branches.

\begin{lemma}[Projection Subtyping\label{lem:proj_sub}]
Let $T = \epp{G}{\role C}$, $T' = \epp{G'}{\role C}$, and $\{\role A,\role B,
\role C\} \subseteq \auxRoles(G)$, $\role C \not \in \{\role A,\role B\}$, then $G
\lto{\gcom{\role A}{\role B}{o}} G'$ implies $T' \prec T$.
\end{lemma}

\begin{proof}
	By induction on the derivation of $G \lto{\gamma} G'$.
\end{proof}

\subsubsection{Typing Environment Reductions}
\label{ssub:typing_environment_reductions}

We define a reduction relation for typing environments. To do so, we first
formalise the writing $k
\not \in \Gamma$, which means that $\Gamma$ has no local typing and buffer
types for session $k$, formally, for some local types $T$ and $T'$
$$
k \not \in \Gamma \iff \nexists\; \role A, \role B \mbox{ s.t. } \pair{k[\role
A]}{T} \in \Gamma \ \vee \pair{\chanto{k}{\role A}{\role B}}{T'} \in
\Gamma
$$

Finally, we formalise the reduction relation for typing environments of the
form $\Gamma \to \Gamma'$, $\to$ being the smallest closed under the rules
below. Note that the annotation labels are a subset of the labels used to
annotate the semantics of FC, ranged over by $\beta$.

$$\begin{array}c	
	\infer[\did{\Gamma}{Send}]
	{
		\Gamma,\Gamma_k
		\lto{\com{k}{\role A}{\role B.o_j}}
		\Gamma,
		\{ \pair{k[\role C]}{\epp{G'}{\role C}} \ | \ k[\role C] \in \Gamma_k \},
		\{\pair{\chanto{k}{\role C}{\role D}}{\epp{G'}{\role C}^{\role D}}
			\ | \ \chanto{k}{\role C}{\role D} \in \Gamma_k \}
	}
	{
		k \not \in \Gamma 
		&
		\Gamma_k \subseteq \epp{G}{k}
		&
		\{\pair{k[\role A]}{T}, \pair{k[\role B]}{T'}\} \in \Gamma_k
		&
		j \in I
		&
		G \lto{\gcom{\role A}{\role B}{o_j}} G'
	}
	\\[10pt]
	\infer[\did{\Gamma}{Recv}]
	{
		\Gamma,\Gamma_k
		\lto{k:{\grecv{\role A} {\role B}{o_j(x)}}}
		\Gamma,
		\{ \pair{k[\role C]}{\epp{G'}{\role C}} \ | \ k[\role C] \in \Gamma_k \},
		\{\pair{\chanto{k}{\role C}{\role D}}{\epp{G'}{\role C}^{\role D}}
			\ |\ \chanto{k}{\role C}{\role D} \in \Gamma_k \},
		\pair{\pid q.x }{U_j}
	}
	{
			k \not \in \Gamma
			&
			\Gamma_k \subseteq \epp{G}{k}
			&
			\{\pair{k[\role A]}{T}, \pair{k[\role B]}{T'}\} \in \Gamma_k
			&
			\Gamma \seq \pair{\pid q}{k[\role B]}
			&
			G \lto{\grecv{\role A}{\role B}{o_j}} G'
	}
	\end{array}
$$
\vspace{2em}

With slight abuse of notation, we also write $\beta_k$ to mark reductions of
$\Gamma$ on session $k$, i.e., $\beta_k \in \{\com{k}{\role A}{\role B.o},\
k:\grecv{\role A}{\role B}{o(x)}\}$.

We define the correspondence operator $G_{\m{act}}(\beta)$ between
$\beta$ and $\gamma$ labels:

$$
G_{\m{act}}(\beta_k) = \begin{cases}
	\gcom{\role A}{\role B}{o} & \mbox{if } \beta_k = \com{k}{\role A}{\role B.o}
	\\
	\grecv{\role A}{\role B}{o} & \mbox{if } \beta_k = k:\grecv{\role A}{\role B.o(x)}
\end{cases} 
$$

In \cref{lemma:session_fidelity} we prove that if a typing environment
$\Gamma$ includes local types that are projection of a global type $G$, then
if the global type can reduce, also the typing environment can reduce. The
reduction preserves the correspondence between the reduced global type and
the reduced local types in $\Gamma$.

\begin{lemma}[Type-Environment Fidelity\label{lemma:session_fidelity}]
	Let $\Gamma = \Gamma_*,\epp{G}{k}$ for some $\Gamma_*$, $k \not \in
	\Gamma_*$, and $G \lto{G_\m{act}(\beta_k)} G'$ then $\Gamma \lto {\beta_k}
	\Gamma'$ and for some $\Gamma_*'$, $k \not \in \Gamma_*'$, $\Gamma' =
	\Gamma_*',\epp{G'}{k}$.
\end{lemma}

\begin{proof}
	Direct by cases on the derivation of $\Gamma$.
\end{proof}

\subsubsection{Proof of Typing Soundness}
\label{ssub:proof_of_typing_soundness}

We also report \cref{lemma:subj_cong,lemma:subj_swap} that prove that typing is
invariant wrt structural equivalence and swapping.

\begin{lemma}[Subject Congruence\label{lemma:subj_cong}]
	$\Gamma \hlseq {\dep,C}$ and $C \equivC C'$ imply $\Gamma \hlseq {\dep,C'}$
	(up to $\alpha$-renaming)
\end{lemma}

\begin{proof}
	By induction on the rules that define $\equivC$.
\end{proof}

\begin{lemma}[Subject Swap\label{lemma:subj_swap}]
	$\Gamma \hlseq{ \dep, C}$ and $C \swapC C'$ imply $\Gamma \hlseq {\dep,C'}$
\end{lemma}

\begin{proof}
	By induction on the derivation of $C \swapC C'$.
\end{proof}
Below we restate the definition of \emph{Deployment Judgements} enriched with
pointers of the kind $\ilab{\Def X.Y}$ for a clearer referencing in the proofs.

\vspace{1em\!}
\def\defDJ{\Def|\ref{def:deployment_judgements}}
\noindent\textbf{\cref{def:deployment_judgements}} (Deployment Judgements)
\\$ \Gamma \hlseq{\env} \iff $
\begin{itemize}
	\step{\defDJ.1} $\forall\ \pid p.x \in \Gamma, \env( \pid p).x : U$

	\step{\defDJ.2} $\forall\ \chanto{k}{\role A}{\role B}: T \in \Gamma
	\wedge \env(\chanto{k}{\role A}{\role B}) = \til m, \ \btypeop(\role A,
	\til m ) = T$
\end{itemize}

Finally, we prove \cref{thm:paper_subj_red} by proving the stronger
result \cref{thm:sub_red}.

In the proof, we use the context over global types $\mathcal{G}[\cdot]$, defined
as
$$
\begin{array}{rl}
\mathcal{G}[ \cdot ] \gram &
  \gcom{\role A}{\role B}{\{ \op{o}_i(U_i) ; \mathcal{G}[\cdot] \}_{i}}
  \\[2pt]
  \Div & \gchoice{\role A}{\role B}{\{\op{o}_i(U_i)\}};\mathcal{G}[\cdot] 
  \\[2pt]
  \Div & \gbranch{\role A}{\role B}{\{\op{o}_i(U_i);\mathcal{G}[\cdot]\}_{i
  \in I}}
  \\[2pt]
  \Div & \grecv{\role A}{\role B}{\op{o}(U)};\mathcal{G}[\cdot] 
\end{array}
$$

We can now proceed to define and prove \cref{thm:sub_red}.

\def\thmSR{\Thm|\ref{thm:sub_red}}
\begin{theorem}[Typing Soundness\label{thm:sub_red}]
	Let $\dep,C$ be an annotated FC and 
	$\ilab{\thmSR.1}$ $\Gamma \hlseq {\dep,C}$ for some $\Gamma$:
	
	\begin{description}
	
		\item if $\ilab{\thmSR.2}$ $\beta \neq \tau$ and $\dep,C \lto{\beta}
			\dep',C'$\, then $\ilab{\thmSR.3}$ $\Gamma \lto \beta \Gamma'$ and
			$\ilab{\thmSR.4}$ $\Gamma' \hlseq {\dep',C'}$;
	
		\item if \ilab{\thmSR.5} $\dep,C \lto{\tau}  \dep',C'$ then, for some
		$\Gamma'$, \ilab{\thmSR.6} $\Gamma' \hlseq {\dep',C'}$.

	\end{description}
\end{theorem}

\begin{proof}
	Proof by induction on the derivation of $\dep,C \lto{\beta} \dep',C'$.
	\begin{itemize}
		\case{$\did{C}{Send}$} The case is:
			$$
			\infer[\did{C}{Send}]
			{
				\dep,\ \eta; C
				\quad \lto {\com{k}{\role A}{\role B.o_j}} \quad
				\dep',\ C
			}{
				\eta = \com{k}{\pid p[\role A].e}{\role B.o_j}
				&
				\renv{\dep}{\eta}{\dep'}
			}
			$$
			Where \ilab{\thmSR.2} has the reductum $C' = C$ and, let $v =
			\evalfn(e,\dep(\pid p))$ and $\til m = \dep(\chanto{k}{\role A}{\role
			B})$, $\dep'= \dep\big[\ \chanto{k}{\role A}{\role B} \mapsto \til m::(o_j,
			v)\ \big]$ by rule $\did {D}{Send}$.

			To prove \ilab{\thmSR.3} we must prove rule $\did{\Gamma}{Send}$ to be
			applicable.

			From \ilab{\thmSR.1} we know that there exists a global type $G$ for
			session $k$ such that $\pco( \Gamma )$ holds. We can partition $\Gamma =
      \Gamma_*,\Gamma_k$ such that $\Gamma_* = \Gamma \setminus {\epp{G}{k}}$
      and $\Gamma_k = \Gamma \setminus {\Gamma_*}$.

			From \ilab{\thmSR.1} we can write the derivation (with $\Gamma = \Gamma_1,
			\pair{k[\role A]}{\lsend{B}{\{o_i(U_i) ; \epp{G_i}{\role A}\}_{i\in
			I}}}$ ) 

			$$
			\infer[\did{T}{\dep C}]
			{
				\Gamma \hlseq{\dep,\gensendj;C}
			}
			{
				\pco( \Gamma )
				&
				\Gamma \hlseq \dep
				&
				\infer[\did{T}{Send}]
				{
					\Gamma_1,\pair{k[\role A]}{\lsend{B}{\{o_i(U_i) ; \epp{G_i}{\role A}\}_{i\in I}}}
					\hlseq{\gensendj; C}
				}
				{
					j \in I
					&
					\Gamma_1 \seq \pair{\pid p}{k[\role A]}
					&
					\Gamma_1 \seq \pair{\pid p.e}{U_j}
					&
					\Gamma_1, \pair{k[\role A]}{\epp{G_j}{\role A}} \hlseq{C }
				}
			}
			$$

			Since $\Gamma \seq \pair{k[\role A]} {\lsend{\role B}{\{o_i(U_i);T_i\}_{i
      \in I}}}$, we can write $G = \mathcal{G}[\gcom{\role A}{\role B}{{o_i
      (U_i);G_i}}]$ where $\forall\ i \in I$, $\epp{G_i} {\role A} = T_i$. Let
      $\pi$ be the reduction of $G$ with rules $\did{G}{Eq}$ and $\did{G}
      {Send}$, we observe the following derivation:

    {\scriptsize
      \vspace{1em}
      \[
      \hspace{-10em}
      \pi = \left\{\hspace{-4em}
      \begin{minipage}{.75\textwidth}
      \[      
      \infer[\did{G}{Eq}]{
          G \lto{\gamma} G'
      }{
        G \equivG G_1
        \hspace{-1em}
        &
        \infer*[\did{G}{Eq}]{
          G_1 \lto{\gamma} G'
        }{
          G_1 \swapG G_2
          \hspace{-1em}
          &
          \infer[\did{G}{Send}]{
            G_2 \lto{\gamma} G'
          }{
            o_i \in U_i\{o_i\}
            &
            G' = \Delta
          }
          &
          \hspace{-3em}
          G' \swapG G'
        }
        &
        \hspace{-3em}
        G \equivG G'
      }
      \]
      \end{minipage}
      \hspace{-3em}
      {\color{lightgray}\rule[-5em]{.5px}{10em}}
      \begin{minipage}{.15\textwidth}
      \[
        \begin{array}l
        \Delta = \gapply{\role A \rangle \role B}{o_i}{
          \mathcal{G}[ \gbranch{\role A}{\role B}{\{o_i(U_i\};G_i} ]}
        \\[1em]
        G_2 = \gchoice{\role A}{\role B}{\{o_i(U_i)\}};
          \mathcal{G}[\gbranch{\role A}{\role B}{\{o_i(U_i\};G_i}]
        \\
        G_1 = \mathcal{G}[\gchoice{\role A}{\role B}{\{o_i(U_i\}};
        \gbranch{\role
        A}{\role B}{\{o_i(U_i\};G_i}]
        \\[2em]
        G' = \mathcal{G}[\grecv{\role A}{\role B}{o_j};G_j]
        \\
        \gamma = \gcom{\role A}{\role B}{o_j}
        \end{array}
      \]
      \end{minipage}
      \vspace{1em}
      \right.\]
    }

    In the reductions, since $C,\env$ reduces with $\beta =
    \com{k}{\role A}{\role B.o_j}$ and $G$ types $C,\env$ in $\Gamma$, there are
    no other exchanges from $\role A$ to $\role B$ in $G$ that could prevent
    from obtaining, after a finite number of derivations on rule $
    \did{G}{Eq}$, the swap-equivalence $G_1 \swapG G_2$. Following a similar
    reasoning, the application $\Delta$ targets the global branching in the
    context, which reduces the continuation $\mathcal{G}[ \gbranch{\role
    A}{\role B}{\{o_i (U_i\};G_i}]$ after the global choice $\gchoice{\role
    A}{\role B}{\{o_i(U_i)\}}$ to $G'$.

    Given $\pi$, we can use it to write the reduction at the level the typing
    environment $\Gamma$, applying rule $\did{\Gamma}{Send}$. Below, we consider
    $\Gamma = \Gamma_*,\Gamma_k$ where $\Gamma_k$ contains all and only typings
    of session $k$ in $\Gamma$.
 
			\[\infer[\did{\Gamma}{Send}]
			{
				\Gamma_*,\Gamma_k
				\lto{\com{k}{\role A}{\role B.o_j}}
				\Gamma_*,
				\{ \pair{k[\role C]}{\epp{G'}{\role C}} \ | \ k[\role C] \in \Gamma_k \},
				\{\pair{\chanto{k}{\role C}{\role D}}{\epp{G'}{\role C}^{\role D}}
					\ | \ \chanto{k}{\role C}{\role D} \in \Gamma_k \}
			}
			{
        k \not \in \Gamma_* 
        &
        \Gamma_k \subseteq \epp{G}{k}
        &
        \{\pair{k[\role A]}{T},\pair{k[\role B]}{T'}\} \in \Gamma_k
        &
        j \in I
        &
				\infer*{
          G \lto{\gcom{\role A}{\role B}{o_j}} G'
        }{
          \pi
		    }
			}\]

			Hence \ilab{\thmSR.3} holds and $\Gamma' = \Gamma_*,
      \{ \pair{k[\role C]}{\epp{G'}{\role C}} \ | \ k[\role C] \in \Gamma_k \},
      \{\pair{\chanto{k}{\role C}{\role D}}{\epp{G'}{\role C}^{\role D}}
      \ | \ \chanto{k}{\role C}{\role D} \in \Gamma_k \}$. We now prove
      $\ilab{\thmSR.4}$ by proving that rule $\did{T}{\dep C}$ applies to
      $\Gamma' \hlseq {\dep',C'}$.

			$$
			\infer[\did{T}{\dep C}]
			{
				\Gamma' \hlseq{\dep',C'}
			}
			{
				\pco( \Gamma' )
				&
				\Gamma' \hlseq C'
				&
				\Gamma' \hlseq \dep'
			}
			$$
			
			Hence we need to prove \pnum{1} $\pco( \Gamma' )$, \pnum{2} $\Gamma' \hlseq
			C'$, and \pnum{3} $\Gamma'\hlseq \dep'$

			\begin{proof}[of \pnum{1}]

				For all sessions $k' \in \Gamma_*$, $\pco( \Gamma' )$ holds as $ \pco(
				\Gamma )$ holds by \ilab{\thmSR.1}. For session $k$, $\pco( \Gamma' )$
				holds by construction.

			\end{proof}

			\begin{proof}[of \pnum{2}]

			From the derivation on $\Gamma \hlseq {\dep, \com{k}{\pid p[\role
			A].e}{\role B.o_j};C}$ we know that $\Gamma_1, \pair{k[\role
			A]}{\epp{G_j} {\role A}}\hlseq{C}$.
      Let $\Gamma'' = \Gamma_1,\pair{k[\role A]}{\epp{G_j} {\role A}}$ and
      $\Gamma_k' = \Gamma_1 \setminus \Gamma_* = \Gamma_k \setminus
      \{\pair{k[\role A]}{\epp {G}{\role A}}\}$. We can write $\Gamma'' =
			\Gamma_*,\Gamma_k',\pair{k[\role A]}{\epp{G_j}{\role A}}$.
      Note that in the premise of rule $\did{T}{Send}$ that types the
      continuation $C$, the buffer types in $\Gamma$ (i.e., those in $\Gamma_1$)
      are unaffected. Therefore $\Gamma''( \chanto{k}{\role A}{\role B})
			\neq \Gamma'(\chanto{k}{\role A}{\role B})$, however from \cref
			{lem:up_to_buffer} we know that we can omit to consider buffer types as
			they are irrelevant for the typing of choreographies.
			For all sessions $k' \neq k$ in $\Gamma''$ their local typings are the
			same in $\Gamma'$. For session $k$, the typing $\Gamma''( k [\role A]) =
			\Gamma'( k[\role A ] ) = \epp{G_j} {\role A}$. From \cref
			{lem:proj_sub}, for all other $k[\role C] \in \Gamma'', \role C \neq
			\role A$ it holds that 
			$\Gamma''( k[\role C] ) = \epp{G}{\role C}$, $\Gamma'( k[\role C] ) =
			\epp{G'}{\role C}$, and $\epp{G'}{\role C} \prec \epp{G}{\role C}$.
			Therefore $\Gamma' \prec \Gamma''$ and \pnum{2} holds by \cref{lem:subsumption}.

			\end{proof}

			\begin{proof}[of \pnum{3}] 
				 
				To prove $\Gamma' \hlseq {D'}$ we need to prove that the conditions of
				\cref{def:deployment_judgements} hold.
				\ilab{\defDJ.1} holds by the application of rule $\did{D}{Send}$, by
				construction of $\Gamma'$, and by \ilab{\thmSR.1}.
				\ilab{\defDJ.2} holds for all sessions $k' \neq k$ by application of
				rule $\did{D}{Send}$ and the construction of $\Gamma'$. 
				The same holds true for session $k$ and any process $\pair{\pid q}{k
				[\role C]} \in \Gamma' \ |\ \role C \neq \role B$.

				Finally, we need to prove that $\Gamma'(\ \chanto{k}{\role A}{\role B}\ ) =
				\btypeop(\ \role A, \dep'(\chanto{k}{\role A}{\role B})\ )$. From \ilab
        {\thmSR.1} we know that \emph{i}) $\Gamma(\chanto{k}{\role A}{\role B})
        = T$ and \emph{ii}) let $\dep(\chanto{k}{\role A}{\role B}) = \til m$,
        that $\btypeop(\role A, \til m) = T$. From \cref{def:bte} we have a
        direct proof that $\btypeop(\ \role A,\ m_1::\dots::m_n\ ) = \btypeop
        (\role A, m_1)\ ;\ \dots\ ;\ \btypeop(\role A,m_n)$. 

        Now, from the reduction on $\did{C}{Send}$ we know that
				\[
          \dep'(\ \chanto{k}{\role A}{\role B}\ ) = m' =	\til m :: (o_j,v)
        \]
        And therefore, $\btypeop(\role A, m') = T;\ \btypeop(\role A,\
        (o_j,v))$. From the reductions on $\Gamma$ and $G$, we observe that the
        reduction on $G$ do not affect the context $\mathcal G$ (which contains
        local type $T$), thus, by the rules of the definition of the Buffer Type
        Projection (\cref{fig:chor_typing_epp_buffer}), we have

        \[
            \epp{G'}{\role B}^\role A = T; \lrecv{\role A}{\op{o_j}(U_j)}
        \]

        Hence, from the reduction on rule $\did{\Gamma}{Send}$, we know that
        $\Gamma'(\ \chanto{k}{\role B}{\role A} \ ) = \Gamma'(\ \epp{G'}{\role
        B}^\role A \ ) = T;\lrecv{\role A}{\op{o_j}(U_j)}$. Finally, from the
        typing rule $\did{T}{Send}$ we know that $\pid p.e \seq U_j$ and from
        reduction rule $\did{C}{Send}$ that $v = \evalfn(\ e,
        \dep(\pid p )\ )$, thus $v$ has type $U_j$. Hence, $\btypeop(\ \role A, 
        (o_j,v) \ ) = \lrecv{\role A}{o_j(U_j)}$ and
        \[
        \Gamma'(\ \chanto{k}{\role A}{\role B}\ ) = T;\lrecv{\role A}{o_j(U_j)} =
        \btypeop(\ \role A, \dep'(\chanto{k}{\role A}{\role B})\ )
        \]
			\end{proof} 

		\case{$\did{C}{Recv}$} The case is:
		$$
			\infer[\did{C}{Recv}]
			{
				\dep,\ \genbranchI
				\quad \lto {k:\grecv{\role A} {\role B}{o_j(x_j)}} \quad
				\dep',\ C_j
			}{
				j \in I
				&
				\renv{\dep}{\com{k}{\role A}{\pid q[\role B].o_j(x_j)}}{\dep'}
			}
		$$

		\ilab{\thmSR.2} has reductum $C' = C_j$. 
		Since we could apply $\did{C}{Recv}$, we know that $\dep(\chanto{k}{\role A}
		{\role B}) = (o_j,v)::\til m$. Let $\dep_1 = \dep\big[ \pid q \mapsto \dep
		(\pid q)[ x \mapsto v ]\big ]$, from the application of rule $\did{D}
		{Recv}$, we know that $\dep' = \dep_1\big[ \chanto{k}{\role A}{\role B}
    \mapsto \til m \big]$. To prove \ilab{\thmSR.3} we must prove that rule 
    $\did{\Gamma}{Recv}$ is applicable.

		Since \ilab{\thmSR.1} holds $\pco( \Gamma )$ and
		$\Gamma \hlseq D$ hold and therefore we know that, by \ilab {\defDJ.2},
		$\Gamma(\chanto{k}{\role A}{\role B}) = \btypeop(\role A,\ (o_j,v)::\
		\til m)$.

    Let $\seq v: U_j$, then $\btypeop(\role A,\ (o_j,v)::\ \til m) = \lrecv {A}
    {o_j(U_j)};T$ where $T = \btypeop(\role A,\ \til m)$ by \cref{def:bte} and
    $\Gamma (\chanto{k}{\role A}{\role B}) = \lrecv{A} {o_j(U_j)};T$. Since
    $\pco(\Gamma)$ holds, there exists a global type $G$ for session $k$ such
    that $G = \mathcal{G}[\grecv{\role A}{\role B}{o_j(U_j)}; G_j]$. Let $\pi$
    be the reduction of $G$ with rules $\did{G}{Eq}$ and $\did{G}{Recv}$, we
    observe the following derivation:

    {\scriptsize
      \vspace{1em}
      \[
      \hspace{-5em}
      \pi = \left\{\hspace{-10em}
      \begin{minipage}{.75\textwidth}
      \[      
      \infer*[\did{G}{Eq}]{
          G \lto{\gamma} G'
      }{
        G \swapG G_1
        &
        \infer[\did{G}{Recv}]{
          G_1
            \lto{\gamma} 
            G'
          }{}
        &
        G \swapG G'
      }
      \]
      \end{minipage}
      \hspace{-9em}
      {\color{lightgray}\rule[-5em]{.5px}{10em}}
      \begin{minipage}{.15\textwidth}
      \[
        \begin{array}l
        G_1 = \grecv{\role A}{\role B}{o_j(U_j)};\mathcal{G}[G_j]
        \\[2em]
        G' = \mathcal{G}[G_j]
        \\
        \gamma = \grecv{\role A}{\role B}{o_j(U_j)}
        \end{array}
      \]
      \end{minipage}
      \vspace{1em}
      \right.\]
    }

    In the reductions, since $C,\env$ reduces with $\beta = k:\grecv{\role A}
    {\role B}{o_j(x_j)}$ and $G$ types $C,\env$ in $\Gamma$, there are no other
    exchanges from $\role A$ to $\role B$ in $G$ that could prevent from
    obtaining, after a finite number of derivations on rule $\did{G}{Eq}$, the
    swap-equivalence $G \swapG G_1$. Then, applying rule $\did{G}{Send}$, $G_1$
    can reduce to $G'$.

    Given $\pi$, we can use it to write the reduction at the level the typing
    environment $\Gamma$, applying rule $\did{\Gamma}{Recv}$. Below, we consider
    $\Gamma = \Gamma_*,\Gamma_k$ where $\Gamma_k$ contains all and only typings
    of session $k$ in $\Gamma$.

		\[
		\infer[\did{\Gamma}{Recv}]
		{
			\Gamma_*,\Gamma_k
			\lto{k:\grecv{\role A} {\role B}{o_j(x)}}
			\Gamma_*,
			\{ \pair{k[\role C]}{\epp{G'}{\role C}} \ | \ k[\role C] \in \Gamma_k \},
			\{\pair{\chanto{k}{\role C}{\role D}}{\epp{G'}{\role C}^{\role D}}
				\ |\ \chanto{k}{\role C}{\role D} \in \Gamma_k \},
			\pair{\pid q.x }{U_j}
		}
		{
        k \not \in \Gamma_*
        &
        \Gamma_k \subseteq \epp{G}{k}
        &
        \{\pair{k[\role A]}{T}, \pair{k[\role B]}{T'}\} \in \Gamma_k
        &
        \Gamma_* \seq \pair{\pid q}{k[\role B]}
        &
        \infer*[]{
          G \lto{\grecv{\role A}{\role B}{o_j}} G'
        }{
          \pi
        }
		}
		\]

		Hence \ilab{\thmSR.3} holds and $\Gamma' = \Gamma_*,\{ \epp{G'}{\role C} \ | \ k[\role C] \in \Gamma_{k} \},\pair{\pid q.x }{U_j}$.

		$\ilab{\thmSR.4}$ holds if we can apply rule $\did {T} {DC}$ on $\Gamma'
		\hlseq {\dep',C'}$

		\[
		\infer[\did{T}{\dep C}]
		{
			\Gamma' \hlseq{\dep',C'}
		}
		{
			\pco( \Gamma' )
			&
			\Gamma' \hlseq C'
			&
			\Gamma' \hlseq \dep'
		}
		\]
			
		and we need to prove \pnum{1} $\pco( \Gamma' )$, \pnum{2} $\Gamma' \hlseq
		C'$, and \pnum{3} $\Gamma'\hlseq \dep'$

		The proof of \pnum{1} for this case is similar to that of \pnum{1} for case
		$\did{C}{Send}$.

		\begin{proof}[of \pnum{2}]
			From \ilab{\thmSR.1}, partitioning $\Gamma = \Gamma_1,\pair{k[\role B]}
			{\lrecv {A} {o_j (U_j) ; \epp {G_j}{\role B}}}$ and since $j \in I$ from
			rule $\did{C}{Recv}$, we can write the derivation

			$$
			\infer[\did{T}{\dep C}]
				{
					\Gamma \hlseq{\dep,\genbranchI}
				}
				{
					\pco( \Gamma )
					&
					\Gamma \hlseq \dep
					&
					\infer[\did{T}{Recv}]
					{
						\Gamma_1, 
						\pair{k[\role B]}{\lrecv{A}{o_j(U_j);\epp{G_j}{\role B}}}
						\hlseq{\genbranchI}
					}
					{
						j \in I
						&
						\Gamma_1 \seq \pair{\pid q}{k[\role B]}
						&
						\Gamma_1, 
						\pair{\pid q.x_j}{U_j}, 
						\pair{k[\role B]}{\epp{G_j}{\role B}}
						\hlseq{C_j}
					}
				}
			$$

			hence we know that $\Gamma_1,\pair{\pid q.x_j}{U_j},\pair{k[\role
			B]}{\epp{G_j}{\role B}}	\hlseq{C_j}$.

      Let $\Gamma'' = \Gamma_1,\pair{\pid q.x_j}{U_j},\pair{k[\role
      B]}{\epp{G_j}{\role B}} \hlseq{C_j}$ and $\Gamma_k' = \Gamma_1 \setminus
      \Gamma_* = \Gamma_k \setminus \{\pair{k[\role B]} {\epp {G}{\role B}}\}$.
      We can write $\Gamma'' =
			\Gamma_*,\Gamma_k',\pair{k[\role B]}{\epp{G_j}{\role B}}$.
			Similarly to \pnum{2} for case $\did{C}{Send}$, $\Gamma''
			(\chanto{k}{\role A}{\role B}) \neq
			\Gamma'( \chanto{k}{\role A}{\role B})$, but we omit to
			consider buffer types as they are irrelevant for the typing of
			choreographies by \cref{lem:up_to_buffer}.
			For all sessions in $\Gamma''$, their local typings are the same as in
			$\Gamma'$. We consider in particular $k$ on which we applied the
			reduction for this case for which it holds
			$$\forall \ k[\role C] \in \Gamma'',\ \Gamma''(k[\role C]) = \Gamma'(k
			[\role C]) = \epp{G'}{\role C}$$
		\end{proof}

		\begin{proof}[of \pnum{3}]
			To prove $\Gamma' \hlseq {D'}$ we prove the conditions in
			\cref{def:deployment_judgements}.
			\ilab{\defDJ.1} holds from the application of rule $\did{D}{Recv}$,
			\ilab{\thmSR.1}, and the construction of $\Gamma'$.
			\ilab{\defDJ.2} holds for all $\pid p.x$ from the application of rule
			$\did{D}{Recv}$, \ilab{\thmSR.1}, and the construction of $\Gamma'$,
			except for $\pid q.x_j$ which is not defined in $\Gamma$. However the
			condition holds by construction of $\Gamma' = \Gamma_1,\pair{\pid q.x_j}{U_j},\pair{k[\role B]}{\epp{G_j}{\role B}}$.
			%
			%
			\ilab{\defDJ.2} holds for all sessions $k' \neq k$ by the application of
			rule $\did{D}{Recv}$ and the construction of $\Gamma'$.
			The same holds true for session $k$ and any process $\pair{\pid p}{k[\role
			C]} \in \Gamma \ |\ \role C \neq \role B$.
			
			For $\pair{\pid q}{k[\role B]}$ and role $\role A$ we know from the
			application of $\did{C}{Send}$ that
			$\dep'( \chanto{k}{\role A}{\role B} )	= \til m$.
			Since we took $G$ such that $\epp{G}{\role B}^\role A = \lrecv{\role
			A}{o_j(U_j);T}$, where $T = \btypeop(\role A, \til m)$, then
			$\epp{G'}{\role B}^\role A = T$.

		\end{proof}

	\case{$\did{C}{Start}$} The case is:
	$$
	\infer[\did{C}{Start}]
	{
		\dep,\ \genstart; C
		\quad \lto {\startlab{k'}{\pid p}{\pids r}} \quad
		\dep',\
		C[k'/k][\pids r/\pids q] 
	}
	{
		\env\fresh{k',\pids r}
		&
		\delta = \deltastart{k'}{l.\prc pA, \wtil{l.\prc rB}}
		&
		\renv{\dep}{\delta}{\dep'}
	}
	$$
		
	Where \ilab{\thmSR.5} has $C' = C[k'/k][\pids r/\pids q]$. $\dep'$ is defined
	non-deterministically but abides the requirements defined in rule
	$\did{D}{Start}$. Let $\wtil {\prc sC} = \prc pA, \wtil{\prc rB }$. Since
	\ilab{\thmSR.1} holds, we can apply rule $\did{T}{Start}$. We partition
	$\Gamma = \Gamma_1,\til l:\initg {G}{\role A}{\til{\role B}}{\til{\role B}}$
	
	$$
	\infer[\did{T}{Start}]
	{
		\Gamma_1, \til l:\initg{G}{\role A}{\til{\role B}}{\til{\role B}}
		\hlseq{\genstart ; C}
	}
	{
		\Gamma_1,\til l:\initg{G}{\role A}{\til{\role B}}{\til{\role
		B}},
		\init{}(\ \wtil{\prc {s'}C}\ ,\ k,\ G\ )
		\hlseq {C}
		&
		\wtil{\prc {s'}C} = \prc pA, \wtil{\prc qB}
		&
		\til{\pid q} \not \in \Gamma_1
	}
	$$

  Coherently with the semantics of rule $\did{C}{Start}$, we take $\Gamma' =
  \Gamma,\init{}(\ \wtil{\prc sC},\ k',\ G\ )$ --- obtainable from
  the typing environment in the left-most premise of rule $\did{T}{Start}$,
  $\alpha$-renaming: \emph{i}) typings on session $k$ to session $k'$ and
  \emph{ii}) process identifies $\pids q$ to $\pids r$ in $\wtil{ \prc
  {s'}C}$ (i.e., such that $\wtil{\prc sC} = [\ \wtil{\prc rC}/\wtil{\prc
  qC}\ ]\wtil{ \prc {s'}C}$) --- and we prove the case by proving that we can
  apply rule $\did{T}{DC}$ on $\Gamma' \hlseq {\dep',C'}$, i.e, that the
  following hold:
	\pnum{1} $\pco(\Gamma')$, \pnum{2} $\Gamma' \hlseq {C'}$, and \pnum {3}
	$\Gamma' \hlseq {\dep'}$.

	\begin{proof}[of \pnum{1}]
		\pnum{1} holds for all session $k'' \in \Gamma', k''\neq k'$ by \ilab
		{\thmSR.1}. For session $k'$ \pnum{1} holds by construction.
	\end{proof}

	\begin{proof}[of \pnum{2}]
    By \ilab{\thmSR.1} we could apply $\did{T}{Start}$ where
    $\Gamma,\init{}(\ \wtil{\prc {s}C},\ k,\ G\ )\  \hlseq C$. Since
    $\Gamma'$ is obtained by $\alpha$-renaming of the left-most premise of
    Rule $\did{T}{Start}$, which types the continuation $C$, $\Gamma'$ types $C[k'/k][\pids r/\pids q]$ and \pnum{2} holds by construction.
	\end{proof}
		
	\begin{proof}[of \pnum{3}]
		To prove \pnum{3} we prove the conditions in \cref{def:deployment_judgements}.
		\ilab{\defDJ.1--\defDJ.2} hold by the application of rule $\did{D}{Start}$
		 and the construction of $\Gamma'$.
	\end{proof}

	\case{$\did{C}{PStart}$} The case is:
	$$
	\infer[\did{C}{PStart}]
	{
	\begin{array}l
		\dep,
		\req{k}{\pid p[\role A]}
		{\wtil{l.\role B}}; C
		\pp
		\prod_i
		\big(
		\genacci; C_i
		\big)
		\lto{\tau}
		\\ \hspace{7em}
		\dep',\ 
		C[k'/k] \pp \prod_i\big(\ C_i[k'/k][\pids r_i /\pids q_i] \ \big)
		\pp
		\prod_i
		\big(
		\genacci; C_i
		\big)
	\end{array}
	}
	{
	\begin{array}c
		i \in \{1,\dots,n\}
		\hfill
		\env\fresh{k',\pids r}
		\hfill
		\{\ \wtil{l.\role B}\ \} = \biguplus_i \{\ \wtil{l_i.\role B_i}\ \}
		\hfill
		\{\pids r\} = \bigcup_i \{ \pids r_i \}
		\\[2pt]
		\pid p \in \dep(l)
		\qquad
		\delta = \deltastart{k'}{l.\pid p[\role A],\wtil{l_1.\pid r_1[\role
		B_1]},\dots,\wtil{l_n.\pid r_n[\role B_n]}}
		\qquad
		\renv{\dep}{\delta}{\dep'}
	\end{array}
	}
	$$
		
  Where \ilab{\thmSR.5} has $C' = C[k'/k] \pp \prod_i\big(\ C_i[k'/k][\pids
  r_i /\pids q_i] \ \big) \pp \prod_i \big( \genacci; C_i \big )$. $\dep'$ is
  defined non-deterministically but abides the requirements defined in rule
  $\did{D}{Start}$.

	We partition $\Gamma$ such that:
	\begin{itemize}
		\item $\Gamma = \Gamma_r, \Gamma_a$
		\item $\Gamma_r \seq \pair{\til l}{\initg{G}
				{\role A} {\til {\role B}}{\emptyset}}$
		\item $\Gamma_a \seq \pair{\til l}{\initg{G}{\role A}{\til
		{\role B}}{\til {\role B}}}$
    \item $\Gamma_a = \Gamma_1,\pair{\til l}{\initg{G}{\role A} {\til {\role
      B}} {\wtil {\role B_1}}}, \cdots, \Gamma_n,\pair{\til
      l}{\initg{G}{\role A} {\til {\role B}} {\wtil {\role B_n}}}$
    \item $\Gamma_a^i = \Gamma_i,\pair{\til l}{\initg{G}{\role A} {\til
    {\role B}}{\wtil {\role B_i}}}, \cdots, \Gamma_n,\pair{\til
    l}{\initg{G}{\role A} {\til {\role B}}{\wtil {\role B_n}}}$
	\end{itemize}
	and we can write the derivation

	$$
		\infer[\did{T}{\dep C}]
		{
			\Gamma \hlseq{\dep,
				\req{k}{\pid p[\role A]}{\wtil{l.\role B}}; C
				\pp \prod_{i \in I} \big( \genacci; C_i \big)}
		}
		{
			\pco( \Gamma )
			&
			\Gamma \hlseq \dep
			&
			\infer[\did{T}{Par}]
			{
				\Gamma \hlseq {
			\req{k}{\pid p[\role A]}{\wtil{l.\role B}}; C
				\pp	\prod_{i \in I}	\big(	\genacci; C_i	\big)}
			}
			{
				\infer[\did{T}{Req}]
				{
					\Gamma_r \hlseq{\genreq ; C}
				}
				{
					\Gamma_r,\pair{\pid p}{k[\role A]},\pair{k[\role A]}{\epp{G}{\role A}}
					\hlseq {C}
					&
					\Gamma_r \seq \pair{\til l}{\initg{G}{\role A}{\til{\role B}}{\emptyset}}
				}
				&
				\Delta_1
			}
		}
	$$

	$$
	\Delta_i = \left\{
	\begin{array}c
		\infer[\did{T}{Par}]
	{
		\Gamma_a^i \hlseq {\acc{k}{\wtil{l_i.\pid q_i[\role B_i]}};C_i
		\pp
		\prod_{j \in I \setminus \{1,\cdots,i\}}
		\big(	\acc{k}{\wtil{l_{j}.\pid q_{j}[\role B_{j}]}}; C_{j}	\big)}
	}
	{
		\infer[\did{T}{Acc}]
		{
			\Gamma_i,\pair{\til l}{\initg{G}{\role A}{\til{\role B}}{\wtil
			{\role B_i}}} \hlseq{
				\acc{k}{\wtil{l_i.\pid q_i[\role B_i]}};C_i	}
		}
		{
			\til l_i \subseteq \til l
			&
      \Gamma_i,\pair{\til l}{\initg{G}{\role A}{\til{\role
      B}}{\emptyset}},\init{}(\ \wtil {\pid q_i [\role B_i]},\ k,\ G\ )
      \hlseq{C_i}
			&
			\pids q_i \not \in \Gamma
		}
		&
		\Delta_{i+1}
	}
	\end{array}\right.
	$$

	Let $\wtil {\prc sC } = \prc pA, \wtil {\pid r_1[\role B_1]}, \cdots, \wtil
	{\pid r_n[\role B_n]}$.

	To prove \ilab{\thmSR.6} we take

	$$\Gamma' = 
		\Gamma,\init(\ \wtil{\prc{s}C},\ k',\ G\ ) =
		\Gamma_r, \Gamma_a, \init(\ \wtil{\prc{s}C},\ k',\ G\ )
	$$

  and we partition $\init(\ \wtil {\prc {s}C},\ k',\ G\ )$ such that
  $$\Gamma' =	\Gamma_r', \Gamma_a', \Gamma_a$$

	Where 
	\begin{itemize}
		\item $\Gamma_r' = \Gamma_r, \init(\ \prc pA,\ k',\ G\ ) $
		\item $\Gamma_a' = \Gamma_1', \cdots, \Gamma_n'$
		\item $\Gamma_i' = \Gamma_i,\pair{\til l}{\initg{G}{\role A}{\til{\role B}}
		{\emptyset}},\init(\ \wtil{\pid r_i [\role B_i]},\ k',\ G\ )$ where $i \in \{1,\dots,n\}$
	\end{itemize}

	To prove \ilab{\thmSR.6} we must prove we can apply rule $\did{T}{\dep C}$
	on $\Gamma' \hlseq {\dep',C'}$.

	{\scriptsize
  \[
	\hspace{-10pt}\infer[\did{T}{\dep C}]
	{
		\Gamma' \hlseq {\dep',C'}
	}{
		\pnum{1}\ \pco( \Gamma' )
    &		
		\ \pnum{3}\ \Gamma' \hlseq {\dep'}
		&
    \pnum{2} \left\{\begin{array}{cc}
		\infer[\did{T}{Par}]
		{
			\Gamma' \hlseq { C[k'/k] \pp \prod_i\big(\ C_i[k'/k][\pids r_i /\pids
			q_i] \ \big)	\pp \prod_i \big( \genacci; C_i \big ) }
		}
		{
			\pnum{2a}\ \Gamma_r' \hlseq {C[k'/k]}
			&
			\infer[\did{T}{Par}]
			{
					\Gamma_a', \Gamma_a \hlseq {\prod_i\big(\ C_i[k'/k][\pids r_i /\pids q_i] \ \big)	\pp \prod_i \big( \genacci; C_i \big ) }
			}
			{
				\begin{array}c
					\pnum{2b}\ \Gamma_a' \hlseq {\prod_i\big(\ C_i[k'/k][\pids r_i /\pids
				q_i] \
				\big)} 
				\\
				\pnum{2c}\ \Gamma_a \hlseq {\prod_i \big( \genacci; C_i \big )}
				\end{array}
			} 
		}
		\end{array}\right.
	}
	\]}

	\begin{proof}[of \pnum{1}]
    \pnum{1} holds by construction.
  \end{proof}

	\begin{proof}[of \pnum{2}]
		\pnum{2} holds as
		\begin{itemize}
      \item \pnum{2a} holds by $\alpha$-renaming $(\Gamma_r,\pair {\pid p} {k
      [\role A]},\pair{k[\role A]}{\epp{G}{\role A}})[k'/k] \hlseq {C[k'/k]}$
      and by omitting to consider buffer types as of \cref{lem:up_to_buffer};
			\item similarly to \pnum{2a}, \pnum{2b} holds by $\alpha$-renaming on the
			derivation of $$(\ \Gamma_i,\pair{\til l}{\initg{G}{\role A}{\til{\role B}}{\emptyset}},\init	(\ \wtil {\pid q_i [\role B_i]},\ k,\ G\ )\ )[k'/k][\pids r_i/\pids q_i]
			\hlseq {C_i[k'/k][\pids r_i/\pids q_i]}$$ and by \cref
			{lem:up_to_buffer};
			\item \pnum{2c} holds by \ilab{\thmSR.1}.
		\end{itemize}
	\end{proof}
		
	\begin{proof}[of \pnum{3}]
		The proof of \pnum{3} of this case is similar to the of \pnum{3} for Case $\did{C}{Start}$.
	\end{proof}

	\case{$\did{C}{Cond}$} The case is:

		$$\infer[\did{C}{Cond}]
		{
			\dep,\ \cond{\pid p.e}{C_1}{C_2}
			\quad\lto{\tau}\quad
			\dep,\ C_i
		}
		{
      i = 1 \text{ if } \evalfn(\ e, \env(\pid p)\ ) = \mathtt{ true },
			i = 2 \text{ otherwise}
		}$$

		In \ilab{\thmSR.5} $\dep' = \dep$ and we have two cases for $C' = C_1$ or
		$C' = C_2$.

		From \ilab{\thmSR.1} we can write

		$$
		\infer[\did{T}{Cond}]
		{\Gamma \hlseq{\gencond}}
		{
			\Gamma \seq \pair{\pid p.e}{\tbool}
			&
			\Gamma \hlseq{C_1}
			&
			\Gamma \hlseq{C_2}
		}
		$$

		The proof of \ilab{\thmSR.6} follows directly from the premises of the
		typing derivation as $\Gamma \hlseq \dep = \dep'$ and in both cases that
		$C' = C_1$ or $C' = C_2$ it holds that $\Gamma
		\hlseq C'$ from the premises of $\did{T}{Cond}$.

	\case{$\did{C}{Ctx}$} The case is:

	$$
	\infer[\did{C}{Ctx}]
	{
		\dep, \recDef{X}{C_2}{C_1} \quad \lto \beta \quad
		\dep^\prime, \recDef{X}{C_2}{C^\prime_1}
	}
	{
		\dep, C_1 \lto \beta \dep^\prime, C^\prime_1
	}
	$$

	From \ilab{\thmSR.1} we know that, $\Gamma = \Gamma_1,\pair{X}{\Gamma_x}$

	$$
	\infer[\did{T}{\dep C}]
	{
		\Gamma \hlseq{\dep, \recDef{X}{C_2}{C_1}}
	}
	{
		\pco( \Gamma )
		&
		\infer[\did{T}{Def}]
		{
			\Gamma \hlseq{\recDef{X}{C_2}{C_1}}
		}
		{
			\Gamma_1,\pair{X}{\Gamma_x} \hlseq{C_1}
			&
			\Gamma_x, \pair{X}{\Gamma_x} \hlseq C_2
			&
			\Gamma_x |_{\locs} \subseteq \Gamma 
		}
		&
		\Gamma \hlseq \dep
	}
	$$

	The proof is divided in two cases on the type of $\beta$.

	\begin{itemize}
	\case{$\beta \neq \tau$} $\dep,C_1$ reduces on some session $k$.
		By the induction hypothesis since $\Gamma \hlseq {\dep,C_1}$ we can find
		$\Gamma'$ such that \ilab {\thmSR.3} holds. We prove \ilab {\thmSR.4}
		by proving that we can apply $\did{T}{\dep C}$ on $\Gamma' \hlseq{\dep',
		\recDef{X}{C_2}{C_1'}}$ and therefore that \pnum{1} $\pco( \Gamma' )$ holds,
		\pnum{2} $\Gamma' \hlseq {\recDef{X}{C_2}{C_1}}$ and \pnum{3}
		$\Gamma'\hlseq \dep'$.

		\pnum{1} holds by the construction of $\Gamma'$ and \pnum{3} holds by the
		induction hypothesis.

		To prove \pnum{2} we have to prove that $\Gamma' \seq \pair{X}{C_2}$ and
		$\Gamma_x|_\locs \subseteq \Gamma'$.

		From the induction hypothesis we have that $\Gamma \lto {\beta} \Gamma'$
		and $\Gamma' \hlseq {\dep',C_1'}$. By construction of $\Gamma'$ it holds
		that $\Gamma' = \Gamma_*',\Gamma_k'$ where $\Gamma' \cap \Gamma = \Gamma_*$
		such that $k \not \in \Gamma_*$ and $\Gamma = \Gamma_*,\Gamma_k$ where
		$\Gamma_k \subseteq \epp{G}{k}$ for some $G$. Therefore it holds that
		$\Gamma_* \seq\pair{X}{\Gamma_x}$ and thus that $\Gamma' \seq
		\pair{X}{\Gamma_x}$. The same applies to $\Gamma_x|_\locs \subseteq
		\Gamma_*$ which proves $\Gamma_x|_\locs \subseteq \Gamma'$. 

	\case{$\beta = \tau$} from the induction hypothesis, for any considered
		derivation we have $\Gamma \subseteq \Gamma'$. We prove \ilab{\thmSR.6} by
		proving that we can apply $\did{T}{\dep C}$ on $\Gamma' \hlseq{\dep',
		\recDef{X}{C_2}{C_1'}}$. \pnum{1}, \pnum{2}, and \pnum{3} hold by
		construction of $\Gamma'$.

	\end{itemize}

	\case{$\did{C}{Par}$} The case is:
		$$\infer[\did{C}{Par}]
		{
			\dep, C_1\pp C_2
			\quad \lto \beta \quad
			\dep', C'_1 \pp C_2
		}
		{
			\dep, C_1
			\quad \lto \beta \quad
			\dep', C'_1
		}$$

		From \ilab{\thmSR.1} we have the derivation below, with $\Gamma$
		partitioned as $\Gamma = \Gamma_1,\Gamma_2$

		$$
		\infer[\did{T}{\dep C}]
		{
			\Gamma \hlseq {\dep, C_1 \pp C_2}
		}
		{
			\pco( \Gamma )
			&
			\infer[\did{T}{Par}]
			{
				\Gamma \hlseq {C_1 \pp C_2}
			}
			{
				\Gamma_1 \hlseq C_1
				&
				\Gamma_2 \hlseq C_2
			}
			&
			\Gamma \hlseq \dep
		}
		$$

		The proof is divided in two cases on the type of $\beta$.

		\begin{itemize}
			
			\case{$\beta \neq \tau$} $\dep,C_1$ reduces on some session $k$.
			By the induction hypothesis and since $\Gamma_1 \hlseq \dep,C_1$ we can
			find $\Gamma_1'$ such that $\Gamma_1 \lto{\beta} \Gamma_1'$ and $\Gamma_1'
			\hlseq {\dep',C_1'}$. Then we take $\Gamma' = \Gamma_1', \Gamma_2$
			which proves \ilab{\thmSR.3} to hold. We prove  \ilab {\thmSR.4} by
			proving that we can apply $\did{T}{\dep C}$ on $\Gamma' \hlseq{\dep',
			C_1' \pp C_2}$ and therefore that \pnum{1} $\pco( \Gamma' )$,
			\pnum{2} $\Gamma' \hlseq {C_1' \pp C_2}$ and \pnum{3}
			$\Gamma'\hlseq \dep'$ hold. \pnum{1}, \pnum{2}, and \pnum{3} hold by
			construction and the induction hypothesis.

			\case{$\beta = \tau$} from the induction hypothesis, for any derivation
			we have that $\Gamma_1' \hlseq {\dep',C_1'}$ and $\Gamma_1 \subseteq
			\Gamma_1'$. Also in this case we take $\Gamma' = \Gamma_1', \Gamma_2$
			and prove \ilab {\thmSR.6} by proving that we can apply $\did{T}{\dep C}$
			on $\Gamma' \hlseq{\dep', C_1' \pp C_2}$. \pnum{1}, \pnum{2}, and
			\pnum{3} hold by construction of $\Gamma'$ and the induction hypothesis.

		\end{itemize}

	\case{$\did{C}{Eq}$} The case is:

	$$
	\infer[\did{C}{Eq}]
	{
		\dep, C \quad \lto \beta \quad \env', C'
	}
	{
		\ctx{R} \in \{\,\equivC\,,\,\swapC\,\}
		&
		C \,\ctx{R}\, C_1
		&
		\env, C_1 \lto \beta \env', C_1'
		&
		C_1' \,\ctx{R}\, C'
	}
	$$

	The proof is divided into two subcases on the type of $\ctx{R}$.
	\begin{itemize}
		\case{$\ctx{R}\ =\ \equivC$} The case is proved by induction hypothesis and
		\cref{lemma:subj_cong}.
		\case{$\ctx{R}\ =\ \swapC$} The case is proved by induction hypothesis
		and \cref{lemma:subj_swap}.
	\end{itemize}

	\end{itemize}
\end{proof}

The proof of \cref{thm:session_fidelity} follows directly from the proof of
\cref{thm:sub_red} and \cref{lemma:session_fidelity}.

\IFSubFileBiblio

\subsection{Proof of Deadlock Freedom}
\label{sec:proof_of_deadlock_freedom}

We report below the statement of 
\cref{thm:deadlock-freedom} enriched with pointers for clearer
referencing in the proof.

\def\thmDF{\Def\ref{thm:deadlock-freedom}}
\noindent{\textbf{\cref{thm:deadlock-freedom}} (Deadlock-freedom)}\\
\ilab{\thmDF.1} $\Gamma \hlseq{\env,C}$ and \ilab{\thmDF.2} $\cofn(\Gamma)$
imply that either \ilab{\thmDF.3} $C \equivC \inact$ or \ilab{\thmDF.4} there
exist $\env'$ and $C'$ such that $\env,C \to \env',C'$.


Like in \cite{CM13,MY13}, frontend choreographies enjoy deadlock freedom,
provided that they \emph{i}) do not contain free variable names and
\emph{ii}) are \emph{well-sorted}, i.e., have no undefined procedure calls.
Notably, well-sortedness is guaranteed by the type system.

\begin{proof}
	Proof by induction on the structure of C.

	\begin{itemize}
		\case{$C \equivC \inact$} trivial.

		\case{$C = \com{k}{\prc pA.e}{\role B.o};C_1$}
			from \ilab{\thmDF.1} and \ilab{\thmDF.2} we know that the requirements
			of $\did{D}{Send}$ hold and we can find $\env'$ such that
			$\renv{\env}{\com{k}{\prc pA.e}{\role B.o}}{\env'}$. We can apply Rule
			$\did{C}{Send}$ for which $C' = C_1$.

		\case{$C = \com{k}{\prc pA.e}{\prc qB.o(x)};C_1$}
			since \ilab{\thmDF.1} holds both receiver and sender are typed by
			$\Gamma$. We apply rule $\did{C}{Eq}$ to split the complete term into respectively a send and a receive partial terms, and similarly to the previous case, we apply rule $\did{C}{Send}$, for which $C' = \com{k}{\role A}{\prc qB.o(x)};C_1$.
			
		\case{$C = \com{k}{\role A}{\prc qB.\{o_i(x_i);C_i\}_{i \in I}}$}
			from \ilab{\thmDF.1} and \ilab{\thmDF.2} we know that the requirements
			of Rule $\did{D}{Recv}$ hold and $\env( \chanto{k}{\role A}{\role B} )
			= (o_j,t_m)::\til m$ for some $j \in I$. We can find $\env'$ such that
			$\renv{\env}{\com{k}{\role A}{\prc qB.o_j( x_j )}}{\env'}$ and apply
			Rule $\did{C}{Recv}$ for which $C' = C_j$.

		\case{$C = \genstart;C_1$}
			from \ilab{\thmDF.1} and \ilab{\thmDF.2} $\did{D}{Start}$ applies and we
			can find $\env'$ such that $\renv{\env}{\deltastart{k'}{l.\prc pA,
			\wtil{l.\prc rB}}}{\env'}$ for some $k'$, $\pids r$ fresh. We can apply
			Rule $\did{C}{Start}$ for which $C' = C_1[k'/k][\pids r/\pids q]$.
			
		\case{$C = \req{k}{\pid p[\role A]}{\wtil{l.\role B}}; C \pp \prod_{i=1}^n
		\big(  \genacci; C_i \big)$} similarly to the previous case, the
		requirements of $\did{D}{Start}$ hold and we can find $\env'$ such that
		$\renv{\env}{\deltastart{k'}{l.\pid p[\role A],\wtil{l_1.\pid r_1[\role
		B_1]},\dots,\wtil{l_n.\pid r_n[\role B_n]}}}{\env'}$ for some $k'$ and
		$\pids r_1, \cdots, \pids r_n$ fresh. We can apply Rule $\did{C}{PStart}$
		for which \\$C' = C[k'/k] \pp \prod_{i=1}^n C_i[k'/k][\pids r_1/\pids q_1]
		\pp \prod_{i=1}^n \big(\genacci; C_i \big)$.

		\case{$C = C_1 \pp C_2$} we can apply the induction hypothesis and Rule
		$\did{C}{Par}$ such that $\env,C_1 \to \env_1,C_1'$ and in \ilab{\thmDF.4}
		$\env'=\env_1$ and $C' = C_1' \pp C_2$.

		\case{$C=\recDef{X}{C_2}{C_1}$} applies the induction hypothesis and Rule
		$\did{C}{Ctx}$ for which $\env,C_1 \to \env',C_1'$, where $C'
		= \recDef{X}{C_2}{C_1'}$.

		\case{$\recDef{X}{C_2}{X;C_1}$} applies Rule $\did{C}{Eq}$ for
		$\recDef{X}{C_2}{X;C_1} \equivC \recDef{X}{C_2}{C_2;C_1}$ and by the
		induction hypothesis $\env,C_2 \to \env',C_2'$ and $C' = \recDef{X}{C_2}{C_2';C_1}$.

		\case{$C=\cond{\pid p.e}{C_1}{C_2}$} from $\ilab{\thmDF.1}$ we know that
		$\Gamma \seq \pair{\pid p.e}{\tbool}$ and therefore we can apply Rule
		$\did{C}{Cond}$ and, according to the evaluation of $e$, we have $C'=C_1$
		or $C'=C_2$.

	\end{itemize}
\end{proof}

\IFSubFileBiblio

\newpage

\subsection{Proof of Endpoint Projection}
\label{sec:proof_of_epp}





To prove our result on the Endpoint Projection we first define the minimal
typing system $\hlseqmin{}$ for FC.

\subsubsection{Minimal Typing} 
\label{sub:minimal_typing}

We recall the definition of subtyping for local and global types (see
\cref{def:local_subtyping,def:global_subtyping}), which we
extend to set inclusion and point-wise to \emph{i}) the typing of services
(i.e., of kind $\pair{\til l} {\serviceTyping{G} {\role A}{\til{\role
B}}{\til{\role C}}}$) and \emph{ii}) the typing of sessions, respectively.
Given two types $G$ and $G'$, we denote their least upper bound wrt $\prec$
with $G \lub G'$ (the same for local types and typing environments).

We define the minimal typing system $\hlseq{\!\!}_{\m{min}}$ on this notion
of subtyping. The minimal typing uses the minimal global and local types for
typing sessions and services such that the projection of the choreography is
still typable. We report the rules for minimal typing in
\cref{fig:minimal_typing}.

\begin{figure}
\small
\[
	\begin{array}c
		\infer[\did{Min}{Start1}]
		{
			\Gamma, 
			\pair{\til l}{\initg{G}{\role A}{\til{ \role B}}{\til{ \role B}}}
			\hlseqmin{
				\genstart;C
			}
		}
		{
			\Gamma,\init(\ \wtil{\prc rC},k,G\ ) \hlseqmin {C}
			&
			\wtil{\prc rC} = \prc pA, \wtil{\prc qB}
			&
			\til{\pid q} \not \in \Gamma
			&
			\til l \not \in \Gamma
		}
		\\[10pt]
		\infer[\did{Min}{Start2}]
		{
			\Gamma, 
			\pair{\til l}{\initg{G\lub G'}{\role A}{\til{ \role B}}{\til{ \role B}}}
			\hlseqmin{
				\genstart;C
			}
		}
		{
			\Gamma,
			\pair{\til l}{\initg{G}{\role A}{\til{ \role B}}{\til{ \role B}}},
			\init(\ \wtil{\prc rC},k,G'\ ) \hlseqmin {C}
			&
			\wtil{\prc rC} = \prc pA, \wtil{\prc qB}
			&
			\til{\pid q} \not \in \Gamma
		}
		\\[10pt]
		\infer[\did{Min}{Req1}]
		{
			\Gamma,
			\pair{\til l}{\initg{G}{\role A}{\til{ \role B}}{\emptyset}}
			\hlseqmin {\genreq ; C}
		}
		{
			\Gamma, \pid p:k[\role A], k[\role A]:\epp{G}{\role A} \hlseqmin {C}
			&
			\til l \not \in \Gamma
		}
		\quad
		\infer[\did{Min}{Req2}] {
			\Gamma,
			\pair{\til l}{\initg{G\lub G'}{\role A}{\til{ \role B}}{\emptyset}}
			\hlseqmin {\genreq ; C}
		}
		{
			\Gamma,
			\pair{\til l}{\initg{G}{\role A}{\til{ \role B}}{\emptyset}},
			 \pid p:k[\role A], k[\role A]:\epp{G'}{\role A} \hlseqmin {C}
		}
	\\[10pt]
	\infer[\did{Min}{Acc}]
	{
		\Gamma, 
		\pair{\til l'}{\initg{G}{\role A}{\til{\role B}}{\til{\role C}}} 
			\hlseqmin { \genaccC ; C	}
	}
	{
		\til l \subseteq \til l'
		&
		\Gamma,
		\init(\ \wtil{\pid q[\role C]},k,G\ )
		\hlseqmin{ C }
		&
		\pids q \not \in \Gamma
		&
		\til l \not \in \Gamma
	}
	\\[10pt]
	\infer[\did{Min}{Cond}]
	{\Gamma_1\lub\Gamma_2 \hlseqmin{\gencond}}
	{
		\Gamma_1\lub\Gamma_2 \seq \pair{\pid p.e}{\tbool}
		&
		\Gamma_1 \hlseqmin{C_1}
		&
		\Gamma_2 \hlseqmin{C_2}
	}
	\\[10pt]
	\infer[\did{Min}{Com}]
	{
		\Gamma, 
		\pair{k[\role A]}{\genlsendBmin},
		\pair{k[\role B]}{\genlrecvPrimemin}
		\hlseqmin{
			\com{k}{\pid p[\role A].e}{\pid q[\role B].o(x)};C}
	}
	{
		\Gamma \seq \pair{\pid p}{k[\role A]},\pair{\pid q}{k[\role B]}
		&
		\Gamma \seq \pair{\pid p.e}{U}
		&
		\Gamma, \pair{\pid q.x}{U}, 
			\pair{k[\role A]}{T}, 
			\pair{k[\role B]}{T'}
			\hlseqmin{C}
	}
	\\[10pt]
	\infer[\did{Min}{Send}]
	{
		\Gamma, 
		\pair{k[\role A]}{\genlsendBmin}
		\hlseqmin{ \com{k}{\pid p[\role A].e}{\role B.o};C }
	}
	{
		\Gamma \seq \pair{\pid p}{k[\role A]}
		&
		\pair{\pid q}{k[\role B]} \not \in \Gamma
		&
		\Gamma \seq \pair{\pid p.e}{U}
		&
		\Gamma, \pair{k[\role A]}{T} \hlseqmin{C }
	}
	\\[10pt]
	\infer[\did{Min}{Recv}]
	{
		\Gamma,
		\pair{k[\role B]}{\lrecv{A}{\{o(U);T\}}}
		\hlseqmin{\com{k}{\role A}{\pid q[\role B].\op{o}(x);C}}
	}
	{
		\Gamma \seq \pair{\pid q}{k[\role B]}
		&
		\pair{\pid p}{k[\role A]} \not \in \Gamma
		&
		\Gamma, 
		\pair{\pid q.x}{U}, 
		\pair{k[\role B]}{T}
		\hlseqmin{C}
	}
	\\[10pt]
	\infer[\did{Min}{Def}]
	{
		(\Gamma\lub\Gamma'),
		\mathsf{solve}( \wtil {k[\role A]:T} \lub
		\wtil {k'[\role	A']:T'}, \genlrecCall_X ) \hlseqmin{\recDef{X}{C'}{C}} 
	} {
	\!\begin{array}{c}
		\Gamma_x(X) = \Gamma_x'(X) 
			\mbox{ if } X \in \dom(\Gamma_x) \cap \dom(\Gamma_x')
		\qquad
		\nexists \ k''[\role A''] \in \dom(\Gamma \lub \Gamma')
		\\[2pt]
		X \not \in \dom( \Gamma \lub \Gamma' )
		\quad
		\Gamma_x' \vartriangleright_X ( \Gamma',\wtil {k'[\role A']:T'} ), C'
		\qquad
		\Gamma_x \vartriangleright_X	( \Gamma,\wtil{k[\role A]:T} ), C
		\qquad
		\Gamma' |_{\locs} \subseteq \Gamma 
	\end{array}
	}
	\\[15pt]
	\infer[\did{Min}{Par}]
	{\Gamma_1, \Gamma_2 \hlseqmin {C_1 \pp C_2}}
	{
		\Gamma_1 \hlseqmin {C_1}
		&
		\Gamma_2 \hlseqmin {C_2}
	}
	\quad
	\infer[\did{Min}{D1}]
	{
		\Gamma_x,\pair{X}{\Gamma_x} \vartriangleright_X \Gamma,C
	}{
		\Gamma, \Gamma_x,\pair{X}{\Gamma_x} \hlseqmin{C}
	}
	\quad
	\infer[\did{Min}{D2}]
	{
		\Gamma_x \vartriangleright_X \Gamma,C
	}{
		X \not \in \dom(\Gamma_x)
		&
		\Gamma, \Gamma_x \hlseqmin{C}
	}
	\\[15pt]
	\infer[\did{Min}{End}]
	{\Gamma \hlseqmin{\inact}}
	{
	\Gamma = \mathsf{ownerships} \cup \mathsf{sessions} \cup \mathsf{vars}
	&
	k[\role A] \in \mathsf{sessions}
	&
	k[\role A]: \gend
	}
	\\[10pt]
	\infer[\did{Min}{Call}]
	{
		\Gamma, \wtil {k[\role A]}:\genlrecCall_X,\wtil {k'[\role A']}: \gend,
		\pair{X}{(\Gamma',\wtil {k[\role A]}:\genlrecCall_X)} \hlseqmin{X}
	}
	{
		\begin{array}c			
		\Gamma = \mathsf{vars} \cup \mathsf{ownerships}
		\qquad
		\wtil {k'[\role A']} = \mathsf{sessions} \setminus \{\wtil {k[\role A]}\}
		\\[2pt]
		\Gamma' = \mathsf{vars}(X) \cup \mathsf{ownerships}(X)
		\quad
		\wtil{k[\role A]} = \mathsf{sessions(X)}
		\qquad
		\Gamma' \subseteq \Gamma
		\end{array}
	}
	\\[1.3em]
	\infer[\did{Min}{\env C}]
	{
		\Gamma, \Gamma' \hlseqmin {\env, C}
	}
	{
		\pco( \Gamma, \Gamma' )
		&
		\Gamma \hlseq{ \env }
		&
		\Gamma' \hlseqmin{ C }
	}
	\end{array}
\]
	\caption{Frontend Choreographies --- Minimal typing rules}
	\label{fig:minimal_typing}
\end{figure}

\begin{proposition}[Existence of Minimal Typing]
	Let $\Gamma \hlseq {\env,C}$, then there exists $\Gamma_0$ such that
	$\Gamma_0 \hlseq{\env,C}$ and for each $\Gamma' \hlseq{\env,C}$ we have
	that $\Gamma_0 \prec \Gamma'$. The environment $\Gamma_0$ can be
	algorithmically calculated from $C$ and is called the minimal typing of
	$C$.
\end{proposition}

\begin{proof}[of Existence of Minimal Typing]
The proof is standard and proceeds by induction on the rules in \cref{fig:minimal_typing}, defining the minimal typing system $\Gamma \hlseqmin{\env,C}$. 

As in \cite{CM13,MY13}, our focus is on the reconstruction of global/local
types, thus we leave the reconstruction of variable types undefined (which it
is entirely standard, e.g., see \cite{PIERCE}).

We give the intuition behind each case corresponding to the derivation on the
rules. $\did{Min}{Start1}$ and $\did{Min}{Start2}$ type the starting of
sessions. The difference between $\did{Min}{Start1}$ and $\did{Min}{Start2}$ is
that, when $\did{Min}{Start1}$ applies, the service typing of $\til l$ is not
used any more in $C$, and thus its typing is dropped to guarantee minimality.
Contrarily, in $\did{Min}{Start2}$ the service typing of $\til l$ is used in the
continuation $C$. In the rule, we consider the minimal global type $G \lub G'$
where $G'$ is minimal in session $k$ and $G$ is minimal in the typing of the
continuation $C$.

Rules $\did{Min}{Req1}$ and $\did{Min}{Req2}$ mirror a similar relationship,
where in the first rule we drop the typing of $\til l$, not used in the
continuation $C$, while in the second we consider $G \lub G'$. Note that Rule
$\did{Min}{Acc}$ directly drops the typing of $\til l$ in the typing of the
continuation. We do this because we assumed (see \cref{sub:fc_syntax}) that
\emph{i}) $\prid{acc}$ terms can only be at the top level (not guarded by
other actions) and \emph{ii}) by rule $\did{T}{Acc}$ no subsequent term
$\prid{start}$ on the same locations $\til l$ is typable (and hence cannot be
present in $C$, well-typed). The same holds for subsequent $\prid{req}$ terms
on $\til l$, which could not be paired with a complementary $\prid{acc}$.

In $\did{Min}{Cond}$ we consider $\Gamma_1 \lub \Gamma_2$ to determine the
least upper bound of receive types. Rules $\did{Min}{Com}$,
$\did{Min}{Send}$, and $\did{Min}{Recv}$ type receptions with a singleton
branching local type. Rule $\did{Min}{Par}$ is standard.

Also in rule $\did{Min}{Def}$ we consider the least upper bound of $\Gamma$ and
$\Gamma'$ respectively typing the continuation $C$ and the body of procedure
$X$. In addition, we also consider the least upper bound of the local typings
$T$ and $T'$, on which we apply function $\mathsf{solve}$. Function
$\mathsf{solve}$ is standard (cf. \cite{CHY12,CM13}) and solves the equations
$\genlrecCall_X = T$ for each $T$ in $\wtil{\pair{k{[\role A]}}{T}}$ where, if
$\genlrecCall_X$ appears in $T$, the corresponding component is $\genlrecDef_X$,
or $T$ otherwise. Rule $\did{Min}{Def}$ uses rules $\did{Min}{D1}$ and $
\did{Min}{D2}$ to determine the content of $\Gamma_x$ and $\Gamma_x'$ to
respectively minimally type the continuation $C$ and the body of procedure $X$.
Indeed, when rule $\did{Min}{D1}$ applies, the choreography $C$ uses the typing
$\pair{X}{\Gamma_x}$, otherwise $\did{Min}{D2}$ applies and the minimal type
does not contain the typing for $X$. Finally, in case both the typing of $C$
and of $C'$ type $X$ (i.e., $X$ in $\dom(\Gamma_x) \cap \dom(\Gamma_x')$),
their judgements coincide.

Rules $\did{Min}{End}$ and $\did{Min}{Call}$ use some auxiliary information,
obtainable by a preliminary top-down visit of the choreography syntax tree (cf.
\cite{CHY12,CM13}). Specifically, $\mathsf{vars}$, $\mathsf{ownerships}$, and
$\mathsf{sessions}$ are respectively the variable, the ownership, and the
session typings of the choreography whose type is being inferred. Similarly,
$\mathsf{vars}(X)$, $\mathsf{ownerships}(X)$, and $\mathsf{sessions}(X)$ yield
respectively the same kind of information regarding the body of procedure $X$ (i.e.,
obtained inspecting the body of the inner-most recursive procedure $X$). In the
rules, in $\did{Min}{End}$ we check that in $\Gamma$ reside only those
ownership, variable, and session typings present in the typed choreography and
that all sessions (i.e., their local types) are terminated. In rule
$\did{Min}{Call}$, \emph{i}) all sessions outside $X$ must be terminated and
those inside $X$ agree on $\genlrecCall_X$ and \emph{ii}) $\Gamma$ and $\Gamma'$
contain only appropriate variable and ownership typings and agree on their
judgement ($\Gamma' \subseteq \Gamma$).

Rule $\did{Min}{DC}$ defines minimal typing for running choreographies.

\end{proof}


\subsubsection{Typing Projection} 
\label{sub:typing_projection}

Here we define the projection of typing environments, which is used to prove
that, given the minimal typing environment $\Gamma$ of a choreography $C$,
from $\Gamma$ we can build the minimal typing environment for the EPP of $C$.

To do that, we have to account for two peculiarities (as defined in
\cref{sec:epp_properties}) of our EPP:

\begin{itemize}

	\item it merges in the output choreography the behaviours of many service
	processes into one process. Hence, to guarantee typing and minimality we
	have to merge typings related to service processes on the same location
	into the same (and only) service process present in $\epp{C}{}$;

	\item it projects recursive definitions of the same procedure on different
	processes, e.g., if in $C$ there are processes $\pid p_1, \ldots, \pid p_n$
	and procedure $X$, in the EPP we will find procedures $X_{\pid p_1},
	\ldots, X_{\pid p_n}$. Thus, we replace the definition typing of any
	procedure $X$ in $\dom(\Gamma)$ with the typings of its projections $X_{\pid
	p_1}$, \ldots, $X_{\pid p_n}$.

\end{itemize}

To indicate the projection of a typing environment $\Gamma$ wrt to its typed
choreography $C$, we write $\gepp{\Gamma}{C}$. To define $\gepp{\Gamma}{C}$
(and also later in this proof) we use the typing environment filtering operator
$\filter{\Gamma}{\pid p}$ defined as
\[
\filter{\Gamma}{\pid p} =
\begin{cases}
\{\ \pair{\pid p.x}{U} \ | \ \pair{\pid p.x}{U} \in \Gamma \ \} & \cup
\\
\{\ \pair{\pid p}{k[\role A]}, \pair{k[\role A]}{T} \} \ | \ \{\ \pair{\pid p}{k[\role A]}, \pair{k[\role A]}{T} \ \} \ \subseteq \Gamma \ \} & \cup
\\
\{\ \geninitg \ | \ \geninitg \in \Gamma \ \} & \cup
\\
\{\ \pair{X_{\pid p}}{\Gamma_x} \ | 
	\ \pair{X_\pid p}{\Gamma_x} \ \in \ \Gamma \ 
\}
\end{cases}
\]

\begin{definition}[Typing Projection\label{def:typing_projection}]
Let $\Gamma \hlseq C$, the projection of $\Gamma$ wrt to $C$, written
$\gepp{\Gamma}{C}$, is defined as:
{\footnotesize\[
\begin{array}{lcl}
\gepp{\Gamma}{C} & = &
\underbrace{\left\{
\bigcup\limits_{\pid q \in \group{C}{l}} 
	\epp{\Gamma}{\pid q}\underbrace{[\pid p/\pid q]}_{i.i)} \ | \ \underbrace{\pid p \in \group{C}{l} \cap \pn(\epp{C}{})}_{i.ii)} \ \wedge \ l \in \{\til l\} \ \wedge \ \til l \in \dom(\Gamma)
\right\}}_{i)}, \underbrace{\big\{ \epp{\Gamma}{\pid r} \ | \ \pid r \ \in \ \fp(\epp{C}{}) \big\}}_{ii)}
\\[25pt]
\epp{\Gamma}{\pid p} & = & 
	\underbrace{\left(\filter{\Gamma}{\pid p} \setminus 
		\left\{ \pair{X}{\Gamma_x} \ | \ \Gamma \hlseq{\pair{X}{\Gamma_x}}
		\right\}\right)}_{iii)},
	\underbrace{\left\{ 
		\pair{X_\pid p}{\epp{\Gamma_x}{\pid p}} \ 
			| \ \Gamma \hlseq {\pair{X}{\Gamma_x}}
	\right\}}_{iv)}
%
\end{array}
\]}

\end{definition}

As mentioned above, in the definition of $\gepp{\Gamma}{C}$ we distinguish
two kinds of projections: the one on service processes \emph{i}) and the one
on active processes \emph{ii}). In the first case, we unify the projection on
service processes at the same location in $C$ (i.e., in $\group{C}{l}$). To
do that in a consistent way, wrt to the EPP of C we:

\begin{itemize}
	\item obtain the identifier of process $\pid p$ \emph{i.i}), the only
	service process at location $l$ that is present in $\epp{C}{}$ (and hence
	the one that merges the behaviours of all service processes in $C$ at $l$);
	\item get the projection of $\Gamma$ on a service process $\pid q$
	($\epp{G}{\pid q}$) in $\group{C}{l}$;
	\item we rename all process-related typings in $\epp{\Gamma}{\pid q}$ to
	correspond to process $\pid p$ (by abusing the notation $\epp{\Gamma}{\pid
	q}[\pid p/\pid q]$) \emph{i.ii});
	\item we merge all the resulting, renamed typing environments into a single typing environment for process $\pid p$.
\end{itemize}

Finally, the projection of typing environment $\Gamma$ on process $\pid p$,
written $\epp{\Gamma}{\pid p}$ corresponds to the union of \emph{iii}) the
typing in $\Gamma$ related to process $\pid p$, from which we remove the
typings of definitions, and \emph{iv}) the projection of the typings of definitions, renamed for process $\pid p$.

Note the definition of $\gepp{\Gamma}{C}$ is coherent with the definition of
process projection (see \cref{def:process_projection}) in which the rule for
projecting $\prid{rec}$ terms is defined as:
\[
	\genproj{\recDef{X}{C^\prime}{C}} \quad = \quad
	\recDef{X_{\pid r}}{
			\proj{\ C'[X_{\pid r}/X]\ }{\pid r}
		}{
			\proj{\ C[X_{\pid r}/X]\ }{\pid r}
		}
\]
Similarly, $\gepp{\Gamma}C$ generates definition typings for each procedure
corresponding to each process in the choreography (assumed to be C). The
typings of definitions are guaranteed minimal (as required in
\cref{thm:EPP_typing_preservation}).

The only remark regards service typings, which are present in all projected
environments, although they might not be used. While having additional,
unused service typings does not compromise type checking, we must consider a
weakened form of minimality of typing where some unused service typings are
allowed. This fact is clearly stated in the definition of the
\cref{thm:EPP_typing_preservation}.


\subsection{Proof of the Well-Typedness property of \cref{thm:epp}} 
\label{sub:proof_of_type_preservation}

To prove the property of well-typedness of \cref{thm:epp} we prove the
stronger result of \cref{thm:EPP_typing_preservation}.

\begin{theorem}[EPP Typing Preservation\label{thm:EPP_typing_preservation}]
Let $\env,C$ be a well-typed running choreography such that $\Gamma
\hlseqmin{\env,C}$, where $\Gamma = \Gamma_d,\Gamma_c$ such that $\Gamma_d
\hlseq{\env}$, then $\gepp{\Gamma_c}C,\Gamma_d \hlseqmin{\env,\epp {C}{}}$ up
to service typings.
\end{theorem}

Intuitively, \cref{thm:EPP_typing_preservation} subsumes the well-typedness
property (1) of \cref{thm:epp}, using the environment projection defined above
to provide a minimal typing environment for $\epp{C}{}$ up to some unused
service typings.

We define some auxiliary lemmas used in the proof of
Theorem~\ref{thm:EPP_typing_preservation}.





\begin{lemma}[Composability of Typing Projections\label{lemma:composability_of_typing_projection}]
	Let $\Gamma \hlseq C$ and $\Gamma = \Gamma',\Gamma''$ then
	$\epp{\Gamma}{}^C = \epp{\Gamma'}{}^C,\epp{\Gamma''}{}^C$.
\end{lemma}

\begin{proof}
	The proof is by contradiction. The projection $\epp{\Gamma}{}^C$ returns
	exactly $\Gamma$ except for the projection of the typings of the
	procedures, as defined in \cref{def:typing_projection}. Hence the
	projection $\epp{\Gamma}{}^C$ can differ from
	$\epp{\Gamma'}{}^C,\epp{\Gamma''}{}^C$ only on definition typings. However,
	it is impossible that $\epp{\Gamma}{}^C \neq
	\epp{\Gamma'}{}^C,\epp{\Gamma''}{}^C$. Indeed, there could be only two
	cases for the partitioning of $\Gamma$ wrt any definition typing $X \in
	\dom(\Gamma)$, either:
	\begin{itemize}

		\item \emph{i}) both $\Gamma'$ and $\Gamma''$ type $X$, in which case, since $\Gamma = \Gamma',\Gamma''$, they must agree on their judgement on $X$;

		\item \emph{ii}) the judgement on $X$ is contained only in $\Gamma'$ or
		$\Gamma''$.

	\end{itemize}
in both cases the projections obtained from $X$ remain the same wrt the one
in $\Gamma$.	  
\end{proof}



We prove Lemma~\ref{lemma:chor_EPP_typing_preservation} that states that
given a well-typed choreography $C$ and a typing environment $\Gamma$ for
which $\Gamma \hlseqmin{C}$ then the projection of $\Gamma$,
$\gepp{\Gamma}{C}$, types minimally the projection of $C$,
$\epp{C}{}$.

\begin{lemma}[Choreography EPP Typing
	Preservation\label{lemma:chor_EPP_typing_preservation}] Let $C$ be a
	well-typed choreography and let $\Gamma \hlseqmin C$ then $\gepp {\Gamma}C
	\hlseqmin{\epp{C}{}}$.
\end{lemma}

\begin{proof}
Like for the proof of Theorem~\ref{thm:deadlock-freedom}, we assume our
choreographies to be well-sorted. The proof is by induction on the typing
derivation of $\Gamma \hlseqmin{C}$.

\begin{itemize}
	\case{$\did{Min}{Start1}$}
	From the premises we have $C = \genstart;C'$. We can partition $\Gamma =
	\pair{\til l}{\serviceTyping{G}{\role A}{\roles B}{\roles B}},\Gamma'$ and we
	can write the derivation
	$$
		\infer[\did{Min}{Start1}]
		{
			\Gamma',\pair{\til l}{\serviceTyping{G}{\role A}{\roles B}{\roles B}} 
			\hlseqmin{ \genstart;C' }
		}
		{
			\Gamma',\init(\wtil{\prc rC},k,G) \hlseqmin{C'}
			&
			\wtil{\prc rC} = \prc pA,\wtil{\prc qB}
			&
			\pids q \not \in \Gamma'
			&
			\til l \not \in \Gamma'
		}
	$$
	Let $\wtil{l.\prc qB} = l_1.\pid q_1[\role B_1],\cdots,l_n.\pid q_n[\role B_n]$.

	Let $\Gamma_c = \Gamma',\init(\wtil{\prc rC},k,G)$, from the induction hypothesis we have that $\Gamma_c \hlseqmin {C'}$ and therefore $\gepp{\Gamma_c}{C'} \hlseqmin {\epp{C'}{}}$. 

	By its definition $\epp{C'}{} \equivC C'_s \pp C''$ where
	$$
		C'_s = \epp{C'}{\pid p} \pp \epp{C'}{\pid q_1} \pp \ldots \pp \epp{C'}{\pid q_n}
	$$
	and
	$$
		C'' = \prod_{\pid r \ \in\ \fp(C') \setminus \{\pid p, \pids q\}} \epp{C'}
		{\pid r}
			\quad \pp \quad
					\prod\limits_{l} \left( 
						\bigsqcup\limits_{\;\;\pid s \ \in \ \group{C'}{l}}
							\epp{C'}{\pid s} \right)
	$$

	We partition $\gepp{\Gamma_c}{C'}$ (as per \cref{lemma:composability_of_typing_projection}) as
	$$
		\gepp{\Gamma_c}{C'} = 
			\Gamma'_{\pid p}
			\ ,\
			\Gamma'_{\pids q}
			\ , \
			\Gamma''
	$$
	where
	$$
		\Gamma'_{\pid p} = \Gamma''_{\pid p},\pair{\pid p}{k[\role
		A]},\pair{k[\role A]}{\epp{G}{\pid p}}
	$$
	and
	$$
	 	\Gamma'_{\pids q} = 
	 		\Gamma'_{\pid q_1} \ , \ 
	 		\ldots \ , \ 
	 		\Gamma'_{\pid q_n}
	$$
	where
	$$
		\Gamma'_{\pid q_i} = \Gamma''_{\pid q_i},\pair{\pid q_i}{k[\role A]},\pair{k[\role A]}{\epp{G}{\pid q_i}}
	$$
	such that we can write the derivation

	{\footnotesize$$
	\hspace{-25pt}
	\infer[\did{Min}{Par}]
	{
		\Gamma'' \e \Gamma'_{\pid p} \e \Gamma'_{\pids q}
		\hlseqmin{C'' \pp C'_s }
	}
	{
		\Gamma'' \hlseqmin C''
		&
		\infer[\did{Min}{Par}]
			{ 
				\Gamma'_{\pid p} \e \Gamma'_{\pids q} \hlseqmin {
					\epp{C'}{\pid p} \pp
					\epp{C'}{\pid q_1} \pp
					\ldots \pp
					\epp{C'}{\pid q_n}
				} 
			}
			{
				\Gamma'_{\pid p} \hlseqmin {\epp{C'}{\pid p}}
				&
				\infer[\did{Min}{Par}]
				{
					\Gamma'_{\pid q_1} \e \Gamma'_{\pid q_2} \e \ldots \e \Gamma'_{\pid q_n} \hlseqmin {
						\epp{C'}{\pid q_1} \pp
						\ldots \pp
						\epp{C'}{\pid q_n}
					}
				}
				{
					\Gamma'_{\pid q_1} \hlseqmin{ \epp{C'}{\pid q_1}}
					&
					\infer[\did{Min}{Par}]
					{
						\Gamma'_{\pid q_2} \e \ldots \e \Gamma'_{\pid q_n} 
							\hlseqmin{ 
								\epp{C'}{\pid q_2} \pp \ldots \pp \epp{C'}{\pid q_n}
						}
					}
					{
						\vdots
					}
				}
			}
	}
	$$}

	Since the ownership and session typings for $k$ in $\Gamma_c$ belong to
	$\init(\wtil {\prc rC},k,G)$ we can write $\Gamma'_\pid p = \Gamma''_\pid
	p, \pair{\pid p}{k[\role A]},\pair{k[\role A]}{T}$ where $\Gamma''_\pid p$
	contains those and only typings (services, ownerships, sessions, etc.) that
	type minimally the projection of continuation $C'$ for process $\pid p$. 

	Since the only difference between $\Gamma$ and $\Gamma_c$ are the typings
	for session $k$, we have that $\Gamma''_\pid p \subseteq
	\gepp{\Gamma}{C}$ and also $\Gamma'' \subseteq
	\gepp{\Gamma}{C}$. The same argument holds for typings
	$\Gamma'_{\pid q_i}$. Indeed, we can partition $\gepp{\Gamma}{C}
	= \Gamma'', \Gamma''_\pid p, \Gamma''_{\pid q_1},\ldots,\Gamma''_{\pid
	q_n}, \pair{\til l}{\serviceTyping{G}{\role A}{\roles B}{\roles B}}$ (as of
	Lemma~\ref{lemma:composability_of_typing_projection}).

	Finally, by the definition of inclusion of service typings in $\Gamma$ (cf
	\cref{sub:environments}), we can write judgement $\pair{\til
	{l}}{\serviceTyping{G}{\role A}{\roles B}{\roles B}}$ as the sequence of
	judgements $\pair{\til l}{\serviceTyping{G}{\role A}{\roles
	B}{\emptyset}},\pair{\til l}{\serviceTyping{G}{\role A}{\roles B}{\role
	B_1}},\ldots,\pair{\til l}{\serviceTyping{G}{\role A}{\roles B}{\role
	B_n}}$. 

	Therefore we write $\gepp{\Gamma'}{C}$ as 
	$$
		\epp{\Gamma'}{C} = \Gamma'' \e 
		\Gamma''_{\pid p}, \Gamma''_{\pid q_1},\ldots,\Gamma''_{\pid
		q_n},\pair{\til l}{\serviceTyping{G}{\role A}{\roles B}{\emptyset}} \e
		\pair{\til l}{\serviceTyping{G}{\role A}{\roles B}{\role B_1}} \e
		\ldots \e	\pair{\til l}{\serviceTyping{G}{\role A}{\roles B}{\role B_n}}
	$$

	Let $\filter{\wtil{l.\pid q[\role B]}}{i} = \{l_i.\pid q_i[\role
	B_i],\ldots, l_n.\pid q_n[\role B_n]\}$, we prove the case by proving the
	typing derivation for $\gepp{\Gamma}{C} \hlseqmin {\epp{C}{}}$.

	From the definition of EPP (Definition~\ref{def:epp}) we can write
	$$\epp{C}{} \equiv C_s \pp C''$$ where, given the shape of $C$, we know
	that $C''$ is the same as the one generated from $\epp{C'}{}$, as seen
	above. $C_s$ is

	$$
		C_s = \genreq;\epp{C'}{\pid p} 
			\quad \pp \quad
					\prod_{l.\prc rC \ \in\ \{\wtil{\scriptstyle l.\prc qB}\}}
						\acc{k}{l.\prc rC};\epp{C'}{\pid r}
	$$

	We now prove we can derive the typing of $\gepp{\Gamma}{C}\hlseqmin
	{\epp{C}{}}$

	{\footnotesize
	$$
		\infer[\did{Min}{Par}]
		{
			\Gamma'' \e 
			\Gamma''_{\pid p},\pair{\til l}{\serviceTyping{G}{\role A}{\roles B}{\emptyset}} \e 
			\Gamma''_{\pid q_1},\pair{\til l}{\serviceTyping{G}{\role A}{\roles B}{\role B_1}} \e 
			\ldots \e 
			\Gamma''_{\pid q_n} \e \pair{\til l}{\serviceTyping{G}{\role A}{\roles B}{\role B_n}}
			\hlseqmin {C_s \pp C''}
		}
		{
			\Gamma'' \hlseqmin{C''}
			&
			\infer[\did{Min}{Par}]
			{
				\Gamma''_{\pid p},\pair{\til l}{\serviceTyping{G}{\role A}{\roles B}{\emptyset}} \e 
				\Gamma''_{\pid q_1},\pair{\til l}{\serviceTyping{G}{\role A}{\roles B}{\role B_1}} \e 
				\ldots \e 
				\Gamma''_{\pid q_n} \e \pair{\til l}{\serviceTyping{G}{\role A}{\roles B}{\role B_n}}
				\hlseqmin {C_s}
			}
			{
				\infer[\did{Min}{Req1}]
				{
					\Gamma''_{\pid p},\pair{\til l}{\serviceTyping{G}{\role A}{\roles B}{\emptyset}} \hlseqmin{\genreq;\epp{C'}{\pid p}}
				}
				{
					\Gamma''_{\pid p},
					\pair{\pid p}{k[\role A]},
					\pair{k[\role A]}{\epp{G}{\role A}}
					\hlseqmin{ \epp{C'}{\pid p}}
					&
					\til l \not \in \Gamma''_\pid p
				}
				&
				\Delta_1
			}
		}
	$$}
	where 
	{\footnotesize$$
	\Delta_i =
	\begin{array}c
		\infer[\did{Min}{Par}]
		{
			\begin{array}l
				\Gamma''_{\pid q_i},\pair{\til l}{\serviceTyping{G}{\role A}{\roles B}{\role B_i}} \e 
				\ldots \e 
				\Gamma''_{\pid q_n} \e \pair{\til l}{\serviceTyping{G}{\role A}{\roles B}{\role B_n}}
				\\
				\hspace{50pt}\hlseqmin{
					\acc{k}{l_i.\pid q_i[\role B_i]};\epp{C'}{\pid q_i}
					\pp
					\prod\limits_{l.\prc rC \ \in\ \filter{\wtil {l.\prc qB}}{i+1}}
						\acc{k}{l.\prc rC};\epp{C'}{\pid r}
				}
			\end{array}
		}
		{
			\Delta_{i+1}
			&
			\infer[\did{Min}{Acc}]
			{
				\Gamma''_{\pid q_i},\pair{\til l}{\serviceTyping{G}{\role A}{\roles B}{\role B_i}}
				\hlseqmin{
					\acc{k}{l_i.\pid q_i[\role B_i]};\epp{C'}{\pid q_i}
				}
			}
			{
				l_i \subseteq \til l
				\qquad
				\Gamma''_{\pid q_i},
				\init(\pid q_i[\role B_i], k, G)
				\hlseqmin{ \epp{C'}{\pid q_i} }
				\qquad
				\pid q_i \not \in \Gamma_{\pid q_i}''
				\qquad
				\til l \not \in \Gamma_{\pid q_i}''
			}
		}
	\end{array}
	$$}

	Note that we are reporting only the derivation terminating with
	$\did{Min}{Req1}$, i.e., the one that applies when $\Gamma''_\pid p$ does
	not contain the typing of $\til l$. The other case is similar and it
	applies rule $\did{Min}{Req2}$.

\begin{itemize}
	\item $\Gamma'' \hlseqmin{C''}$;
	\item $\Gamma''_{\pid p},
		\pair{\pid p}{k[\role A]},\pair{k[\role A]}{\epp{G}{\role A}} \hlseqmin{ \epp{C'}{\pid p}}$;
	\item $\Gamma''_{\pid q_i},\init(\pid q_i[\role B_i], k, G) \hlseqmin{
		\epp{C'}{\pid q_i} }$.
\end{itemize}

	  hold by the induction hypothesis.

\case{$\did{Min}{Start2}$} Similar to case $\did{Min}{Start1}$.

\case{$\did{Min}{Req1}$} and \textbf{\emph{Case}.} $\did{Min}{Req2}$ follow the
proof of case $\did{Min}{Start1}$, focussing on the request branch.

\case{$\did{Min}{Acc}$} Follows the proof of case
$\did{Min}{Start1}$, following the accept branch.

\case{$\did{Min}{Cond}$} By induction hypothesis on $C_1$ or $C_2$.

\case{$\did{Min}{Com}$} 
	From the premises we have $C = \gencom;C'$ on which we can apply the typing derivation

	$$
	\infer[\did{Min}{Com}]
	{
		\Gamma',
		\pair{k[\role A]}{\lsend{\role B}{o(U);T}},
		\pair{k[\role B]}{\lrecv{\role A}{o(U);T'}}
		\hlseqmin{
			\gencom;C'
		}
	}
	{
		\Gamma' \seq \pair{\pid p}{k[\role A]},\pair{\pid q}{k[\role B]}
		&
		\Gamma' \seq \pair{\pid p.e}{U}
		&
		\Gamma',
		\pair{\pid q.x}{U},
		\pair{k[\role A]}{T},
		\pair{k[\role B]}{T'}
		\hlseqmin{
			C'
		}
	}
	$$

	Hence we consider $\Gamma = \Gamma', \pair{k[\role A]}{\lsend{\role B}{o
	(U);T}},\pair{k[\role B]}{\lrecv{\role A}{o(U);T'}}$. From the definition of
	EPP (Definition~\ref{def:epp}) we have $\epp{C}{} \equiv C_c \pp C''$ where

	$$
		C_c = \gensend;\epp{C'}{\pid p} \ \pp \ \genrecv;\epp{C'}{\pid q}
	$$
	$$
		C'' = 
			\prod_{\pid r \ \in\ \{\fp(C') \setminus \{\pid p, \pid q\}\}} \epp{C'}{\pid
			r}
			\ \pp \
			\prod_{l} \left( 
				\bigsqcup_{\pid s \ \in\  \group{C'}{l}} \epp{C'}{\pid s}
			\right)
	$$

	From the definition of $\gepp{\Gamma}{C}$ we can write

	$$
		\gepp{\Gamma}{C} = 
			\gepp{\Gamma'}{C},
			\pair{k[\role A]}{\lsend{\role B}{o(U);T}},
			\pair{k[\role A]}{\lrecv{\role A}{o(U);T'}}
	$$

	from the induction hypothesis we have that, let $\Gamma_c = \Gamma',\pair{\pid q.x}{U},\pair{k[\role A]}{T},\pair{k[\role B]}{T'}$, $\Gamma_c \hlseqmin{C'}$ and therefore $\gepp{\Gamma_c}{C'}\hlseqmin{ \epp{C'}{}	}$. We can partition $\gepp{\Gamma_c}{C'}$ as

	$$
		\gepp{\Gamma_c}{C'} = 
			\Gamma'',
			\Gamma_{\pid p},\pair{k[\role A]}{T} \e 
			\Gamma_{\pid q},\pair{\pid q.x}{U},\pair{k[\role B]}{T'}
	$$

	such that

	$$
		\infer[\did{Min}{Par}]
		{
			\Gamma'' \e \Gamma_{\pid p},\pair{k[\role A]}{T} \e 
			\Gamma_{\pid q},\pair{\pid q.x}{U},\pair{k[\role B]}{T'}
			\hlseqmin{
				\epp{C'}{\pid p} \pp \epp{C'}{\pid q} \pp C''
			}
		}
		{
			\Gamma'' \hlseqmin{ C'' }
			&
			\infer[\did{Min}{Par}]
			{
				\Gamma_{\pid p},\pair{k[\role A]}{T} \e 
				\Gamma_{\pid q},\pair{\pid q.x}{U},\pair{k[\role B]}{T'}
					\hlseqmin{
						\epp{C'}{\pid p} \pp \epp{C'}{\pid q}
					}
			}
			{
				\Gamma_{\pid p},\pair{k[\role A]}{T} \hlseqmin{ \epp{C'}{\pid p} }
				&
				\Gamma_{\pid q},\pair{\pid q.x}{U},\pair{k[\role B]}{T'} \hlseqmin{
			\epp{C'}{\pid q} } } }
	$$

	From the derivation on rule $\did{Min}{Com}$ we know that

	$$
	\gepp{\Gamma'}{C'} = \Gamma'' \e \Gamma_{\pid p} \e \Gamma_{\pid q}
	$$

	and therefore that 

	$$
	\gepp{\Gamma}{C} = \Gamma'' \e 
		\Gamma_{\pid p},\pair{k[\role A]}{\lsend{\role B}{o(U);T}} \e
		\Gamma_{\pid q},\pair{k[\role B]}{\lsend{\role A}{o(U);T'}}
	$$

	To prove $\gepp{\Gamma}{C} \hlseqmin{\epp{C}{}}$ we prove that
	we can apply rule $\did{Min}{Par}$.

	{\footnotesize$$\hspace{-25pt}
	\infer[\did{Min}{Par}]
	{
		\begin{array}l
			\Gamma'' \e
			\Gamma_{\pid p},\pair{k[\role A]}{\lsend{\role B}{o(U);T}} \e
			\Gamma_{\pid q},\pair{k[\role B]}{\lsend{\role A}{o(U);T'}}
			\\ \hspace{50pt}
			\hlseqmin{
				\gensend;\epp{C'}{\pid p} \pp \genrecv;\epp{C'}{\pid q} \pp C'' 
			}
		\end{array}
	}
	{
		\Gamma'' \hlseqmin {C''}
		&
		\infer[\did{Min}{Par}]
		{
			\begin{array}l
				\Gamma_{\pid p},\pair{k[\role A]}{\lsend{\role B}{o(U);T}} \e
				\Gamma_{\pid q},\pair{k[\role B]}{\lsend{\role A}{o(U);T'}}
				\\ \hspace{50pt}
				\hlseqmin{
					\gensend;\epp{C'}{\pid p} \pp \genrecv;\epp{C'}{\pid q}
				}
			\end{array}
		}
		{
			\infer[\did{Min}{Send}]
			{
				\begin{array}l
				\Gamma_{\pid p},\pair{k[\role A]}{\lsend{\role B}{o(U);T}}
				\\ \hspace{25pt}
				\hlseqmin{ \gensend;\epp{C'}{\pid p}}	
				\end{array}
			}
			{
				\begin{array}{c}
					\Gamma_{\pid p} \seq \pair{\pid p}{k[\role A]}
					\quad
					\pair{\pid q}{k[\role B]} \not \in \Gamma_{\pid p}
					\\
					\Gamma_{\pid p} \seq \pair{\pid p.e}{U}
					\\
					\Gamma_{\pid p},
					\pair{k[\role A]}{T} \hlseqmin{\epp{C'}{\pid p}}
				\end{array}
			}
			&
			\infer[\did{Min}{Recv}]
			{
				\begin{array}l
					\Gamma_{\pid p},\pair{k[\role B]}{\lrecv{\role A}{o(U);T'}}
					\\ \hspace{25pt}
					\hlseqmin{ \genrecv;\epp{C'}{\pid p}}	
				\end{array}
			}
			{
				\begin{array}{c}
					\Gamma_{\pid q} \seq \pair{\pid p}{k[\role B]}
					\qquad
					\pair{\pid p}{k[\role A]} \not \in \Gamma_{\pid q}
					\\
					\Gamma_{\pid q},\pair{\pid q.x}{U},\pair{k[\role B]}{T'}
					\hlseqmin{
						\epp{C'}{\pid q}
					}
				\end{array}
			}
		}
	}
	$$}

\case{$\did{Min}{Send}$}
	Analogous to case $\did{Min}{Com}$

\case{$\did{Min}{Recv}$}
	Analogous to case $\did{Min}{Com}$.

\case{$\did{Min}{Par}$}
	From the premises we know that $C = C_1 \pp C_2$ on which we can apply the typing derivation

	$$
	\infer[\did{Min}{Par}]
	{
		\Gamma_1 \e \Gamma_2 \hlseqmin {C_1 \pp C_2}
	}
	{
		\Gamma_1 \hlseqmin{C_1}
		&
		\Gamma_2 \hlseqmin{C_2}
	}
	$$

	the case is proved applying the induction hypothesis.

\case{$\did{Min}{Def}$}
	From the premises we know that $C = \recDef{X}{C''}{C'}$ on which we can apply
	the typing derivation, with $\Gamma = (\Gamma'\lub\Gamma''),
		\mathsf{solve}( \wtil {k[\role A]:T} \lub
		\wtil {k'[\role	A']:T'}, \genlrecCall_X )$

\[
\infer[\did{Min}{Def}]{
		(\Gamma'\lub\Gamma''),
		\mathsf{solve}( \wtil {k[\role A]:T} \lub
		\wtil {k'[\role	A']:T'}, \genlrecCall_X ) \hlseqmin{\recDef{X}{C''}{C'}} 
	} {
	\!\begin{array}{c}
		\Gamma_x(X) = \Gamma_x'(X) 
			\mbox{ if } X \in \dom(\Gamma_x) \cap \dom(\Gamma_x')
		\qquad
		\nexists \ k''[\role A''] \in \dom(\Gamma \lub \Gamma')
		\\[2pt]
		X \not \in \dom(\Gamma \lub \Gamma')
		\quad
		\Gamma_x' \vartriangleright_X ( \Gamma'',\wtil {k'[\role A']:T'} ), C''
		\qquad
		\Gamma_x \vartriangleright_X	( \Gamma',\wtil{k[\role A]:T} ), C'
		\qquad
		\Gamma' |_{\locs} \subseteq \Gamma'
	\end{array}
	}
	\]

To prove $\gepp{\Gamma}{C} \hlseqmin{\epp{C}{}}$, we consider the processes
$\pid p \in \pids p = \pn(\epp{C}{})$ with cardinality $[1,n]$ and we let

\begin{itemize}
	\item $\epp{C}{} = \prod_\pid p C_\pid p$
 
 \item $C_\pid p = \recDef{X_\pid p}{\epp{C''[X_\pid p/X]}{\pid p}}{\epp{C'
 [X_\pid p/X]}{\pid p}}$

 \item $\Gamma_c = \gepp{\Gamma}{C}$

	\item $\wtil{ \pair{k_{\pid p}[\role A]}{T}} = \left\{
	\pair{k[\role A]}{T} \ | \ 
		\left\{ 
			\pair{\pid p}{k[\role A],\pair{k[\role A]}{T}} 
		\right\} \subseteq \Gamma_c
		\wedge
		\pair{k[\role A]}{T} \in \wtil{\pair{k[\role A]}{T}}
	\right\}$

	\item $\wtil{ \pair{k'_{\pid p}[\role A']}{T'}} = \left\{
	\pair{k'[\role A']}{T'} \ | \ 
		\left\{ 
			\pair{\pid p}{k'[\role A'],\pair{k'[\role A']}{T'}} 
		\right\} \subseteq \Gamma_c
		\wedge
		\pair{k'[\role A']}{T'} \in \wtil{\pair{k'[\role A']}{T'}}
	\right\}$

\end{itemize}

The case is proved by the derivation $\Delta_1$ where

\[
\Delta_i = 
\begin{array}c
	\infer[\did{Min}{Par}]
	{
		\filter{\Gamma_c}{\pid p_i}, 
		\bigcup\limits_{\pid p \ \in\ \{\pid p_{i+1},\ldots,\pid p_n\}}
	 	\filter{\Gamma_c}{\pid p}
		\hlseqmin{
			C_{\pid p_i} \pp 
			\prod\limits_{\pid p \in \{\pid p_{i+1}, \ldots,\pid p_n\}} C_\pid p}
	}{
		\pi_{\pid p_i}
		&
		\infer[\did{Min}{Par}]
		{
			\bigcup\limits_{\pid p \ \in\ \{\pid p_{i+1},\ldots,\pid p_n\}}
				 	\filter{\Gamma_c}{\pid p}
					\hlseqmin{
						\prod\limits_{\pid p \in \{\pid p_{i+1}, \ldots,\pid p_n\}} C_\pid p}
		}{
			\Delta_{i+1}
		}
	}
\end{array}
\]
and
{\footnotesize
\[
\hspace{-3em}\pi_\pid p = 
\begin{array}c
\infer[\did{Min}{Def}]
{
\begin{array}l
	\filter{\gepp{\Gamma'}{C}}{\pid p} \lub 	
	\filter{\gepp{\Gamma''}{C}}{\pid p},
	\mathsf{solve}\left(
		\wtil{\pair{k_\pid p[\role A]}{T}}
		\lub
		\wtil{\pair{k'_\pid p[\role A']}{T'}},
		\genlrecCall_{X_{\pid p}}
	\right) 
	\hlseqmin{
		\recDef{X_\pid p}{\epp{C''[X_\pid p/X]}{\pid p}}{\epp{C'[X_\pid p/X]}{\pid p}}
	}
\end{array}
}{
	\begin{array}c
	\Gamma_x( X_\pid p ) = \Gamma_x'( X_\pid p ) \mbox{ if } X_\pid p \in \dom
	(\Gamma_x) \cap \dom(\Gamma_x')
	\\[5pt]
	\nexists \pair{k''[\role A'']} \in \dom\left(\filter{\gepp{\Gamma'}{C}}{\pid p} \lub
	\filter{\gepp{\Gamma''}{C}}{\pid p}\right)
	\qquad
	\Gamma_x' \vartriangleright \left(\filter{\gepp{\Gamma''}{C}}{\pid
	p},\wtil{\pair{k'_\pid p[\role A']}{T'}}\right),\epp{C''[X_\pid p/X]}{\pid p}
	\\[2pt]
	\Gamma_x \vartriangleright \left(\filter{\gepp{\Gamma'}{C}}{\pid p},
	\wtil{\pair{k_\pid p[\role A]}{T}}\right),\epp{C''[X_\pid p/X]}{\pid p}
	\qquad
	\filter{\filter{\gepp{\Gamma''}{C}}{\pid p}}{\mathsf{locs}}
	\subseteq
	\filter{\gepp{\Gamma'}{C}}{\pid p} 
	\end{array}
}	
\end{array}
\]}

Essentially, using the filtrations $\filter{\Gamma}{\pid p}$ and the
partitions $\wtil{ \pair{k_{\pid p}[\role A]}{T}}$ and $\wtil{ \pair{k'_{\pid
p}[\role A']}{T'}}$ in $\Delta_i$, we shape $\gepp{\Gamma}{C}$ in such a way
that its partitions contain all and only the typings (variable, ownership,
definitions) that minimally type the endpoint choreography $C_\pid p$, with
the exception of service typings, which are duplicated in all filtrations (as
per its definition). However, this is not a problem, as we consider a
weakened form of minimal typing that allows for additional, unused service
typings.

Such a partitioning of $\gepp{\Gamma}{C}$ is possible by the definitions of
$\gepp{\Gamma}{C}$ and $\lub$ (and $\prec$ by extension):

\[
\begin{array}{lll}
\gepp{\Gamma}{C} 
& = &
\gepp{(\Gamma'\lub\Gamma''),
		\mathsf{solve}( \wtil {k[\role A]:T} \lub
		\wtil {k'[\role	A']:T'}, \genlrecCall_X )}{C}
= \\ & = &
\bigcup\limits_{\pid p \in \pids p}
\filter{
\left(
\left(\gepp{\Gamma'}{C}\lub
\gepp{\Gamma''}{C}\right),
		\gepp{\mathsf{solve}( \wtil {k[\role A]:T} \lub
		\wtil {k'[\role	A']:T'}, \genlrecCall_X )}{C}
\right)
}{\pid p}
= \\ & = &
\bigcup\limits_{\pid p \in \pids p}
\filter{
\left(
\left(\gepp{\Gamma'}{C}\lub
\gepp{\Gamma''}{C}\right),
		\mathsf{solve}( \wtil {k[\role A]:T} \lub
		\wtil {k'[\role	A']:T'}, \genlrecCall_{X} )
\right)
}{\pid p}
\end{array}
\]

Finally, we simply rename $\genlrecCall_{X}$ to $\genlrecCall_{X_{\pid p}}$ (in
each filtration $\pid p \in \pids p$).

Then in $\pi_\pid p$ we prove the partition $\filter{\Gamma_c}{\pid p}$ to
minimally type the endpoint choreography $C_\pid p$. All preconditions in
$\pi_\pid p$ hold as the environments $\gepp{\Gamma'}{C}$,
$\gepp{\Gamma''}{C}$, $\wtil{\pair{k_\pid p[\role A]}{T}}$, and
$\wtil{\pair{k'_\pid p[\role A']}{T'}}$, contain those and only definition,
ownership, variable, and session types related to process $\pid p$ (with the
exception of duplicated service typings) and originally contained in
$\Gamma'$, $\Gamma''$,$\wtil{\pair{k[\role A]}{T}}$, and $\wtil{
\pair{k'[\role A']}{T'}}$. Definition typing identifiers are properly
renamed to be unique for $\pid p$ (i.e., from $X$ to $X_\pid p$).

\case{$\did{Min}{End}$}
	Trivial.

\case{$\did{Min}{Call}$}
	From the premises we know that 
	$C = X$ , on which we can apply the typing derivation

	$$
		\infer[\did{Min}{Call}]
	{
		\Gamma', \wtil {k[\role A]}:\genlrecCall_X,\wtil {k'[\role A']}: \gend,
		\pair{X}{(\Gamma'',\wtil {k[\role A]}:\genlrecCall_X)} \hlseqmin{X}
	}
	{
		\begin{array}c			
		\Gamma' = \mathsf{vars} \cup \mathsf{ownerships}
		\qquad
		\wtil {k'[\role A']} = \mathsf{sessions} \setminus \{\wtil {k[\role A]}\}
		\\[2pt]
		\Gamma'' = \mathsf{vars}(X) \cup \mathsf{ownerships}(X)
		\quad
		\wtil{k[\role A]} = \mathsf{sessions(X)}
		\qquad
		\Gamma'' \subseteq \Gamma'
		\end{array}
	}
	$$

	Thus, in the case, $\Gamma = \Gamma', \wtil {k[\role
	A]}:\genlrecCall_X,\wtil {k'[\role A']}: \gend,\pair{X}{(\Gamma'',\wtil
	{k[\role A]}:\genlrecCall_X)}$. Given our assumption of well sortedness, we
	can consider as EPP of $X$ the composition

	\[
		\epp{X}{} = \prod_{\pid p \in \pids p} X_\pid p
	\]

	Where processes $\pids p$ are a subset of the processes present both in the
	prefix of procedure call $X$ in $C$ and in the typing environment $\Gamma$
	(we recall, $\Gamma$ contains typings that are coalesced in
	$\gepp{\Gamma}{C}$). From the definition of $\gepp{\Gamma}{C}$, we can write 

	\[
		\Gamma_c = \gepp{\Gamma}{C} = \gepp{\Gamma'}{C}, \wtil {k_\pid p[\role
		A]}:\genlrecCall_X,\wtil {k_\pid p'[\role A']}: \gend,\bigcup_{\pid p \in
		\pids p}\pair{X_\pid p}{(\epp{\Gamma''}{\pid p},\wtil {k_\pid p[\role
		A]}:\genlrecCall_{X_{\pid p}})}
	\]

	where
	
	\begin{itemize}
	
	\item $\pair{\wtil{k_{\pid p}[\role A]}}{\genlrecCall_{X_{\pid p}}} = \left\{
	\pair{k_{\pid p}[\role A]}{\genlrecCall_{X_{\pid p}}} \ | \ 
		\left\{ 
			\pair{\pid p}{k[\role A],\pair{k[\role A]}{\genlrecCall_{X}}} 
		\right\} \subseteq \Gamma
	\right\}$

	\item $\pair{\wtil{k'_{\pid p}[\role A']}}{\gend} = \left\{
	\pair{k'_\pid p[\role A']}{\gend} \ | \ 
		\left\{ 
			\pair{\pid p}{k'[\role A'],\pair{k'[\role A']}{\gend}} 
		\right\} \subseteq \Gamma
	\right\}$
	
	\end{itemize}

	Finally, let the cardinality of $\pids p$ be $[1,n]$. The case is proved by
	the derivation $\Delta_1$ where

\[
\Delta_i = 
\begin{array}c
	\infer[\did{Min}{Par}]
	{
		\filter{\Gamma_c}{\pid p_i}, 
		\bigcup\limits_{\pid p \ \in\ \{\pid p_{i+1},\ldots,\pid p_n\}}
	 	\filter{\Gamma_c}{\pid p}
		\hlseqmin{
			X_{\pid p_i} \pp 
			\prod\limits_{\pid p \in \{\pid p_{i+1}, \ldots,\pid p_n\}} X_\pid p}
	}{
		\pi_{\pid p_i}
		&
		\infer[\did{Min}{Par}]
		{
			\bigcup\limits_{\pid p \ \in\ \{\pid p_{i+1},\ldots,\pid p_n\}}
				 	\filter{\Gamma_c}{\pid p}
					\hlseqmin{
						\prod\limits_{\pid p \in \{\pid p_{i+1}, \ldots,\pid p_n\}} X_\pid p}
		}{
			\Delta_{i+1}
		}
	}
\end{array}
\]
and
{\footnotesize
\[
\hspace{-3em}\pi_\pid p = 
\begin{array}c
\infer[\did{Min}{Call}]
{
\begin{array}l
	\filter{\gepp{\Gamma'}{C}}{\pid p},
	\pair{\wtil{k_\pid p[\role A]}}{\genlrecCall_{X_{\pid p}}},
	\pair{\wtil{k'_\pid p[\role A']}}{\gend},
	\pair{X_\pid p}{(\epp{\Gamma''}{\pid p},\wtil {k_\pid p[\role A]}:\genlrecCall_{X_{\pid
	p}})} \hlseqmin{
		X_\pid p
	}
\end{array}
}{
		\begin{array}c			
		\filter{\gepp{\Gamma'}{C}}{\pid p} = \mathsf{vars} \cup \mathsf{ownerships}
		\qquad
		\wtil {k'_\pid p[\role A']} = \mathsf{sessions} \setminus \{\wtil
		{k_\pid p[\role A]}\}
		\\[2pt]
		\filter{\gepp{\Gamma'}{C}}{\pid p} = \mathsf{vars}(X_\pid p) \cup \mathsf{ownerships}(X_\pid p)
		\quad
		\wtil{k_\pid p[\role A]} = \mathsf{sessions(X_\pid p)}
		\qquad
	 \epp{\Gamma''}{\pid p} \subseteq \filter{\gepp{\Gamma'}{C}}{\pid p}
		\end{array}
}	
\end{array}
\]}

Where in $\pi_p$ we consider the usage of auxiliary functions
$\mathsf{vars}$, $\mathsf{owenerships}$, and $\mathsf{sessions}$ on the
projection $\epp{C}{\pid p}$.

\end{itemize}
\end{proof}

\noindent We finally prove Theorem~\ref{thm:EPP_typing_preservation}.

\begin{proof}[of EPP Typing Preservation]
	From \cref{thm:EPP_typing_preservation}, we have that $\Gamma = \Gamma_d,
	\Gamma_c$ and we need to prove that we can apply rule $\did{Min}{\dep C}$
	on $\Gamma_d,\gepp{\Gamma_c}{C} \hlseqmin {\dep, \epp{C}{}}$

	$$
	\infer[\did{Min}{DC}]
	{\Gamma_d,\gepp{\Gamma_c}{C} \hlseqmin { \dep, \epp{C}{} } }
	{
		\pco( \Gamma_d,\gepp{\Gamma_c}{C} )
		&
		\Gamma_d \hlseq { \dep }	
		&
		\gepp{\Gamma_c}{C} \hlseqmin { \epp{C}{} }
	}
	$$
	
	where
	\begin{itemize}
		\item $\pco(\Gamma_d,\gepp{\Gamma}{C})$ holds as, regarding session
		typings, $\gepp{\Gamma}{C}$ just coalesces session typings and their
		related ownerships of service processes;
		\item $\Gamma_d \hlseqmin{ \dep }$ holds as per premises of
		\cref{thm:EPP_typing_preservation};
		\item $\epp{\Gamma}{} \hlseqmin{ \epp{C}{} }$ holds from
		Lemma~\ref{lemma:chor_EPP_typing_preservation} and the assumption of
		well-sortedness on $C$ (if $C$ is well-sorted also $\epp{C}{}$ is
		well-sorted and typable by $\gepp{\Gamma}{C}$).
	\end{itemize}
\end{proof}

\subsection{EPP Theorem}

Before proving Theorem~\ref{thm:epp} we define some auxiliary concepts to
establish a correspondence between a choreography and its projection.

\begin{lemma}[EPP Swap Invariance\label{lemma:swap_invariance}]
	Let $C \swapC C'$ then $\epp{C}{} \swapC \epp{C'}{}$.
\end{lemma}

\begin{proof}[Sketch]
	In the proof we show that the projection is invariant under the rules for the
	swapping relation $\swapC$ defined in \cref{fig:chor_swap}.
	$\did{CS}{EtaEta}$ is trivial. For rule $\did{CS}{EtaCnd}$ we need to check
	that the projections of the processes in the swapped interaction $\eta$ do
	not change, which holds by the definition of EPP for $\prid{cond}$ terms and
	the merging operator (merging the same $\eta$ returns $\eta$). The same
	reasoning on the EPP and the merging operator applies to all other cases.
\end{proof}

\begin{lemma}[EPP under $\equiv$\label{lemma:epp_under_equiv}] Let $C \equivC
C'$ then $\epp{C}{} \equivC \epp{C'}{}$.
\end{lemma}

\begin{proof}
	Easy by cases on the rules of $\equivC$.
\end{proof}

\begin{lemma}[Compositional EPP\label{lemma:compositional_EPP}]
	Let $C$ be well-typed and $C = C_1 \pp C_2$ then
	$\epp{C}{} \equivC \epp{C_1}{} \pp \epp{C_2}{}$.
\end{lemma}

\begin{proof}
	By definition of EPP

	$$
		\epp{C}{} = \prod_{\pid p \ \in \ \fp(C)} \epp{C}{\pid p}
			\pp \prod_{l}\left(
				\bigsqcup_{\pid s \in \group{C}{l} } \epp{C}{\pid s}
			\right)
	$$

	Since $C$ is well-typed and $C = C_1 \pp C_2$, rule $\did{T}{Par}$ applies and
	by definition of $\Gamma_1 \e \Gamma_2$ there cannot be a process $\pid p$
	such that $\pid p \in \fp(C_1) \cap \fp(C_2)$. Therefore we can write

	$$
		\epp{C}{} \equivC 
		\prod_{\pid p \ \in \ \fp(C_1)} \epp{C_1}{\pid p}
		\pp
		\prod_{\pid q \ \in \ \fp(C_2)} \epp{C_2}{\pid q}
			\pp \prod_{l}\left(
				\bigsqcup_{\pid s \in \group{C}{l} } \epp{C}{\pid s}
			\right)
	$$

	By the definition of service typing we know that \emph{i}) locations can
	implement only one role in a choreography and \emph{ii}) a location can appear
	only in one service typing. Therefore there cannot be two service processes at
	the same location in $C_1$ and $C_2$. Thus we can write

	$$
		\epp{C}{} \equivC
		\underbrace{
			\prod_{\pid p \ \in \ \fp(C_1)} \epp{C_1}{\pid p}
		}_{C_1^a}
		\pp
		\underbrace{
			\prod_{\pid q \ \in \ \fp(C_2)} \epp{C_2}{\pid q}
		}_{C_2^a}
		\pp
		\underbrace{
		\prod_{l}\left(
			\bigsqcup_{\pid r \ \in \ \group{C_1}{l} } \epp{C_1}{\pid r}
		\right)
		}_{C_1^s}
		\pp
		\underbrace{
			\prod_{l'}\left(
					\bigsqcup_{\pid s \ \in \ \group{C_2}{l'} } \epp{C_2}{\pid s}
			\right)
		}_{C_2^s}
	$$

	where $\epp{C_1}{} = C_1^a \pp C_1^s$ and $\epp{C_2}{} = C_2^a \pp C_2^s$ by definition of EPP.
\end{proof}

\subsubsection{Pruning}\label{sub:pruning}

Following our definition of EPP, the projection of $\prid{start}$ terms on
service processes yield a parallel composition of $\prid{acc}$ terms on the
locations subject of the $\prid{start}$. However, the reduction of a
$\prid{start}$ term might remove the availability to start new processes on the
locations subject of the $\prid{start}$ (i.e., if the reductum does not contain
another $\prid{start}$ term on the same locations). Contrarily, $\prid{acc}$
terms remain always available.

A similar observation can be drawn between conditional branches that contain
$\prid{com}$ terms whose projection merges all possible communications into
$\prid{recv}$ and $\prid{send}$ terms. Also in this case, reducing the condition
and projecting the result we obtain a subset of all possible branches for the
considered communication.

Similarly to~\cite{MY13} and~\cite{CHY12}, we deal with these asymmetries by
introducing the \emph{pruning relation} (see \cref{def:pruning}), which allows
us to ignore unused \emph{i}) endpoint services and \emph{ii}) input branches.

Before continuing with the last auxiliary results and the proof of
Theorem~\ref{thm:epp} we need to extend the labels of the semantics of annotated
Frontend Choreographies (see \cref{sub:ac_annotated_semantics}) with the
identifiers of the processes involved in a reduction
$$
\beta \gram \com{k}{\pid p[\role A]}{\role B.o} \; \Div \; \grecv{\role
A}{\pid q[\role B]}{o(x)} \; \Div \; \tau@\pid p \; \Div \; \tau
$$
 and the annotation of the reduction with rule $\did{C}{Cond}$ as

$$
\infer[\did{C}{Cond}]
{
	\dep,\ \cond{\pid p.e}{C_1}{C_2}
	\quad\lto{\tau@\pid p}\quad
	\dep,\ C_i
}
{
	i = 1 \text{ if } \auxfn{eval}(e, \dep(\pid p)) = \text{ true, }
	i = 2 \text{ otherwise}
}
$$ 
Let also 
$\pn(\com{k}{\pid p[\role A]}{\role B.o}) = \{\pid p\}$,
$\pn(\grecv{\role A}{\pid q[\role B]}{o(x)}) = \{\pid q\}$, 
$\pn(\tau@\pid p) = \{\pid p\}$, and $\pn(\tau) = \emptyset$

\begin{lemma}[Passive Processes Pruning
	Invariance\label{lemma:passive_processes_pruning_invariance}] $\dep, C
	\lto{\beta} \dep',C'$ implies that for all $\pid p \in
	\fp(C)\setminus\pn(\beta)$, $\epp{C'}{\pid p} \prec \epp{C}{\pid p}$.
\end{lemma}

\begin{proof}[Sketch]
	By cases on the derivation of $C$. The only interesting case is
	$\did{C}{Cond}$ in which the projection of the processes receiving
	selections are merged. The thesis follows directly from \cref{def:pruning}
	and \cref{lemma:swap_invariance,lemma:epp_under_equiv}.

\end{proof}

\subsection{Proof of Theorem \ref{thm:epp}} 
\label{sub:proof_of_theorem_thm:epp}

We restate items (2) and (3) of Theorem~\ref{thm:epp} to include annotated
reductions.
\newline

\noindent\textbf{Theorem~\ref{thm:epp}}~(EPP Operational Correspondence)

\textit{Let $\dep,C$ be well-typed and well-annotated. Then,
\begin{enumerate}
\item (Completeness) $\dep,C \lto{\beta} \dep',C'$ implies $\dep,\epp{C}{} \lto{\beta} \dep',C''$ and $\epp{C'}{} \prec C''$.
\item (Soundness) $\dep,\epp{C}{} \lto{\beta} \dep',C''$ implies $\dep,C \lto{\beta} \dep',C'$ and $\epp{C'}{} \prec C''$.
\end{enumerate}}

We report below the respective proofs of \emph{(Completeness)} and \emph{(Soundness)} separately.

\begin{proof}[(Completeness)]~\\

	Proof by induction on the derivation of $\dep,C \lto{\beta} \dep',C'$.
	
	\begin{description}
		
		\case{$\did{C}{Send}$} we know that $C = \gensend;C_c$ and we can
		write the derivation

			$$
				\infer[\did{C}{Send}]
				{
					\dep,\ \eta; C
					\quad \lto {\com{k}{\prc pA}{\role B.o}} \quad
					\dep',\ C_c
				}{
					\eta = \gensend &
					\renv{\dep}{\gensend}{\dep'}
				}
			$$

			and $C' = C_c$.

			From the definition of EPP we have that $\epp{C}{} = C_{\mathit{act}} \pp
			C_s$ such that

			$$
				C_{\mathit{act}} = \com{k}{\prc pA.e}{\role B.o;\epp{C_c}{\pid p}}
				\pp \prod_{\pid r \ \in\ \fp(C)\setminus\{\pid p\}}
				\epp{C_c}{\pid r}
			$$

			and

			$$
				C_s = \prod_l \left( \ 
					\bigsqcup_{\pid s \ \in \ \group{C}{l}}
				 	\epp{C_c}{\pid s}
				 \ \right)
			$$

			While $\epp{C'}{} \equivC C'_{\mathit{act}} \pp C_s$

			$$
				C'_{\mathit{act}} = 
					\epp{C_c}{\pid p}
					\pp \prod_{\pid r \ \in\ \fp(C')\setminus\{\pid p\}} \epp{C_c}{\pid r}
			$$

			We can apply Rules $\did{C}{Par}$, $\did{C}{Eq}$, and $\did{C}{Send}$ on $\dep,\epp{C}{}$ such that

			$$
			\infer*[\did{C}{Par}]
			{
				\dep,\ \epp{C}{}
				\quad \lto {\com{k}{\prc pA}{\role B.o}} \quad
				\dep'',\ C''
			}{
				\eta = \gensend
				&
				\renv{\dep}{\eta}{\dep''}
			}
			$$

			for which it holds that $\dep' = \dep''$ by rule $\did{D}{Send}$.

			$$
				C'' = \epp{C_c}{\pid p}
					\pp \prod_{\pid r \ \in\ \fp(C')\setminus\{\pid p\}} \epp{C_c}{\pid r}
					\pp
					C_s
			$$

			for which it holds that $\epp{C'}{} \prec C''$.

		\case{$\did{C}{Recv}$}

		we know that $\dep,C = \dep,\ \genbranchI$  and we can write the
		derivation

		$$
			\infer[\did{C}{Recv}]
			{
				\dep,\ \genbranchI
				\quad \lto {k:\grecv{\role A} {\prc qB}{o_j(x_j)} } \quad
				\dep',\ C_j
			}{
				j \in I
				&
				\renv{\dep}{\com{k}{\role A}{\pid q[\role B].o_j(x_j)}}{\dep'}
			}
		$$

		for $\beta = k:\grecv{\role A} {\prc qB}{o_j(x_j)}$ and
		$C' = C_j$.

		By the definition of EPP we have 
		\[
		\epp{C}{} \equivC
			\com{k}{\role A}{\pid q[\role B].\left\{o_i(x_i);
				\epp{C_i}{\pid q}\right\}_ {i\in I}} 
			\pp 
				\prod_{\pid p \ \in \ \fp(C)\setminus\{\pid q\}}
					\left(\bigsqcup_{i \	\in \ I} \epp {C_i}{\pid p}\right)
			\pp
				\prod_l \left(
					\bigsqcup_{\pid r \ \in \ \group{C}{l} } \epp{C}{\pid r}
				\right)
		\]
		Then we can apply rules $\did{C}{Par}$, $\did{C}{Eq}$, and
		$\did{C}{Recv}$ such that

		{\footnotesize
		$$
			\infer*[\did{C}{Par}]
			{
				\dep,\ \epp{C}{}
				\quad \lto {k:\grecv{\role A} {\prc qB}{o_j(x_j)} } \quad
				\dep'',\ \epp{C_j}{\pid q} 
			\pp 
				\prod\limits_{\pid p \ \in \ \fp(C)\setminus\{\pid q\}}
					\left(\bigsqcup\limits_{i \	\in \ I} \epp {C_i}{\pid p}\right)
			\pp
				\prod_l \left(
					\bigsqcup\limits_{\pid r \ \in \ \group{C}{l} } \epp{C}{\pid r}
				\right)
			}{
				j \in I
				&
				\renv{\dep}{\com{k}{\role A}{\pid q[\role B].o_j(x_j)}}{\dep''}
			}
		$$}

		and 

		$$
			C'' = \epp{C_j}{\pid q} 
			\pp 
				\prod_{\pid p \ \in \ \fp(C)\setminus\{\pid q\}}
					\left(\bigsqcup_{i \	\in \ I} \epp {C_i}{\pid p}\right)
			\pp
				\prod_l \left(
					\bigsqcup_{\pid r \ \in \ \group{C}{l} } \epp{C}{\pid r}
				\right)
		$$

		From rule $\did{D}{Recv}$ we know that $D'' = D'$. Finally $\epp{C'}{} \prec
		C''$ by \cref{def:pruning} and
		\cref{lemma:passive_processes_pruning_invariance}.

		\case{$\did{C}{Start}$}

			we know that $C = \genstart; C_c$ and we can write the derivation

			\[
				\infer[\did{C}{Start}]
				{
					\env,\ \genstart; C
					\quad \to \quad
					\env',\
					C[k'/k][\pids r/\pids q] 
				}
				{
					\env\fresh{k',\pids r}
					&
					\delta = \start{k'}{\pid p[\role A]}{\wtil{l.\pid q[\role B]}}
					&
					\renv{\env}{\delta}{\env'}
				}
			\]

			and $C' = C_c[k'/k][\pids r/\pids q]$.

			From the definition of EPP we have

			$$
				\epp{C'}{} =
					\prod_{\pid q \ \in \ \fp(C') } \epp{C'}{\pid q}
				\pp
					\prod_{l}\left(
						\bigsqcup_{\pid s \ \in \ \group{C'}{l} } \epp{C'}{\pid s}
					\right)
			$$

			and

			\[
				\epp{C}{} \equivC
				\begin{cases}
						& \genreq;\epp{C_c}{\pid p}
						\\
						\pp &
						\prod\limits_{l.\prc qB \ \in \ \wtil{l.\prc qB}}
							\acc{k}{l.\prc qB};\epp{C_c}{\pid	q}
						\\
						\pp &	
							\prod\limits_{\pid r \ \in \ \fp(C)\setminus\{\pid p\}} \epp{C}{\pid r}
						\\
						\pp &
						\prod\limits_{l' \not \in \ \til l}
							\left(
								\prod\limits_{\pid s \ \in \ \group{C}{l'}}
								\epp{C}{\pid s}
							\right)
				\end{cases}
			\]

			we can apply rules $\did{C}{Par}$, $\did{C}{Eq}$, $\did{C}{PStart}$
			such that

			\[
			\infer*[\did{C}{Par}]
			{
				\begin{array}{l}
				\env, \epp{C}{} \lto{\tau} \env'', C''
				\end{array}
			}
			{
				\begin{array}c
				i \in \{1,\dots,n\}
				\qquad
				\env\fresh{k'',\pids r'}
				\qquad
				\{ \wtil{l.\role B} \} = \biguplus_i \{\wtil{l_i.\role B_i}\}_i
				\qquad
				\{\pids r'\} = \bigcup_i \{ \pids r'_i \}
				\\[2pt]
				\delta = \start{k''}{\pid p[\role A]}{\wtil{l_1.\pid r'_1[\role
				B_1]},\dots,\wtil{l_n.\pid r'_n[\role B_n]}}
				\qquad
				\renv{\env}{\delta}{\env''}
				\end{array}
			}
			\]

			where

			\[
			C'' \equivC
			\begin{cases}
				& \epp{C_c}{\pid p}[k''/k]
				\\
				\pp &
				\prod\limits_{(\pid q, \pid r') \ \in \ 
				\big\{
					(\pid q_1, \pid r'_1),\ldots,(\pid q_n,\pid r'_n) \big\}}
					\epp{C_c}{\pid q}[k''/k][\pid q/\pid {r'}]
				\\
				\pp &
				\prod\limits_{\pid r \ \in \ \fp(C_c) \setminus \{\pid p, \pids q\}}
				\epp{C_c}{\pid r}
				\\
				\pp &
				\prod\limits_{l.\prc qB \ \in \ \wtil{l.\prc qB}}
						\acc{k}{l.\prc qB};\epp{C_c}{\pid	q}
				\\
				\pp &
				\prod\limits_{l' \ \not \in \ \til l}
					\left(
						\prod\limits_{\pid s \ \in \ \group{C_c}{l'}}
						\epp{C_c}{\pid s}
					\right)
				\end{cases}
			\]

			Observe that we can $\alpha$-rename $k''$ to $k'$ and $\pids r'$ to
			$\pids r$ as $k''$, $k'$, $\pids r'$, and $\pids r$ are all fresh wrt
			$\dep,C$.

			From the application of rule $\did{D}{Start}$ we can find $\Gamma$ such
			that $$\Gamma \hlseqmin{( \ \dep'',C'' \ )[k'/k''][\pids r/ \pids r']}$$
			and $$\Gamma \hlseqmin{(\dep',C'')[k'/k''][\pids r/ \pids r']}$$

			and by $\alpha$-renaming we have that 

			$$\dep,\epp{C}{} \lto{\tau} \dep',C''[k'/k''][\pids r/ \pids r']$$

			Finally $\epp {C'} {} \prec C''[k'/k''][\pids r/ \pids r']$ by
			\cref{lemma:passive_processes_pruning_invariance}.

		\case{$\did{C}{PStart}$}

		Similar to (in particular the second part of) the proof of case
		$\did{C}{Start}$.

		\case{$\did{C}{Cond}$}

			we know that $C \equivC \cond{\pid p.e}{C_1}{C_2}$ and we can write the
			derivation

			$$
			\infer[\did{C}{Cond}]
			{
				\dep,\ \cond{\pid p.e}{C_1}{C_2}
				\quad\lto{\tau@\pid p}\quad
				\dep,\ C_i
			}
			{
				i = 1 \text{ if } \auxfn{eval}(e, \dep(\pid p)) = \text{ true, }
				i = 2 \text{ otherwise}
			}
			$$

			We only consider the case for $\auxfn{eval}(e, \dep(\pid p)) =
			\text{true}$ as $\auxfn{eval}(e, \dep(\pid p)) = \text{false}$
			is similar.

			$C' = C_1$ and by the definition of EPP

			$$
			\epp{C}{} \equivC
			\cond{\pid p.e}{\epp{C_1}{\pid p}}{\epp{C_2}{\pid p}}
				\pp 
				\prod_{\pid q \ \in \ \fp(C')\setminus\{\pid p\}}
					\epp{C_1}{\pid q} \sqcup \epp{C_2}{\pid q} 
				\pp
				\prod_{l}\left(\bigsqcup_{\pid r \ \in \ \group{C}{l}} \epp{C}{\pid r}\right)
			$$

			and

			$$
			\epp{C'}{} \equivC
				\epp{C_1}{\pid p} 
				\pp 
				\prod_{\pid q \ \in \ \fp(C')\setminus\{\pid p\}}	\epp{C_1}{\pid q} 
				\pp
				\prod_{l}\left(\bigsqcup_{\pid r \ \in \ \group{C_1}{l}} \epp{C_1}{\pid r}\right)
			$$

			We can apply rules $\did{C}{Par}$, $\did{C}{Eq}$, and $\did{C}{Cond}$
			such that $\dep,\epp{C}{} \lto{\tau@\pid p} \dep,C''$ where

			$$
				C'' = \epp{C_1}{\pid p}
				\pp
				\prod_{\pid q \ \in \ \fp(C')\setminus\{\pid p\}}
					\epp{C_1}{\pid q} \sqcup \epp{C_2}{\pid q} 
				\pp
				\prod_{l}\left(\bigsqcup_{\pid r \ \in \ \group{C}{l}} \epp{C}{\pid r}\right)
			$$

			and $\epp{C'}{} \prec C''$ by \cref{lemma:passive_processes_pruning_invariance}.

		\case{$\did{C}{Ctx}$ and \textbf{Case} $\did{C}{Par}$} 

			proved by the definition of EPP and the induction hypothesis.
		
		\case{$\did{C}{Eq}$}

			We can write the derivation
			$$
			\infer[\did{C}{Eq}]
			{
				\dep, C_1 \quad \lto \beta \quad \dep^\prime, C_2
			}
			{
				\ctx{R} \in \{\,\equivC\,,\,\swapC\,\}
				&
				C_1 \,\ctx{R}\, C^\prime_1
				&
				\dep, C^\prime_1 \lto \beta \dep^\prime, C^\prime_2
				&
				C^\prime_2 \,\ctx{R}\, C_2
			}
			$$

			For $\ctx{R}\ = \ \equivC$, proved by the definition of EPP,
			\cref{lemma:epp_under_equiv}, and the induction hypothesis.

			For $\ctx{R} \ = \ \swapC$, proved by the definition of EPP,
			\cref{lemma:swap_invariance}, and the induction hypothesis.

	\end{description}
\end{proof}

\begin{proof}[(Soundness)]
	Proof by induction on the structure of $C$.
		
	\begin{description}
		
		\case{$C = \gencom;C_c$}
			From the definition of EPP we have 

			{\footnotesize
			$$
				\epp{C}{} \equivC
					\gensend;\epp{C_c}{\pid p} 
					\pp
					\genrecv;\epp{C_c}{\pid q}
					\pp
					\prod_{\pid r \ \in \ \fp(C)} \epp{C_c}{\pid r}
					\pp
					\prod_{l}\left(
						\bigsqcup_{\pid s \ \in \ \group{C}{l}} \epp{C}{\pid s}
					\right)
			$$}

			we proceed by subcases on the last applied rule in the derivation of
			$\dep,\epp{C}{} \lto{\beta}	\dep',C''$.

			\begin{description}
				\case{$\did{C}{Send}$}

					Divided into subcases whether $\beta = \com{k}{\prc pA}{\role B.o}$
					holds or not.

					\begin{description}
						\case{$\beta=\com{k}{\prc pA}{\role B.o}$} $\dep,\epp{C}{}$
							reduces to $\dep',C''$ with rules $\did{C}{Par}$,
							$\did{C}{Eq}$, ending with rule $\did{C}{Send}$ such that

							$$
							C'' = 
								\epp{C_c}{\pid p} 
								\pp
								\genrecv;\epp{C_c}{\pid q}
								\pp
								\prod_{\pid r \ \in \ \fp(C)\setminus\{\pid p, \pid q\}} \epp{C_c}{\pid
								r}
								\pp
								\prod_{l}\left(
									\bigsqcup_{\pid s \ \in \ \group{C}{l}} \epp{C}{\pid s}
								\right)
							$$

							$\dep, C$ mimics $\dep,\epp{C}{}$ with rules $\did{C}{Eq}$ and
							$\did{C} {Send}$ for which $\dep,C \lto{\beta}
							\dep'',C'$, $\dep' = \dep''$ by rule $\did{D}{Send}$,

							$$
							\epp{C'}{} \equivC
								\epp{C_c}{\pid p} 
								\pp
								\genrecv;\epp{C_c}{\pid q}
								\pp
								\prod_{\pid r \ \in \ \fp(C)\setminus\{\pid p, \pid q\}} \epp{C_c}{\pid r}
								\pp
								\prod_{l}\left(
									\bigsqcup_{\pid s \ \in \ \group{C}{l}} \epp{C}{\pid s}
								\right)
							$$

						and $\epp{C'}{} \prec C''$.

						\case{$\beta \neq \com{k}{\prc pA}{\role B.o}$}

							In this case $\dep,C$ can mimic $\dep,\epp{C}{}$ with the
							application of rules $\did{C}{Eq}$, $\did{C}{Par}$, and
							$\did{C}{Send}$ and the thesis follows by the induction
							hypothesis.
 
					\end{description}

				\case{$\did{C}{Recv}$, $\did{C}{PStart}$, or $\did{C}{Cond}$}
					In this case $\dep,\epp{C}{}$ reduces with rules $\did{C}{Eq}$,
					$\did{C}{Par}$, and respectively ends the derivation with either
					$\did{C}{Recv}$, $\did{C}{PStart}$, or $\did{C}{Cond}$, i.e., some
					process $\pid r \in \fp(C)$ ($\pid p$ and $\pid q$ included) either
					receives a message, starts a new session with some service
					processes, or reduces to some branch. $\dep,C$ can mimic
					$\dep,\epp{C}{}$ applying rules $\did{C}{Eq}$, $\did{C}{Par}$ and
					terminates the derivation with either rules $\did{C}{Recv}$,
					$\did{C}{PStart}$ (or $\did{C}{Start}$, depending on the form of
					$C$) or $\did{C}{Cond}$. The thesis follows by the induction
					hypothesis.

			\end{description}

		\case{$C = \gensend;C_c$}

			Similar to case $C = \gencom;C_c$.

		\case{$C = \genbranchI$}

			From the definition of EPP we have

			$$
				\epp{C}{} \equivC
					\com{k}{\role A}{\pid q[\role B].\{o_i(x_i);\epp{C_i}{\pid q}\}_{i
					\in I}}
					\pp
					\prod_{i \ \in \ I}\left(
						\bigsqcup_{\pid p \ \in \ \fp(C_i)} \epp{C_i}{\pid p}
					\right)
					\pp
					\prod_{k}\left(
						\bigsqcup_{\pid r \ \in \ \group{C}{l}} \epp{C}{\pid r} 
					\right)
			$$
			we proceed by subcases on the last applied rule in the derivation of
			$\dep,\epp{C}{} \lto{\beta} \dep',C''$.

			\begin{description}
				\case{$\did{C}{Recv}$} 

					Divided into subcases whether $\beta = \pair{k}{\grecv{\role
					A}{\prc qB}{o_j}}$, $j \in I$ or not.

					\begin{description}
						\case{$\beta = \pair{k}{\grecv{\role A}{\prc qB}{o_j}}$, $j \in
						I$} $\dep,\epp{C}{}$ reduces to $\dep',C''$ with rules
						$\did{C}{Par}$, $\did{C}{Eq}$, and terminates with rule
						$\did{C}{Recv}$ such that

						$$
							C'' =
								\epp{C_j}{\pid q}
								\pp
								\prod_{i \ \in \ I}\left(
									\bigsqcup_{\pid p \ \in \ \fp(C_i)\setminus\{\pid q\}} \epp{C_i}{\pid p}
								\right)
								\pp
								\prod_{k}\left(
									\bigsqcup_{\pid r \ \in \ \group{C}{l}} \epp{C}{\pid r} 
								\right)
						$$

						$\dep,C$ mimics $\dep,\epp{C}{}$ with rule $\did{C}{Recv}$ for
						which $\dep,C \lto{\beta} \dep'',C'$ where $\dep'' = \dep'$ by rule $\did{D}{Recv}$ and 

						$$
						\epp{C'}{} =
							\epp{C_j}{\pid q}
							\pp
							\prod_{\pid p \ \in \ \fp(C_j)\setminus\{\pid q\}} \epp{C_j}{\pid p}
							\pp
							\prod_{k}\left(
								\bigsqcup_{\pid r \ \in \ \group{C_j}{l}} \epp{C_j}{\pid r} 
							\right)
						$$

						and $\epp{C'}{} \prec C''$ by Lemma~\ref{lemma:passive_processes_pruning_invariance}.

						\case{$\beta \neq \pair{k}{\grecv{\role A}{\prc qB}{o_j}}$}
							For any $\beta$ of this case $\dep,C$ can mimic
							$\dep,\epp{C}{}$ with the application of rules $\did{C}{Eq}$
							and $\did{C}{Par}$, terminating with rule $\did{C}{Recv}$ and the thesis follows by the induction hypothesis.

 					\end{description}

				\case{$\did{C}{Send}$, $\did{C}{PStart}$, or $\did{C}{Cond}$} is
				similar to subcase \textbf{Case} $\did{C}{Recv}$, $\did{C}{PStart}$, or
				$\did{C}{Cond}$ of \\\textbf{Case} $C = \gencom;C_c$.

			\end{description}

		\case{$C = \genstart;C_c$}

			$$
				\epp{C}{} \equivC
					\genreq;C_c 
					\pp
					\prod_{\pid r \ \in \ \fp(C_c)\setminus\{\pid p\}} \epp{C_c}{\pid r}
					\pp
					\prod_{l}\left(
						\bigsqcup_{\pid s \ \in \ \group{C}{l}}\epp{C}{\pid s}
					\right)
			$$

			we proceed by subcases on the last applied rule in the derivation of
			$\dep,\epp{C}{} \lto{\beta} \dep,C''$.

			\begin{description}
				\case{$\did{C}{PStart}$}
					$\dep,\epp{C}{}$ can reduce to $\dep',C''$ with a process $\pid r$
					(including $\pid p$) that starts a new session with some service
					processes. $\dep,C$ can reduce to $\dep'',C'$ mimicking
					$\dep,\epp{C}{}$ by applying rules $\did{C}{Eq}$, $\did{C}{Par}$,
					terminating with either rule $\did{C}{PStart}$ or $\did{C}{Start}$.

				\case{$\did{C}{Send}$, $\did{C}{Recv}$, and $\did{C}{Cond}$} are
				similar to the corresponding proof for the previous cases.

			\end{description}

		\case{$C = \gencond$}

			From the definition of EPP we have

			{\footnotesize
			$$\epp{C}{} \equivC
				\cond{\pid p.e}{\epp{C_1}{\pid p}}{\epp{C_2}{\pid p}}
				\pp
				\prod_{\pid q \ \in \ \fp(C_1) \ \cup \ \fp(C_2) \setminus \{\pid p\}}
					\epp{C_1}{\pid q} \sqcup \epp{C_2}{\pid q}
				\pp
				\pp
				\prod_{l}\left(
					\bigsqcup_{\pid r \ \in \ \group{C}{l}} \epp{C}{\pid r}
				\right)
			$$}

			we proceed by subcases on the derivation of $\dep,\epp{C}{} \lto{\beta}
			\dep',C''$.

			\begin{description}

				\case{$\did{C}{Cond}$} 
					$\dep,\epp{C}{}$ can reduce to $\dep',C''$ with:
					\begin{description}
						
						\case{$\beta = \tau@\pid p$} that reduces to a branch. $\dep,C$ can
						mimic $\dep, \epp{C}{}$ applying rules $\did{C}{Eq}$,
						$\did{C}{Par}$, and terminating the derivation with rule
						$\did{C}{Cond}$. The case is proved by
						Lemma~\ref{lemma:passive_processes_pruning_invariance}.

						\case{$\beta = \tau@\pid r,\ \pid r \neq \pid p$} where process
						$\pid r$ reduced to a branch. The case follows the proof of the
						previous case and the thesis follows by the induction hypothesis.

					\end{description}

				\case{$\did{C}{Recv}$, $\did{C}{Send}$, $\did{C}{PStart}$}
					are similar to the corresponding proof for the previous cases.

			\end{description}

		\case{$C = \genreq;C_c$}
			Case not allowed by the hypothesis that $\dep,\epp{C}{} \lto \beta
			\dep,C''$.

		\case{$C = \genacc;C_c$}
			Case not allowed by the hypothesis that $\dep,\epp{C}{} \lto \beta
				\dep,C''$.

		\case{$C = \recDef{X}{C''}{C'}$}
			proved by Lemma~\ref{lemma:epp_under_equiv} and the induction hypothesis.

		\case{$C = \recCall{X}$}
			Case not allowed by the hypothesis that $C$ is well-sorted.

		\case{$C = C_1 \pp C_2$}

			$\epp{C}{} \equivC \epp{C_1}{} \pp \epp{C_2}{}$ by
			Lemma~\ref{lemma:compositional_EPP}.

			we proceed by subcases for $n$ equal to the length of the derivation of
			$\dep,\epp{C}{} \lto{\beta} \dep',C''$

			\begin{description}

				\case{$n=1$}
					In this case the only applicable rule is $\did{C}{PStart}$ where,
					Since both $\epp{C_1}{}$ and $\epp{C_2}{}$ reduce, we can infer, let

					$$\wtil{l.\prc qB}
					= l_1.\pid q_1[\role B_1], \ldots, l_i.\pid q_i[\role
					B_i],l_{i+1}.\pid q_{i+1}[\role B_{i+1}] \ldots, l_n.\pid q_n[\role
					B_n]$$

					that

					$$
						C_1 \equivC \genreq;C_1^r
							\pp
							\prod_{j=1}^i \acc{k}{l_j.\pid q_j[\role B_j]};C_1^{j}
							\pp C_c^1
					$$

					$$
						C_2 \equivC 
							\prod_{j=i+1}^n \acc{k}{l_j.\pid q_j[\role B_j]};C_2^{j}
							\pp C_c^2
					$$

					and by the definition of EPP that

					$$
						\epp{C_1}{} \equivC \genreq;\epp{C_1^r}{\pid p}
							\pp
							\prod_{j=1}^i 
								\acc{k}{l_j.\pid q_j[\role B_j]};\epp{C_1^{j}}{\pid q_j}
							\pp \epp{C_c^1}{}
					$$

					$$
						\epp{C_2}{} \equivC
							\prod_{j=i+1}^n \acc{k}{l_j.\pid q_j[\role B_j]};\epp{C_2^{j}}{\pid q_j}
							\pp \epp{C_c^2}{}
					$$

					Observe that we can proceed without loss of generality as the
					symmetric case (with $\pid p \in \fp(C_2)$) follows the same
					structure.
					
					\[
					\infer[\did{C}{PStart}]
					{
						\begin{array}{l}
						\env, \epp{C_1}{} \pp \epp{C_2}{}
						\quad \lto{\tau} \quad
						\env'',C''
						\end{array}
					}
					{
						\begin{array}c
						i \in \{1,\dots,n\}
						\qquad
						\env\fresh{k',\pids r}
						\qquad
						\{ \wtil{l.\role B} \} = \biguplus_i \{\wtil{l_i.\role B_i}\}_i
						\qquad
						\{\pids r\} = \bigcup_i \{ \pids r_i \}
						\\[2pt]
						\delta = \start{k'}{\pid p[\role A]}{\wtil{l_1.\pid r_1[\role
						B_1]},\dots,\wtil{l_n.\pid r_n[\role B_n]}}
						\qquad
						\renv{\env}{\delta}{\env''}
						\end{array}
					}
					\]

					where

					$$
						C'' \equivC 
						\left\{\begin{array}l
							\epp{C_1^r}{\pid p}[k'/k]
								\pp
								\left(
									\begin{array}c
										\prod_{j=1}^i \epp{C_1^{j}}{\pid q_j}
										\\[8pt]
										\pp
										\prod_{j=i+1}^n \epp{C_2^{j}}{\pid q_j}
									\end{array}
								\right)[k'/k][\pids r/\pids q]
								\\[5pt]
								\pp
								\left(
									\begin{array}c
									\prod_{j=1}^i 
									\acc{k}{l_j.\pid q_j[\role B_j]};\epp{C_1^{j}}{\pid q_j}
									\\[8pt]
									\pp
									\prod_{j=i+1}^n \acc{k}{l_j.\pid q_j[\role
										B_j]};\epp{C_2^{j}}{\pid q_j}
									\end{array}
								\right)
									\pp \epp{C_c^1}{} \pp \epp{C_c^2}{}
						\end{array}\right.
					$$

					Then $\dep,C$ can mimic $\dep,\epp{C}{}$ applying rule
					$\did{C}{PStart}$ with reduction

					\[
					\infer[\did{C}{PStart}]
					{
						\begin{array}{l}
						\env, C_1 \pp C_2
						\quad \lto{\tau} \quad
						\env'',C'
						\end{array}
					}
					{
						\begin{array}c
						i \in \{1,\dots,n\}
						\qquad
						\env\fresh{k'',\pids r'}
						\qquad
						\{ \wtil{l.\role B} \} = \biguplus_i \{\wtil{l_i.\role B_i}\}_i
						\qquad
						\{\pids r'\} = \bigcup_i \{ \pids r'_i \}
						\\[2pt]
						\delta = \start{k''}{\pid p[\role A]}{\wtil{l_1.\pid r'_1[\role
						B_1]},\dots,\wtil{l_n.\pid r'_n[\role B_n]}}
						\qquad
						\renv{\env}{\delta}{\env''}
						\end{array}
					}
					\]

					where

					{\footnotesize$$
						C' \equivC
						C_1^r[k''/k]
								\pp
						\left(
							\begin{array}c
								\prod_{j=1}^i C_1^{j} \pp
								\\[8pt]
								\prod_{j=i+1}^n C_2^{j}
							\end{array}
						\right)[k''/k][\pids r'/\pids q]
						\pp
						\left(
							\begin{array}c
							\prod_{j=1}^i 
							\acc{k}{l_j.\pid q_j[\role B_j]};C_1^{j}
							\\[8pt]
							\pp
							\prod_{j=i+1}^n \acc{k}{l_j.\pid q_j[\role
								B_j]};C_2^{j}
							\end{array}
						\right)
						\pp \epp{C_c^1}{}
						\pp \epp{C_c^2}{}
					$$}

					Following the structure of the second part of the proof of
					\textbf{Case} $\did{C}{Start}$ for the proof of \emph{Completeness} of
					Theorem~\ref{thm:epp}, by $\alpha$-renaming we have $\dep'' = \dep'$
					and $\epp{C'}{} \prec C''$.

				\case{$n>1$}

					For $n > 1$ we have a derivation similar to

					$$
						\infer[\did{C}{Par}]
						{
							\dep,\epp{C_1}{} \pp \epp{C_2}{}
								\quad \lto{\beta} \quad
							\dep',C_1'' \pp \epp{C_2}{}
						}
						{
							\hspace{115pt}
							\begin{array}l
								R
								\\
								\; \vdots
								\quad 
								\begin{array}l
								\mbox{\footnotesize $n-1$ times, each either}
								\\
								\mbox{$\did{C}{Par}$ or $\did{C}{Eq}$}
								\end{array}
							\end{array}
						}
					$$

					where $R$ is the last applied rule, $R \in \{\did{C}{Send},
					\did{C}{Recv}, \did{C}{PStart}, \did{C}{Cond}\}$. The thesis follows
					from the induction hypothesis.

					The proof for the mirror case $\dep,\epp{C_1}{} \pp \epp{C_2}{}
					\quad \lto{\beta} \quad \dep',\epp{C_1}{} \pp C_2''$ follows the same
					structure.

			\end{description}

		\case{$C = \inact$}
			trivial.

	\end{description}
\end{proof}

\IFSubFileBiblio

\newpage

\subsection{Proof of Compilation from Frontend Choreographies to DCC Networks}
\label{sub:proof_of_fc_to_dcc_compilation}
 
We first define some auxiliary results used in the proof of
\cref{thm:applied_choreographies}.




We provide some results on DCC variable substitution. We remind that the
only bound names in $DCC$ are the variables in $\prid{accept}$ terms (e.g., $x$
in $!(x);B$). However, the following lemmas prove that renaming free variables
with fresh names in processes (and, by extension, in services) preservers
bisimilarity.

In the following, we abuse the notation for $\alpha$-renaming to denote
variable renaming in running processes.
%
%
%
%
%
%
%
We define the variable renaming operator for DCC processes $P[x'/x]$.

\begin{definition}[DCC Variable Renaming Operator\label{def:variable_renaming}]
 Let $\jpr{B}{t}$ be a DCC process, then $(\jpr{B}{t})[x'/x] =
 \jpr{B[x'/x]}{t\tcopy{x'}{x(t)}\tcopy{x}{\emptyset}}$ where $B[x'/x]$
 substitutes every occurrence of $x$ with $x'$.
\end{definition}

\begin{lemma}[DCC Process Variable Renaming\label{lemma:variable_renaming}]
	Let $\jsrv{\strBhv,P \pp P_c,M}{l}$ be a DCC service where $P=\jpr{B}{t}$. Let $P'
	= P[x'/x]$ where $x'$ is fresh in $B$. Then $\jsrv{\strBhv,P \pp P_c,M}{l} \to
	\jsrv{\strBhv,P'' \pp P_c,M}{l} \iff \jsrv{\strBhv,P' \pp P_c,M}{l} \to
	\jsrv{\strBhv,P''[x'/x] \pp P_c,M}{l}$.
\end{lemma}

\begin{proof}
The proof is by induction on the form of $P$. We report the most interesting
cases. Below we consider $t' = t\tcopy{x'}{x(t)} \tcopy{x}{\emptyset}$.

\begin{itemize}
	\ecase{$P = \jpr{\oneway{o}{y} \qfrom{e};B'}{t}$}
	The only applicable rule is $\did{DCC}{Recv}$, hence we consider the
	interesting case in which $M$ contains a message for the queue defined by $e$.
	In the other case the Lemma trivially holds as services cannot reduce on $P$
	and $P'$. The case unfolds on the combinations of whether \emph{i}) $y
	\neq x$ and \emph{ii}) expression $e$ contains $x$. Below we consider the
	comprehensive case for $y = x$ and $e$ that contains $x$. The proof of the
	other cases is either trivial or a slight modification of the reported one.

	Since we assume we can apply rule $\did{DCC}{Recv}$ we take $t_c = \evalfn(e,
	t)$ and $M( t_c ) = ( o, t' ) :: \til m$. From \cref{def:variable_renaming} we
	have that $t_c = \evalfn(e[x'/x],t')$.

	Meaningful reductions on $P$ and $P'$ are of the form $P \to
	\jpr{B'}{t\tcopy{x}{t_m}}$ and $P' \to \jpr{B'[x'/x]}{t'\tcopy{x'}{t_m}}$ and
	the thesis follow by induction hypothesis.

	\ecase{$P = \jpr{\sum_{i \in I}\choice{\oneway{o_i}{x_i}\qfrom{e}}{B_i}}{t}$}
	The only applicable rule on both $P$ and $P'$ is $\did{DCC}{Recv}$. The most
	comprehensive case is for $M$ that contains a message for operation $o_j$, $j
	\in I$ where $x_j = x$ and expression $e$ contains $x$. The remainder of the
	proof follows that of the previous case.
	
	\ecase{$P = \jpr{\cond{e}{B_1}{B_2}}{t}$}
	Trivial by \cref{def:variable_renaming} for which $\evalfn(e,t) =
	\evalfn(e[x'/x],t')$.

	\ecase{$P = \jpr{y = e;B}{t}$}
	The only applicable rule on both $P$ and $P'$ is $\did{DCC}{Assign}$. The most
	comprehensive case is for $y = x$ and expression $e$ that contains $x$. The
	case is proved considering that, by \cref{def:variable_renaming}, it holds
	that $\evalfn(e,t) = \evalfn(e[x'/x],t')$.

	\ecase{$P = \jpr{\recDef{X}{B_1}{B}}{t}$} The thesis follows from the
	application of rule $\did{DCC}{Ctx}$ and the induction hypothesis.

	\ecase{$P = \jpr{\cq{x};B'}{t}$}
	Let $t_c \not \in M$. We have the reduction on rule $\did{DCC}{Newque}$
	\[
	S \to	\jsrv{\strBhv,\jpr{B'}{t\tcopy{x}{t_c}}\pp P_c,M[t_c \mapsto \emptyseq]}{l}
	\]
	Let service $S'$ be equal to $S$ with $P$ replaced with $P'$. $S'$ can mimic
	the behaviour of $S$ by taking the fresh value $t'_c =t_c$, obtaining the
	reduction
	\[
	S' \to \jsrv{\strBhv,\jpr{B'[x'/x]}{t'\tcopy{x'}{t'_c}} \pp P_c, M[t'_c
	\mapsto \emptyseq]}{l}
	\]
	The same holds if we let $S'$ reduce and prove that $S$ can mimic it.
	
	\ecase{$P = \jpr{\notify{o}{e_1}{e_2}\qto e_3; B'}{t}$}
	We consider the comprehensive case in which expressions $e_1$, $e_2$ and $e_3$
	contain $x$. From \cref{def:variable_renaming} we know that $\auxfn{eval}(e_1, t)
	= \auxfn{eval}(e_1[x'/x], t')$. Similarly the couples $e_2$ and $e_2[x'/x]$ and
	$e_3$ and $e_3[x'/x]$ enjoy the same property when evaluated respectively on
	$t$ and $t'$.

	We analyse the case in which $P$ moves and $P[x'/x]$ mimics it. The other
	case, for $P[x'/x]$ that reduces and $P$ that mimics it, follows the same
	structure.

	\[
	\infer[\did{DCC}{InSend}]
		{
			\jsrv{\strBhv,\ \jpr{B}{t} \pp P, M}{l}
			\quad \to \quad 
			\jsrv{\strBhv,\ \jpr{B'}{t} \pp P, M[t_c \mapsto M(t_c)::(o,t_m)]}{l}
		}
		{
		\begin{array}c
			B = \notify{\op{o}}{e_1}{e_2} \qto e_3;B'
			\qquad
			\evalfn(e_1, t ) = l
			\\[2pt]
			\evalfn( e_3, t ) = t_c
			\qquad
			\evalfn( e_2, t ) = t_m
			\qquad
			t_c \in \dom(M)
		\end{array}
		}
	\]

	and

	\[
	\infer[\did{DCC}{InSend}]
		{
			\jsrv{\strBhv,\ \jpr{B[x'/x]}{t'} \pp P, M}{l}
			\quad \to \quad 
			\jsrv{\strBhv,\ \jpr{B'[x'/x]}{t'} \pp P, M[t_c \mapsto M(t_c)::(o,t_m)]}{l}
		}
		{
		\begin{array}c
			B[x'/x] = \notify{\op{o}}{e_1[x'/x]}{e_2[x'/x]} \qto e_3[x'/x];B'[x'/x]
			\qquad
			\evalfn(e_1[x'/x], t' ) = l
			\\[2pt]
			\evalfn( e_3[x'/x], t' ) = t_c
			\qquad
			\evalfn( e_2[x'/x], t') = t_m
			\qquad
			t_c \in \dom(M)
		\end{array}
		}
	\]
	
	\ecase{$\jpr{\notify{?}{e_1}{e_2};B''}{t}$}
	We consider the comprehensive case where expressions $e_1$ and $e_2$ contain
	$x$. From \cref{def:variable_renaming} we know that $\auxfn{eval}(e_1,
	t) =\auxfn{eval}(e_1[x'/x], t')$. Similarly $e_2$ and $e_2[x'/x]$ enjoy the
  same property when evaluated respectively on $t$ and $t'$.

	Below we describe the case in which $P$ moves and $P[x'/x]$ mimics it. The
	other case, for $P[x'/x]$ that reduces and $P$ that mimics it, follows the
	same structure. We assume the start behaviour $\strBhv = !(y);B'$.

	\[
	\infer[\did{DCC}{InStart}]
		{
			\jsrv{!(y);B',\ \jpr{B}{t} \pp P_c,\ M}{l}
			\quad \to \quad
			\jsrv{!(y);B',\ Q \pp \jpr{B''}{t} \pp P_c,\ M}{l}
		}
		{
			B = \notify{?}{e_1}{e_2};B''
			&
			Q = \jpr{B'}{\emptyset\tcopy{y}{\evalfn( e_2, t )} }
		}
	\]

	and

		\[
		\infer[\did{DCC}{InStart}]
			{
				\jsrv{!(y);B',\ \jpr{B[x'/x]}{t'} \pp P_c,\ M}{l}
				\quad \to \quad
				\jsrv{!(y);B',\ Q \pp \jpr{B''[x'/x]}{t'} \pp P_c,\ M}{l}
			}
			{
				B[x'/x] = \notify{?}{e_1[x'/x]}{e_2[x'/x]};B''[x'/x]
				&
				Q = \jpr{B'}{\emptyset\tcopy{y}{\evalfn( e_2[x'/x], t' )} }
			}
	\]

\end{itemize}
\end{proof}

\begin{lemma}[DCC Network Variable Renaming\label{lemma:network_variable_renaming}]
	Let $S$ and $S'$ be two DCC networks such that $S = \jsrv{\strBhv,P \pp Q,M}{l} \pp
	S_*$ and $S' = \jsrv{\strBhv,P[x'/x] \pp Q,M}{l} \pp S_*$ then \[S \to \jsrv{\strBhv,P' \pp
	Q',M'}{l} \pp S_*'\iff S' \to \jsrv{\strBhv,P'[x'/x] \pp Q',M'}{l} \pp S_*'\].
\end{lemma}

\begin{proof}[Sketch]
	The proof is by induction on the derivation of S. The main observation is that
	the most part of cases are already considered in
	\cref{lemma:variable_renaming}. The cases not considered in
	\cref{lemma:variable_renaming} regard derivations on rules:

\begin{itemize}
	\item $\did{DCC}{Send}$ whose proof follows the same steps of case $P =
	\jpr{\notify{o}{e_1}{e_2}\qto e_3; B'}{t}$ in \cref{lemma:variable_renaming};

	\item $\did{DCC}{Start}$ proved following the same steps of 
	case $P = \jpr{\notify{?}{e_1}{e_2};B''}{t}$ in
	\cref{lemma:variable_renaming};

	\item $\did{DCC}{Eq}$ and $\did{DCC}{Par}$ where the thesis follows from the
	application of the induction hypothesis.

\end{itemize}
\end{proof}

We report below the statement of Theorem~\ref{thm:applied_choreographies},
enriched with annotation on the transitions of $\dep,C$.

\noindent\textbf{Theorem~\ref{thm:applied_choreographies}} \emph{(Applied Choreographies)} \newline 
Let $\env,C$ be a Frontend choreography where $C$ is projectable and $\Gamma
\hlseq{\env, C}$ for some $\Gamma$. Then:
\begin{enumerate}[itemsep=1em,partopsep=1ex,parsep=1ex]
 \item (Completeness)
 $\env, C \lto{\beta} \env',C'$ implies \\
	\begin{enumerate}[itemsep=.5em,partopsep=1ex,parsep=1ex]
		\item $\genenc{\geneenc{\env},\epp{C}{}} \to^+ \enc{\eenc{\env'}{\Gamma'},C''}^{\Gamma'}$
		\item $\epp{C'}{} \prec C''$
		\item for some $\Gamma',\ \Gamma' \hlseq{\env',C'}$
	\end{enumerate}
 \item (Soundness)
 $\genenc{\geneenc{\env},\epp{C}{}} \to^* S$ implies 
 \begin{enumerate}[itemsep=.5em,partopsep=1ex,parsep=1ex]
 	\item $\env,C \to^* \env', C'$
 	\item $S \to^* \enc{\eenc{\env'}{\Gamma'},C''}^{\Gamma'}$
 	\item $\epp{C'}{} \prec C''$
 	\item for some $\Gamma',\ \Gamma' \hlseq{\env',C'}$
\end{enumerate}
\end{enumerate}

\begin{proof}[(Completeness)] We proceed by induction on the derivation of
	$\dep,C \lto{\beta} \dep',C'$. The general strategy is to:
	\begin{itemize}
		\item apply \cref{thm:encoding_operational_correspondence} from which, let
		$\aenv = \geneenc{\env}$, we have that $\aenv,C \lto{\beta} \aenv',C'$,
		$\aenv' = \eenc{\env}{\Gamma'}$;
		\item since $C$ is \emph{projectable}, we can always apply \cref{thm:epp},
		from which, $\env, \epp{C}{} \lto{\beta} \env',C''$ and $\epp{C'}{} \prec
		C''$;
		\item we compile the Backend Endpoint choreography $\aenv,\epp{C}{}$ into the DCC network $\genenc{\aenv,\epp{C}{}}$ and prove that we can reduce it in such a way that its reductum is $\equivD$-equivalent to the compilation of the reductum $\enc{\eenc{\env'}{\Gamma'},C''}^{\Gamma'}$.
 	\end{itemize}

	\begin{itemize}
		\ecase{$\did{C}{Send}$} We know that
			\begin{itemize}
				
				\item $\epp{C}{} \equivC C_{\pid p} \pp C_c$ with $C_{\pid p} =
				\gensend;C_{p}'$;
				
				\item $\dep,\epp{C}{} \lto{\beta} \dep',C''$ with $\did {C}{Send}$ being
				the last applied rule, where $\beta = \gensend$ and $C'' = C_{\pid p}' \pp
				C_c$;

				\item let $\til m = \env(\chanto{k}{\role A}{\role B})$ and $v =
				\auxfn{eval}(e, \env(\pid p))$ we have, by rule $\did{D}{Send}$,

				\[
					\dep' = \dep\big[\ \chanto{k}{\role A}{\role B} \mapsto
						\til m::(o, v) \ \big]
				\]

				which, by \cref{thm:encoding_operational_correspondence}, corresponds to $\aenv' = \aenv\big[\ l^*:t_c \mapsto	\aenv(l^*:t_c)::(o, t_m) \ \big]$ by $\did{\aenv}{Send}$ where $l^*$ is the location of the receiving process playing role $\role B$ and $t_c$ is the correlation key used by the process playing $\role A$ to send to the process playing role $\role B$. The tree $t_m$ corresponds to value $v$ exchanged in rule $\did{D}{Send}$.

			\end{itemize}

			We have two cases, whether the receiving process $\pid q$ is in the same
			location of the sender $\pid p$ or not. Formally, let $\pid p \in
			\aenv(l)$ we consider the exhaustive cases:

			\begin{itemize}
				\ecase{$\pid q \in \aenv( l )$}

					From Definition~\ref{def:compilation} we have that
					$
						\genenc{\aenv,\epp{C}{}} \equivD S \pp S_c
					$
					where, let $t_\pid p = \aenv( \pid p )$ and $M = \filter{\aenv}{l}$
					\begin{itemize}
						\item $S = \jsrv{\genenc{ \filter{C_c}{l} }, \	P \pp Q, M\ }{l}$
						\item $
						P = \jpr{%
							\notify{o}{\jpath{k.B.l}}{e} \qto \jpath{k.A.B};
							\genenc{C_{\pid p}'}}{t_{\pid p}}
						$
					
						\item $Q = \prod\limits_{\pid q \ \in \ \aenv(l)\setminus\{\pid p\}}
							\jpr{\genenc{\filter{C_c}{\pid q}}}{\aenv(\pid q)}$

						\item $
							S_c = \prod\limits_{l' \ \in \ \Gamma \setminus \{l\}}
								\jsrv{
									\genenc{ \filter{C_c}{l'} }, \
										\prod\limits_{\pid r \ \in \ \aenv(l')} 
											\jpr{\genenc{\filter{C_c}{\pid r}}}
												{\aenv(\pid s)}, \filter{\aenv}{l'}
								}{l'}
						$

					\end{itemize}

					In this case $\genenc{\aenv,\epp{C}{}}$ can mimic $\dep,C$ applying rules
					$\did{DCC}{Eq}$, $\did{DCC}{SPar}$, and $\did{DCC}{InSend}$ where
					$S \pp S_c \to S' \pp S_c$ with $\did{DCC}{SPar}$ and $S \to S'$ with

					\[
						\infer[\did{DCC}{InSend}]
						{
							\jsrv{\strBhv,\ P \pp Q, M}{l}
							\quad \to \quad 
							\jsrv{\strBhv,\ P' \pp Q, M[t_c \mapsto M(t_c)::(o,t_m)]}{l}
						}
						{
						\begin{array}c
							P = \jpr{\notify{o}{\jpath{k.B.l}}{e} \qto \jpath{k.A.B};
								\genenc{C_{\pid p}'}}{t_{\pid p}}
							\qquad
							\evalfn(\jpath{k.B.l}, t_{\pid p} ) = l
							\\[2pt]
							\evalfn( \jpath{k.A.B}, t_{\pid p} ) = t_c
							\qquad
							\evalfn( e, t_{\pid p} ) = t_m
							\qquad
							t_c \in \dom(M)
						\end{array}
						}
					\]

					where $P' = \jpr{\genenc{C_{\pid p}'}}{t_{\pid p}}$. Since by
					\cref{def:compilation} $l$, $t_c$, and $t_m$ result from the
					evaluation of the state of process $\pid p$, $\aenv(\pid p)$, we have
					that $M[t_c \mapsto M(t_c)::(o,t_m)] = \filter{\aenv'}{l}$.

					This corresponds to the compilation of the reduction $\env', C'$, i.e,

					{\footnotesize\[\hspace{-4em}
						\enc{\eenc{\env'}{\Gamma'},C''}^{\Gamma'} \equivD
						\begin{array}l
							\left.\jsrv{
								\genencp{\filter{C_c}{l}},
								\overbrace{
									\jpr{\genencp{C_{\pid p}'}}{\aenv'(\pid p)}}^{P'}
								\pp
								\overbrace{
									\prod\limits_{\pid q \ \in \ \aenv'(l) \setminus \{\pid p\}}
										\jpr{\genencp{\filter{C_c}{\pid q}}}{\aenv'(\pid q)}}^{Q} 
								, \filter{\aenv'}{l}
							}{l}\right\}S'
							\\\pp\\
							\left.\prod\limits_{l' \ \in \ \Gamma' \setminus \{l\}}
								\vphantom{\raisebox{14pt}{C}}
								\jsrv{
									\genencp{ \filter{C_c}{l'} }, \
										\prod\limits_{\pid r \ \in \ \aenv'(l')}
											\jpr{\genencp{\filter{C_c}{\pid r}}} {\aenv'(\pid r)}
											,\filter{\aenv'}{l'}}{l'}\right\}S_c
						\end{array}
					\]}

					Where the changes in $\env'$ and $\Gamma'$ affect only the compilation
					of the queue in $\filter{\aenv'}{l}$ identified by $t_c$, while for
					all other terms $\genenc{\cdot} = \genencp{\cdot}$ and
					$\filter{\aenv'}{l'}=\filter{\aenv}{l'}$.

				\ecase{$\pid q \not \in \aenv( l )$}

					Similar to \textbf{Case} $\pid q \in \aenv( l )$ except the last
					applied rule in the reduction of $\genenc{\aenv,\epp{C}{}}$ is
					$\did{DCC}{Send}$.

			\end{itemize}
		
		\ecase{$\did{C}{Recv}$}
			We know that

			\begin{itemize}
				\item $\epp{C}{} \equivC C_{\pid q} \pp C_c$ with $C_{\pid q} = \genbranchI$

				\item $\dep,\epp{C}{} \lto{\beta} \dep',C''$ with rule $\did{C}{Recv}$ where
				$\beta = \pair{k}{\grecv{\role A}{\prc qB}{o_j(x_j)}}$, $C' \equivC C_j
				\pp C_c$. Let $\aenv = \geneenc{\env}$ and $\env(\chanto{k}{\role
				A}{\role B}) = (o_j,v)::\til m$, we have
				\[
					\dep' =	\dep
						\big[\ \pid q \mapsto \env(\pid q)[ x_j \mapsto v ]\ \big]
						\big[\ \chanto{k}{\role A}{\role B} \mapsto \til m\ \big]
				\]

				By \cref{thm:encoding_operational_correspondence}, let $\aenv(t_c:l^*)
				= (o_j,t_m)::\til m^*$ we have 

				\[
				\aenv' = \aenv
				\big[\ \pid q \mapsto \aenv(\pid q)[x_j \to t_m]\big]
				\big[\ l^*:t_c \mapsto \til m^* \ \big]
				\]

				by $\did{\aenv}{Recv}$ where $l^*$ is the location of the receiving
				process playing role $\role B$ and $t_c$ is the correlation key used by
				the process playing $\role A$ to send to the process playing role $\role
				B$. The tree $t_m$ corresponds to the encoding of value $v$ in the
				queue.

			\end{itemize}

			Let $\pid q@l \in \Gamma$, $t_\pid q = \aenv(\pid q)$, and $M =
			\filter{\aenv}{l}$, from \cref{def:compilation} we have
			$\genenc{\aenv,\epp{C}{}} \equivD S \pp S_c$ where

			\begin{itemize}
				\item $S = \jsrv{ \genenc{\filter{C_c}{l}}, Q \pp R, M }{l}$
				\item $Q = \jpr{\sum_{i \in I}\choice{\oneway{o_i}{x_i}
						\qfrom{\jpath{k.A.B}}}{\ \genenc{C_i} \ }}{t_{\pid q}}$
				\item $R = \prod\limits_{\pid r \in \aenv(l)\setminus\{\pid q\}}
					\jpr{\genenc{\filter{C_c}{\pid r}}}{\aenv(\pid p)}$
				\item $S_c = \prod\limits_{l' \ \in \ \Gamma\setminus\{l\}}
					\jsrv{\genenc{\filter{C_c}{l'}},
					\prod\limits_{\pid s \in \aenv(l')}
						\jpr{\genenc{\filter{C_c}{\pid s}}}{t_{\pid s}}, \filter{\aenv}{l'}
						}{l'}
				$
			\end{itemize}
			
			In this case $\genenc{\aenv,\epp{C}{}}$ can mimic $\dep,C$ applying rules
			$\did{DCC}{Eq}$, $\did{DCC}{SPar}$, and $\did{DCC}{Recv}$.

			{\footnotesize
			\[
			\infer[\did{DCC}{SPar}] { S \pp S_c \quad \to \quad S' \pp S_c } {
				\infer[\did{DCC}{Recv}]
				{
					\jsrv{ \genenc{\filter{C_c}{l}}, Q \pp R, M }{l}
					\quad \to \quad 
					\jsrv{ \genenc{\filter{C_c}{l}}, \jpr{\genenc{C_j}}{t_\pid
					q\tcopy{x_j} {t_m}} \pp R ,\ M[t_c \mapsto \til m^*]}l } {
					\begin{array}c
					Q = \jpr{\sum_{i \in I}\choice{\oneway{o_i}{x_i}
					\qfrom{\jpath{k.A.B}}}{\ \genenc{C_i} \ }}{t_\pid q}
				  \qquad
				  j \in I
				  \qquad
				  t_c = \evalfn(e,t_\pid q)
				  \qquad
				  M(t_c) = (o_j,t_m)::\til m^*
					\end{array}
				}
				}
			\]}

			Where $S' = \jsrv{ \genenc{\filter{C_c}{l}}, \jpr{\genenc{C_j}}{t_\pid
			q\tcopy{x_j} {t_m}} \pp R ,\ M[t_c \mapsto \til m^*]}l$. Let $t_{\pid q}'
			= t_\pid q\tcopy{x_j}{t_m}$, $Q' = \jpr{\genenc{C_j}}{t_\pid q'}$, and
			$M' = M [t_c \mapsto \til m]$.

			Since by \cref{def:compilation} $t_c$ and $t_m$ respectively result from
			the evaluation of the state of process $\pid q$, $\aenv(\pid q)$ and the
			encoding of value $v$, we have that $M' = \filter{\aenv'}{l}$ and $t'_\pid
			q = \aenv'(\pid q)$.

			This corresponds to the compilation of the reduction $\env', C''$, i.e,

			{\footnotesize
			$$
				\hspace{-20pt}\enc{\eenc{\dep'}{\Gamma'},C''}^{\Gamma'} \equivD
					\overbrace{\jsrv{
						\genencp{\filter{C_c}{l}}, 
						\overbrace
						{
							\jpr{\genencp{\filter{C_j}{\pid q}}}{t'_{\pid q}}
						}^{Q'}
						\pp
						\overbrace{
							\prod\limits_{\pid r \ \in \ \aenv'(l)\setminus\{\pid q\}} 
								\jpr{\genencp{\filter{C_c}{\pid r}}}{\aenv'(\pid r)}
						}^{R}, M'
					}{l}}^{S'}
					\pp
					\underbrace{\prod\limits_{l' \ \in \ \Gamma \setminus \{l\}}
						\jsrv{
							\genencp{ \filter{C_c}{l'} }, \
								\prod\limits_{\pid s \ \in \ \dep(l')} 
									\jpr{\genencp{\filter{C_c}{\pid s}}}
										{t_{\pid s}}, \filter{\aenv'}{l}
						}{l'}}_{S_c}
			$$}

			Where the changes in $\env'$ and $\Gamma'$ affect only the compilation of
			the queue in $\filter{\aenv'}{l}$ identified by $t_c$ and the state of
			$\pid q$; while for all other terms $\genenc{\cdot} = \genencp{\cdot}$ and
			$\filter{\aenv'}{l'}=\filter{\aenv}{l'}$.
		
		\ecase{$\did{C}{PStart}$}
			We know that

			\begin{itemize}
				\item $\epp{C}{} \equivC C_r \pp C_a \pp C_c$ where, let $\pair{\til l}
				{\serviceTyping{G}{\role A}{\roles B}{\roles B}} \in \Gamma$
				\item $C_r = \genreq;C_r'$
				\item let $l_1.\role B_1,\ldots,l_n.\role B_n = \wtil{l.\role B}$,
				$C_a = \prod\limits_{i=1}^{n} \acc{k}{l_i.\pid q_i[\role B_i];C_{\pid
				q_i}}$
			\end{itemize}

			We can apply rules $\did{C}{Par}$ and $\did{C}{Eq}$ and lastly rule
			$\did{C}{PStart}$ such that

			\[
			\infer[\did{C}{PStart}]
			{
				\env,
				C_r
				\pp
				C_a
				\to
				\env',\ 
				C_r'[k'/k] 
				\pp 
				\prod_i\big(\ C'_{\pid q_i}[k'/k][\pid r_i /\pid q_i] \ \big)
				\pp
				C_a
			}
			{
				\begin{array}c
				i \in \{1,\dots,n\}
				\qquad
				\env\fresh{k',\pids r}
				\qquad
				\{ \wtil{l.\role B} \} = \biguplus_i \{\wtil{l_i.\role B_i}\}_i
				\qquad
				\{\pids r\} = \bigcup_i \{ \pids r_i \}
				\\[2pt]
				\delta = \start{k'}{\pid p[\role A]}{\wtil{l_1.\pid r_1[\role
				B_1]},\dots,\wtil{l_n.\pid r_n[\role B_n]}}
				\qquad
				\renv{\env}{\delta}{\env'}
				\end{array}
			}
			\]

			and 

			$$\dep, C_r \pp C_a \pp C_c \quad \lto{\tau} \quad
			\dep',C_r'[k'/k] \pp 
			\prod_i\big(\ C_{\pid q_i}[k'/k][\pid r_i /\pid q_i] \ \big) \pp
			C_a \pp C_c
			$$

			thus $C'' = C_r'[k'/k] \pp 
			\prod_i\big(\ C_{\pid q_i}[k'/k][\pid r_i /\pid q_i] \ \big) \pp
			C_a \pp C_c$

			We can find $\Gamma' = \Gamma,\auxfn{init}(k',(\prc pA,\wtil{\prc qB}),G)$
			and $\Gamma' \hlseq{\dep',C'}$.

			\begin{remark}
				We have two cases for, let $\pid p@l \in \Gamma$, whether $l \in
				\{\til l\}$ or not. For a clearer treatment of the case we proceed
				considering that $l \not \in \{\til l\}$ (i.e., no service process is
				created in the same location --- service --- of the requester $\pid p$).
				The other case follows the same structure of $l \not \in \{\til l\}$
				although the service located at $l$ has $\genenc{\filter{C_a}{l}}$ as
				start behaviour and $\genenc{\aenv,\epp{C}{}}$ applies rule $\did{DCC}
				{InStart}$ in place of the $\did{DCC}{Start}$ for starting the DCC
				process located at $l$.

				Henceforth we proceed analysing the case for $l \not \in \{\til l\}$.
			\end{remark}				

			From \cref{def:compilation} we have, let $\aenv^* = \eenc{\env'}{\Gamma'}$
			and $M^* = \filter{\aenv^*}{l}$ and $M^*_i = \filter{\aenv^*}{l_i}$

			$$
				\genencp{\aenv^*,C''} = 
					\jsrv{\genencp{\filter{C_c}{l}}, P'' \pp R',M^*}{l} \pp
					\prod\limits_{i=1}^{n} \jsrv{Q_i'', Q_i^{*} \pp R_{l_i}',M^*_{i}}{l_i}
					\pp
					S_c'
			$$

			In the following, we use the abbreviation $t_{\pid s}^* = \aenv^*(\pid
			s)$ for process $\pid s$ in $\aenv^*$.

			\begin{itemize}
				
				\item $P'' = \jpr{\genencp{C_r'[k'/k]}}{t_{\pid p}^*}$

				\item $ R' = \prod\limits_{\pid p' \ \in \ \aenv^*(l)\setminus\{\pid p\}}
					\jpr{\genencp{\filter{C_c}{\pid p'}}}{t_{\pid p'}^{*}}
				$

				\item $Q_i'' = \m{accept}(k,\role B_i,\serviceTyping{G}{\role A}{\roles
				B}{\roles B});\genencp{C_{\pid q_i}}$

				\item $Q_i^{*} = 
				\jpr{\genencp{C_{\pid q_i}[k'/k][\pid r_i/\pid q_i]}}{t_{q_i}^{*}}$
			
				\item $ R'_{l_i} = \prod\limits_{\pid s \ \in \ \aenv^*(l_i)}
					\jpr{\genencp{\filter{C_c}{\pid s}}}{t_{\pid s}^{*}}
				$

				\item $ S_c' = \prod\limits_{l' \ \in \ \Gamma \setminus \{ l, \til l
				\}} \jsrv{ \genencp{\filter{C_c}{l'}} ,
					\prod\limits_{\pid s' \ \in \ \aenv^*(l')}
						\jpr{\genencp{\filter{C_c}{\pid s'} }}{t_{\pid
						s'}^{*}},\filter{\aenv^*} {l'} }{l'}
				$
			\end{itemize}

			From \cref{thm:encoding_operational_correspondence} we can apply rule
			$\did{\aenv}{Start}$ on $\aenv,\epp{C}{} \to \aenv^*,C''$ such that we
			know that

			$$
				\jpath{k'}(t_{\pid p}^{*}) = \jpath{k'}(t_{\pid q_1}^{*}) = \ldots =
				\jpath{k'} (t_{\pid q_n}^{*}) = t_{k'}
			$$

			for some $t_{k'}$ session descriptor of session $k'$.

			We proceed by proving that we can reduce $\genenc{\aenv,\epp{C}{}} \to^{+} S$.

			From \cref{def:compilation} we have, let $t_\pid p = \aenv(\pid p)$, $M =
			\filter{\aenv}{l}$, and $M_i = \filter{\aenv}{l_i}$

			$$
				\genenc{\aenv,\epp{C}{}} \equivD
					\jsrv{\genenc{\filter{C_c}{l}}, P \pp R, M }{l}
					\pp
					\prod\limits_{i = 1}^{n} \jsrv{Q_i, R_{l_i}, M_i}{l_i}
					\pp S_c
			$$

			where

			\begin{itemize}
				
				\item $
			\begin{array}{rl} P = & \jpr{\m{start}(\ k, \ (l.\role A,\wtil{ l.\role
				B }) \ ); \genenc {C_r'}}{t_{\pid p}} =
				\\ = &
				\jpr{
					\left(\begin{array}{l}
						\Seq\limits_{
							\substack{ \jpath{I} \in \{\role A, \til{\role B}\}}}
								\jpath{k.I.l} = l_{\role I}\ ;
						\\
						\Seq\limits_{\jpath{I} \in \{\til{\role B}\}}
							\Big(
								\cq{\jpath{k.I.A}};
								\notify{?}{\jpath{k.I.l}}{\jpath{k}};
								\oneway{\mathit{sync}}{\jpath{k}} \qfrom \jpath{k.I.A}
							\Big);
						\\
						\Seq\limits_{\jpath{I} \in \{\til{\role B}\}}
							\notify{\mathit{start}}{\jpath{k.I.l}}{\jpath{k}} \qto \jpath{k.A.I};
						\genenc{C_r'}
					\end{array}
					\right) }{t_{\pid p}}
			\end{array}
			$

			\item $ Q_i = \m{accept}(k, \role B_i,\initg{G}{\role A}{\til{\role
				B}}{\roles B}) ; \genenc{C_{\pid q_i}} =
				\begin{array}l
					\oneway{!}{\jpath{k}};
					\Seq\limits_{ \jpath{I} \in \{\role A,\roles B\}\setminus\{\role B_i\} }
						\cq{\jpath{k.I.B_i}} \ ;
				\\
					\notify{\mathit{sync}}{\jpath{k.A.l}}{\jpath{k}} \qto{\jpath{k.B_i.A}}\ ;
				\\
					\oneway{\mathit{start}}{\jpath{k}} \qfrom{\jpath{k.A.B_i}} \ ; \
					\genenc{C_{\pid q_i}}
				\end{array}
			$

			\item $ R = \prod\limits_{\pid p' \ \in \ \aenv(l)\setminus\{\pid p\}}
					\jpr{\genenc{\filter{C_c}{\pid p'}}}{t_{\pid p'}}
			$

			\item $ R_{l_i} = \prod\limits_{\pid s \ \in \ \aenv(l_i)}
					\jpr{\genenc{\filter{C_c}{\pid s}}}{t_{\pid s}}
			$

			\item $ S_c = \prod\limits_{l' \ \in \ \Gamma \setminus \{ l, \til l \}}
					\jsrv{ \genenc{\filter{C_c}{l'}} ,
					\prod\limits_{\pid s' \ \in \ \aenv(l')}
						\jpr{\genenc{\filter{C_c}{\pid s'} }}{t_{\pid s'}},\filter{\aenv}{l'}
					}{l'}
			$

			\end{itemize}

			$\genenc{\aenv,\epp{C}{}}$ can mimic $\dep, C$ with the following sequence
			of reductions. Note that we make use of renaming on $\prid{accept}$ terms
			in $Q_1,\ldots, Q_n$ and variable renaming on $P$ (as of
			Definition~\ref{def:variable_renaming}) to align the evolution of
			$\genenc{\aenv,\epp{C}{}}$ with the evolution of $\dep,C$, in which $k$
			has been renamed with the fresh name $k'$. Since the renamed DCC network
			and the original one are bisimilar, as per
			\cref{lemma:network_variable_renaming}, we can proceed to prove our
			results on the original DCC network using the DCC renamed network as a
			proxy.

			Therefore we take $S^{*}_{0} \sim \genenc{\aenv,\epp{C}{}}$

			$$
			S^{*}_{0} =
				\jsrv{
					\genenc{\filter{{C_c}}{l}}, P[\jpath{k'}/\jpath{k}] \pp R,M}{l}
				\pp
				\prod\limits_{i = 1}^{n}
					\jsrv{Q_i[\jpath{k'}/\jpath{k}], R_{l_i}, M_i }{l_i}
				\pp S_c
			$$

			$$
			S^{*}_{0}
			\to
			\begin{array}{cl}
				\left.
					\lto{\did{DCC}{SEq} \quad \did{DCC}{SPar} \quad
					\did{DCC}{PPar} \quad \did{DCC}{Assign} }
				\right\}_{\mbox{\footnotesize$n+1$ times}}\hspace{-42pt}\pnum{1}
				\\
				\left. \begin{array}l
						\smallpnum{2.1} \lto{\did{DCC}{SEq} \quad \did{DCC}{SPar} \quad
						\did{DCC}{Newque} }
						\\
						\smallpnum{2.2} \lto{\did{DCC}{SEq} \quad \did{DCC}{SPar} \quad
						\did{DCC}{Start} }
						\\
								\smallpnum{2.3}
								\left.\lto{\did{DCC}{SEq} \quad \did{DCC}{SPar} \quad \did{DCC}{Newque} }
								\right\}_{\mbox{\footnotesize$n$ times}}
						\\
						\smallpnum{2.4}\lto{\did{DCC}{SEq} \quad \did{DCC}{SPar} \quad
						\did{DCC}{Send} }
						\\
						\smallpnum{2.5}\lto{\did{DCC}{SEq} \quad \did{DCC}{SPar} \quad
						\did{DCC}{Recv} }
				\end{array}\right\} &
					\hspace{-5pt}\pnum{2}
					\hspace{-10pt}\raisebox{-40pt}[0pt][0pt]{\footnotesize\mbox{n times}}
				\\
				\left.
					\begin{array}l
					\smallpnum{3.1} \lto{\did{DCC}{SEq} \quad \did{DCC}{SPar} \quad
					\did{DCC}{Send} }
					\\
					\smallpnum{3.2} \lto{\did{DCC}{SEq} \quad \did{DCC}{SPar} \quad
					\did{DCC}{Recv} }
				\end{array}
				\right\}_{\mbox{\footnotesize$n$ times}}
				\hspace{-25pt}\pnum{3}
			\end{array}
			\to S^{*}_1
			$$

			We briefly comment the numbered transitions.

			\begin{itemize}
				\item In $\pnum{1}$ $P[k'/k]$ proceeds to store (for $n+1$ times, $l$
				plus $l_i, i \in \{1,\ldots,n\}$) the locations of all roles under
				$\jpath {k'}$.

				\item In $\pnum{2}$, for each location $l_i, i \in \{1,\ldots,n\}$ (for
				each service process):
					\begin{itemize}
						\item $P$ creates its receiving queue for the service process
						\smallpnum{2.1};
						\item in \smallpnum{2.2} $P$ synchronises  with the service at location $l_i$ starting
						($\did{DCC}{Start}$) a new service process;
						\item in \smallpnum{2.3} the service process creates its own queues
						for all other roles in the session (hence $n$ times);
						\item in \smallpnum{2.4} the service process sends the correlation
						values to $P$;
						\item finally $P$ receives the message in \smallpnum{2.5}.
					\end{itemize}

				\item In \pnum{3} for each service process ($n$ times)
					\smallpnum{3.1} the starter sends a message to the service process to
					start the session and \smallpnum{3.2} the addressee receives it.
			\end{itemize}

			Finally we have

			$$
				S^{*}_{1} =
					\jsrv{
						\genenc{\filter{C_c}{l}} \pp P' \pp R, M'}{l}
					\pp
						\prod_{i=1}^{n} \jsrv{Q_i[\jpath{k'}/\jpath{k}], Q'_i \pp R_{l_i}
						,M'_i}{l_i}
					\pp S_c
			$$

			where

			\begin{itemize}
				\item $P' =
				\jpr{\genenc{C'_r} \ [\jpath{k'}/\jpath{k}]}{t_{p}'}$, and
				\item $Q_i' = \jpr{\genenc{C_{\pid q_i}} \ [\jpath{k'}/\jpath{k}]}{t_
				{k'}}$
			\end{itemize}

			From the transitions presented above we know that there exists $t_{k'}'$
			such that $t_{\pid p}' = t_{\pid p} \tcopy{\jpath{k'}}{t_{k'}'}$, where
			$t_{k'}'$ is a session descriptor for session $k'$ (i.e., it contains all
			the locations and correlation keys used by the processes in session
			$k'$). In this case, we take $t_{k}'=t_{k'}$ obtained from the derivation
			$\aenv,C \to \aenv^*,C'$.

			Similarly, $M'$ and $M_1',\ldots,M_n'$ contain
			the necessary (empty) queues to support communication in session $k'$.

			$$
				M' = M[\jpath{k'.B_1.A}(t_{k'}) \mapsto \emptyseq ]
					\ \ldots \ [\jpath{k'.B_n.A}(t_{k'}) \mapsto \emptyseq ]
			$$

			and ($\emptyfunc$ being a totally undefined function on $\Val
			\rightharpoonup \Queues$)

			$$
				M_{i} = \emptyfunc
					\begin{array}l [\jpath{k'.A.B_i}(t_k) \mapsto \emptyseq ]
						[\jpath{k'.B_1.B_i}(t_k) \mapsto \emptyseq ]
						\ \ldots \ [\jpath{k'.B_{i-1}.B_i}(t_k) \mapsto \emptyseq ]
						\ \ldots \
						\\
						\ \ldots \ [\jpath{k'.B_{i+1}.B_i}(t_k) \mapsto \emptyseq ]
						\ \ldots \ [\jpath{k'.B_{n}.B_i}(t_k) \mapsto \emptyseq ]
					\end{array}
			$$

			We proceed with the proof taking $S \sim S^{*}_{1}$ as $S$ is simply the
			renaming of $\jpath{k'}$ to $\jpath{k}$ on start behaviours $Q_i, i \in
			\{1,\ldots,n\}$ (trivially
			$Q_i[\jpath{k'}/\jpath{k}][\jpath{k}/\jpath{k'}] = Q_i$)

			$$
				S =
					\jsrv{
						\genenc{\filter{C_c}{l}} \pp P' \pp R,M'}{l}
					\pp
						\prod_{i=1}^{n} \jsrv{Q_i, Q'_i \pp R_{l_i}, M_i' }{l_i}
					\pp S_c
			$$

			We now proceed to prove that $\genenc{\aenv, \epp{C}{}} \to^{+}
			\enc{\aenv^*,C''}^{\Gamma'}$, i.e. that $\enc{\aenv^*,C''}^{\Gamma'} = S$
			with $\Gamma' \hlseq {\dep',C'}$.

			We prove that 

			{\footnotesize\[
			\begin{array}l
				\overbrace{\jsrv{\genencp{\filter{C_c}{l}}, P'' \pp R',M^*}{l} \pp
					\prod\limits_{i=1}^{n} \jsrv{Q_i'', Q_i^{*} \pp R_{l_i}', M_i^*}{l_i}
					\pp	S_c'
				}^{\enc{\aenv^*,C''}^{\Gamma'}}
					=
				\overbrace{
					\jsrv{
							\genenc{\filter{C_c}{l}} \pp P' \pp R,M'}{l}
						\pp
							\prod_{i=1}^{n} \jsrv{Q_i, Q'_i \pp R_{l_i}, M_i' }{l_i}
						\pp S_c
				}^{\mbox{\normalsize $S$}}
			\end{array}
			\]}

			\begin{itemize}
				\item $M^*$ and $M'$ are equal and similarly $M_i^*$ and $M_i$ are
				pair-wise equal by construction and rule $\did{\aenv}{Start}$;
				\item $\genenc{\filter{C_c}{l}} = \genencp{\filter{C_c}{l}}$
					as $\filter{\Gamma}{\m{locs}} = \filter{\Gamma'}{\m{locs}}$ by
					construction;
				\item $P'' = P'$ is proved by 
				$$\jpr{\genencp{C_r'[k'/k]}}{t_{\pid p}^*} = 
					\jpr{\genenc{C'_r} \ [\jpath{k'}/\jpath{k}]}{t_{p}'}$$ 
					which holds as
					\begin{enumerate}[label=\emph{\roman*})]
						\item $\genencp{C_r'[k'/k]} = \genenc{C'_r} \
						[\jpath{k'}/\jpath{k}] $ since 
							\begin{enumerate}
							\item $\Gamma'$ does not contain any new process used in $C'_r$;
							\item by renaming, and \cref{lemma:network_variable_renaming}.
							\end{enumerate}
					\item $t_{\pid p}^{*} = t_{\pid p}'$ by construction and
					rule $\did{\aenv}{Start}$.
					\end{enumerate}

				\item $Q_i'^{*} = Q_i'$ proved by
					$$\jpr{\genencp{C_{\pid q_i}[k'/k][\pid r_i/\pid q_i]}}
					{t_{q_i}^{*}} = \jpr{\genenc{C_{\pid q_i}} \ [\jpath
					{k'}/\jpath{k}]}{t_{k'}}$$ whose proof of equivalence
					follows that of $P'' = P'$, except that $\Gamma'$ contains the
					location of the process ($\pid r_i$) used in $C_{\pid q_i}[k'/k][\pid
					r_i/\pid q_i]$.

				\item $Q_i'' = Q_i$ proved by

				$$
					\m{accept}(k,\role B_i,\serviceTyping{G}{\role A}{\roles B}{\roles
				B});\genencp{C_{\pid q_i}} = \m{accept}(k, \role B_i,\initg{G}{\role
				A}{\til{\role B}}{\roles B}) ; \genenc{C_{\pid q_i}}
				$$
				which holds as $\genencp{C_{\pid q_i}} = \genenc{C_{\pid q_i}}$ because
				$\Gamma$ and $\Gamma'$ contain the same service typings.

				\item $R' = R$ is proved by

				$$
					\prod\limits_{\pid p' \ \in \ \aenv^*(l)\setminus\{\pid p\}}
					\jpr{\genencp{\filter{C_c}{\pid p'}}}{t_{\pid p'}^{*}}
					=
					\prod\limits_{\pid p' \ \in \ \aenv(l)\setminus\{\pid p\}}
					\jpr{\genenc{\filter{C_c}{\pid p'}}}{t_{\pid p'}}
				$$

				for which 

				\begin{enumerate}[label=\emph{\roman*})]
					\item $\genencp{\filter{C_c}{\pid p'}} = \genenc{\filter{C_c}{\pid
					p'}}$ as $\Gamma'$ does not contain any new process used in $C_c$.
					\item $t_{\pid p'}^{*} = t_{\pid p'}$ unchanged by the reductions of
					$\aenv,C$ and $\genenc{\aenv,\epp{C}{}}$.
				\end{enumerate}

				\item $R_{l_i}' = R_{l_i}$ whose proof follows that of $R' = R$.

				\item $S_c' = S_c$ following the proof of $\genenc{\filter{C_c}{l}} =
				\genencp{\filter{C_c}{l}}$ and $R_{l_i}' = R_{l_i}$.

			\end{itemize}

		\ecase{$\did{C}{Start}$}
			While the original FC program reduces applying rule $\did{C}{Start}$, the
			endpoint projection $\env,\epp{C}{}$ will mimic it applying rule $\did{C}
			{PStart}$, as per \cref{thm:epp}. Hence, to prove this case, we can follow
			the same proof of case $\did{C}{PStart}$.

		\ecase{$\did{C}{Cond}$}
			We have $\epp{C}{} = C_{\pid p} \pp C_c$ where $C_{\pid p} = \gencond$. 
			Let $\pid p@l \in \Gamma$ and
			\begin{itemize}
				\item $t_\pid p = \aenv(\pid p)$;
				\item $P = \jpr{\cond{e}{\ \genenc{C_1} \ }{ \ \genenc{C_2} \
				}}{t_{\pid p}}$; 
				\item $R = \prod\limits_{\pid r \ \in \
				\aenv(l)\setminus\{\pid p\}}
					\jpr{\genenc{\filter{C_c}{\pid r}}}{t_{\pid r}}$
				\item $S_c = \prod\limits_{l' \ \in \ \Gamma\setminus\{l\}}
					\jsrv{ \genenc{\filter{C_c}{l'}} ,
						\prod\limits_{\pid r \ \in \ \aenv(l')}
							\jpr{\genenc{\filter{C_c}{\pid r}}}{t_{\pid r}},\filter{\aenv}{l'}
					}{l'}$
			\end{itemize}

			From Definition~\ref{def:compilation} we have, let $M = \filter{\aenv}{l}$

			$$
				\genenc{\aenv,\epp{C}{}} \equivD
					\jsrv{\genenc{\filter{C_c}{l}}, P \pp R,M}{l} \pp S_c
			$$

			we reduce $\aenv,\epp{C}{}$ applying rules $\did{C}{Par}$, $\did{C}{Eq}$
			and lastly rule $\did{C}{Cond}$. We analyse only the case for
			$\evalfn(e,t_{\pid p}) = \m{true}$ as the other case for
			$\evalfn(e,t_{\pid p}) = \m{false}$ follows the same structure.

			$$\aenv, \epp{C}{} \quad  \lto{\tau} \quad  \aenv',C''$$

			and $C'' = C_1 \pp C_c$ and $\aenv'=\aenv$ by the definition of $\did{C}
			{Cond}$. We can choose $\Gamma = \Gamma'$, for which it holds that $\Gamma
			\hlseq{\dep',C'}$.

			From \cref{def:compilation} we have

			$$
				\genencp{\aenv',C''} = \genenc{\aenv,C''} = 
					\jsrv{
						\genenc{\filter{C_c}{l}}, 
						\jpr{\genenc{C_1}}{t_{\pid p}} \pp R,M
					}{l} \pp S_c
			$$

			$\genenc{\aenv,\epp{C}{}}$ can mimic $\dep,C$ applying rules
			$\did{DCC}{Eq}$, $\did{DCC}{SPar}$, $\did{DCC}{PPar}$, and lastly
			$\did{DCC}{Cond}$ for which

			\[
			\genenc{\aenv,\epp{C}{}} \to \jsrv{\genenc{\filter{C_c}{l}},
			\jpr{\genenc{C_1}}{t_{\pid p}} \pp R,M}{l} \pp S_c
			\]

		\ecase{$\did{C}{Ctx}$}
			The thesis follows from the induction hypothesis as $\dep,C$ applies rule
			$\did{C}{Ctx}$ and $\genenc{\aenv,\epp{C}{}}$ can mimic it with rule
			$\did{DCC}{Ctx}$.

		\ecase{$\did{C}{Par}$}
			The thesis follows from the induction hypothesis.

		\ecase{$\did{C}{Eq}$}
			The thesis follows from the induction hypothesis. Starting from any
			configuration of $\dep,C$, $\genenc{\aenv,\epp{C}{}}$ can always mimic the
			evolution of $\dep,C$ when it applies rule $\did{C}{Eq}$: in both cases
			that $\ctx{R} = \ \equiv$ or $\ctx{R} =  \ \swapC$,
			$\genenc{\aenv,\epp{C}{}}$ can apply $\did{DCC}{Eq}$, $\did{DCC}{SPar}$,
			and $\did{DCC}{PPar}$ to mimic $\dep,C$.

	\end{itemize}
\end{proof}

\newpage

Before proceeding with the proof of (Soundness) of
Theorem~\ref{thm:applied_choreographies}, we extend the semantics of DCC by
annotating its transitions with the variable paths (of the kind $x =
\jpath{x.y.z}$) on which DCC operations execute. We range over DCC transition
labels with $\lambda$.
\[
\lambda \gram x \ \Div \ \cq{x} \ \Div \ ?(x) \ \Div \ o \qfrom{x} \ \Div \
o \qto x \ \Div \ \tau
\]
We report in Figure~\ref{fig:js_semantics_annotated} the annotated semantics of
DCC.

\begin{figure}
\scalebox{.85}[.85]{%
\(
\hspace{-15pt}\begin{array}{c}
	\infer[\did{DCC}{Assign}]
	{ \jpr{x = e\ ;B}{t} \quad \lto{x} \quad \jpr{B}{t\tcopy{x}{t'}}}
	{ t' = \evalfn(x,t) }
	\qquad
	\infer[\did{DCC}{Ctx}]
	{
		\jpr{\recDef{X}{B_1}{B}}{t}
		\quad \lto{\lambda} \quad
		\jpr{\recDef{X}{B_1}{B'}}{t'}
	}
	{
		\jpr{B}{t} \lto{\lambda} \jpr{B'}{t'}
	}
	\\[15pt]
	\infer[\did{DCC}{Cond}]
	{
		\jpr{\cond{e}{B_1}{B_2}}{t}
			\quad \lto{\tau} \quad
		\jpr{B_i}{t}
	}
	{
		i = 1 \mbox{ if } \evalfn( e, t ) = \mbox{true}, i = 2 \mbox{ otherwise}
	}
	\qquad
	\infer[\did{DCC}{PEq}]
	{ \jsrv{\strBhv,\ P,\ M}l \quad \lto{\lambda} \quad \jsrv{\strBhv,\ P',\ M}l }
	{ P \equivD P_1 \pp P_2 & P_1 \lto{\lambda} P_1' & P_1' \pp P_2 \equivD P'}
	\\[15pt]
	\infer[\did{DCC}{Newque}]
	{
		\jsrv{\strBhv,\ \jpr{B}{t} \pp P,\ M}l
			\quad \lto{\cq{x}} \quad 
		\jsrv{\strBhv,\ \jpr{B}{t \tcopy{x}{t_c}} \pp P,\ M'}l
	}
	{
		B = \cq{x};B
		&
		t_c \not \in \dom(M)
		&
		M' = M[t_c \mapsto \emptyseq]
	}
	\\[15pt]
	\infer[\did{DCC}{Recv}]
	{
		\jsrv{\strBhv,\ \jpr{B}{t} \pp P,\ M}l 
		\quad \lto{o_j \qfrom e} \quad 
		\jsrv{\strBhv,\ \jpr{B_j}{t\tcopy{x_j}{t_m}} \pp P ,\ M[t_c \mapsto \til m]}l
	}
	{
		\begin{array}c
		B \in \{\; o_j(x_j) \qfrom e;B_j 
		\; , \; 
						  \sum_{i \in I}\choice{o_i(x_i) \qfrom e}{B_i}  
	  \; \}
	  \\[2pt]
	  j \in I
	  \qquad
	  t_c = \evalfn(e,t)
	  \qquad
	  M(t_c) = (o_j,t_m)::\til m	
		\end{array}
	}
	\\[15pt]
	\infer[\did{DCC}{InSend}]
	{
		\jsrv{\strBhv,\ \jpr{B}{t} \pp P, M}{l}
		\quad \lto{o \qto e_3} \quad 
		\jsrv{\strBhv,\ \jpr{B'}{t} \pp P, M[t_c \mapsto M(t_c)::(o,t_m)]}{l}
	}
	{
	\begin{array}c
		B = \notify{\op{o}}{e_1}{e_2} \qto e_3;B'
		\qquad
		\evalfn(e_1, t ) = l
		\\[2pt]
		\evalfn( e_3, t ) = t_c
		\qquad
		\evalfn( e_2, t ) = t_m
		\qquad
		t_c \in \dom(M)
	\end{array}
	}
	\\[15pt]
	\infer[\did{DCC}{InStart}]
	{
		\jsrv{!(x);B',\ \jpr{B}{t} \pp P,\ M}{l}
		\quad \lto{?(e_2)} \quad
		\jsrv{!(x);B',\ Q \pp \jpr{B''}{t} \pp P,\ M}{l}
	}
	{
		B = \notify{?}{e_1}{e_2};B''
		&
		Q = \jpr{B'}{\emptyset\tcopy{x}{\evalfn( e_2, t )} }
	}
	\\[15pt]
	\infer[\did{DCC}{Send}]
	{
		\jsrv{\strBhv, \jpr{B}{t} \pp P,\ M}{l} 
		\pp 
		\jsrv{\strBhv', P',\ M'}{l'}
		\quad \lto{o \qto e_3} \quad
		 \jsrv{\strBhv, \jpr{B''}{t} \pp P,\ M}{l} \pp 
		 \jsrv{\strBhv', P', M''}{l'}
	}
	{
		\begin{array}c
		B = \notify{\op{o}}{e_1}{e_2} \qto e_3; B''
		\qquad
		\evalfn( e_1, t ) = l'
		\qquad
		\evalfn( e_3, t ) = t_c
		\\[2pt]
		\evalfn(	e_2, t ) = t_m
		\qquad
		t_c \in \dom(M')
		\qquad
		M'' = M'[t_c \mapsto M'(t_c)::(o,t_m)]	
		\end{array}
	}
	\\[15pt]
	\infer[\did{DCC}{Start}]
	{
		\begin{array}l
		\jsrv{\strBhv,\ \jpr{B}{t} \pp P, M}{l}
		\pp 
		\jsrv{\strBhv',\ P',\ M'}{l'} 
		 \quad \lto{?(e_2)} \quad 
		\jsrv{\strBhv,\ \jpr{B''}{t} \pp P,\ M}{l}	
		\pp 
		\jsrv{\strBhv', Q \pp P',\ M'}{l'} 
		\end{array}
	}
	{
		B = \notify{?}{e_1}{e_2};B''
		&
		\strBhv' = !(x);B'
		&
		\evalfn( e_1, t ) = l'
		&
		Q = \jpr{B'}{\emptyset\tcopy{x}{\evalfn( e_2, t )}}
	}
	\\[15pt]
	\infer[\did{DCC}{SPar}]
	{
		S \pp S_1 \quad \lto{\lambda} \quad S' \pp S_1
	}
	{
		S  \lto{\lambda}  S'
	}
	\qquad
	\infer[\did{DCC}{SEq}]
	{
		S \quad \lto{\lambda} \quad S'
	}
	{
		S \equivD S_1
		&
		S_1  \lto{\lambda}  S_1'
		&
		S_1' \equivD S'
	}
\end{array}
\)%
}
\caption{Dynamic Correlation Calculus, annotated semantics.}
\label{fig:js_semantics_annotated}
\end{figure}

We also introduce two operators on sequences of DCC transition labels. Let
$\lambda,\til \lambda$ be a sequence of DCC labels, the filtering of
$\lambda,\til \lambda$ on $k$ , written $\filter{(\lambda,\til
\lambda)}{k}$ is defined as
$$
	\filter{(\lambda,\til \lambda)}{k} =
		\begin{cases}
			\lambda,(\filter{\til \lambda}{k}) & \mbox{if }
				\lambda \in \left\{ 
					\begin{array}c
						\jpath{k.x.y},\ \cq{\jpath{k.x.y}},\ ?(\jpath{k}),\\
						\mathit{sync}\qfrom{\jpath{k.x.y}},\ \mathit{sync}@\jpath{k.x.y}, \\
						\mathit{start}	\qfrom{\jpath{k.x.y}},\ \mathit{start}@\jpath{k.x.y} 		
					\end{array}
				\right\}
			\\
			\filter{\til \lambda}{k} & \mbox{otherwise}
		\end{cases}
$$

Let $\lambda_1,\til \lambda_1$ and $\lambda_2,\til \lambda_2$ be two sequences
of DCC labels, the complement of $\lambda_1,\til \lambda_1$ on $\lambda_2,\til
\lambda_2$ , written $(\lambda_1,\til \lambda_1)\setminus(\lambda_2,\til
\lambda_2)$ is defined as
\[
	(\lambda_1,\emptyseq)\setminus(\lambda_2,\til \lambda_2) = \begin{cases}
		\emptyseq & \mbox{if } \lambda_1 = \lambda_2
		\\
		\lambda_1 & \mbox{otherwise}
	\end{cases}
\]
\[
	(\lambda_1,\til \lambda_1)\setminus(\lambda_2,\til \lambda_2) =
	\begin{cases}
		\til \lambda_1\setminus\til \lambda_2 & \mbox{if } \lambda_1 = \lambda_2
		\\
		\lambda_1,(\til \lambda_1\setminus(\lambda_2,\til \lambda_2)) & \mbox{otherwise}
	\end{cases}
\]

Below we state Lemma~\ref{lemma:dcc_permutations} that proves that, given a DCC system $S$ and a sequence of reductions $\til \lambda$ for which $S \lto{\til \lambda} S'$, if the first action is the initiation of a session $k$, then we can reorder the execution of the subsequent actions in $\til \lambda$ such that we first execute all transitions related to the start of $k$ and then all the remaining actions, obtaining the same final system $S'$.

\begin{lemma}[DCC Start Permutation\label{lemma:dcc_permutations}]
	Let S be a composition of DCC services such that $S \lto{\til \lambda} S'$
	where $\til \lambda = \jpath{k.C.l},\til \lambda'$, then let $\til\lambda_k = \filter{\til \lambda}{k}$ and $\til\lambda_* = \til\lambda\setminus\til\lambda_k$ we have $S \lto{\til \lambda_k} S_1$ and $S_1 \lto{\til \lambda_*} S'$.
\end{lemma}

\begin{proof}[Sketch]
	The proof is by induction on the length of $\til \lambda$. The main intuition
	is that, since the first action is the start of the new session $k$, all
	other actions in $\til \lambda'$ either are related to the initiation of $k$
	or do not affect it. Hence, we can reorder the execution of actions in $\til
	\lambda$ such that first we execute all actions regarding the start of
	the session\footnote{Note, this does not imply nor require that $\lambda$
	contains all actions needed to start session $k$.} contained in $\til
	\lambda_k$ and then all the other actions in $\til \lambda_*$.
\end{proof}

Next we state \cref{lemma:start_completion} that proves that, given

\begin{itemize}
	\item a well-typed FC endpoint choreography $\env,C$
	\item its DCC compilation $S$
	\item the DCC system $S'$ that results from an arbitrary number of steps of
	reduction belonging to the start of a session $k$ in $S$
\end{itemize}

we can execute the remaining steps of reduction in $S'$ to complete the start
of session $k$, obtaining the final system $S''$ and prove that $S''$ is the
same DCC system as the one obtained from the compilation of $\env',C'$, the
reductum of the source FC choreography $\env,C$ after the step of reduction to
start session $k$.

\begin{lemma}[DCC Start Completion\label{lemma:start_completion}]
Let $\Gamma \hlseq{\dep,C}$, $C$ a endpoint choreography $$C= \req{k}{\prc
pA}{l_1.[\role B_1],\ldots,l_n.[\role B_n]};C_r \pp
\prod_{i=1}^{n}\acc{k}{l_i.\pid q_i[\role B_i];C_{\pid q_i}}$$ and
$\genenc{\geneenc{\dep},C}=S$ such that $S \lto{\til
\lambda} S'$ where $\filter{\til \lambda}{k} = \til \lambda$ then \emph{i}) $S'
\lto{\til \lambda'} S''$, \emph{ii}) $\dep,C \to \dep',C'$, and \emph{iii})
there exists some $\Gamma'$ s.t. $\Gamma' \hlseq{\dep',C'}$ and
$\genencp{\eenc{\dep'}{\Gamma'},C'} = S''$.
\end{lemma}

\begin{proof}
Proof by case analysis on the length of $\til \lambda$. 

Let $\pid p@l \in \Gamma$. To proceed, we have two subcases whether $l \in
\{l_1,\ldots,l_n\}$, i.e., whether one of the service processes is at the same
location of $\pid p$. Since the subcases follow the same structure, we detail
only the proof for $l \not \in \{l_1,\ldots,l_n\}$ which allows for a uniform
treatment. In the other case, \emph{i}) we should account for transitions on
the same service of $\pid p$ with rules $\did{DCC}{InStart}$ and
$\did{DCC}{InSend}$ and \emph{ii}) we would have a newly created process in
parallel with $\pid p$ in $\dep,C$ and in the correspondent DCC system $S''$.

Provided $n$ is the number of service processes involved in the start of the
session $k$, from \cref{def:compilation} we can count the number of transitions
needed to complete the start of a session. Indeed, given a $\dep,C$ with

$$
C = \req{k}{\prc pA}{l_1.[\role B_1],\ldots,l_n.[\role B_n]};C_r \pp
\prod_{i=1}^{n}\acc{k}{l_i.\pid q_i[\role B_i];C_{\pid q_i}}
$$

and $\genenc{\geneenc{\dep},C} = S$ then we can write the sequence of
transitions of the compiled DCC system

$$
\begin{array}l
	\hspace{-30pt} S \hspace{30pt} \overbrace{\lto{\jpath{k.I.l}}
	}^{\pnum{1}}
	\\[5pt]
	\underbrace{
		\overbrace{\lto{\cq{\jpath{k.A.I}}}}^{\smallpnum{2.1}}
		\overbrace{\lto{?(\jpath{k})}}^{\smallpnum{2.2}}
		\overbrace{\lto{\cq{\jpath{k.I.I'}}}}^{\smallpnum{2.3}}
		\overbrace{\lto{\mathit{sync}@\jpath{k.I'.A}}}^{\smallpnum{2.4}}
		\overbrace{\lto{\mathit{sync}
	\qfrom{\jpath{k.I'.A}}}}^{\smallpnum{2.5}}
	}_{\raisebox{-10pt}{\pnum{2}}}
	\\
	\underbrace{
		\overbrace{\lto{\mathit{start}@\jpath{k.I'.A}}}^{\smallpnum{3.1}}
		\overbrace{\lto{\mathit{start}
	\qfrom{\jpath{k.I'.A}}}}^{\smallpnum{3.2}}
	}_{\raisebox{-10pt}{\pnum{3}}} \qquad S''
\end{array}
$$

and count the number of all the transitions to complete the start, let it be
$m$, as the sum of:

\begin{description}
	\item[\pnum{1}] n+1 times, for $\jpath{I} \in \{\role A,\roles B\}$, with
	last Rule $\did{DCC}{Assign}$;
	\item[\pnum{2}] n times, for $\jpath{I} \in \roles B$:
		\begin{description}
			\item [\smallpnum{2.1}] reduces with last applied rule $\did{DCC}{Newque}$;
			\item [\smallpnum{2.2}] reduces with last applied rule $\did{DCC}{Start}$;
			\item [\smallpnum{2.3}] n times for $\jpath{I'} \in \{\role A,\roles B\}\setminus\{\jpath{I}\}$, reduces with last applied rule $\did{DCC}{Newque}$;
			\item [\smallpnum{2.4}] reduces with last applied rule $\did{DCC}{Send}$;
			\item [\smallpnum{2.5}] reduces with last applied rule $\did{DCC}{Recv}$;
		\end{description}
	\item[\pnum{3}] n times, for $\jpath{I} \in \roles B$:
		\begin{description}
			\item [\smallpnum{3.1}] reduces with last applied rule $\did{DCC}{Send}$;
			\item [\smallpnum{3.2}] reduces with last applied rule $\did{DCC}{Recv}$;
		\end{description}
\end{description}

and $m = n^2 + 7n + 1$.
We proceed unfolding the proof on the length of $\til \lambda$.

\begin{itemize}

	\ecase{$|\{\til \lambda\}| = 1$}

	Since the cardinality of $\til \lambda$ is one and that from the premises we
	know that $\til \lambda$ contains only transitions belonging to the start of
	session $k$, we can infer that $\til \lambda = \jpath{k.C.l}$ where $\role C
	\in \{\role A, \roles B\}$.

	To prove the thesis we let $S'$ do all the remaining transitions to start the
	session and show that $\dep,C$ can  mimic it. Let $\wtil{l.\role B} =
	l_1.\role B_1, \ldots, l_n.\role B_n$ and $\pair{\til l}{\serviceTyping{G}{\role A}{\roles B}{\roles B}} \in \Gamma$.

		From \cref{def:compilation} and \cref{thm:encoding_operational_correspondence} we have, let $\aenv = \geneenc{\env}$,
		$M = \aenv(l)$, $M_i = \aenv(l_i)$, and $t_\pid p = \aenv(\pid p)$

		$$
			\genenc{\geneenc\dep,C} \equivD
				\jsrv{\genenc{\filter{C_c}{l}}, P \pp R,M}{l} \pp
				\prod\limits_{i=1}^n \jsrv{Q_i, R_{l_i},M_i}{l_i} \pp S_c
		$$

		where

		\begin{itemize}

		\item $
		\begin{array}{rl} P = & 
		\jpr{\m{start}(\ k, \ (l.\role A,\wtil{l.\role B})\ ); \genenc
		{C_r}}{t_{\pid p}} =
			\\ = &
			\jpr{
				\left(\begin{array}{l}
					\Seq\limits_{
						\substack{ \jpath{I} \in \{\role A, \til{\role B}\}}}
							\jpath{k.I.l} = l_{\role I}\ ;
					\\
					\Seq\limits_{\jpath{I} \in \{\til{\role B}\}}
						\Big(
							\cq{\jpath{k.I.A}};
							\notify{?}{\jpath{k.I.l}}{\jpath{k}};
							\oneway{sync}{\jpath{k}} \qfrom \jpath{k.I.A}
						\Big);
					\\
					\Seq\limits_{\jpath{I} \in \{\til{\role B}\}}
						\notify{start}{\jpath{k.I.l}}{\jpath{k}} \qto \jpath{k.A.I};
				\end{array}
				\right);\genenc{C_r} }{t_{\pid p}}
		\end{array}
		$

		\item $ Q_i = \m{accept}(k, \role B_i,\initg{G}{\role A}{\til{\role
			B}}{\roles B}) ; \genenc{C_{\pid q_i}} =
			\begin{array}l
				\oneway{!}{\jpath{k}};
				\Seq\limits_{ \jpath{I} \in \{\role A,\roles B\}\setminus\{\role B_i\} }
					\cq{\jpath{k.I.B_i}} \ ;
			\\
				\notify{sync}{\jpath{k.A.l}}{\jpath{k}} \qto{\jpath{k.B_i.A}}\ ;
			\\
				\oneway{start}{\jpath{k}} \qfrom{\jpath{k.A.B_i}} \ ; \
				\genenc{C_{\pid q_i}}
			\end{array}
		$

		\item $ R = \prod\limits_{\pid p' \ \in \ \aenv(l)\setminus\{\pid p\}}
				\jpr{\genenc{\filter{C_c}{\pid p'}}}{t_{\pid p'}}
		$

		\item $ R_{l_i} = \prod\limits_{\pid s \ \in \ \aenv(l_i)}
				\jpr{\genenc{\filter{C_c}{\pid s}}}{t_{\pid s}}
		$

		\item $ S_c = \prod\limits_{l' \ \in \ \Gamma \setminus \{ l, \til l \}}
				\jsrv{ \genenc{\filter{C_c}{l'}} ,
				\prod\limits_{\pid s' \ \in \ \dep(l')}
					\jpr{\genenc{\filter{C_c}{\pid s'} }}{t_{\pid s'}},\filter{\aenv}{l'}
				}{l'}
		$

		\end{itemize}

		The first transition, $\lambda = \jpath{k.C.l}$ consumed the first
		assignment of location and assigned the location of role $\role C$ to
		$\jpath{k.C.l}$ in the state of the starter $t_\pid p$.

		Let us suppose, without loss of generality, that $\role C =
		\role A$, then we have

		{\footnotesize 
		$$\begin{array}{rl} P' = & 
			\jpr{
				\left(\begin{array}{l}
					\Seq\limits_{
						\substack{ \jpath{I} \in \{\til{\role B}\}}}
							\jpath{k.I.l} = l_{\role I}\ ;
					\\
					\Seq\limits_{\jpath{I} \in \{\til{\role B}\}}
						\Big(
							\cq{\jpath{k.I.A}};
							\notify{?}{\jpath{k.I.l}}{\jpath{k}};
							\oneway{\mathit{sync}}{\jpath{k}} \qfrom \jpath{k.I.A}
						\Big);
					\\
					\Seq\limits_{\jpath{I} \in \{\til{\role B}\}}
						\notify{\mathit{start}}{\jpath{k.I.l}}{\jpath{k}} \qto \jpath{k.A.I};
				\end{array}
				\right);\genenc{C_r} }{t_{\pid p}\tcopy{\jpath{k.A.l}}{l}}
		\end{array}
		$$}

		and $\genenc{\geneenc\dep,C} \lto{\jpath{k.A.l}} S'$ where
		
		$$
			S' = \jsrv{\genenc{\filter{C_c}{l}}, P' \pp R,M}{l} \pp
			\prod\limits_{i=1}^n \jsrv{Q_i, R_{l_i},M_i}{l_i} \pp S_c
		$$

		Since in its reduction $\dep,C$ renames the new session with a
		fresh name, we first rename session $k$, in $P$ and the service
		processes $Q_i$, to $k'$, which is fresh. We take 

		{\footnotesize 
		$$\begin{array}{rl}
			P'' = P'[k'/k] = &
			\jpr{
				\left(\begin{array}{l}
					\Seq\limits_{
						\substack{ \jpath{I} \in \{\til{\role B}\}}}
							\jpath{k'.I.l} = l_{\role I}\ ;
					\\
					\Seq\limits_{\jpath{I} \in \{\til{\role B}\}}
						\Big(
							\cq{\jpath{k'.I.A}};
							\notify{?}{\jpath{k'.I.l}}{\jpath{k'}};
							\notify{?}{\jpath{k'.I.l}}{\jpath{k'}};
						\Big);
					\\
					\Seq\limits_{\jpath{I} \in \{\til{\role B}\}}
							\oneway{\mathit{sync}}{\jpath{k'}} \qfrom \jpath{k'.I.A}
				\end{array}
				\right); \genenc{C_r}\ [\jpath{k'}/\jpath{k}] }{t_{\pid p}''}
		\end{array}
		$$}

		where, let $t_{\pid p}' = t_{\pid p}\tcopy{\jpath{k.A.l}}{l}$,
		 $
			t_{\pid p}'' = 
				t_{\pid p}' \tcopy{ \jpath{k'} } 
													{ \jpath{k} (t_{\pid p}') }
										\tcopy{ \jpath{k} }
													{ \emptyset }
		$.

		We take

		$$
			S^*_0 = \jsrv{\genenc{\filter{C_c}{l}}, P'' \pp R,M}{l} \pp
				\prod\limits_{i=1}^n \jsrv{Q_i[\jpath{k'}/\jpath{k}], R_{l_i},M_i}{l_i}
				\pp S_c
		$$

		and by \cref{lemma:network_variable_renaming} we have $S^*_0
		\sim S'$.

		Now we can proceed with the rest of the transitions of the start procedure as
		defined at the beginning of the proof, so that $S^*_0 \to^+ S_1^*$. Finally
		we have

	$$
			S_1^* \sim S'' =
				\jsrv{
					\genenc{\filter{C_c}{l}}, P''' \pp R,M'}{l}
				\pp
					\prod_{i=1}^{n} \jsrv{Q_i[\jpath{k'}/\jpath{k}], Q'_i \pp R_{l_i},M_i'
					}{l_i}
				\pp S_c
		$$

		where
		$P''' = \jpr{\genenc{C_r} \ [\jpath{k'}/\jpath{k}]}{t_{p}'}$ and 
		$Q_i' = \jpr{\genenc{C_{\pid q_i}} \ [\jpath{k'}/\jpath{k}]}{t_{k'}}$

		From the transitions presented above we know that there exists $t_{k'}'$
		such that $t_{\pid p}' = t_{\pid p} \tcopy{\jpath{k'}}{t_{k'}'}$, where
		$t_{k'}'$ is a session descriptor for session $k'$ (i.e., it contains all
		the locations and correlations keys used by the processes in session $k'$).

		We proceed by proving that $\dep,C$ can mimic
		$\genenc{\geneenc\dep,C}$.

		We can apply rules $\did{C}{Par}$ and $\did{C}{Eq}$ and lastly
		rule $\did{C}{PStart}$ such that

		\[
			\infer[\did{C}{PStart}]
			{
				\begin{array}{l}
				\env,
				\req{k}{\pid p[\role A]}
				{\wtil{l.\role B}}; C
				\pp
				\prod_i
				\big(
				\genacci; C_i
				\big)
				\quad \to
				\\ \hspace{50pt}
				\env',\ 
				C[k'/k] \pp \prod_i\big(\ C_i[k'/k][\pids r_i /\pids q_i] \ \big)
				\pp
				\prod_i
				\big(
				\genacci; C_i
				\big)
				\end{array}
			}
			{
				\begin{array}c
				i \in \{1,\dots,n\}
				\qquad
				\env\fresh{k',\pids r}
				\qquad
				\{ \wtil{l.\role B} \} = \biguplus_i \{\wtil{l_i.\role B_i}\}_i
				\qquad
				\{\pids r\} = \bigcup_i \{ \pids r_i \}
				\\[2pt]
				\delta = \start{k'}{\pid p[\role A]}{\wtil{l_1.\pid r_1[\role
				B_1]},\dots,\wtil{l_n.\pid r_n[\role B_n]}}
				\qquad
				\renv{\env}{\delta}{\env'}
				\end{array}
			}
		\]

		and

		$$
			\dep, C \pp C_c \quad \to \quad
			\dep',C_r[k'/k] \pp 
			\prod_i\big(\ C_{\pid q_i}[k'/k][\pid r_i /\pid q_i] \ \big) 
			\pp
			\prod_{i=1}^{n}\acc{k}{l_i.\pid q_i[\role B_i];C_{\pid q_i}}
			\pp 
			C_c
		$$

		thus $C' = C_r[k'/k] \pp 
			\prod_{i=1}^{n}\big(\ C_{\pid q_i}[k'/k][\pid r_i /\pid q_i]
			\ \big)
			\pp
			\prod_{i=1}^{n}\acc{k}{l_i.\pid q_i[\role B_i];C_{\pid q_i}}
			\pp C_c$.

		From the hypothesis we know that $\Gamma \hlseq{\dep,C}$ and
		therefore that $\Gamma = \Gamma_1,\pair{\til
		l}{\serviceTyping{G}{\role A}{\roles B}{\roles B}}$. We can
		find $\Gamma' = \Gamma,\auxfn{init}(k',(\prc pA,
		\wtil{\prc qB}),G)$ and $\Gamma' \hlseq{\dep',C'}$.



		Finally, we need to prove that $S_1^* =
		\genencp{\eenc{\dep'}{\Gamma'},C'}$.

		From \cref{def:compilation} we have

		$$
			\genencp{\eenc{\dep'}{\Gamma'},C'} = 
				\jsrv{\genencp{\filter{C_c}{l}}, P^* \pp R',M^*}{l} \pp
				\prod\limits_{i=1}^{n} \jsrv{Q_i'', Q_i^{*} \pp R_{l_i}',M^*_i}{l_i}
				\pp
				S_c'
		$$

		Let $\aenv^* = \eenc{\env'}{\Gamma'}$ we use the abbreviations $t_{\pid
		s}^* =\aenv^*(\pid s)$, for $\pid s$ process in $\aenv^*$, and  $M^{*} = 
		\filter{\aenv}{l}$, and $M^{*}_i = \filter{\aenv}{l_i}$, in $
		\genencp{\aenv^*,C'}$ we have

		\begin{itemize}
			
			\item $P^* = \jpr{\genencp{C_r[k'/k]}}{t_{\pid p}^*}$

			\item $R' = \prod\limits_{\pid p' \ \in \ \aenv^*(l)\setminus\{\pid p\}}
				\jpr{\genencp{\filter{C_c}{\pid p'}}}{t_{\pid p'}^{*}}
			$

			\item $Q_i'' = \m{accept}(k,\role B_i,\serviceTyping{G}{\role A}{\roles
			B}{\roles B});\genencp{C_{\pid q_i}}$

			\item $Q_i^{*} = \jpr{\genencp{C_{\pid q_i}[k'/k][\pid r_i/\pid q_i]}}
			{t_{q_i}^{*}}$
		
			\item $R'_{l_i} = \prod\limits_{\pid s \ \in \ \aenv^*(l_i)}
				\jpr{\genencp{\filter{C_c}{\pid s}}}{t_{\pid s}^{*}}
			$

			\item $ S_c' = \prod\limits_{l' \ \in \ \Gamma \setminus \{ l, \til l \}}
				\jsrv{ \genencp{\filter{C_c}{l'}} ,
				\prod\limits_{\pid s' \ \in \ \aenv^*(l')}
					\jpr{\genencp{\filter{C_c}{\pid s'} }}{t_{\pid
					s'}^{*}},\filter{\aenv^*} {l'}} {l'}
			$
		\end{itemize}

		From Rule $\did{D}{Start}$ we know that

		$$
			\jpath{k'}(t_{\pid p}^{*}) = \jpath{k'}(t_{\pid q_1}^{*}) = \ldots =
			\jpath{k'} (t_{\pid q_n}^{*}) = t_{k'}
		$$

		for some $t_{k'}$ session descriptor of session $k'$.

		We prove the case by taking $t_{k'}=t'_{k'}$, $t'_{k'}$ obtained from the
		derivation of $\genenc{\geneenc\dep,C}$ and $M^* = M'$ and
		$M_{i}^* = M'_{i}$, $i \in \{1,\ldots,n\}$.

	\ecase{$1 < |\{\til  \lambda\}| < m-1$}
		The case follows the same structure of the previous case. We
		rename $k$ to $k'$ on $\pid p$ and all the newly created service
		processes. Then we let the system complete all the transitions and prove
		that the reductum corresponds to the compilation of $\dep',C'$.

	\ecase{$|\{\til  \lambda\}| = m$}
		Since $|\{\til \lambda\}| = m$ then $S = S'$ where $S'$ has terminated all
		the transitions to start the session. Here we only have to rename $k$ to
		$k'$, as per \cref{lemma:network_variable_renaming}, for all the involved
		processes, proving $S' = \genencp{\eenc{\dep'}{\Gamma'},C'}$.

\end{itemize}
\end{proof}

We now proceed to prove the \emph{(Soundness)} of
Theorem~\ref{thm:applied_choreographies}, restated here below to
consider annotated transitions: 

\begin{itemize}
	\item \emph{(Soundness)} $\enc{\dep,C}^{\Gamma} \lto{\til \lambda} S$
	implies \emph{i}) $\dep,C \to^* \dep', C'$ and \emph{ii}) $S
	\to^* \enc{\dep',C'}^{\Gamma'}$ for some $\dep'$,$C'$, and $\Gamma'$ such that
	\emph{iii}) $\Gamma' \hlseq{\dep',C'}$
\end{itemize}

In the following we use the shortcut
	
	$$C_{\mathit{start}} =
		\req{k}{\prc pA}{l_1.[\role B_1],\ldots,l_n.[\role B_n]};C_r \pp
		\prod_{i=1}^{n}\acc{k}{l_i.\pid q_i[\role B_i];C_{\pid q_i}}
	$$

\begin{proof}[(Soundness)]
	
We proceed by induction on the cardinality of $\til \lambda$. Then we consider
sub-cases on the shape of $C$ and the shape of $\til \lambda$.

\begin{itemize}
	
	\ecase{$|\{\til \lambda\} | = 0$}

	 	Trivial, $\genenc{\dep,C} = S = \genencp{\dep',C'}$, $\dep,C =
		\dep',C'$, and $\Gamma \hlseq{\dep',C'}$.

	\ecase{$|\{\til \lambda\}| = 1$} Since the cardinality of $\til \lambda$ is
	one, we can directly consider the single annotated transition $\lambda = \til
	\lambda$. In the sub-cases of this case we omit to consider impossible cases
	for $\lambda = \cq{x}$ and $\lambda =\ ?(x)$ since these transitions
	(corresponding respectively to rules $\did{DCC}{Newque}$, and $\did{DCC}
	{InStart}$ or $\did{DCC} {Start}$) can happen only within of a start session
	sequence (i.e., not at the first position).

	
	\newcommand{\seeCase}[1]{\emph{follows} (#1)}

	In the following we use the abbreviation \seeCase{\#} to indicate that the
	case unfolds following the proof of \textbf{Case} \emph{\#} for the same subcase for $\lambda$, with the thesis following by applying the induction
	hypothesis.

	\begin{enumerate}

			\ecase{$C = \genbranchI;C_\pid q \pp C_c$}

			\hspace{1.9em}\textbf{Case} \hl{\lambda = x} \seeCase{$C = C_{\mathit{start}} \pp C_c$}.
				
			\noindent\hspace{1.6em} \textbf{Case} \hl{\lambda = o \qto x} \seeCase{$C = \gensend;C_\pid p \pp C_c$}.
			
			\noindent\hspace{1.6em} \textbf{Case} \hl{\lambda = \tau} \seeCase{$C = \cond{\pid p.e}{C_1}{C_2} \pp C_c$}.

			\begin{enumerate}

				 \ecase{$\lambda = o \qfrom{x}$}
					 Since receptions in compiled DCC systems can only happen on
					correlating queues within sessions, without loss of generality we can
					assume that $\lambda = o  \qfrom{\jpath{k.A.B}}$ where $o_j \not \in
					\{\mathit{start},
					\mathit{sync}\}$, indeed these operation names are reserved for session
					initiation and cannot appear as first (in this case, only) reduction
					of a compiled system.

					Let $\pid q@l \in \Gamma$, from Definition~\ref{def:compilation} and \cref{thm:encoding_operational_correspondence} we have

						$$
							\genenc{\dep,C} \equivD
								\jsrv{ \genenc{\filter{C_c}{l}}, Q \pp R, M}{l}
								\pp S_c
						$$

						where, let $\aenv = \geneenc{\env}$, $M = \filter{\aenv}{l}$ and
						$t_\pid q = \aenv(\pid q)$,

						\begin{itemize}

							\item $Q =\jpr{\sum_{i \in I}\choice{ o_i(x_i)
									\qfrom{\jpath{k.A.B}}}{\genenc{C_i}}}{t_{\pid q}}$
							
							\item $R = \prod\limits_{\pid r \ \in \ \aenv(l)\setminus\{\pid
							q\}}\jpr{\genenc{\filter{C_c}{\pid r}}}{t_{\pid r}}$
							
							\item $S_c = \prod\limits_{l'
							\in\Gamma\setminus\{l\}}\jsrv{\genenc{\filter{C_c}
							{l'}},\prod\limits_{\pid s
							\ \in \ D(l')}\jpr{\genenc{\filter{C_c}{\pid s}}}{t_{\pid
							s}}}{l'}$

						\end{itemize}

						and we can apply rules $\did{DCC}{Eq}$, $\did{DCC}{SPar}$ and
						$\did{DCC}{Recv}$ such that, let $t_c = \evalfn(\jpath{k.A.B},
						t_{\pid q})$, $t_m = \evalfn(e, t_{\pid q})$, and $M(t_c)
						= (o_j,t_m)::\til m$

						$$\genenc{\dep,C} \quad \lto{o_j \qfrom{\jpath{k.A.B}}} \quad S$$

						where

						$$S = S' \pp S_c$$

						and $S' = \jsrv{ \genenc{\filter{C_c}{l}},
						\jpr{\genenc{C_{j}}}{t_{\pid q}\tcopy{x_j}{t_m}} \pp R, M[t_c
						\mapsto \til m] }{l}$.

						$\dep,C$ can mimic $\genenc{\dep,C}$ with rules $\did{C}{Eq}$,
						$\did{C}{Par}$, and $\did{C}{Recv}$ for which $$\dep,C
						\quad \to \quad \dep',C_{\pid p} \pp C_c$$ where, let
						$D(\chanto{k}{\role A}{\role B}) = (o_j,v_m)::\til m'$, we have
						$$\dep' = \dep[\pid q \mapsto t_{\pid q}\tcopy{x_j}{v_m}][
						\chanto{k}{\role A}{\role B} \mapsto \til m']$$

						Since from the premises $\Gamma \hlseq{\dep,C}$ then \\ $\Gamma =
						\Gamma_1, \pair{k[\role A]}{\lrecv{\role A}{\{o_i(U_i);T_i\}_{i \in
						I}}}, \pair{\chanto{k}{\role A}{\role B}}{
						\lrecv{A}{o_j(U_j);T'}}$	and we can find \\$\Gamma' = \Gamma_1, \pair{k[\role A]}{T_j},
						\pair{\chanto{k}{\role A}{\role B}}{T'}$ such that $\Gamma'
						\hlseq{\dep', C'}$.

						At the level of choreographies, since the changes in $\env'$ and
						$\Gamma'$ and the related $\aenv' = \enc{\env'}^{\Gamma'}$ affect
						only the queue related to $\filter{\aenv'}{l}$ and the state of
						$\pid q$, for all other terms $\genenc{\cdot} =
						\genencp{\cdot}$ and $\filter{\aenv'}{l'}=\filter{\aenv}{l'}$.

						Hence we can write $\genencp{\env',C'} = S'' \pp S_c$ where $S'' = S'$ by \cref{thm:encoding_operational_correspondence}.

				\end{enumerate}

			\ecase{$C = \gensend;C_\pid p \pp C_c$}

					\hspace{1.6em} \textbf{Case} \hl{\lambda = x} \seeCase{$C = C_{\mathit{start}} \pp C_c$}.
					
					\noindent\hspace{1.6em} \textbf{Case} \hl{\lambda = o \qfrom{x}} \seeCase{$C = \genbranchI;C_\pid q \pp C_c$}.
					
					\noindent\hspace{1.6em} \textbf{Case} \hl{\lambda = \tau} \seeCase{$C = \cond{\pid p.e}{C_1}{C_2} \pp C_c$}.
				
					\begin{enumerate}

					\ecase{$\lambda = o \qto x$}
					\label{case:lambda_send}

					As for Case $C = \genbranchI;C_\pid q \pp C_c$, we know that all send
					actions in DCC systems compiled from FC programs happen on a
					session-related queues, hence we can assume $\lambda =
					o@\jpath{k.A.B}$. Also, we know that $o \not  \in \{\mathit{start},
					\mathit{sync}\}$ for the reasons explained in Case $C =
					\genbranchI;C_\pid q \pp C_c$.

					From \cref{thm:encoding_operational_correspondence}, let $\aenv = \geneenc{\env}$, $t_\pid p = \aenv(\pid p)$, and $M =
					\filter{\aenv}{l}$. Now we consider two cases for which, let $\pid p@l
					\in \Gamma$, whether the location of the receiving process (stored
					under path $\jpath{k.B.l}$ in the state of $\pid p$) equals $l$, we
					either reduce the compiled DCC system by means of rule
					$\did{DCC}{InSend}$ or rule $\did{DCC}{Send}$. For brevity we just
					consider the case for $\did{DCC}{InSend}$ as the other case follows
					similarly.

					Since $\did{DCC}{InSend}$ applies, we can infer that

						$$
							\genenc{\dep,C} \equivD
								\jsrv{ \genenc{\filter{C_c}{l}}, P \pp Q \pp R, M}{l}
								\pp S_c
						$$

						where

						\begin{itemize}
							\item $P =
								\jpr{o@\jpath{k.B.l} \qto{\jpath{k.A.B}};
									\genenc{C_{\pid p}}}{t_{\pid p}}$
							\item $Q = \jpr{\genenc{\filter{C_c}{\pid q}}}{t_{\pid
							q}}$
							\item $R = \prod\limits_{\pid r \ \in \ D(l)\setminus\{\pid p, \pid q\}}
								\jpr{\genenc{\filter{C_c}{\pid r}}}{t_{\pid r}}
							$
							\item $S_c = \prod\limits_{l' \in \Gamma\setminus\{l\}}\jsrv{
							\genenc{\filter{C_c}{l'}}, \prod\limits_{\pid s \ \in
								\ \aenv(l')}\jpr{\genenc{\filter{C_c}{\pid s}}}{t_{\pid
								s}},\filter{\aenv}{l'}}{l'}$
						\end{itemize}

						Let $t_c = \evalfn(\jpath{k.A.B}, t_{\pid p})$, $t_m = \evalfn(e,
						t_{\pid p})$, and $M(t_c) = \til m$

						$$\genenc{\dep,C} \quad \lto{o@\jpath{k.A.B}} \quad S$$

						where

						$$S = S' \pp S_c$$

						and $S' = \jsrv{\genenc{\filter{C_c}{l}},
							\jpr{\genenc{C_{\pid p}}}{t_{\pid p}}
							\pp
							\jpr{\genenc{\filter{C_c}{\pid q}}}{t_{\pid q}} \pp R, M[t_c
							\mapsto \til m :: (o, t_m)]}{l}$ 

						$\dep,C$ can mimic
						$\genenc{\dep,C}$ with rules $\did{C}{Eq}$, $\did{C}{Par}$, and
						$\did{C}{Send}$ for which $$\dep,C
						\quad \to \quad \dep',C_{\pid p} \pp C_c$$ where, let $v_m =
						\evalfn(e, \env(\pid p))$ and $\til m' = \env(\chanto{k}{\role A}{\role
						B})$, $\dep'= \dep[\chanto{k}{\role A}{\role B} \mapsto \til m' :: (o,
						v_m)]$.

						Since from the premises $\Gamma \hlseq{\dep,C}$ then $\Gamma =
						\Gamma_1, \pair{k[\role A]}{\lsend{\role B}{o(U);T}},
						\pair{\chanto{k}{\role A}{\role B}}{T'}$ and we can find $\Gamma' =
						\Gamma_1, \pair{k[\role A]}{T},
						\pair{\chanto{k}{\role A}{\role B}}{T';\lrecv{\role A}{o(U)}}$
						such that $\Gamma' \hlseq{\dep', C'}$.

						At the level of choreographies, since the changes in $\env'$ and
						$\Gamma'$ and the related $\aenv' = \enc{\env'}^{\Gamma'}$ affect
						only the queue related to $\filter{\aenv'}{l}$, for all other terms
						$\genenc{\cdot} =
						\genencp{\cdot}$ and $\filter{\aenv'}{l'}=\filter{\aenv}{l'}$.

						Hence we can write $\genencp{\env',C'} = S'' \pp S_c$ where $S'' = S'$ by \cref{thm:encoding_operational_correspondence}.

				\end{enumerate}

			\ecase{$C = C_{\mathit{start}} \pp C_c$}

					\noindent\hspace{1.6em} \textbf{Case} \hl{\lambda = o \qfrom{x}} \seeCase{$C = \genbranchI;C_\pid q \pp C_c$}.
					
					\noindent\hspace{1.6em} \textbf{Case} \hl{\lambda = o \qto	x} \seeCase{$C = \gensend;C_\pid p \pp C_c$}
					
					\noindent\hspace{1.6em} \textbf{Case}	\hl{\lambda = \tau} \seeCase{$C = \cond{\pid p.e}{C_1}{C_2} \pp C_c$}.

				\begin{enumerate}

					\ecase{$\lambda = x$}
						From \cref{def:compilation} we know that assignments in DCC systems
						that are compiled from FC programs appear only within the starting
						of a session. In this case, since $\til \lambda$ contains only one
						action which corresponds to the first reduction of the compiled DCC
						system, it must be the first assignment for the creation of the
						session descriptor for some session $k'$ in $C$.

						Let $\role C \in \{\role A, \roles B\}$, we have two subcases
						whether $\til \lambda = \lambda = \jpath{k.C.l}$ or $\til \lambda =
						\lambda = \jpath{k''.C.l}$, i.e., whether we are starting session
						$k$ or we are starting another session $k''$.

							\begin{itemize}

								\ecase{$\lambda = \jpath{k.C.l}$}
								In this case $\genenc{\dep,C}$ is starting a new session on $k$.
								The case is proved applying Lemma~\ref{lemma:start_completion}.

								\ecase{$\lambda \neq \jpath{k'.C.l}$} In this case we are
								starting a session on $k'' \neq k$. The case unfolds following
								the proof of case $C = C_{\mathit{start}} | C_c$ where $C_c$
								contains the endpoint choreographies for the starter process and
								the service processes for session $k''$. The thesis follows by
								applying the induction hypothesis.

						\end{itemize}

				\end{enumerate}

			\ecase{$C = \cond{\pid p.e}{C_1}{C_2} \pp C_c$}

				\noindent\hspace{1.6em} \textbf{Case} \hl{\lambda = x} \seeCase{$C = C_{\mathit{start}} \pp C_c$}.

				\noindent\hspace{1.6em} \textbf{Case} \hl{\lambda = o \qfrom{x}} \seeCase{$C = \genbranchI;C_\pid q \pp C_c$}.

				\noindent\hspace{1.6em} \textbf{Case} \hl{\lambda = o \qto x} \seeCase{$C = \gensend;C_\pid p \pp C_c$}.

				\begin{enumerate}
					
					\ecase{$\lambda = \tau$}
					\label{case:lambda_cond}

					In this case, the label does not allow us to establish a correspondence
					between the considered shape of C and the actual reduction
					annotated by the label.

					However, since \(\tau\) labels only correspond to the reduction of
					conditionals, without loss of generality, we can consider here only the
					case where the reduction acts on the considered term. The other case
					follows the unfolding of this case for the term reduced by the action and
					the induction hypothesis.

					Let $\pid p@l \in \Gamma$. From \cref{def:compilation} we
					have

					$$
						\genenc{\dep,C} \equivD \jsrv{\genenc{\filter{C_c}{l}}, P \pp R, M}{l}
						\pp S_c
					$$ 

					where, let $\aenv = \geneenc{\env}$, $t_\pid p = \aenv( \pid p)$, and $M =
					\filter{\aenv}{l}$

					\begin{itemize}
						\item $P = \jpr{\cond{\pid
						p.e}{\genenc{C_1}}{\genenc{C_2}}}{t_{\pid p}}$

						\item $R = \prod\limits_{\pid r \in \aenv(l)\setminus\{\pid p\}}
							\jpr{\genenc{\filter{C_c}{\pid r}}}{t_{\pid r}}$

						\item $S_c = \prod\limits_{l' \in \Gamma \setminus \{l\}}
							\jsrv{\genenc{\filter{C_c}{l'}},
								\prod\limits_{\pid s \in \aenv(l')}
									\jpr{\genenc{\filter{C_c}{\pid s}}}{t_{\pid s}},
									\filter{\aenv}{l'}} {l'}$
					\end{itemize}

					The case unfolds into two cases, on whether
					$\evalfn(e,\aenv(\pid p)) = \m{true}$. Here we proceed with the
					positive case. The other case follows the same structure.

					We proceed considering that $\evalfn(e,\aenv(\pid p)) =
					\m{true}$. $\genenc{\dep,C}$ reduces with rules $\did{DCC}{Eq}$,
					$\did{DCC}{SPar}$, $\did{DCC}{Par}$, and $\did{DCC}{Cond}$ such that

					$$
						\genenc{\dep,C} \quad \to \quad \jsrv{\genenc{\filter{C_c}{l}},
							\jpr{\genenc{\filter{C_c}{\pid p}}}{t_{\pid p}}\pp R, M
							}{l} \pp S_c
					$$

					where $S =
					\jsrv{\genenc{\filter{C_c}{l}},\jpr{\genenc{\filter{C_c}{l}}}{t_{\pid
					p}} \pp R,M}{l} \pp S_c$. $\dep,C$ can mimic $\genenc{\dep,C}$ with
					rules $\did{C}{Eq}$, $\did{C}{Par}$, and $\did{C}{Cond}$ such that

					$$
						\dep,C \quad \to \quad \dep,C_1 \pp C_c
					$$

					We choose $\Gamma' = \Gamma$ for which it holds that $\Gamma \hlseq
					{\dep,C_1 \pp C_c}$.

					Finally, $\genenc{\dep,C_1 \pp C_c} = S$ by
					Definition~\ref{def:compilation}.

				\end{enumerate}

			Finally, \textbf{Case}s \hl{C = C_1 \pp C_2}, \hl{C = \recDef{X}{C^\prime}
			{C_\pid p} \pp C_c}, \hl{C = \recCall{X} \pp C_c} and \hl{C = \inact \pp
			C_c} unfold applying the induction hypothesis on the respective sub-cases
			\textbf{Case} \hl{\lambda = x} \seeCase{$C = C_{\mathit{start}} \pp C_c$},
			\textbf{Case} \hl{\lambda = o \qfrom{x}} \seeCase{$C = \genbranchI;C_\pid q \pp C_c$},
			\textbf{Case} \hl{\lambda = o \qto x} \seeCase{$C = \gensend;C_\pid p \pp C_c$},
			\textbf{Case} \hl{\lambda = \tau} \seeCase{$C = \cond{\pid p.e}{C_1}{C_2} \pp C_c$}.

		\end{enumerate}

	\ecase{$|\{\til \lambda\}| > 1 $}

		The case unfolds considering $\lambda$ as the first action in $\til \lambda
		= \lambda, \til \lambda'$. For any shape of $C$ and label $\lambda \neq x$
		we can \emph{i}) apply the same steps followed in the related case for the
		same $C$ with $|\{\til \lambda\}| = 1$, $\til \lambda =
		\lambda$ and \emph{ii}) inductively unfold the case on the remaining part
		$\til \lambda'$.

		For $\lambda = x$ and $C$ of shape $C_{\mathit{start}} \pp C_c$ (the case
		for other shapes of $C$ can be re-conducted to this case), let $x =
		\jpath{k.A.l}$ (other cases for $x=\jpath{k'.B.l}$ are similar) and the
		thesis follows by applying \cref{lemma:dcc_permutations},
		\cref{lemma:start_completion} and the induction hypothesis on the remaining
		transitions in $\til\lambda \setminus \filter{\til \lambda}{k}$.

\end{itemize}
\end{proof}

\IFSubFileBiblio

\end{document}